\documentclass[fleqn,usenatbib,useAMS]{mnras}

\usepackage{adjustbox}
\usepackage{multirow}
\usepackage{graphicx}	
\usepackage{amsmath}	
\usepackage{amssymb}	
\usepackage{multicol}        
\usepackage{bm}		
\usepackage{pdflscape}	

\usepackage[T1]{fontenc}
\usepackage{ae,aecompl}

\usepackage{newtxtext,newtxmath}

\title[Shock--multicloud systems I]{Shock--multicloud interactions in galactic outflows - I. Cloud layers with log-normal density distributions}

\author[W.~E.~Banda-Barrag\'{a}n et al.]{W.~E.~Banda-Barrag\'{a}n,$^{1}$\thanks{E-mail: wlady.bsc@gmail.com (WBB)}
M.~Br\"uggen,$^{1}$
C.~Federrath,$^{2}$
A.~Y.~Wagner,$^{3}$\newauthor
E.~Scannapieco,$^{4}$ and
J.~Cottle$^{4}$
\\
$^{1}$Hamburger Sternwarte, Universit\"{a}t Hamburg, Gojenbergsweg 112, D-21029 Hamburg, Germany\\
$^{2}$Research School of Astronomy and Astrophysics, Australian National University, Canberra, ACT 2611, Australia\\
$^{3}$Center for Computational Sciences, University of Tsukuba, 1-1-1 Tennodai, Tsukuba, Ibaraki 305-8577, Japan\\
$^{4}$School of Earth and Space Exploration, Arizona State University, Tempe AZ, USA\\
}

\date{Accepted XXX. Received YYY; in original form ZZZ}

\pubyear{2020}

\begin{document}
\label{firstpage}
\pagerange{\pageref{firstpage}--\pageref{lastpage}}
\maketitle

\begin{abstract}
We report three-dimensional hydrodynamical simulations of shocks (${\cal M_{\rm shock}}\geq 4$) interacting with fractal multicloud layers. The evolution of shock-multicloud systems consists of four stages: a shock-splitting phase in which reflected and refracted shocks are generated, a compression phase in which the forward shock compresses cloud material, an expansion phase triggered by internal heating and shock re-acceleration, and a mixing phase in which shear instabilities generate turbulence. We compare multicloud layers with narrow ($\sigma_{\rho}=1.9\bar{\rho}$) and wide ($\sigma_{\rho}=5.9\bar{\rho}$) log-normal density distributions characteristic of Mach $\approx 5$ supersonic turbulence driven by solenoidal and compressive modes. Our simulations show that outflowing cloud material contains imprints of the density structure of their native environments. The dynamics and disruption of multicloud systems depend on the porosity and the number of cloudlets in the layers. `Solenoidal' layers mix less, generate less turbulence, accelerate faster, and form a more coherent mixed-gas shell than the more porous `compressive' layers. Similarly, multicloud systems with more cloudlets quench mixing via a shielding effect and enhance momentum transfer. Mass loading of diffuse mixed gas is efficient in all models, but direct dense gas entrainment is highly inefficient. Dense gas only survives in compressive clouds, but has low speeds. If normalised with respect to the shock-passage time, the evolution shows invariance for shock Mach numbers $\geq10$ and different cloud-generating seeds, and slightly weaker scaling for lower Mach numbers and thinner cloud layers. Multicloud systems also have better convergence properties than single-cloud systems, with a resolution of $8$ cells per cloud radius being sufficient to capture their overall dynamics.

\end{abstract}

\begin{keywords}
hydrodynamics -- turbulence -- methods: numerical -- galaxies: starburst -- galaxies: ISM -- ISM: clouds
\end{keywords}



\section{Introduction}
\label{sec:Introduction}
Multi-phase flows are ubiquitous in the Universe. They are found in the interstellar medium surrounding supernova remnants (e.g., see \citealt{2006ApJ...636..859F,2020MNRAS.491.2855V}), in the circumgalactic medium as inflowing gas streams (e.g., see \citealt{2018ApJ...861..148M,2019NatAs...3..822M,2020MNRAS.494.2641M}), outflowing winds (e.g., see \citealt{2013ApJS..204...17W,2016ApJ...833...54W}) and fountains (e.g., see \citealt{2015ApJ...814...83L}), and also in the intracluster medium (e.g., see \citealt{2019ApJ...883..145J}). Galactic winds are examples of such outflows as multi-wavelength observations of several galaxies, mainly starburst galaxies, reveal the presence of a cold, dense gas component embedded in a much hotter and more diffuse gas component (e.g., see \citealt{1998ApJ...493..129S}; \citealt*{2005ARA&A..43..769V}; \citealt{2011Sci...334..952T,2016ApJ...826..215L,2018ApJ...855...33D,2018ApJ...856...97S,2019ApJ...881...43K,2019ApJ...885L..32D}; \citealt*{2020ApJ...888...51L}). The prevalence of dense gas in galactic outflows poses challenges to current theoretical models as analytical considerations and numerical simulations of wind-cloud interactions show that dynamical instabilities and evaporation can make the acceleration and survival of dense gas difficult over a wide range of the parameter space (e.g., see \citealt{2017MNRAS.468.4801Z} and \citealt{2018Galax...6..114Z} for a recent review).\par

In recent years, there have been efforts to understand both shock-cloud and wind-cloud interactions by studying the roles of radiative cooling (\citealt*{2010ApJ...722..412Y,2019MNRAS.482.5401S}; \citealt{2020MNRAS.492.1841L}), thermal conduction (\citealt{2005MNRAS.362..626M,2017MNRAS.470..114A}), self-gravity (\citealt{2014MNRAS.444.2884L}), turbulence (\citealt{2017ApJ...834..144S,2018MNRAS.473.3454B,2020MNRAS.491.5056L}), and magnetic fields (\citealt{2017ApJ...845...69G,2018ApJ...865...64G}) in such models. Radiative cooling and thermal conduction generally prolong the lifetimes of wind-swept clouds, but they impede acceleration (e.g., \citealt{2015ApJ...805..158S,2016ApJ...822...31B}). Magnetic fields can shield the clouds by stabilising shear layers and preventing the emergence of Kelvin-Helmholtz instabilities (hereafter KH instabilities, see e.g., \citealt{2015MNRAS.449....2M,2016MNRAS.455.1309B}). Turbulent densities favour cloud disruption but the initial dissipation of supersonic turbulence aids acceleration (e.g., \citealt{2018MNRAS.473.3454B}).\par

Despite the progress made towards understanding wind-cloud models, most of the above works focused solely on the interplay between a single, isolated cloud with either a shock or a wind (e.g., see \citealt{2015ApJS..217...24S,2016MNRAS.457.4470P,2017ApJ...839..103D,2018ApJ...864...96C}; and \citealt{2017MNRAS.470.2427G,2018MNRAS.476.2209G} for a recent comparison between shock- and wind-cloud problems). Studies of shocks/winds interacting with multicloud systems are, however, more scarce, even though, in most astrophysical situations, interstellar clouds are not isolated but rather are part of larger multi-cloud complexes\footnote{The community studying outflows driven by Active Galactic Nuclei (AGN) activity has also investigated non-uniform cloud systems, e.g., see \citealt*{2007ApJS..173...37S,2012ApJ...757..136W,2013ApJ...763L..18W}; \citealt{2016MNRAS.461..967M,2017MNRAS.464.1854B}.}. In the case of galactic winds, such complexes can be found both at the base and along the outflowing gas (e.g., see \citealt{2017ApJ...835..265W,2017ApJ...849...90S,2019MNRAS.488.3904L,2019ApJ...881...43K}). Cloud complexes are the birthplace of the starburst outflows as they host active star-forming regions, which drive turbulence and promote the vertical circulation of gas in the host galaxy (e.g., see \citealt{2008ApJ...674..157C,2012ApJ...750..104H,2013MNRAS.430.3235M,2018ApJ...853..173K}). Thus, investigating how multicloud systems evolve and disentangling the collective effects of different cloud distributions when they are overrun by a shock is essential to deepening our understanding of dense gas entrainment and mass loading into multi-phase galactic outflows.\par

The problem of shocks interacting with multicloud systems has been studied both analytically and numerically by previous authors. \cite*{1996ApJ...468L..59J} showed that vortical motions produced by supernova ejecta interacting with clumpy media can enhance Rayleigh-Taylor instabilities (hereafter RT instabilities). \cite*{2002ApJ...576..832P} studied the adiabatic interaction between shocks and cylindrical clouds. They found that the evolution of a shocked multicloud system depends primarily on the thickness of the cloud layer, and showed that mass loading is inefficient as dense gas travels marginal distances and reaches $<10\,\%$ of the flow speed before destruction. In addition, \cite{2005MNRAS.361.1077P} studied how a collection of mass sources embedded in transonic and supersonic flows affect their dynamics. They showed that the spatial separation among different mass sources determines whether or not a wind can percolate through a clumpy medium. The tails of clouds in close proximity interacted effectively with each other, while increasing the distance between them created multiple bow shocks that favoured the acceleration of gas in between the clouds.\par

Later, \cite{2012MNRAS.425.2212A,2014MNRAS.444..971A} studied the 2D hydrodynamic (HD) and magnetohydrodynamic (MHD) interactions, respectively, of supersonic shocks interacting with multiclump media. In their HD study they found that the ablation of clouds in a clumpy medium overrun by a shock leads to the formation of dense shells as a result of mass loading. These shells are highly turbulent and speed up the destruction of downstream clouds (see also \citealt{2009MNRAS.394.1351P}). In their MHD study they showed that the role of magnetic fields depends on their orientation and how the clouds are arranged in the shocked multiclump system. While fields aligned with the flow prevent adjacent clouds from expanding and mixing, transverse magnetic fields are effective at drawing nearby clouds together (and even inducing cloud mergers if they were on the same field line). Similarly, \citealt{2019AJ....158..124F} showed that upstream clouds can effectively shield downstream clouds when they are initially placed along a stream in close proximity.\par

Owing to the complexity of the problem, most of the above studies on multicloud systems investigated models with either cylindrical clouds (in 2D) or spherical clouds (in 3D) in purely adiabatic configurations. However, clouds in the interstellar medium (ISM) at the base of outflows are turbulent (e.g., see \citealt{2004RvMP...76..125M,2004ARA&A..42..211E,2007ARA&A..45..565M,2014prpl.conf...77P,2019FrASS...6....7K}) and mass-loaded gas along the outflow is subjected to radiative processes (e.g., see \citealt{2008ApJ...674..157C,2015ApJ...803....6M,2016MNRAS.455.1830T,2018MNRAS.473.5407M,2018ApJ...862...56S}). In this paper we relax the first assumption and present, for the first time, shock-multicloud models including clouds with log-normal density distributions of the type that arises from supersonic turbulence (e.g., see \citealt{2008ApJ...688L..79F}). In subsequent papers in this series, we will present models with radiative heating and cooling, magnetic fields, and other source terms. In this context, this paper broadens the parameter space by investigating the adiabatic interactions between shocks with different Mach numbers and layers of clouds with different density distributions (i.e., compact versus porous systems) and varying cloud population densities (i.e., systems with few clouds versus systems with many).\par

This paper is organised as follows. In Section \ref{sec:Method} we describe the computational set-up and the set of diagnostics and time-scales we use for the analysis of the simulations. In Section \ref{sec:Results} we analyse the effects of changing the density structure of the multicloud system, the shock Mach number, the cloud layer thickness, and the numerical resolution on both the shock and the clumpy medium. In this section we also comment on the implications for cloud entrainment and mass loading into galactic outflows. In Section \ref{sec:FutureWork} we discuss the limitations of this work and the content of the next papers in this series. In Section \ref{sec:Conclusions} we summarise our findings. 

\section{Method}
\label{sec:Method}

\subsection{Simulation code}
\label{subsec:SimulationCode}
For the simulations reported in this paper we solve the equations of hydrodynamics using the \verb#HLLC# approximate Riemann solver \citep*{Toro:1994} with a Courant-Friedrichs-Lewy (CFL) number of $C_{\rm a}=0.3$, implemented in the {\sevensize PLUTO v4.3} code (\citealt{2007ApJS..170..228M}). The mass, momentum, and energy conservation laws we solve are:

\begin{equation}
\frac{\partial \rho}{\partial t}+\bm{\nabla\cdot}\left[{\rho \bm{v}}\right]=0,
\label{eq:MassConservation}
\end{equation}

\begin{equation}
\frac{\partial \left[\rho \bm{v}\right]}{\partial t}+\bm{\nabla\cdot}\left[{\rho\bm{v}\bm{v}}+{\bm{I}}P\right]=0,
\label{eq:MomentumConservation}
\end{equation}

\begin{equation}
\frac{\partial E}{\partial t}+\bm{\nabla\cdot}\left[\left(E+P\right)\bm{v}\right]=0,
\label{eq:EnergyConservation}
\end{equation}

\begin{equation}
\frac{\partial\left[\rho C\right]}{\partial t}+\bm{\nabla\cdot}\left[{\rho C \bm{v}}\right]=0,
\label{eq:tracer}
\end{equation}

\noindent where $\rho$ is the mass density, $\bm{v}$ is the velocity, $P=\left(\gamma-1\right)\rho\epsilon$ is the gas thermal pressure, $E=\rho\epsilon+\frac{1}{2}\rho\bm{v^2}$ is the total energy density, $\epsilon$ is the specific internal energy, and $C$ is a Lagrangian scalar that allows us to track gas originally in the multicloud system (at time $t=0$, $C=1$ inside the multicloud layer, and $C=0$ everywhere else).

\subsection{Scale-free models and normalisation}
\label{subsec:Normalisation}
The adiabatic simulations in this paper lack source terms, so they are scale-free shock-multicloud models. This means the reader can normalise the results to their target systems by adequately following the relevant scaling relations (including the equation of state). In addition, we report some of the results normalised with physical units relevant for the galactic wind in galaxy M82 (e.g., see \citealt{1998ApJ...493..129S,2009ApJ...697.2030S}) and the nuclear wind in our own Galaxy (e.g., see \citealt{2003ApJ...582..246B,2013ApJ...770L...4M}), which are the main motivation of our paper.  Reporting the results in both scale-free units and a set of fiducial physical units allows us to readily compare these results with those reported in the next papers of this series, for which we use the same units, but where scaling is limited by the inclusion of radiative cooling and magnetic fields.

\subsection{Computational set-up}
\label{subsec:Initial and Boundary Conditions}
\begin{figure}
\begin{center}
  \begin{tabular}{c c}
       \hspace{-1cm}1a) Solenoidal cloud model & \hspace{-2cm}1b) Compressive cloud model\\
    \hspace{+0.2cm}\resizebox{38mm}{!}{\includegraphics{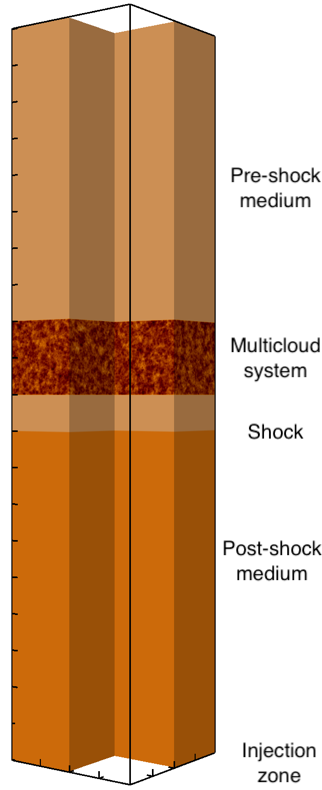}} & \hspace{-0.15cm}\resizebox{38mm}{!}{\includegraphics{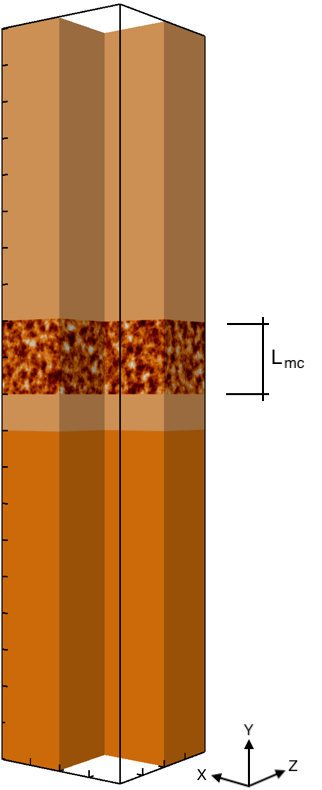}}\vspace{+0.2cm}\\
     \multicolumn{2}{c}{\resizebox{70mm}{!}{\includegraphics{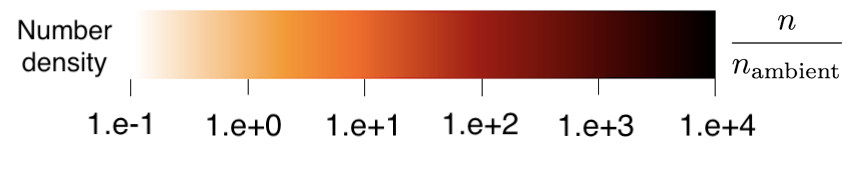}}}\vspace{-0.1cm}\\
  \end{tabular}
  \caption{3D computational setup of two shock-multicloud simulations with a solenoidal layer (panel 1a; model sole-k8-M10) and a compressive layer (panel 1b; comp-k8-M10); see table \ref{Table1}. These computational setups are representative of the whole simulation sample. A quarter of the volume has been clipped to show the interior of the computational domain. The colour bar indicates the gas number densities for all panels in normalised units. To convert to our fiducial physical units, the reader can fix the value of the ambient number density to $n_{\rm ambient}=10^{-2}\,\rm cm^{-3}$.} 
  \label{Figure1}
\end{center}
\end{figure}

The simulation set-up consists of a multicloud system, a pre-shock ambient medium, and a post-shock ambient medium (see Figure \ref{Figure1}). The multicloud system is a rectangular prism (layer) with thickness $L_{\rm mc}$ that contains fractal clouds with an initial average density $\bar{\rho}_{\rm cloud,0}$ (corresponding to an average number density $\bar{n}_{\rm cloud,0}=\bar{\rho}_{\rm cloud,0}/{\mu m_u}=1\,\rm cm^{-3}$ in our fiducial example, where $\mu$ is the mean particle mass and $m_u$ is the atomic mass unit). The multicloud system is embedded in a pre-shock ambient medium with a constant density, $\rho_{\rm ambient}$ (corresponding to a constant number density $n_{\rm ambient}={\rho_{\rm ambient}}/{\mu m_u}=10^{-2}\,\rm cm^{-3}$). Both the pre-shock ambient medium and the multicloud system are initially at rest, and they are swept by a supersonic shock characterised by a Mach number,
\begin{equation}
{\cal M_{\rm shock}}=\frac{v_{\rm shock}}{c_{\rm ambient}}=10,\:\rm or\:4,\:\rm or\:30,
\label{eq:MachNumber}
\end{equation}

\noindent depending on the model. In this equation, $v_{\rm shock}$ and $c_{\rm ambient}=\sqrt{\gamma {P_{\rm ambient}}/{\rho_{\rm ambient}}}$ are the shock speed and the sound speed of the pre-shock ambient medium, respectively. In all models the initial density contrast between the cloud layer and the pre-shock ambient material is
\begin{equation}
\chi=\frac{\bar{\rho}_{\rm cloud,0}}{\rho_{\rm ambient}}=\frac{\bar{n}_{\rm cloud,0}}{n_{\rm ambient}}=10^2,
\label{eq:DensityContrast}
\end{equation}

\noindent while the density contrast between the post-shock ("psh") and pre-shock ambient media is determined by the Rankine-Hugoniot jump conditions (\citealt{1987flme.book.....L}), which also determine the pressure and velocity jumps across the shock. For ${\cal M_{\rm shock}}=10$, ${\rho_{\rm psh}}\approx 4\,{\rho_{\rm ambient}}$ and ${P_{\rm psh}}\approx 125\,{P_{\rm ambient}}$. For ${\cal M_{\rm shock}}=4$, ${\rho_{\rm psh}}\approx 3.4\,{\rho_{\rm ambient}}$ and ${P_{\rm psh}}\approx 20\,{P_{\rm ambient}}$. For ${\cal M_{\rm shock}}=30$, ${\rho_{\rm psh}}\approx 4\,{\rho_{\rm ambient}}$ and ${P_{\rm psh}}\approx 1125\,{P_{\rm ambient}}$. The shock-multicloud systems are evolved in the rest frame of the pre-shock medium (i.e., $v_{\rm ambient}=0$), so ${v_{\rm psh}}\approx 0.75\,v_{\rm shock}$ for ${\cal M_{\rm shock}}=10,30$ and ${v_{\rm psh}}\approx 0.70\,v_{\rm shock}$ for ${\cal M_{\rm shock}}=4$. The lower Mach numbers, ${\cal M_{\rm shock}}\leq10$, represent the conditions expected in the inner region of galactic winds driven by stellar feedback (e.g., see \citealt{2008ApJ...674..157C,2017ApJ...834..144S}), while the high Mach number, ${\cal M_{\rm shock}}=30$, is chosen solely for theoretical purposes as it allows us to study Mach scaling in the strong-shock regime. 

\subsubsection{Log-normal density fields}
\label{subsec:Log-normal}
Following \cite{2019MNRAS.486.4526B}, the initial probability density function (PDF) of the density fields assigned to the fractal multicloud systems is log-normal,
\begin{equation}
{\cal P}(\rho_{\rm cloud,0})=\frac{1}{s_0\sqrt{2\pi}\rho_{\rm cloud,0}}\,\exp\left({-\frac{[\ln(\rho_{\rm cloud,0})-\bar{m}_0]^2}{2s_0^2}}\right),
\label{eq:PDF}
\end{equation}

\noindent where $\rho_{\rm cloud,0}$ is the cloud density, $\bar{m}_0$ and $s_0$ are the mean and the standard deviation of the logarithm of the density at $t=0$ (\citealt{2007ApJS..173...37S}). Accordingly, the mean and the variance of the density are $\bar{\rho}_{\rm cloud,0}={\exp{(\bar{m}_0+s_0^2/2)}}$ and $\sigma_{\rho_{\rm cloud,0}}^2=\bar{\rho}_{\rm cloud,0}^2(\exp{[s_0^2]}-1)$, respectively. Using this parametrisation, the normalised standard deviation of the initial log-normal PDF is
\begin{equation}
\sigma_{\rm cloud,0}=\frac{\sigma_{\rho_{\rm cloud,0}}}{\bar{\rho}_{\rm cloud,0}}=1.9\:\rm or\: 5.9,
\label{eq:PDFsigma}
\end{equation}

\noindent depending on whether the clouds are solenoidal (i.e., consistent with divergence-free supersonic turbulence) or compressive (i.e., consistent with curl-free supersonic turbulence), respectively (see \citealt{2010A&A...512A..81F}). Note that we will, henceforth, use the terms ``solenoidal clouds'' to refer to the former, i.e., to more compact, more uniform systems; and ``compressive clouds'' to refer to the latter, i.e., to more permeable, more porous systems (see Figure \ref{Figure1}). The log-normal density fields for the cloud layers are generated with the pyFC library (available at \url{https://bitbucket.org/pandante/pyfc}), which constructs randomly-generated, periodic scalar fields that follow pre-defined power-law spectra, $D(k)$, in Fourier space. Solenoidal clouds have $D(k)\propto k^{-0.78}$ and compressive clouds have $D(k)\propto k^{-1.44}$ (see \citealt{2009ApJ...692..364F}). The largest spatially-correlated scales in the two-point fractal distribution are determined by a minimum wavenumber, $k_{\rm min}$, which we change depending on the desired number of cloudlets inside the cloud layer, while the smallest-correlated scales are given by the Nyquist limit, $k_{\rm max}$. In solenoidal models this process creates smoothly-varying density fields with larger fractal dimensions and less pronounced density contrasts than in compressive models.\par

The relation between the normalised standard deviation of the density PDF, $\sigma_{\rm cloud,0}$, and the turbulence Mach number, ${\cal M}_{\rm turb}$, is $\sigma_{\rm cloud,0}\approx b\,{\cal M}_{\rm turb}$ (\citealt{1997ApJ...474..730P,1998PhRvE..58.4501P}). Therefore, the solenoidal density fields correspond to an rms Mach number of ${\cal M}_{\rm turb}\approx 5.3$ with $b\approx0.36$, and the compressive density fields correspond to ${\cal M}_{\rm turb}\approx5.6$ with $b\approx1.05$ (see \citealt{2008ApJ...688L..79F}). Our choice of Mach number reflects the turbulent properties of interstellar gas at the boundary between the cold and warm phases (e.g., see \citealt{2014A&A...564A.106T,2015ApJ...811L..28B}). We will explore other rms Mach numbers in future studies. Note also that, owing to the fractal nature of our multicloud systems, some cloudlets/cores inside the fractal multicloud systems are $\gtrsim 10^4$ times denser than the ambient medium, with $n_{\rm cloud}\gtrsim 10^2\,\rm cm^{-3}$; while, diffuse regions in the layers have number densities of $n_{\rm cloud}\sim 10^{-3}-10^{-1}\,\rm cm^{-3}$, thus reflecting more realistic density gradients than previous uniform-cloud models.

\begin{table*}\centering
\caption{Initial conditions for the 3D shock-multicloud models. Column 1 indicates the model name. Columns 2 and 3 indicate the type of density field in the multicloud system and the seed ID used for generating the log-normal density field. Columns 4 and 5 report the normalised standard deviation of the initial density field in the multicloud system, $\sigma_{\rm cloud,0}=\sigma_{\rho_{\rm cloud,0}}/\bar{\rho}_{\rm cloud,0}$, and the normalised wavenumber, $k_{\rm min}\equiv k$, of the multicloud density field, respectively. Column 6 shows the shock Mach number. Columns 7, 8, and 9 show the scale-free, $L_{\rm mc}$-normalised domain size, the number of grid cells in the computational volume, and the size of the domain in our fiducial physical model, respectively. Column 10 indicates the length of the cloud layer in the streaming direction, $L_{\rm mc}$. Columns 11 and 12 report the cloudlet sizes, $r_{\rm cloud}$, in the multicloud system and the number of grid cells covering a cloudlet radius in the traditional notation. In all models, the adiabatic index is $\gamma=\frac{5}{3}$, the turbulence Mach number is ${\cal M}_{\rm turb}\approx 5$, the $L$-normalised domain is $(L\times 5L \times L)$, and the initial density contrast between the multicloud system and the ambient medium is $\chi=10^2$.}
\begin{adjustbox}{max width=\textwidth}
 \hspace*{-0.35cm}\begin{tabular}{c c c c c c c c c c c c}
\hline
\textbf{(1)} & \textbf{(2)} & \textbf{(3)} & \textbf{(4)} & \textbf{(5)} & \textbf{(6)} & \textbf{(7)} & \textbf{(8)} & \textbf{(9)} & \textbf{(10)}  & \textbf{(11)} & \textbf{(12)}\\
\textbf{Model} & \textbf{Density} & $S_d$ & $\sigma_{\rm cloud,0}$ & $k$ & $\cal M_{\rm shock}$ & \textbf{Domain} & \textbf{Number of cells} & \textbf{Fiducial domain}& $L_{\rm mc}$ & $r_{\rm cloud}$ & $\frac{{\rm cells}}{r_{\rm cloud}}$\\
 & & & & & & & & $[\rm pc^3]$ & $[\rm pc]$ & $[\rm pc]$ & \\ \hline
sole-k4-M10 & Solenoidal & 1 & $1.9$ & $4$ & $10$ & $(2\times 10 \times 2)\,L_{\rm mc}$ & $(256\times1280\times256)$ & $(100\times 500 \times 100)$ & $50$ & $12.5$ & $32$ \\
sole-k8-M10 & Solenoidal & 1 & $1.9$ & $8$ & $10$ & $(2\times 10 \times 2)\,L_{\rm mc}$ & $(256\times1280\times256)$ & $(100\times 500 \times 100)$ & $50$ & $6.3$ & $16$ \\
sole-k16-M10 & Solenoidal & 1 & $1.9$ & $16$ & $10$ & $(2\times 10 \times 2)\,L_{\rm mc}$ & $(256\times1280\times256)$ & $(100\times 500 \times 100)$ & $50$ & $3.1$ & $8$ \\
comp-k4-M10 & Compressive & 1 & $5.9$ & $4$ & $10$ & $(2\times 10 \times 2)\,L_{\rm mc}$ & $(256\times1280\times256)$ & $(100\times 500 \times 100)$ & $50$ & $12.5$ & $32$ \\
comp-k8-M10 & Compressive & 1 & $5.9$ & $8$ & $10$ & $(2\times 10 \times 2)\,L_{\rm mc}$ & $(256\times1280\times256)$ & $(100\times 500 \times 100)$ & $50$ & $6.3$ & $16$ \\
comp-k16-M10 & Compressive & 1 & $5.9$ & $16$ & $10$ & $(2\times 10 \times 2)\,L_{\rm mc}$ & $(256\times1280\times256)$ & $(100\times 500 \times 100)$ & $50$ & $3.1$ & $8$ \\\hline
sole-k8-M10-th & Solenoidal & 1 & $1.9$ & $8$ & $10$ & $(4\times 20 \times 4)\,L_{\rm mc}$ & $(256\times1280\times256)$ & $(100\times 500 \times 100)$ & $25$ & $6.3$ & $16$ \\
comp-k8-M10-th & Compressive & 1 & $5.9$ & $8$ & $10$ & $(4\times 20 \times 4)\,L_{\rm mc}$ & $(256\times1280\times256)$ & $(100\times 500 \times 100)$ & $25$ & $6.3$ & $16$ \\\hline
sole-k8-M4 & Solenoidal & 1 & $1.9$ & $8$ & $4$ & $(2\times 10 \times 2)\,L_{\rm mc}$ & $(256\times1280\times256)$ & $(100\times 500 \times 100)$ & $50$ & $6.3$ & $16$ \\
comp-k8-M4 & Compressive & 1 & $5.9$ & $8$ & $4$ & $(2\times 10 \times 2)\,L_{\rm mc}$ & $(256\times1280\times256)$ & $(100\times 500 \times 100)$ & $50$ & $6.3$ & $16$ \\
sole-k8-M30 & Solenoidal & 1 & $1.9$ & $8$ & $30$ & $(2\times 10 \times 2)\,L_{\rm mc}$ & $(256\times1280\times256)$ & $(100\times 500 \times 100)$ & $50$ & $6.3$ & $16$ \\
comp-k8-M30 & Compressive & 1 & $5.9$ & $8$ & $30$ & $(2\times 10 \times 2)\,L_{\rm mc}$ & $(256\times1280\times256)$ & $(100\times 500 \times 100)$ & $50$ & $6.3$ & $16$ \\\hline
sole-k8-M10-sd & Solenoidal & 2 & $1.9$ & $8$ & $10$ & $(2\times 10 \times 2)\,L_{\rm mc}$ & $(256\times1280\times256)$ & $(100\times 500 \times 100)$ & $50$ & $6.3$ & $16$ \\
comp-k8-M10-sd & Compressive & 2 & $5.9$ & $8$ & $10$ & $(2\times 10 \times 2)\,L_{\rm mc}$ & $(256\times1280\times256)$ & $(100\times 500 \times 100)$ & $50$ & $6.3$ & $16$ \\\hline
sole-k8-M10-hr & Solenoidal & 1 & $1.9$ & $8$ & $10$ & $(2\times 10 \times 2)\,L_{\rm mc}$ & $(512\times2560\times512)$ & $(100\times 500 \times 100)$ & $50$ & $6.3$ & $32$ \\
sole-k8-M10-lr & Solenoidal & 1 & $1.9$ & $8$ & $10$ & $(2\times 10 \times 2)\,L_{\rm mc}$ & $(128\times640\times128)$ & $(100\times 500 \times 100)$ & $50$ & $6.3$ & $8$ \\
comp-k8-M10-hr & Compressive & 1 & $5.9$ & $8$ & $10$ & $(2\times 10 \times 2)\,L_{\rm mc}$ & $(512\times2560\times512)$ & $(100\times 500 \times 100)$ & $50$ & $6.3$ & $32$ \\
comp-k8-M10-lr & Compressive & 1 & $5.9$ & $8$ & $10$ & $(2\times 10 \times 2)\,L_{\rm mc}$ & $(128\times640\times128)$ & $(100\times 500 \times 100)$ & $50$ & $6.3$ & $8$ \\\hline
\end{tabular}
\end{adjustbox}
\label{Table1}
\end{table*}

\subsubsection{3D domain and grid resolution}
\label{subsec:3Ddomainandresolution}
The 3D computational domain consists of a rectangular prism with a volume $L_{\rm X}\times L_{\rm Y}\times L_{\rm Z}$, where $L_{\rm X}=L_{\rm Z}=\frac{1}{5}L_{\rm Y}=L$. In our fiducial configuration, we choose $L=100\,\rm pc$, so that the corresponding physical size of the domain is $100\,\rm pc\times 500\,\rm pc\times100\,\rm pc$. The grid in all models is uniform and has a standard resolution of $(N_{\rm X}\times N_{\rm Y}\times N_{\rm Z})=(256\times1280\times256)$. Our high-resolution simulations have twice that number of cells, i.e., $(N_{\rm X}\times N_{\rm Y}\times N_{\rm Z})=(512\times2560\times512)$, and our low-resolution simulations have half that number, i.e., $(N_{\rm X}\times N_{\rm Y}\times N_{\rm Z})=(128\times640\times128)$. The shock is placed at $Y=-L/4$ in all models, so there is a time delay, $\Delta t_{\rm ini}$, until it reaches the multicloud layer. The 3D multicloud system occupies the region between the planes $Y=0$ and $Y=L_{\rm mc}$ (see Figure \ref{Figure1}). We choose $L_{\rm mc}=50\,\rm pc$ in our standard models, and $L_{\rm mc}=25\,\rm pc$ in our thin-layer models (see table \ref{Table1}). Note that our models represent an idealised vertical section of an outflow, so these thicknesses are chosen to be lower than the estimated disc scale heights of $\sim 300\,\rm pc$ in our main targets, namely galaxy M82 and our own Galaxy.\par

Note that: 1) Assigning a large volume for the bottom half of our computational domain is needed because shocks reflected from the multicloud system travel upstream and need to be kept within the computational domain at all times to ensure that the post-shock gas conditions are not altered; 2) in all the figures henceforth we will crop the bottom part of the computational domain to highlight the region of interest, where the multicloud system is located; and 3) employing uniform grids, instead of adaptive ones, allows us to capture the evolution of dense cores in the multicloud layer, shock-cloud interfaces where vorticity is deposited, reflected and refracted shocks, and the diffuse mixed gas at the same resolution in all models.\par

\subsubsection{Boundary conditions}
\label{subsec:Boundaries}
In all models we set up a diode boundary condition on the upper side of the simulation domain, periodic boundary conditions on the four lateral sides, and an inflow boundary condition on the bottom side. A constant supply of gas with post-shock gas properties is injected into the computational domain from the latter zone.\par

\subsubsection{Models}
\label{subsec:models}
Our simulation sample comprises $18$ models in total (see table \ref{Table1}). We initialise the multicloud systems with log-normal density fields characteristic of solenoidal and compressive fractal clouds. We set up $9$ models with solenoidal cloud layers and $9$ models with compressive cloud layers. Within each sample, we vary the minimum wavenumber of the cloud distribution ($k_{\rm min}\equiv k$), the cloud layer thickness ($L_{\rm mc}$), the shock Mach number (${\cal M}_{\rm shock}$), the cloud-generating seed ($S_d$), and the numerical resolution of the computational domain.\par

To insert the pyFC-generated fractal clouds, we follow our standard four-step procedure (see \citealt{2018MNRAS.473.3454B}), i.e., 1) we mask regions in the cloud layer outside a length of $L_{\rm mc}$, 2) we scale the average density to $\bar{\rho}_{\rm cloud,0}$ in the multicloud system, 3) we interpolate the resulting density data cube into the 3D domain, and 4) we initialise the simulations with the multicloud systems in thermal pressure equilibrium with the ambient medium. This process allows us to compare the evolution of solenoidal and compressive multicloud models by ensuring that all of them contain clouds with the same initial average density. All the multicloud systems have an initially stationary velocity field, and turbulence forcing is also excluded from the models. While including a turbulent velocity field would be more consistent, it would also broaden the parameter space as subsonically- and supersonically-turbulent clouds evolve differently (see a comparison in Section 4.2 of \citealt{2018MNRAS.473.3454B}).\par

The standard model names indicate the type of density field (sole/comp), the minimum wavenumber in the cloud layer (k4/8/16), and the shock Mach number (M10/4/30). Models with thinner cloud layers (i.e., with smaller $L_{\rm mc}$) are labeled with a "$\rm th$" subscript, and their cloud layers contain half the mass of the other models. Models with clouds generated with different random seeds are labelled with a "$\rm sd$" subscript, and models with higher and lower numerical resolutions are labeled with "$\rm hr$" and "$\rm lr$" subscripts, respectively.\par

\subsubsection{Cloud layer porosity, cloudlet population density, and individual resolutions}
\label{subsec:CloudSizes}
The porosity of a multicloud system is determined by the standard deviation of the log-normal density distributions. Compressive multicloud models are therefore more porous than solenoidal multicloud models as the mass in them is concentrated in higher-density cores (cloudlets) and there are larger voids of low-density gas surrounding them.\par

On the other hand, the cloud population density in a particular multicloud system is given by the number of individual cloudlets inside the cloud layer volume. The number of cloudlets in a layer is given by the normalised, dimensionless wavenumber of the density distribution, $k=\frac{k_{l}\,L}{2\pi}$, where $k_l$ is the wavenumber in units of $1$/length. Thus, the number of cloudlets, $N_{\rm cloudlets}$, and their typical size, $r_{\rm cloudlet}=\frac{\pi}{k_{l}}$, in each multicloud system is given by the wavenumber as,
\begin{equation}
N_{\rm cloudlets}\approx k^3\frac{L_{\rm mc}}{L},\: \rm and,
\label{eq:NumberCloudlets}
\end{equation}

\begin{equation}
r_{\rm cloudlet}\approx \frac{L}{2\,k},
\label{eq:RadCloudlets}
\end{equation}

\noindent respectively. Therefore, in our standard models $N_{\rm cloudlets}\approx 32,\:256,\:2048$ for $k=4\:,8,\:16$ models, respectively; and in our thin-layer, $k=8$, models $N_{\rm clouds}\approx128$. In our fiducial models with physical units, the above equation implies typical cloud sizes of $r_{\rm cloudlet}\approx\rm 12.5\,pc,\:6.3\,pc,\:3.1\,pc$ for $k=4\:,8,\:16$ models, respectively. Similarly, the numerical resolutions (in terms of number of grid cells per cloudlet radius) are 32, 16, 8 ($R_{32}$, $R_{16}$, $R_{8}$ in the conventional notation) for $k=4\:,8,\:16$ models, respectively, in our standard-resolution simulations.\par

\subsection{Diagnostics}
\label{subsec:Diagnostics}
To investigate how varying the initial conditions affect the evolution of shock-multicloud systems, we use the following set of diagnostic quantities.\par

\noindent a) First, we measure the volumetric averages of the thermal pressure in cloud gas as
\begin{equation}
\left[~P_{{\rm cloud}}~\right]=\frac{\int P\,C\,dV}{\int C\,dV}.
\label{eq:mean_prs}
\end{equation}

\noindent b) Second, we calculate the volumetric filling factor of cloud material in the computational domain (see \citealt{2002ApJ...576..832P}) as
\begin{equation}
F_{v}=\frac{\int C\,dV}{\int dV}.
\label{eq:vol_filling}
\end{equation}

\noindent c) Third, we detect and track shocks inside the computational domain using an algorithm that searches for cells where there are large pressure gradients and ${\bm \nabla} \cdot {\bm v}<0$ (our algorithm is based on the methods described in \citealt{2011MNRAS.418..960V} and \citealt*{2016MNRAS.463.1026L}). The Mach number in each cell, $i$, is  ${\cal M}_{i}=\sqrt{{\cal M}^2_{x_i}+{\cal M}^2_{y_i}+{\cal M}^2_{z_i}}$, where each component is obtained from the local directional speed gradients, $\Delta v_{x_i,y_i,z_i}\approx|\partial v_{i}/\partial x_i,y_i,z_i|\,(2\,\Delta x_i,y_i,z_i)$, where the derivatives are calculated using a central difference method. Thus,
\begin{equation}
{\cal M}_{x_i,y_i,z_i}\approx \left|\frac{4}{3}\frac{\Delta v_{x_i,y_i,z_i}}{c_{\rm sound}}\right|, 
\label{eq:Mach_local}
\end{equation}

\noindent where $c_{\rm sound}=c_{\rm psh}$ for the reflected shock (``rs'') and $c_{\rm sound}=c_{\rm ambient}$ for the forward transmitted shock (``ts''). We assume both sound speeds are constant for simplicity. Then, we calculate the Mach numbers along the streaming direction ($Y$), averaged over the $X$ and $Z$ axes,
\begin{equation}
\left[~{\cal M}~\right]_{y}=\frac{\int {\cal M}_i\,dxdz}{\int \,dxdz},
\label{eq:Mach_2D}
\end{equation}

\noindent and identify the Mach numbers for the reflected shock and the transmitted forward shock as ${\cal M}_{\rm rs}$ and ${\cal M}_{\rm ts}$, respectively. We use a local-maximum detection algorithm and the direction of the $Y$ speed gradients to isolate them from the 1D Mach numbers, $\left[~{\cal M}~\right]_{y}$.

\noindent d) In addition, we measure the degree of mixing between cloud and ambient gas by using a mixing fraction expressed as
\begin{equation}
f_{{\rm mix}}=\frac{\int \rho\,C_{\rm mix}\,dV}{M_{{\rm mc},0}},
\label{eq:MixingFraction}
\end{equation}

\noindent where the numerator is the mass of mixed gas. $C_{\rm mix}$ tracks material in mixed cells, so $C_{\rm mix}=C$ if $0.1\leq C \leq 0.9$ and $C_{\rm mix}=0$ otherwise. The denominator, $M_{{\rm mc},0}$, represents the total mass of the multicloud layer at time $t=0$ (see also \citealt{1995ApJ...454..172X,2005A&A...444..505O,2015ApJ...805..158S}).\par

\noindent e) Next, we define the velocity dispersion along $\rm j=X,Z$, transverse to the direction of shock propagation (see also \citealt{2016MNRAS.455.1309B}),
\begin{equation}
\delta_{{\rm v}}\equiv|\bm{\delta_{{\rm v}}}|=\sqrt{\sum_{\rm j}\delta_{{\rm v}_{{\rm j}}}^2},
\label{eq:rmsVelocity}
\end{equation}

\noindent where the corresponding dispersion of the $\rm j$-component of the velocity (see also \citealt{1994ApJ...433..757M}), $\delta_{{\rm v}_{{\rm j}}}$, reads
\begin{equation}
\delta_{{\rm v}_{{\rm j}}}=\left(\langle~v^2_{{\rm j}}~\rangle-\langle~v_{{\rm j}}~\rangle^2\right)^{\frac{1}{2}}.
\label{eq:rmsVelocityComponent}
\end{equation}

\noindent f) We define the displacement of the centre of mass of the multicloud layer along the streaming axis, $Y$, as
\begin{equation}
\langle~d_{{\rm y}}~\rangle=\frac{\int \rho\,Y CdV}{\int \rho\,C\,dV}=\frac{\int \rho\,Y C\,dV}{M_{\rm mc}}.
\label{eq:cm}
\end{equation}

\noindent where $M_{\rm mc}$ is the time-dependent mass in the multicloud layer.\par

\noindent g) Similarly, we define the average mass-weighted velocity of the cloud layer along the streaming axis, $Y$,
\begin{equation}
\langle~v_{{\rm y}}~\rangle=\frac{\int \rho\,v_{{\rm y}}\,C\,dV}{\int \rho\,C\,dV}=\frac{\int \rho\,v_{{\rm y}}\,C\,dV}{M_{\rm mc}}.
\label{eq:vm}
\end{equation}

\noindent h) In general, the cloud layer mass is $M_{\rm mc}=\int \rho\,C\,dV$, but we also define the mass of cloud gas denser than $\bar{\rho}_{\rm cloud, 0}/3$ as
\begin{equation}
M_{\rm mc_{1/3}}=\int [\rho\,C]_{\rho_{\rm cloud} \geq \bar{\rho}_{\rm cloud, 0}/3}\,dV.
\label{eq:mlo}
\end{equation}

We note that some of the above diagnostics can become affected at late times when either the shock front or cloud material leave the computational domain. Thus, in the diagnostic plots presented in Section \ref{sec:Results}, we only show the curves up to the times when comparisons are still meaningful. These times vary with the model and diagnostic under consideration, but, in general, volume-weighted diagnostics are affected earlier than mass-weighted diagnostics.

\subsection{Transmitted shock speed and dynamical time-scales}
\label{subsec:DynamicalTime-Scales}
The dynamical time-scales relevant for our shock-multicloud models (see table \ref{Table2}) depend on the speed of the internal shock transmitted to the cloud layer after the initial collision, $v_{\rm ts}$. Usually, this speed is taken as $v_{\rm ts}\approx\chi^{-\frac{1}{2}}{\cal M}_{\rm shock}\,c_{\rm ambient}$, which provides a good approximation for most shock-cloud systems. However, for this study we will utilise a more precise definition, introduced by \cite*{1994ApJ...420..213K}, as we find that it provides a better match to our models,
\begin{equation}
v_{\rm ts}=\chi^{-\frac{1}{2}}\,(F_{\rm c1}F_{\rm st})^{\frac{1}{2}}{\cal M}_{\rm shock}\,c_{\rm ambient},
\label{eq:TransmittedShockSpeed}
\end{equation}

\noindent where $F_{\rm st}\approx 1+2.16/(1+6.55\chi^{-\frac{1}{2}})$ and $F_{\rm c1}\approx 1.3$ are dimensionless factors that relate the postshock ambient pressure with the stagnation pressure, and the latter with the pressure behind the transmitted shock, respectively (see also \citealt{2002ApJ...576..832P}). We note that in our models, (1) $\chi=100$, so the factor $(F_{\rm c1}F_{\rm st})^{\frac{1}{2}}$ in equation (\ref{eq:TransmittedShockSpeed}) is $\approx 1.73$; and (2) the speed of the transmitted shock is not homogeneous across the multicloud system as it moves faster in low-density regions and slower in high-density regions than what equation (\ref{eq:TransmittedShockSpeed}) predicts.\par

\begin{table}\centering
\caption{Same as table \ref{Table1}, but here we show the dynamical time-scales relevant for our simulations. Column 1 indicates the model name. Columns 2, 3, and 4 show the shock-passage, cloud-crushing, and simulation time-scales, respectively, in physical units, assuming the fiducial set of physical parameters described in Section \ref{subsec:Initial and Boundary Conditions}. Columns 5 and 6 show the scale-free simulation time normalised with respect to the shock-passage time (note that it is the same in all models) and the cloud-crushing time, respectively.}
\begin{adjustbox}{max width=\textwidth}
\begin{tabular}{c c c c c c}
\hline
\textbf{(1)} & \textbf{(2)} & \textbf{(3)} & \textbf{(4)} & \textbf{(5)} & \textbf{(6)}\\
\textbf{Model} & $t_{\rm sp}$ & $t_{\rm cc}$ & $t_{\rm sim}$ & $\frac{t_{\rm sim}}{t_{\rm sp}}$ & $\frac{t_{\rm sim}}{t_{\rm cc}}$\\
 & $[\rm Myr]$ & $[\rm Myr]$ & $[\rm Myr]$ & &\\ \hline
sole-k4-M10 & 0.20 & 0.09 & 0.60 & 3 & 7\\
sole-k8-M10 & 0.20 & 0.04 & 0.60 & 3 & 14\\
sole-k16-M10 & 0.20 & 0.02 & 0.60 & 3 & 28\\
comp-k4-M10 & 0.20 & 0.09 & 0.60 & 3 & 7\\
comp-k8-M10 & 0.20 & 0.04 & 0.60 & 3 & 14\\
comp-k16-M10 & 0.20 & 0.02 & 0.60 & 3 & 28\\\hline
sole-k8-M10-th & 0.10 & 0.04 & 0.30& 3 & 7\\
comp-k8-M10-th & 0.10 & 0.04 & 0.30 & 3 & 7\\\hline
sole-k8-M4 & 0.50 & 0.11 & $1.50$ & 3 & 14\\
comp-k8-M4 & 0.50 & 0.11 & $1.50$  & 3 & 14\\
sole-k8-M30 & 0.07 & 0.01 & $0.20$ & 3 & 14\\
comp-k8-M30 & 0.07 & 0.01 & $0.20$  & 3 & 14\\\hline
sole-k8-M10-sd & 0.20 & 0.04 & 0.60 & 3 & 14\\
comp-k8-M10-sd & 0.20 & 0.04 & 0.60 & 3 & 14\\\hline
sole-k8-M10-hr & 0.20 & 0.04 & 0.60 & 3 & 14\\
sole-k8-M10-lr & 0.20 & 0.04 & 0.60 & 3 & 14\\
comp-k8-M10-hr & 0.20 & 0.04 & 0.60 & 3 & 14\\
comp-k8-M10-lr & 0.20 & 0.04 & 0.60 & 3 & 14\\\hline
\end{tabular}
\end{adjustbox}
\label{Table2}
\end{table} 

Based on the transmitted shock speed, we can now define global time-scales to characterise the evolution of our multicloud systems. First, we define the shock-passage time, which is the approximate time for the transmitted internal shock to travel from the upstream end to the downstream end of the multicloud layer,
\begin{equation}
t_{\rm sp}=\frac{L_{\rm mc}}{v_{\rm ts}}=\frac{L_{\rm mc}\,\chi^{\frac{1}{2}}}{(F_{\rm c1}F_{\rm st})^{\frac{1}{2}}{\cal M}_{\rm shock} c_{\rm ambient}},
\label{eq:ShockPassageTime}
\end{equation}

\noindent which in our fiducial standard models is $t_{\rm sp}=0.20\,\rm Myr$, in our thin-layer models is $t_{\rm sp}=0.10\,\rm Myr$, in our Mach-4 models is $t_{\rm sp}=0.50\,\rm Myr$, and in our Mach-30 models is $t_{\rm sp}=0.07\,\rm Myr$. The shock-passage time has a similar definition as the more widely-used cloud-crushing time, as defined in \cite{1994ApJ...420..213K}, for shock-cloud models,
\begin{equation}
t_{\rm cc}=\frac{r_{\rm cloudlet}}{v_{\rm ts}}=\frac{r_{\rm cloudlet}\,\chi^{\frac{1}{2}}}{(F_{\rm c1}F_{\rm st})^{\frac{1}{2}}{\cal M}_{\rm shock} c_{\rm ambient}}=\frac{r_{\rm cloudlet}}{L_{\rm mc}}\,t_{\rm sp},
\label{eq:CloudCrushing}
\end{equation}

\noindent where $r_{\rm cloudlet}$ is the cloudlet size (see equation \ref{eq:RadCloudlets}). This time-scale is relevant for describing the evolution of individual cloudlets within multicloud systems, but, since the radius of individual cloudlets differs for multicloud models with different wavenumbers, $t_{\rm cc}$ also varies substantially from model to model. Therefore, we use $t_{\rm sp}$ as our standard normalisation time-scale.\par

The total simulation time is $t_{\rm totsim}=\Delta t_{\rm ini}+t_{\rm sim}$, where $\Delta t_{\rm ini}$ is measured from $t_0=-0.09\,t_{\rm sp}$ to the time when the shock arrives at the multicloud layer (which we define as $t=0$), and $t_{\rm sim}$ is the actual shock-multicloud interaction time. The interaction time is the same in all our models and is given by
\begin{equation}
t_{\rm sim}=3\,t_{\rm sp}.
\label{eq:SimulationTime}
\end{equation}

\noindent Finally, we define the destruction time, $t_{\rm des}$, of a multicloud system as the time when only $25$ per cent of the initial cloud mass in the system has densities above $1/3$ of the original average density in the cloud, $\bar{\rho}_{\rm cloud,0}$ (see \citealt{2015ApJ...805..158S} and our previous study \citealt{2019MNRAS.486.4526B}).


\section{Results}
\label{sec:Results}

\subsection{Evolution of shock-multicloud systems}
\label{subsec:Evolution}
The interaction between supersonic shocks and multicloud systems consists of four phases. Figures \ref{Figure2} and \ref{Figure3} show 2D slices at $Z=0$ of the gas number density, $n$, normalised with respect to the ambient number density, $n_{\rm ambient}$, in three solenoidal and three compressive multicloud models, respectively. The upper panels of these figures correspond to models with a normalised wavenumber $k=4$, the middle panels to $k=8$, and the bottom panels to $k=16$. Independently of whether the multicloud system is initially solenoidal or compressive, the global evolution of the clouds and the shock can be characterised in the following stages:

\begin{figure*}
\begin{center}
  \begin{tabular}{c c c c c c c}
       \multicolumn{1}{l}{\hspace{-2mm}2a) sole-k4-M10 \hspace{+2.5mm}$t_0$} & \multicolumn{1}{c}{$0.5\,t_{\rm sp}=0.10\,\rm Myr$} & \multicolumn{1}{c}{$1.1\,t_{\rm sp}=0.22\,\rm Myr$} & \multicolumn{1}{c}{$1.8\,t_{\rm sp}=0.36\,\rm Myr$} & \multicolumn{1}{c}{$2.4\,t_{\rm sp}=0.48\,\rm Myr$} & \multicolumn{1}{c}{$3.0\,t_{\rm sp}=0.60\,\rm Myr$} & $\frac{n}{n_{\rm ambient}}$\\   
       \hspace{-0.00cm}\resizebox{27mm}{!}{\includegraphics{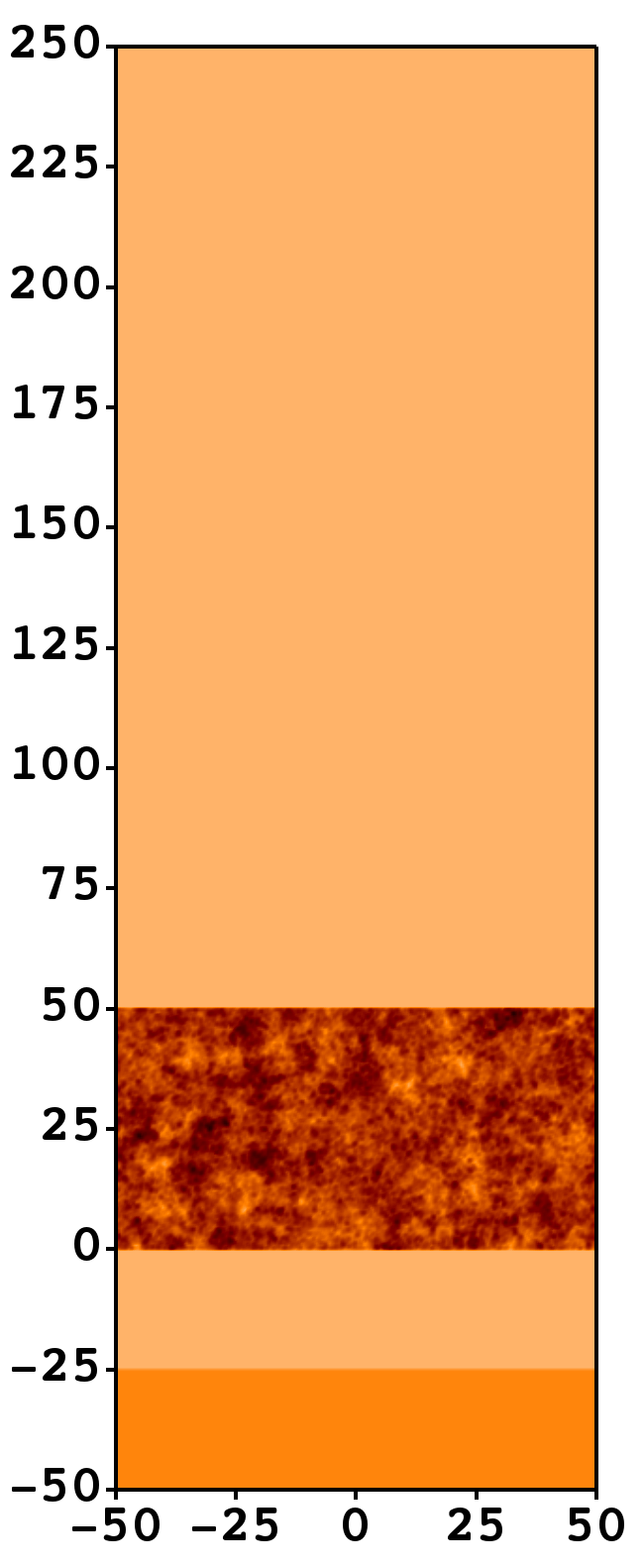}} & \hspace{-0.4cm}\resizebox{27mm}{!}{\includegraphics{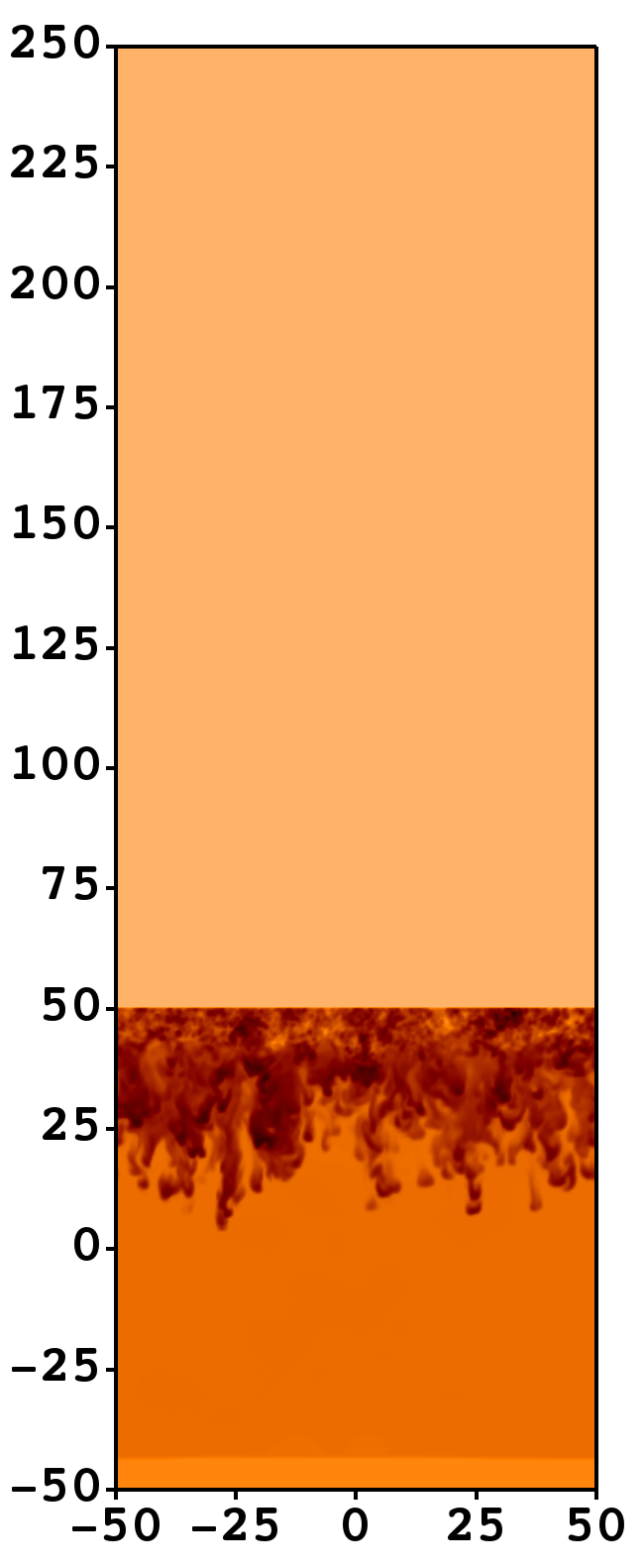}} & \hspace{-0.4cm}\resizebox{27mm}{!}{\includegraphics{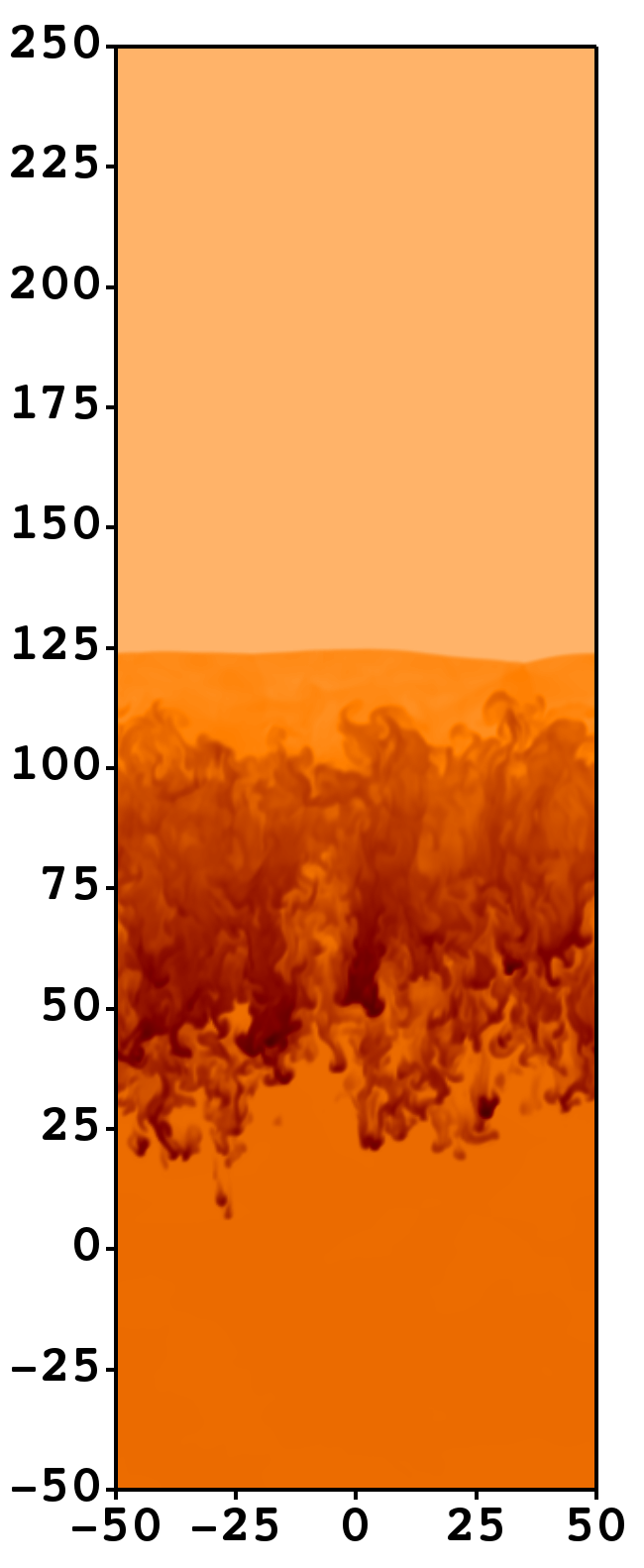}} & \hspace{-0.4cm}\resizebox{27mm}{!}{\includegraphics{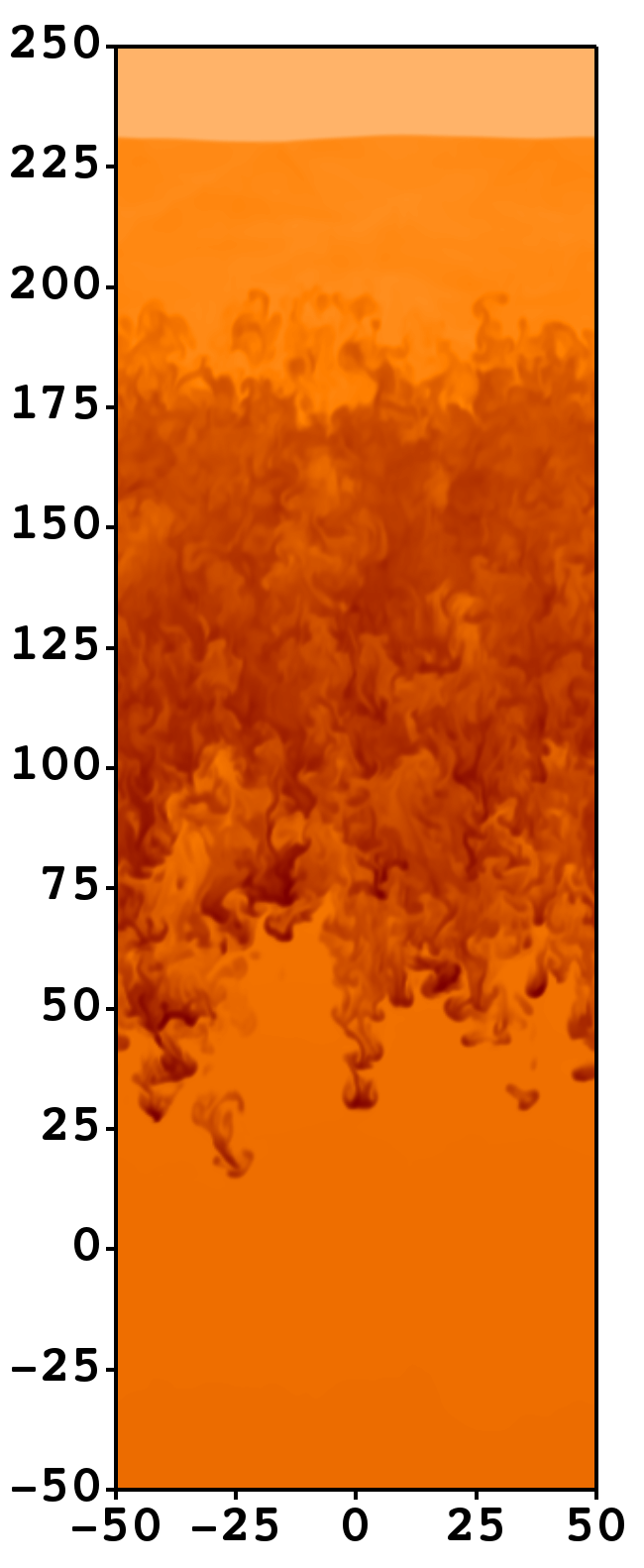}} & \hspace{-0.4cm}\resizebox{27mm}{!}{\includegraphics{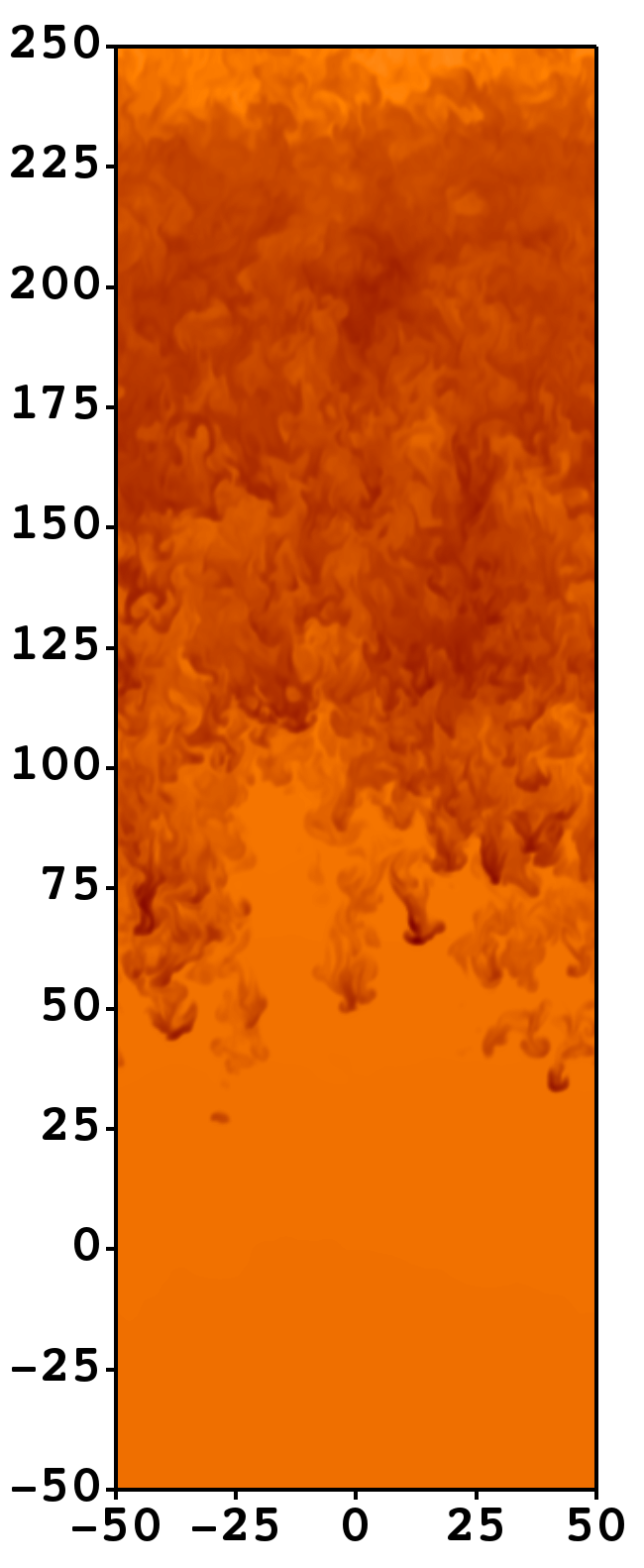}} & \hspace{-0.4cm}\resizebox{27mm}{!}{\includegraphics{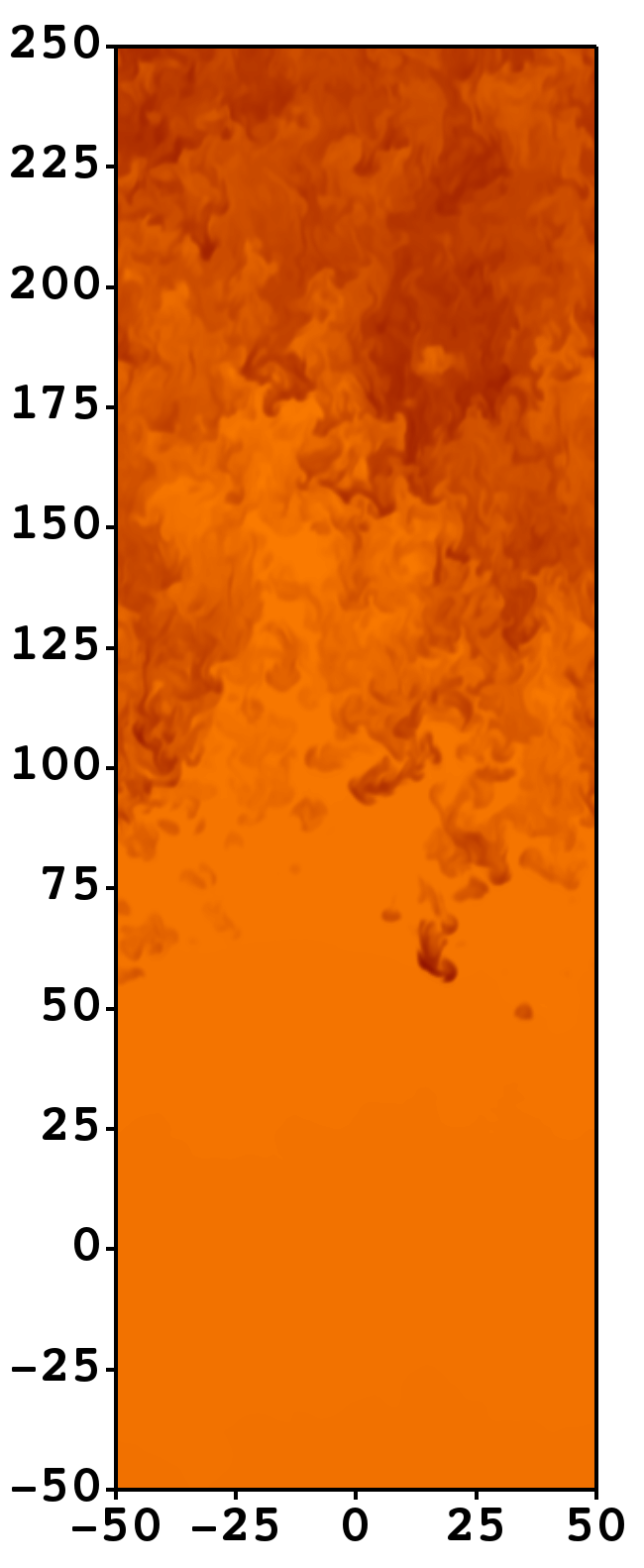}} &
\hspace{-0.2cm}\resizebox{12.8mm}{!}{\includegraphics{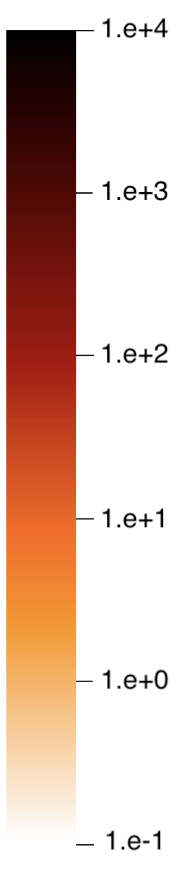}}\\
       \multicolumn{1}{l}{\hspace{-2mm}2b) sole-k8-M10 \hspace{+3.5mm}$t_0$} & \multicolumn{1}{c}{$0.5\,t_{\rm sp}=0.10\,\rm Myr$} & \multicolumn{1}{c}{$1.1\,t_{\rm sp}=0.22\,\rm Myr$} & \multicolumn{1}{c}{$1.8\,t_{\rm sp}=0.36\,\rm Myr$} & \multicolumn{1}{c}{$2.4\,t_{\rm sp}=0.48\,\rm Myr$} & \multicolumn{1}{c}{$3.0\,t_{\rm sp}=0.60\,\rm Myr$} & $\frac{n}{n_{\rm ambient}}$\\  
       \hspace{-0.00cm}\resizebox{27mm}{!}{\includegraphics{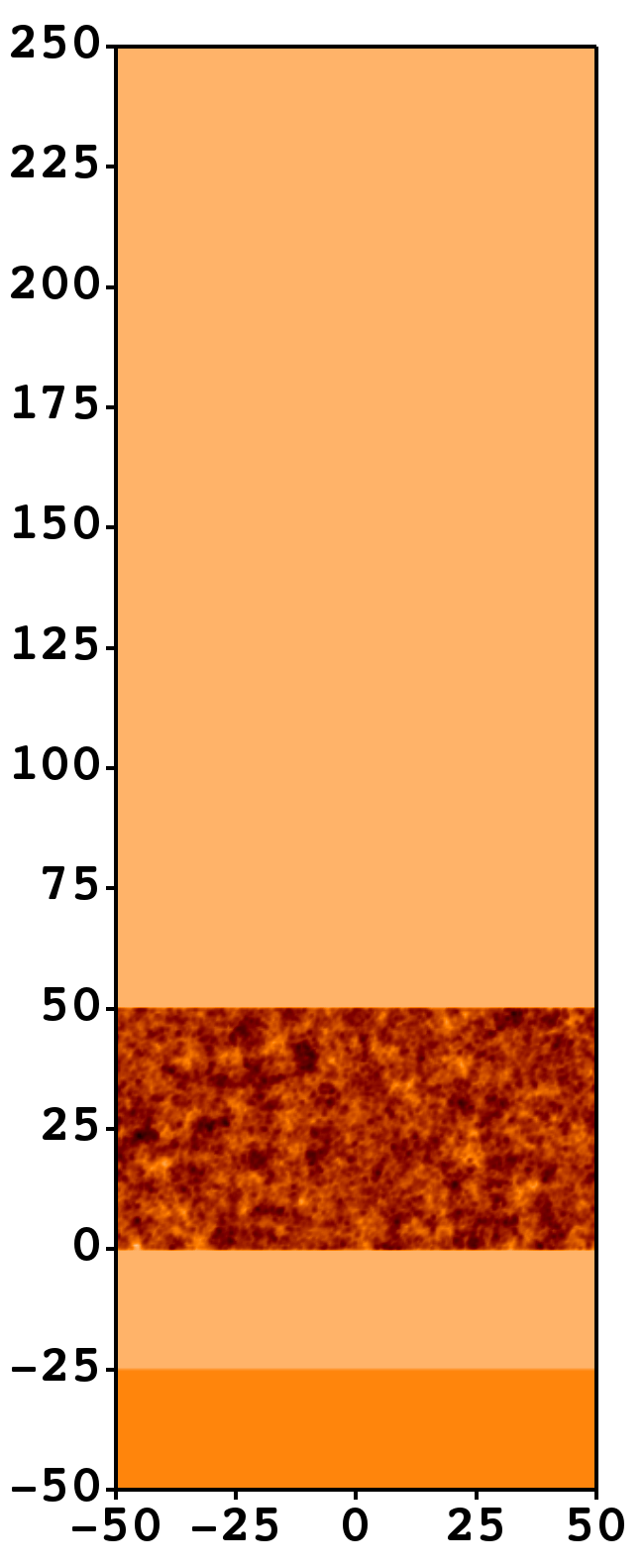}} & \hspace{-0.4cm}\resizebox{27mm}{!}{\includegraphics{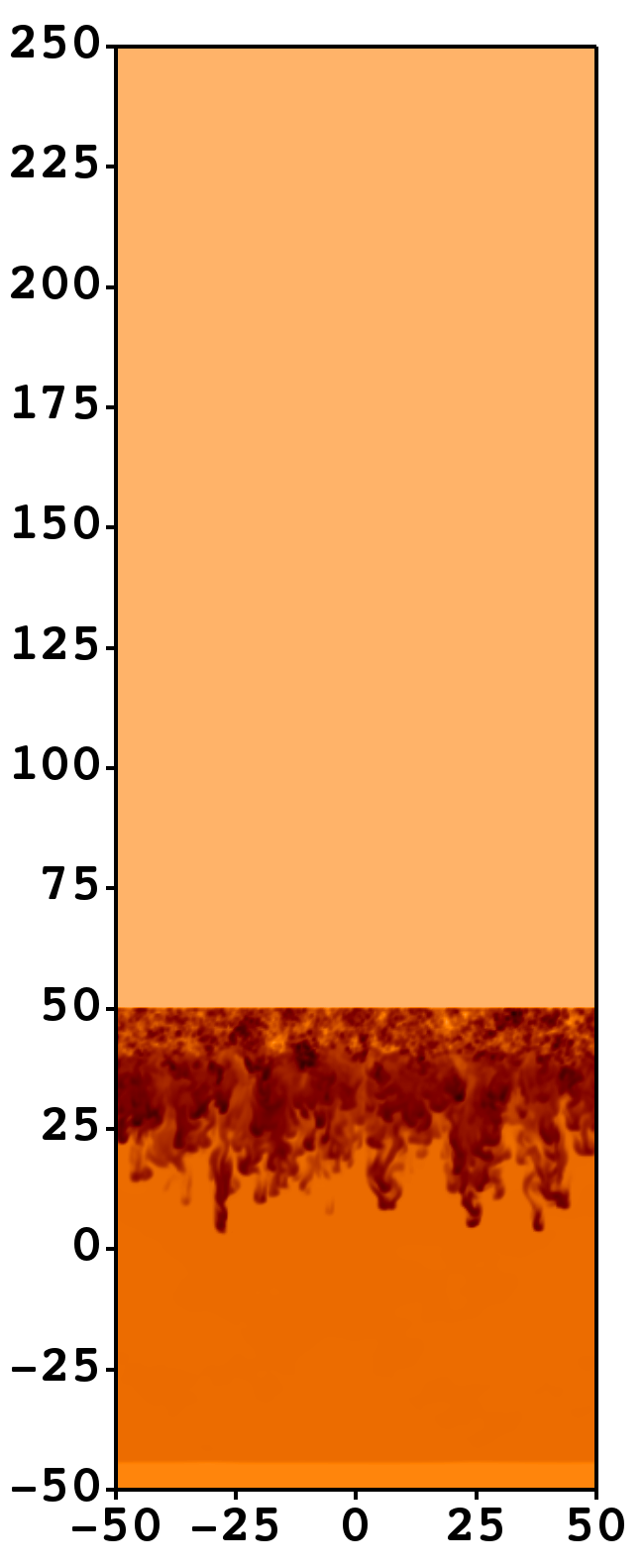}} & \hspace{-0.4cm}\resizebox{27mm}{!}{\includegraphics{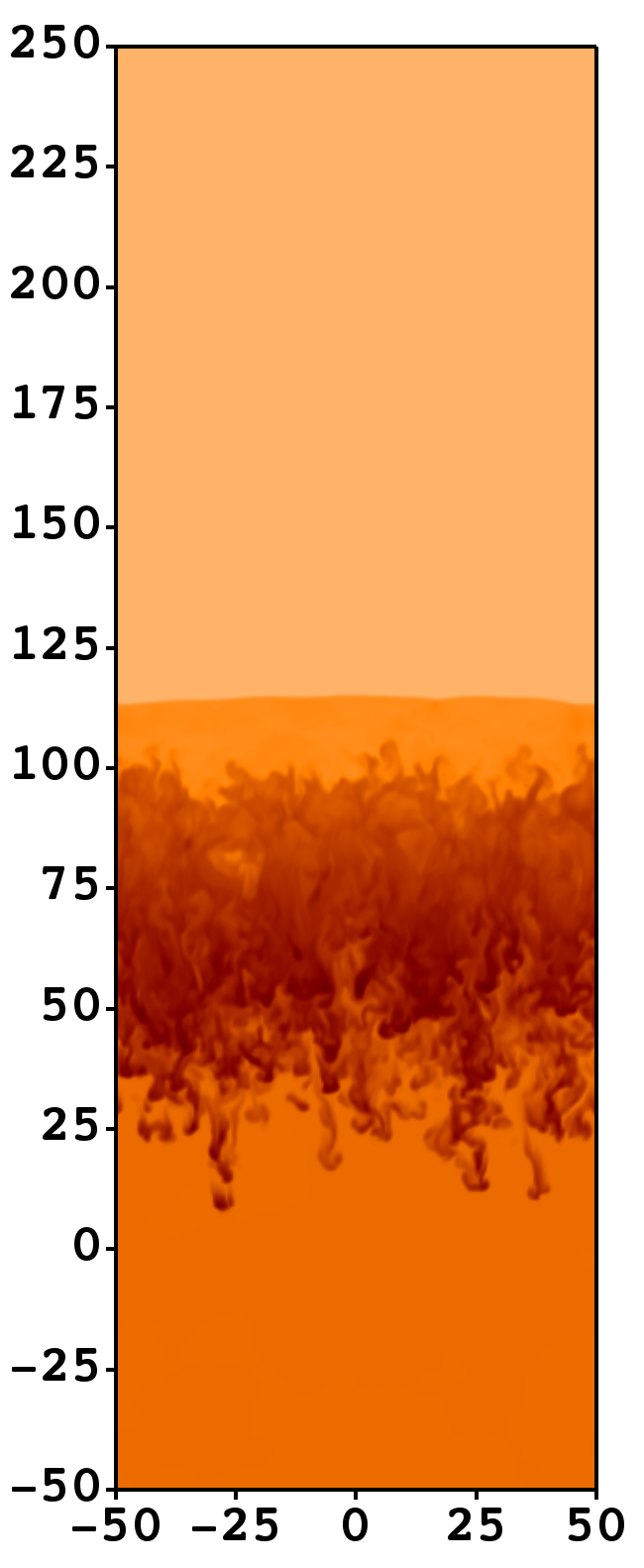}} & \hspace{-0.4cm}\resizebox{27mm}{!}{\includegraphics{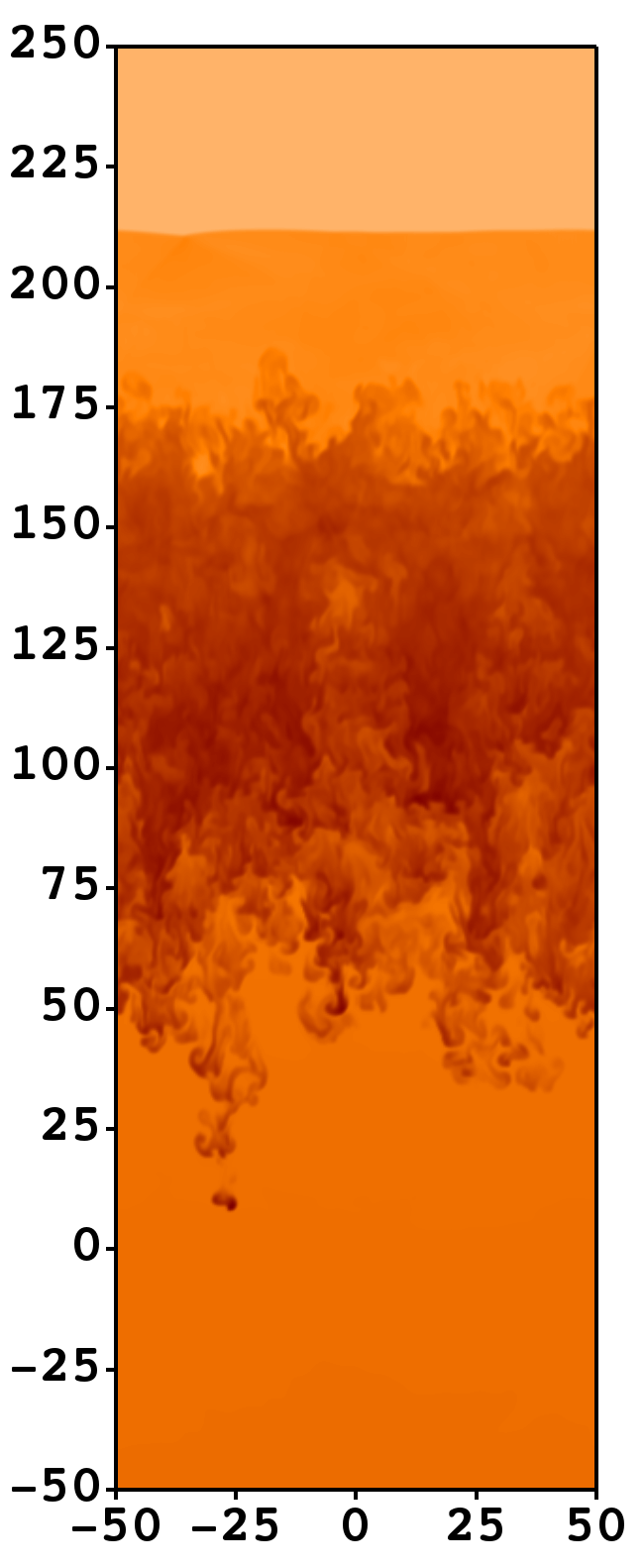}} & \hspace{-0.4cm}\resizebox{27mm}{!}{\includegraphics{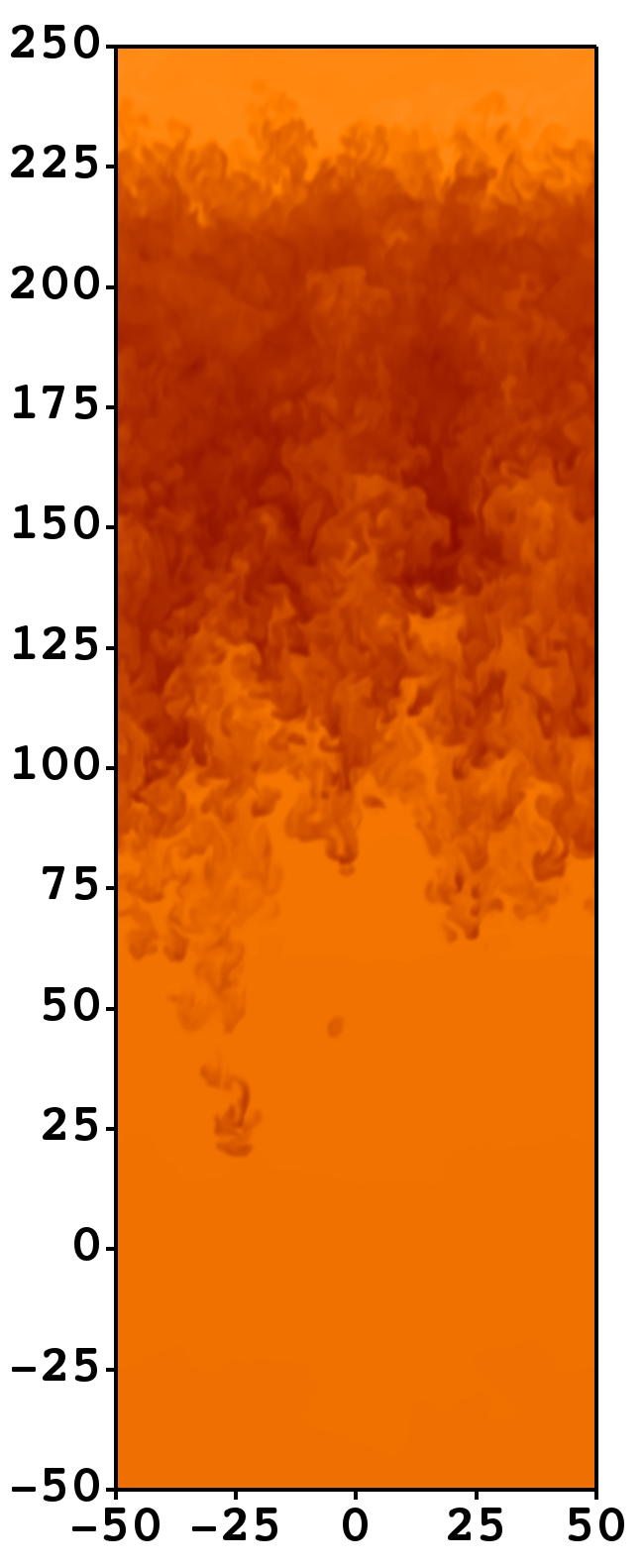}}  & \hspace{-0.4cm}\resizebox{27mm}{!}{\includegraphics{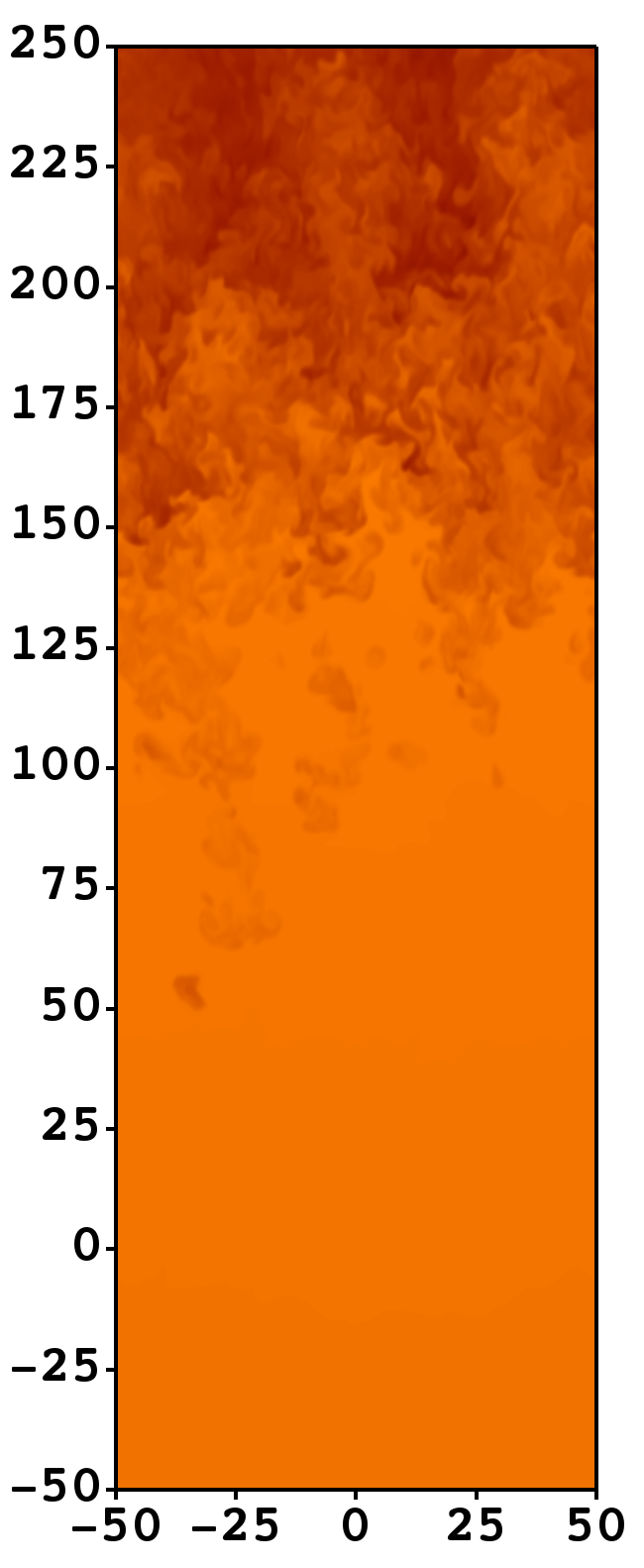}} &
\hspace{-0.2cm}\resizebox{12.8mm}{!}{\includegraphics{bar_vert_2.png}}\\
       \multicolumn{1}{l}{\hspace{-2mm}2c) sole-k16-M10 \hspace{+2.5mm}$t_0$} & \multicolumn{1}{c}{$0.5\,t_{\rm sp}=0.10\,\rm Myr$} & \multicolumn{1}{c}{$1.1\,t_{\rm sp}=0.22\,\rm Myr$} & \multicolumn{1}{c}{$1.8\,t_{\rm sp}=0.36\,\rm Myr$} & \multicolumn{1}{c}{$2.4\,t_{\rm sp}=0.48\,\rm Myr$} & \multicolumn{1}{c}{$3.0\,t_{\rm sp}=0.60\,\rm Myr$} & $\frac{n}{n_{\rm ambient}}$\\
       \hspace{-0.00cm}\resizebox{27mm}{!}{\includegraphics{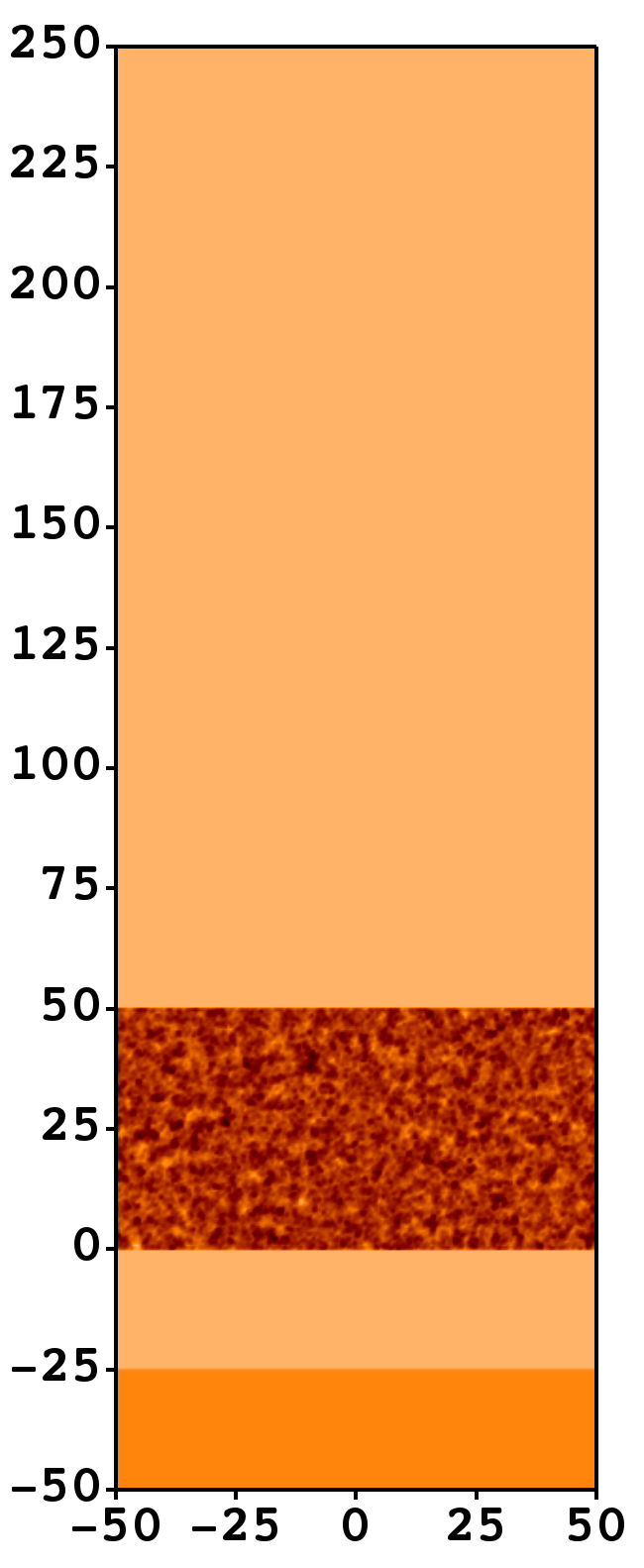}} & \hspace{-0.4cm}\resizebox{27mm}{!}{\includegraphics{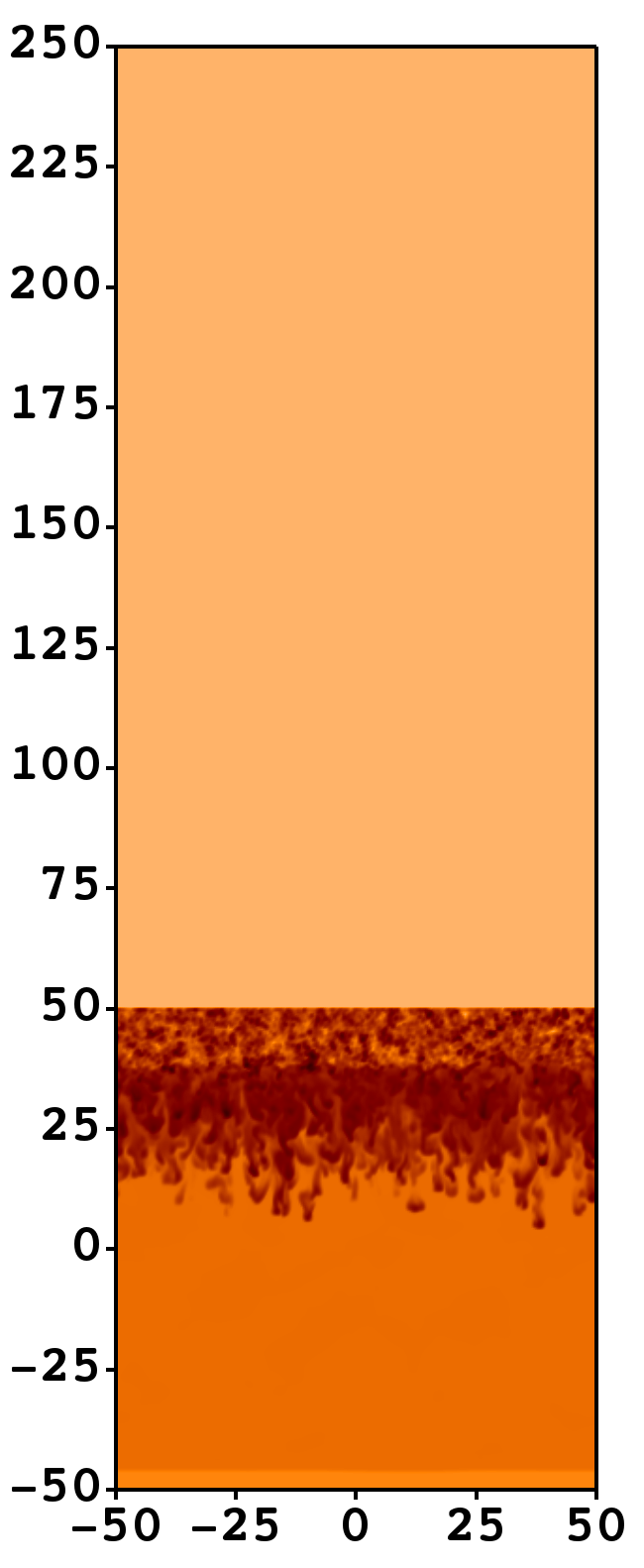}} & \hspace{-0.4cm}\resizebox{27mm}{!}{\includegraphics{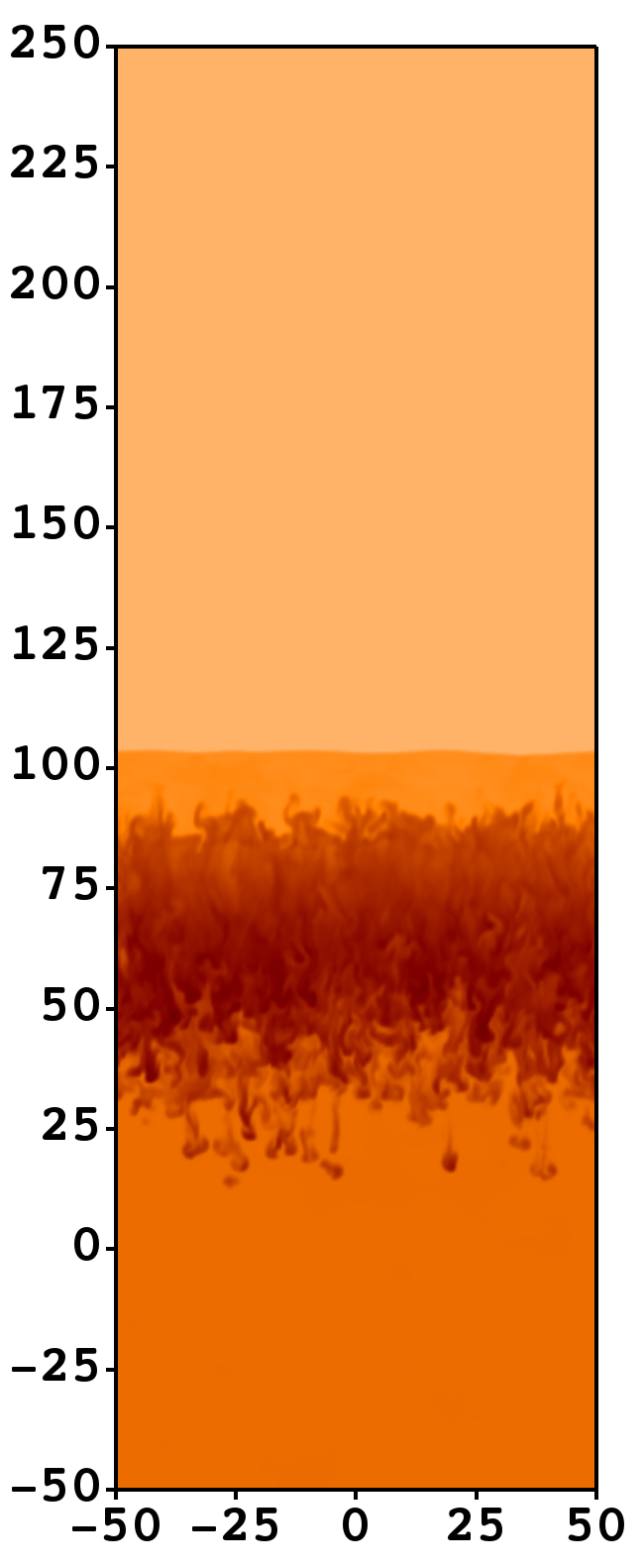}} & \hspace{-0.4cm}\resizebox{27mm}{!}{\includegraphics{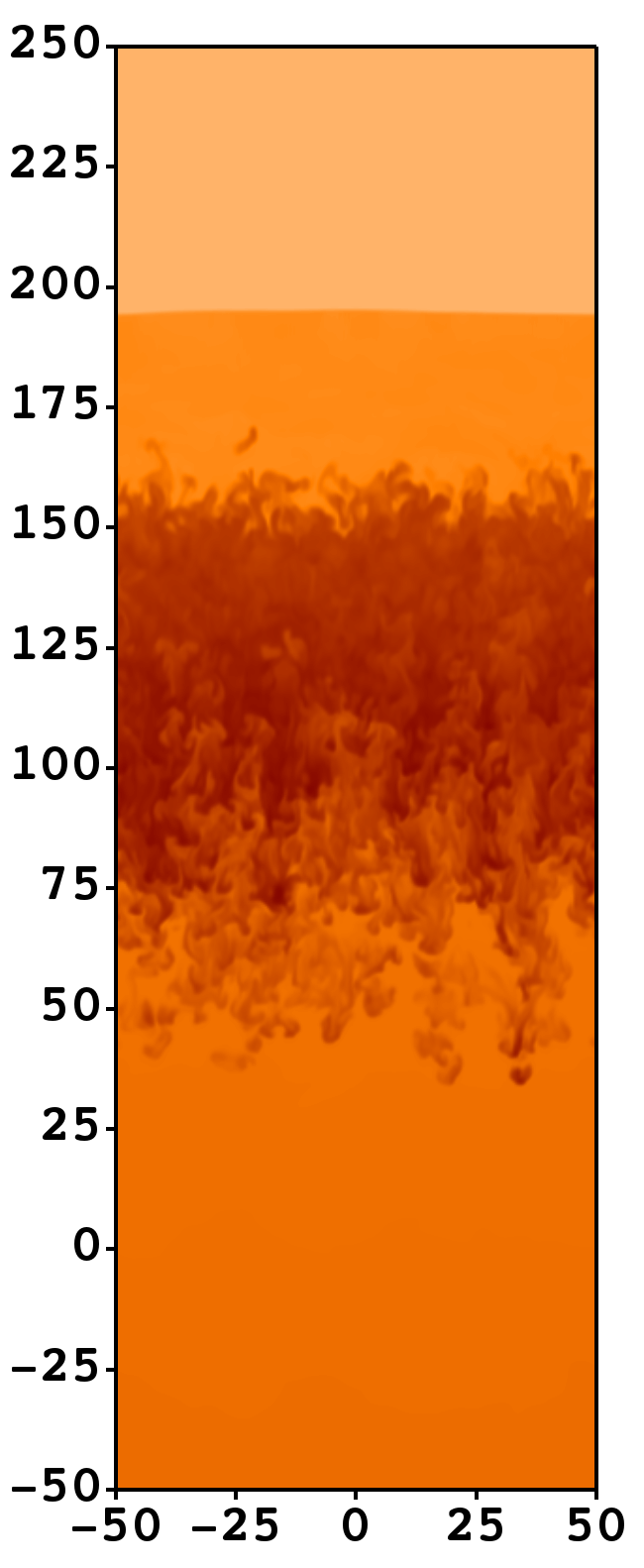}} & \hspace{-0.4cm}\resizebox{27mm}{!}{\includegraphics{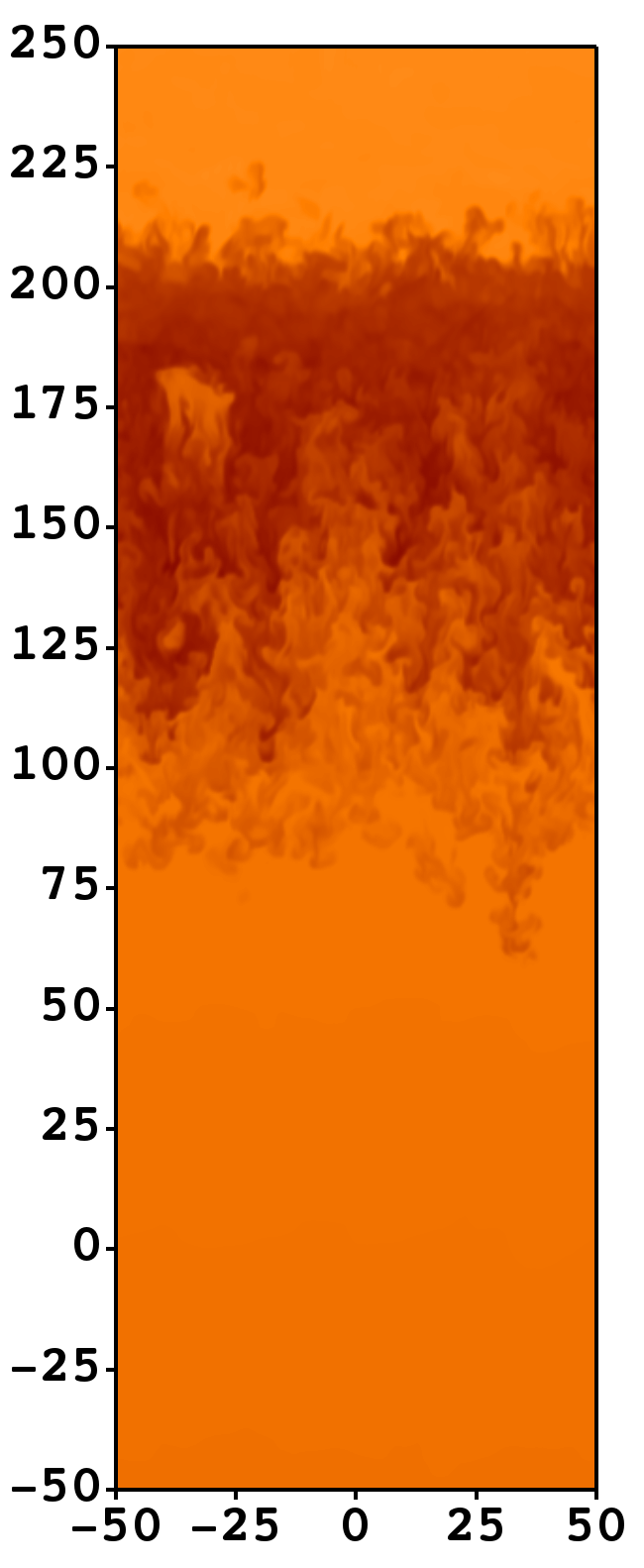}} & \hspace{-0.4cm}\resizebox{27mm}{!}{\includegraphics{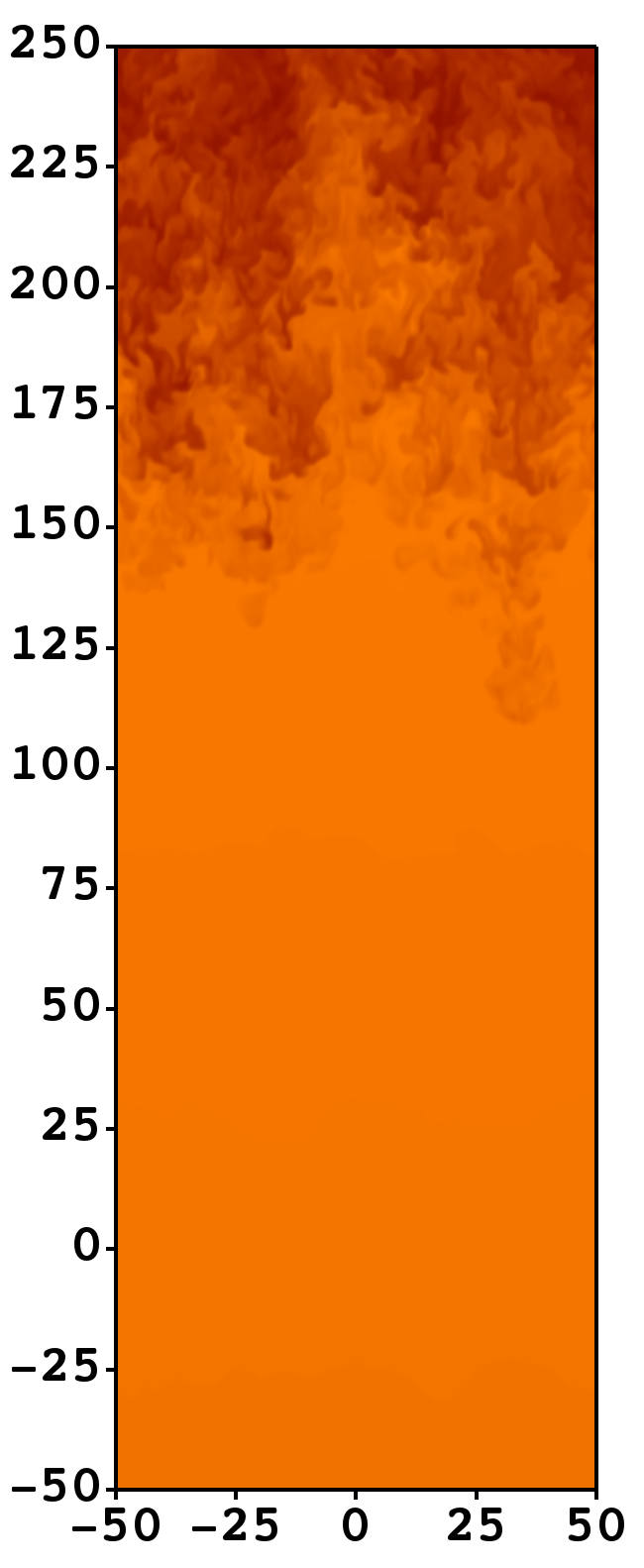}} &
\hspace{-0.2cm}\resizebox{12.8mm}{!}{\includegraphics{bar_vert_2.png}}\\
  \end{tabular}
  \caption{2D slices at $Z=0$ showing the evolution for $t/t_{\rm sp}\leq3.0$ of the gas number density ($n$), normalised with respect to the ambient number density ($n_{\rm ambient}$). We show three solenoidal multicloud models, sole-k4-M10 (panel 2a), sole-k8-M10 (panel 2b), and sole-k16-M10 (panel 2c), which correspond to fractal multicloud layers with normalised wavenumbers $k=4$, $k=8$, and $k=16$, respectively, and the same shock Mach number, ${\cal M_{\rm shock}}=10$. The spatial ($X,Y$) extent is ($L\times3L$)$\equiv$($2\,L_{\rm mc}\times6\,L_{\rm mc}$) as we cropped the bottom part of the domain to zoom into the multicloud region. In our fiducial physical units, $t_{\rm sp}=0.2\,\rm Myr$, so the time range corresponds to $t\leq0.6\,\rm Myr$, and the $X$ and $Y$ axes are given in $\rm pc$, so they cover a spatial extent of ($100\,\rm pc\times300\,\rm pc$).} 
  \label{Figure2}
\end{center}
\end{figure*}

\begin{figure*}
\begin{center}
  \begin{tabular}{c c c c c c c}
       \multicolumn{1}{l}{\hspace{-3mm}3a) comp-k4-M10 \hspace{+2.5mm}$t_0$} & \multicolumn{1}{c}{$0.5\,t_{\rm sp}=0.10\,\rm Myr$} & \multicolumn{1}{c}{$1.1\,t_{\rm sp}=0.22\,\rm Myr$} & \multicolumn{1}{c}{$1.8\,t_{\rm sp}=0.36\,\rm Myr$} & \multicolumn{1}{c}{$2.4\,t_{\rm sp}=0.48\,\rm Myr$} & \multicolumn{1}{c}{$3.0\,t_{\rm sp}=0.60\,\rm Myr$} & $\frac{n}{n_{\rm ambient}}$\\   
       \hspace{-0.25cm}\resizebox{27mm}{!}{\includegraphics{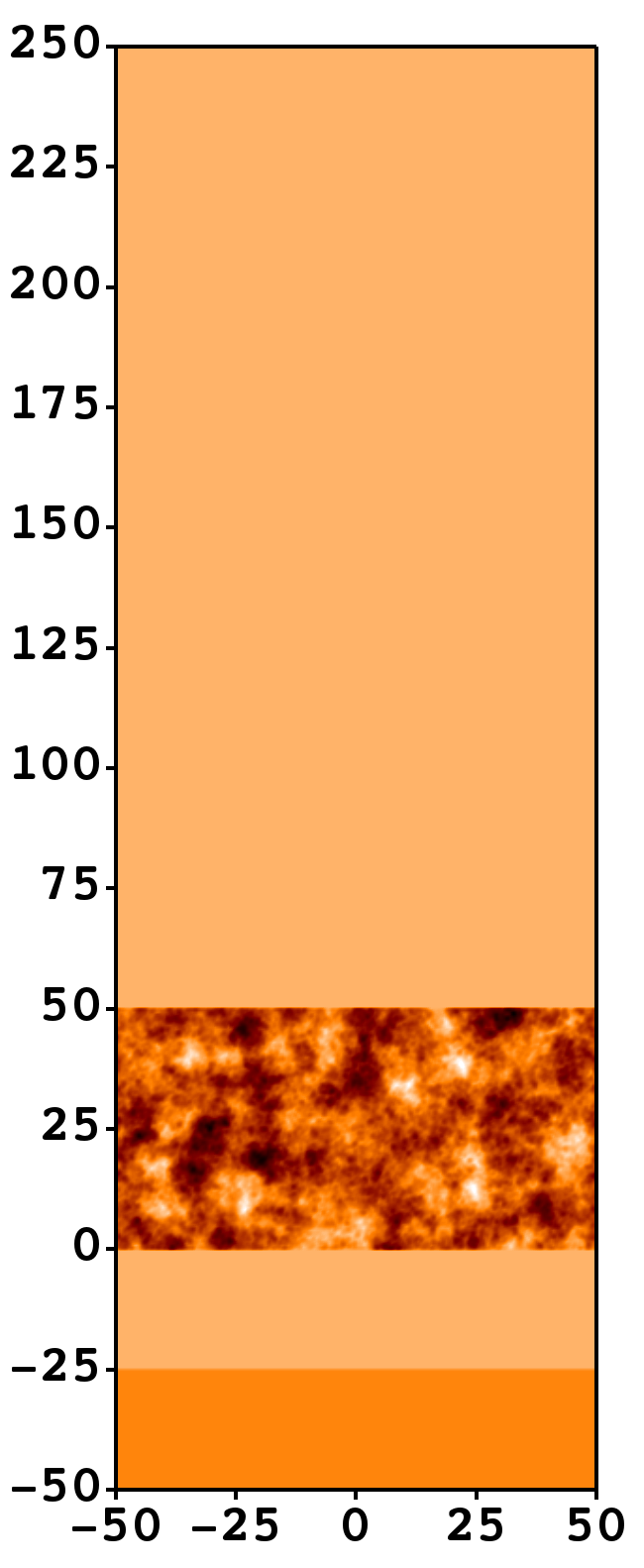}} & \hspace{-0.4cm}\resizebox{27mm}{!}{\includegraphics{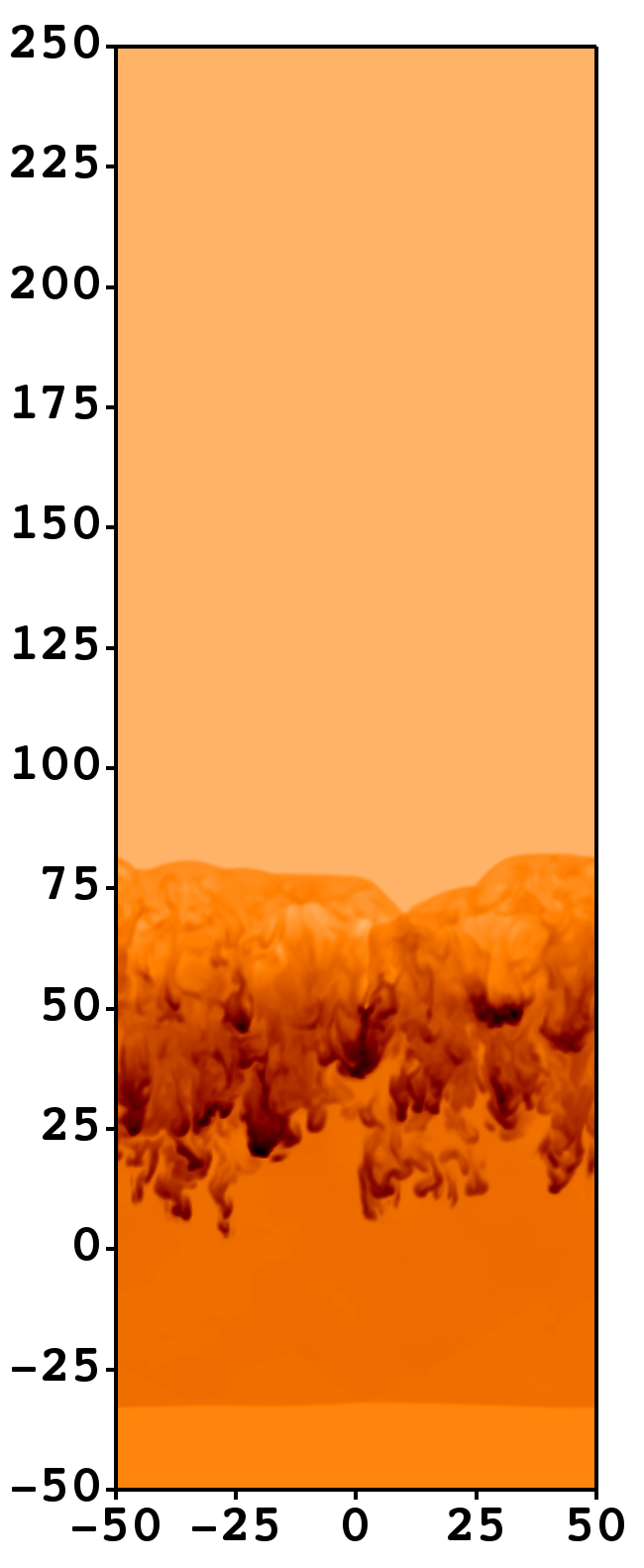}} & \hspace{-0.4cm}\resizebox{27mm}{!}{\includegraphics{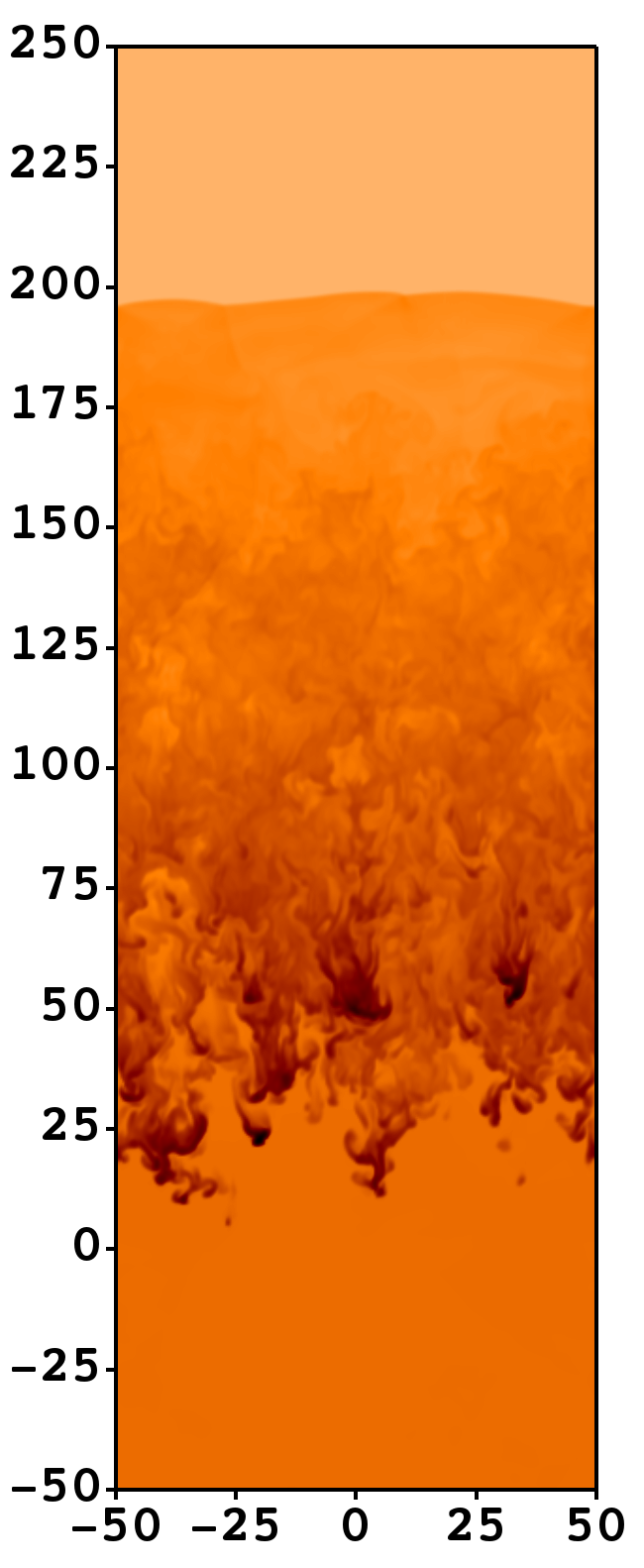}} & \hspace{-0.4cm}\resizebox{27mm}{!}{\includegraphics{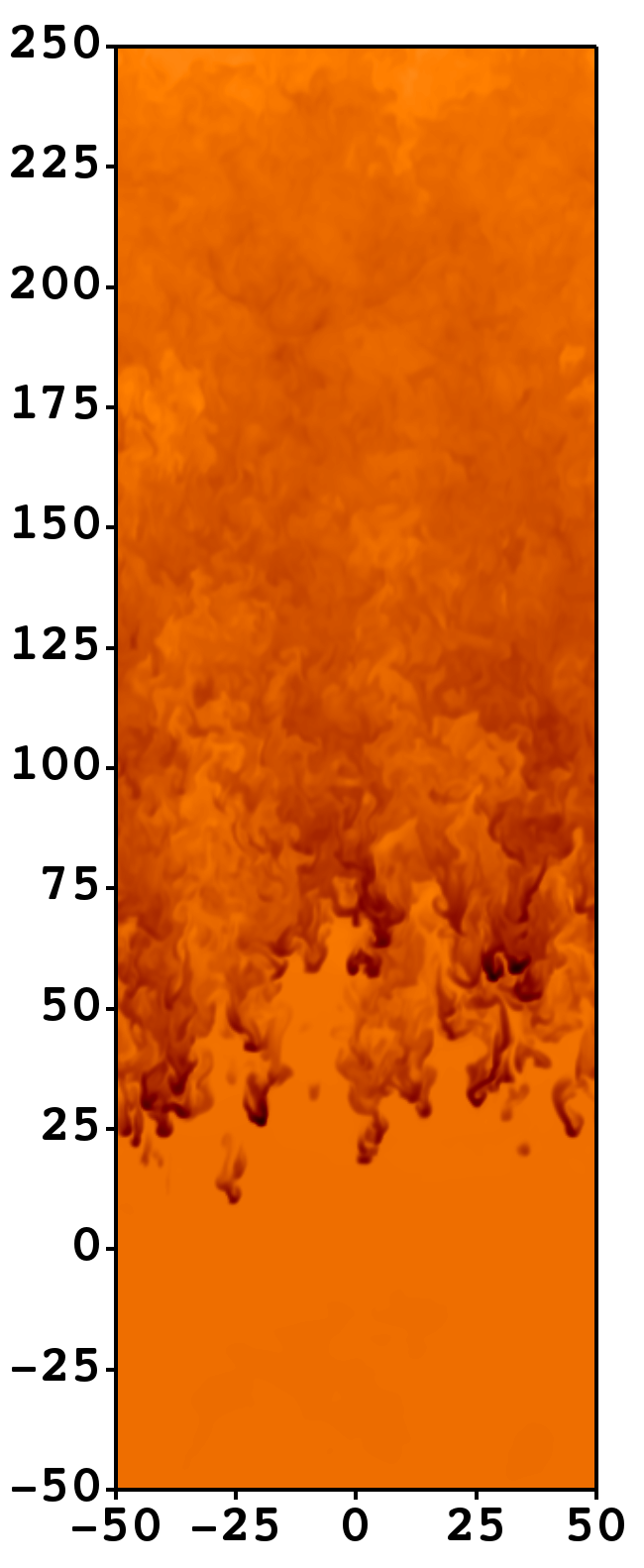}} & \hspace{-0.4cm}\resizebox{27mm}{!}{\includegraphics{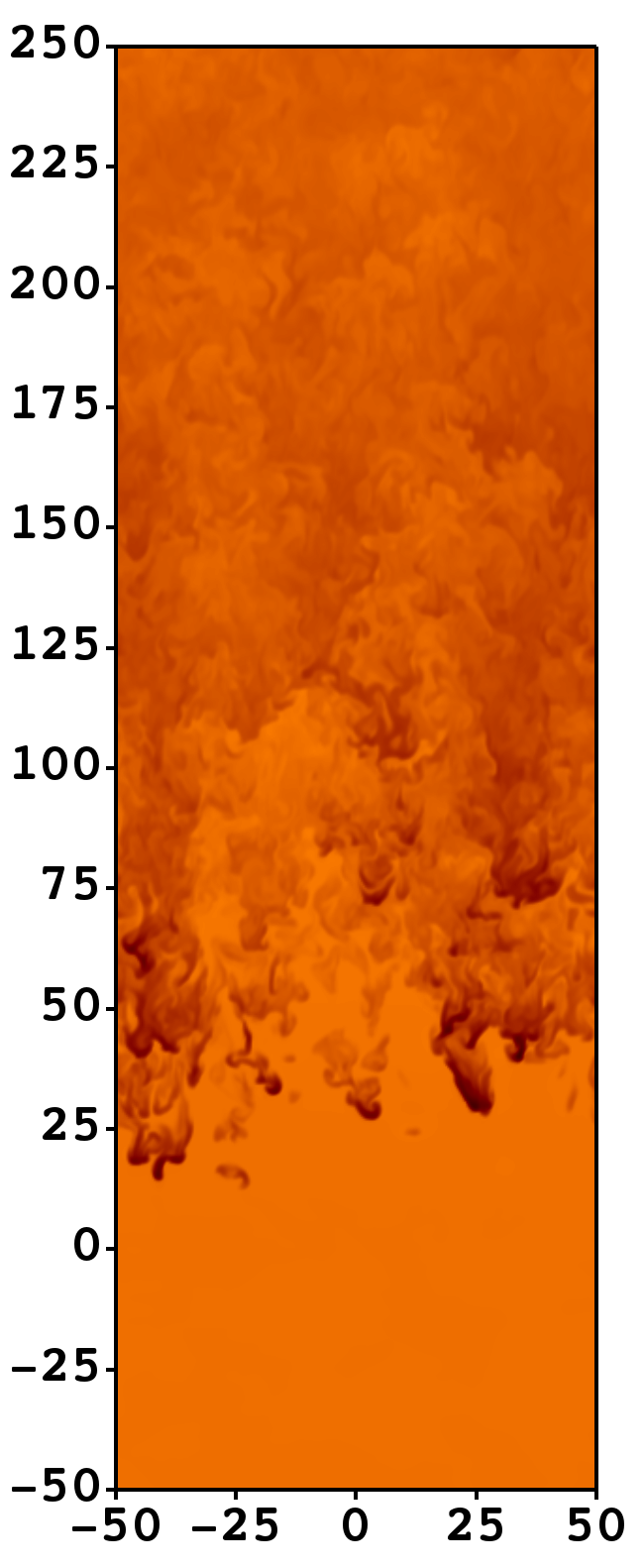}}  & \hspace{-0.4cm}\resizebox{27mm}{!}{\includegraphics{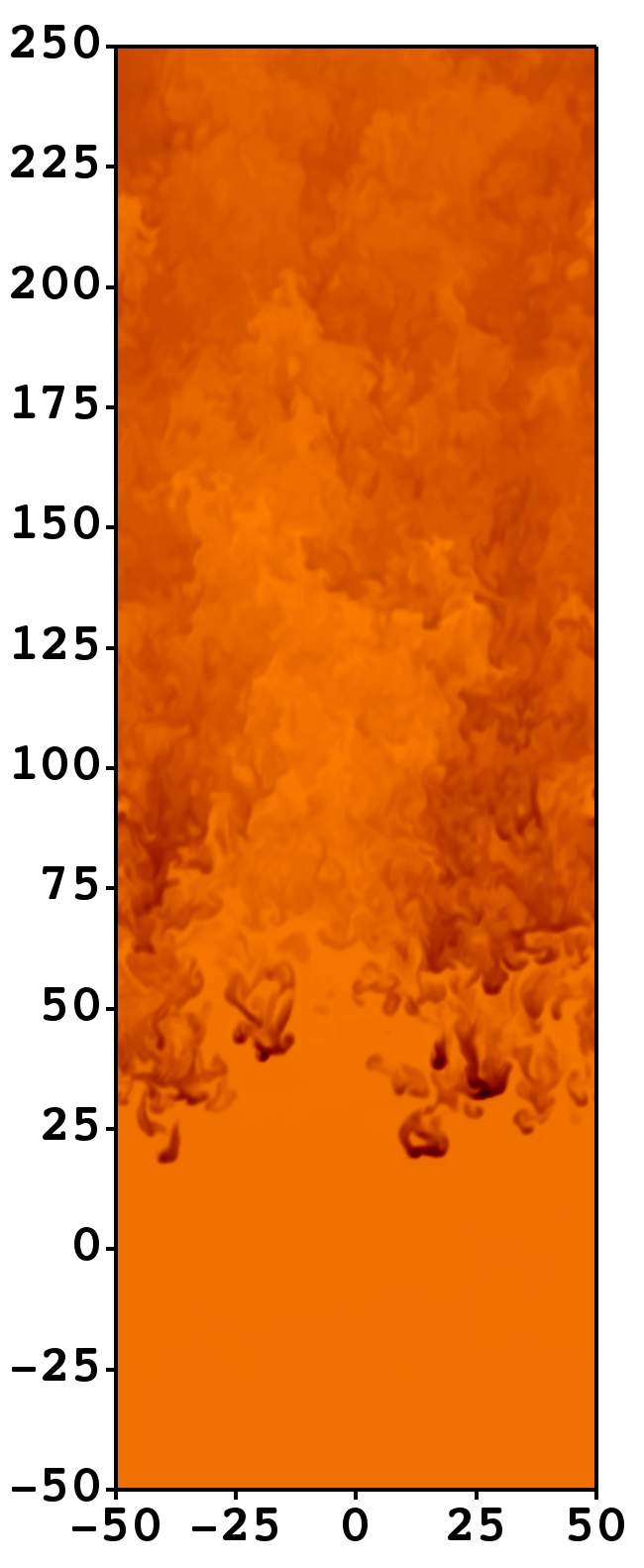}} &
\hspace{-0.2cm}\resizebox{12.8mm}{!}{\includegraphics{bar_vert_2.png}}\\
       \multicolumn{1}{l}{\hspace{-3mm}3b) comp-k8-M10 \hspace{+2.5mm}$t_0$} & \multicolumn{1}{c}{$0.5\,t_{\rm sp}=0.10\,\rm Myr$} & \multicolumn{1}{c}{$1.1\,t_{\rm sp}=0.22\,\rm Myr$} & \multicolumn{1}{c}{$1.8\,t_{\rm sp}=0.36\,\rm Myr$} & \multicolumn{1}{c}{$2.4\,t_{\rm sp}=0.48\,\rm Myr$} & \multicolumn{1}{c}{$3.0\,t_{\rm sp}=0.60\,\rm Myr$} & $\frac{n}{n_{\rm ambient}}$\\    
       \hspace{-0.25cm}\resizebox{27mm}{!}{\includegraphics{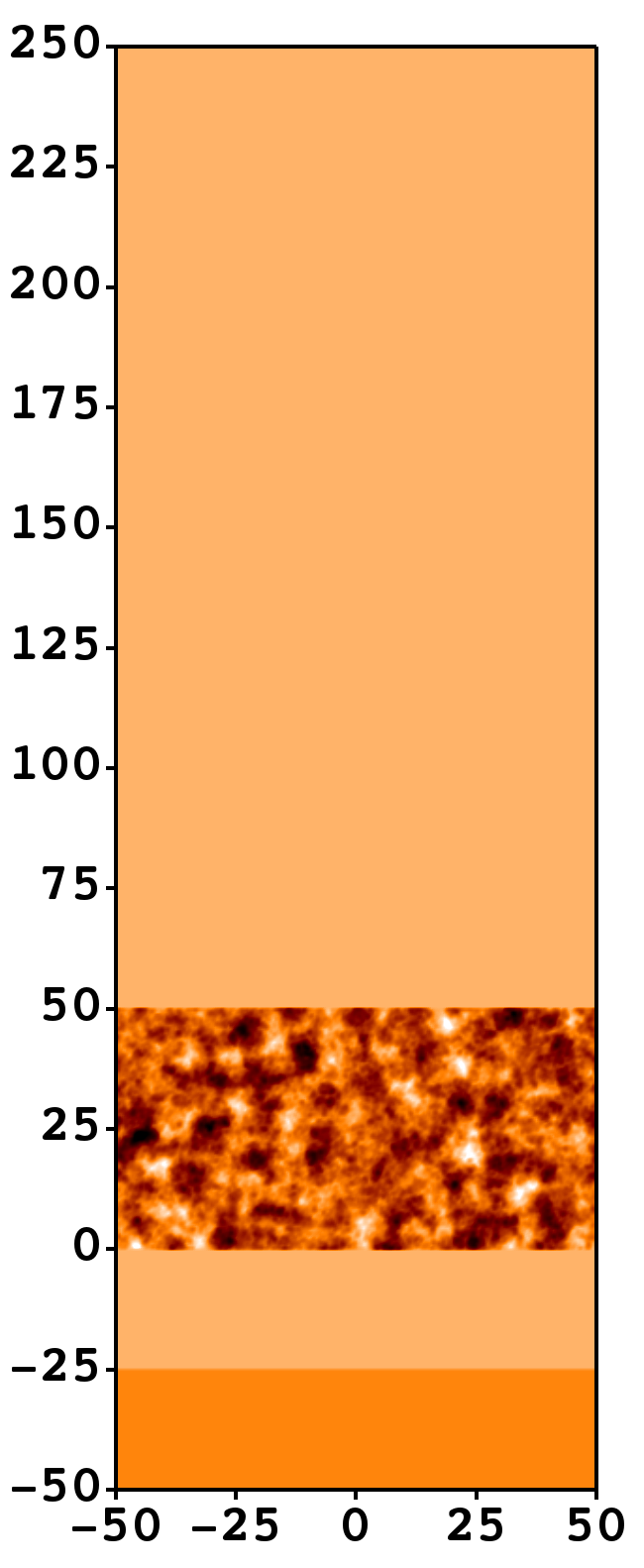}} & \hspace{-0.4cm}\resizebox{27mm}{!}{\includegraphics{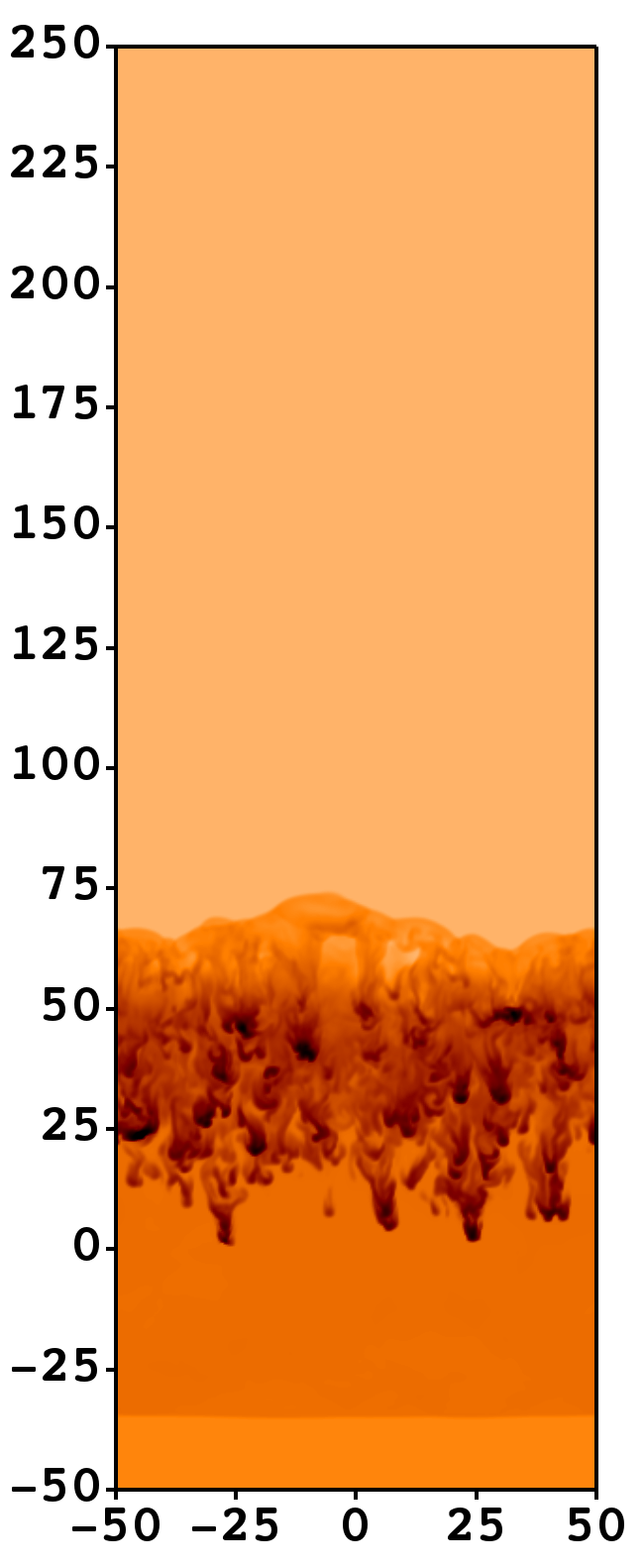}} & \hspace{-0.4cm}\resizebox{27mm}{!}{\includegraphics{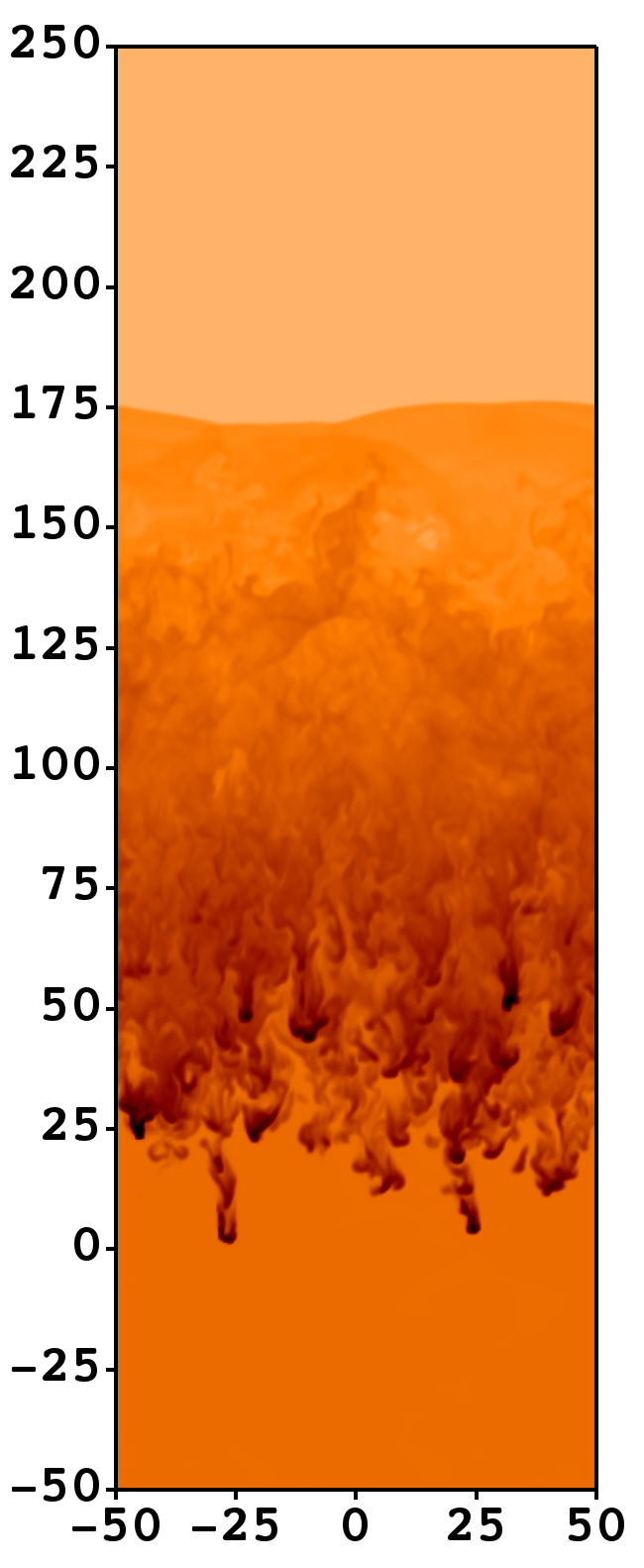}} & \hspace{-0.4cm}\resizebox{27mm}{!}{\includegraphics{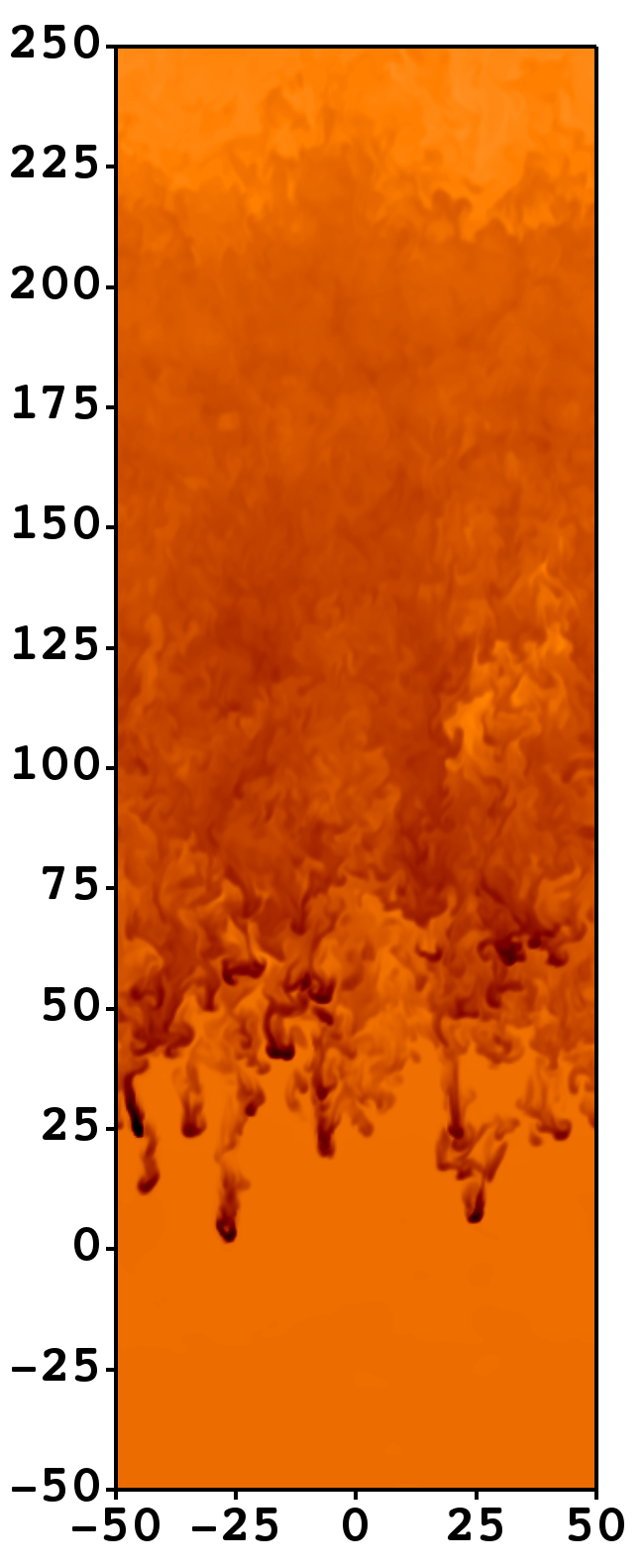}} & \hspace{-0.4cm}\resizebox{27mm}{!}{\includegraphics{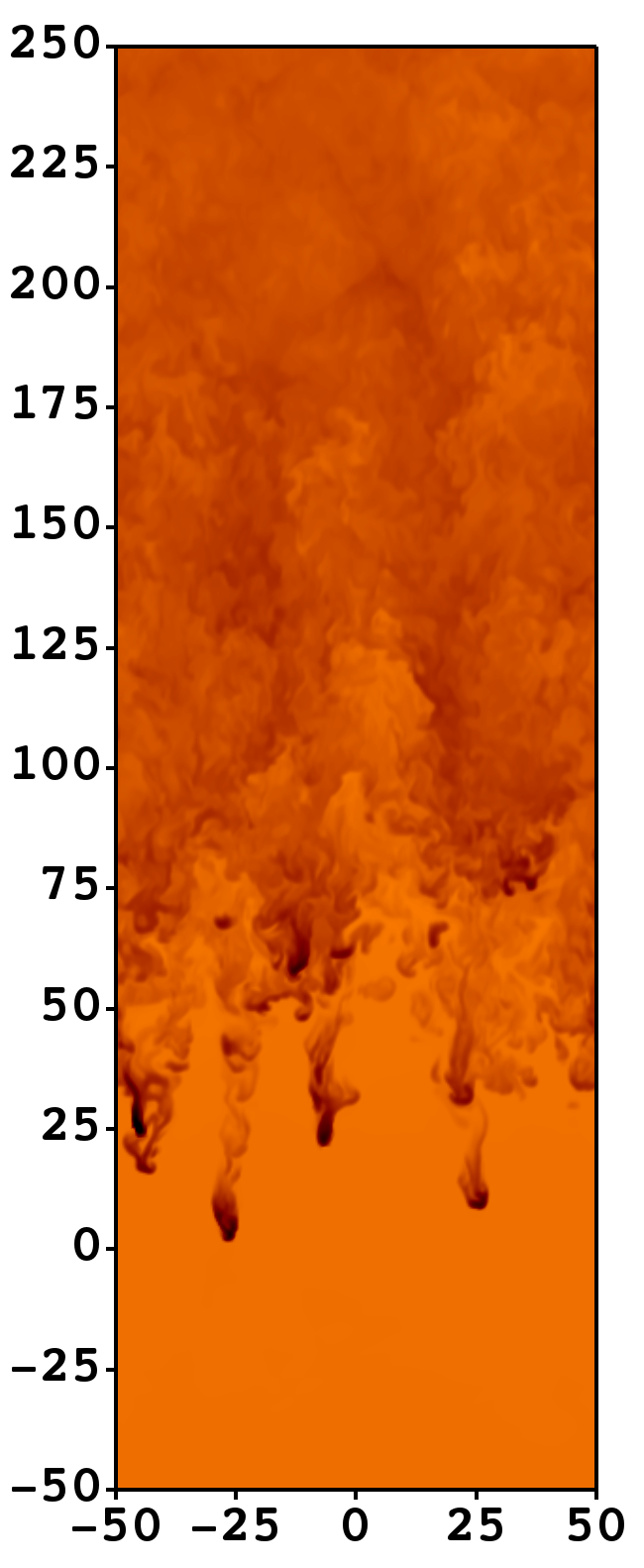}}  & \hspace{-0.4cm}\resizebox{27mm}{!}{\includegraphics{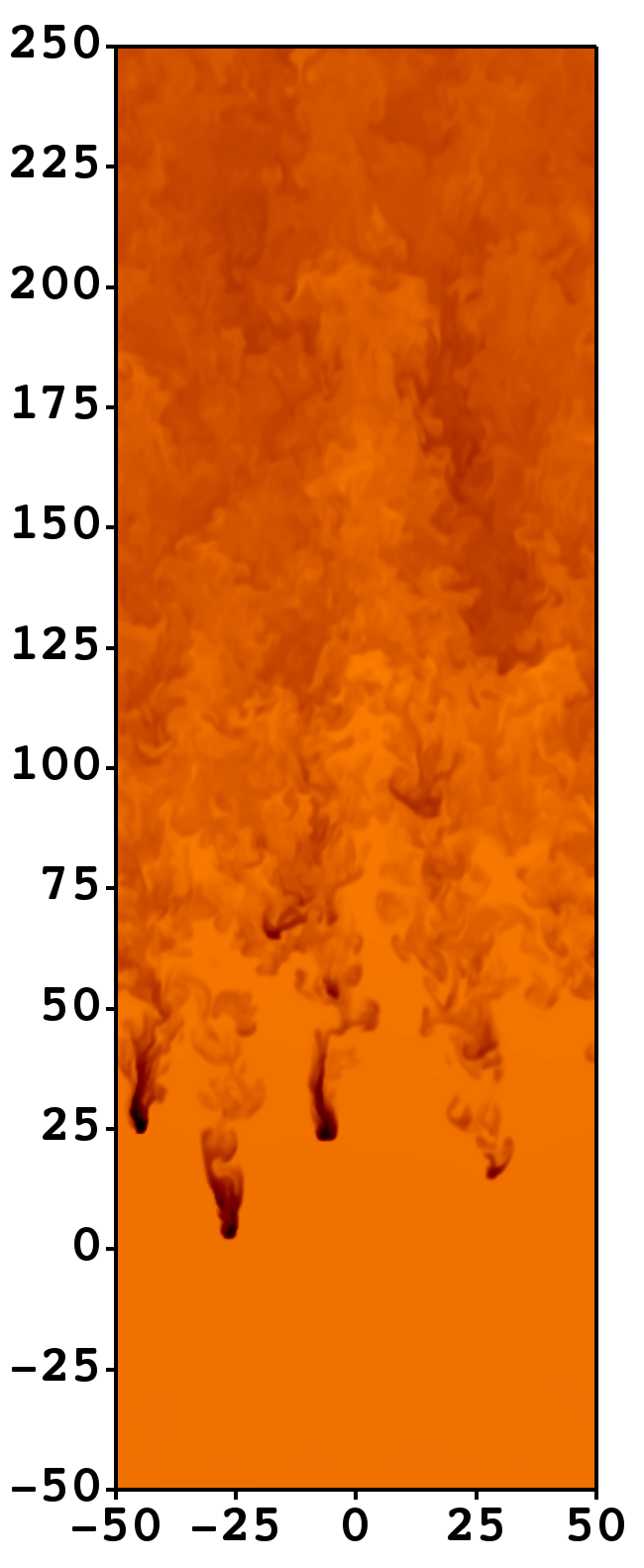}}&
\hspace{-0.2cm}\resizebox{12.8mm}{!}{\includegraphics{bar_vert_2.png}}\\
       \multicolumn{1}{l}{\hspace{-3mm}3c) comp-k16-M10 \hspace{+2.5mm}$t_0$} & \multicolumn{1}{c}{$0.5\,t_{\rm sp}=0.10\,\rm Myr$} & \multicolumn{1}{c}{$1.1\,t_{\rm sp}=0.22\,\rm Myr$} & \multicolumn{1}{c}{$1.8\,t_{\rm sp}=0.36\,\rm Myr$} & \multicolumn{1}{c}{$2.4\,t_{\rm sp}=0.48\,\rm Myr$} & \multicolumn{1}{c}{$3.0\,t_{\rm sp}=0.60\,\rm Myr$} & $\frac{n}{n_{\rm ambient}}$\\
       \hspace{-0.25cm}\resizebox{27mm}{!}{\includegraphics{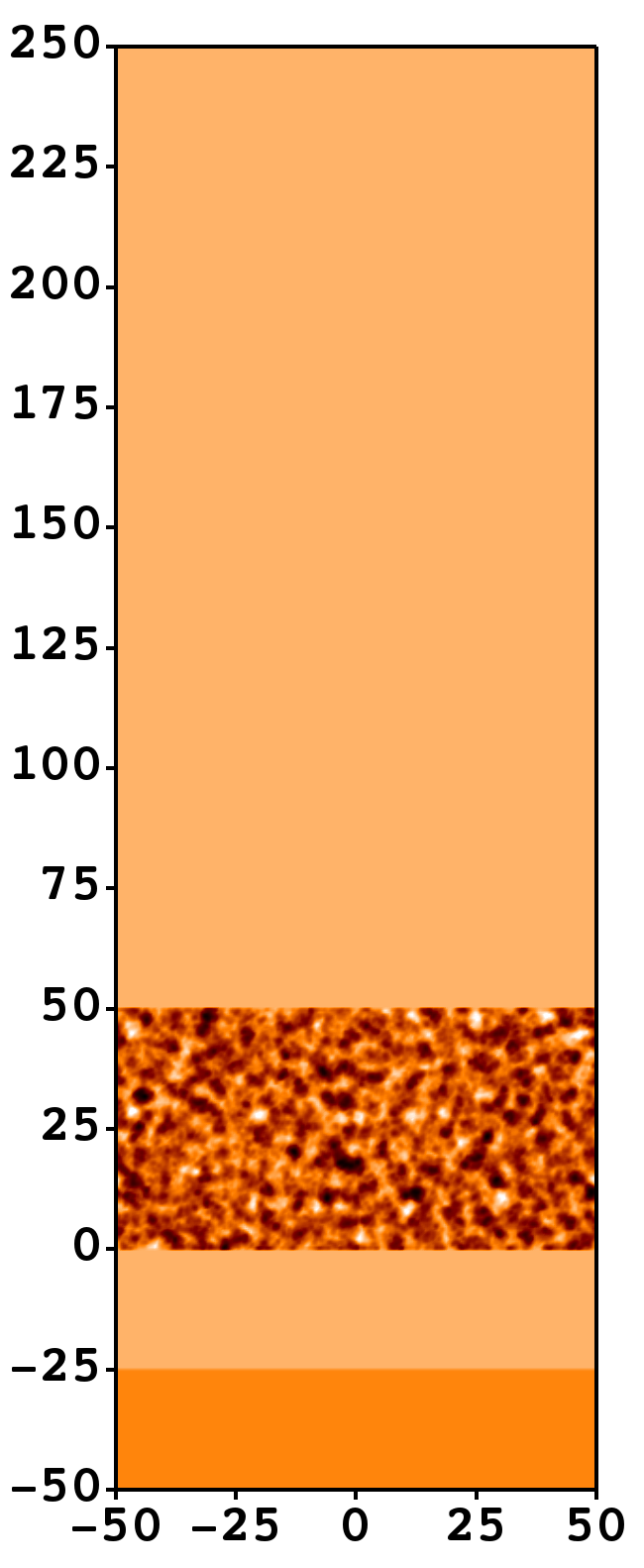}} & \hspace{-0.4cm}\resizebox{27mm}{!}{\includegraphics{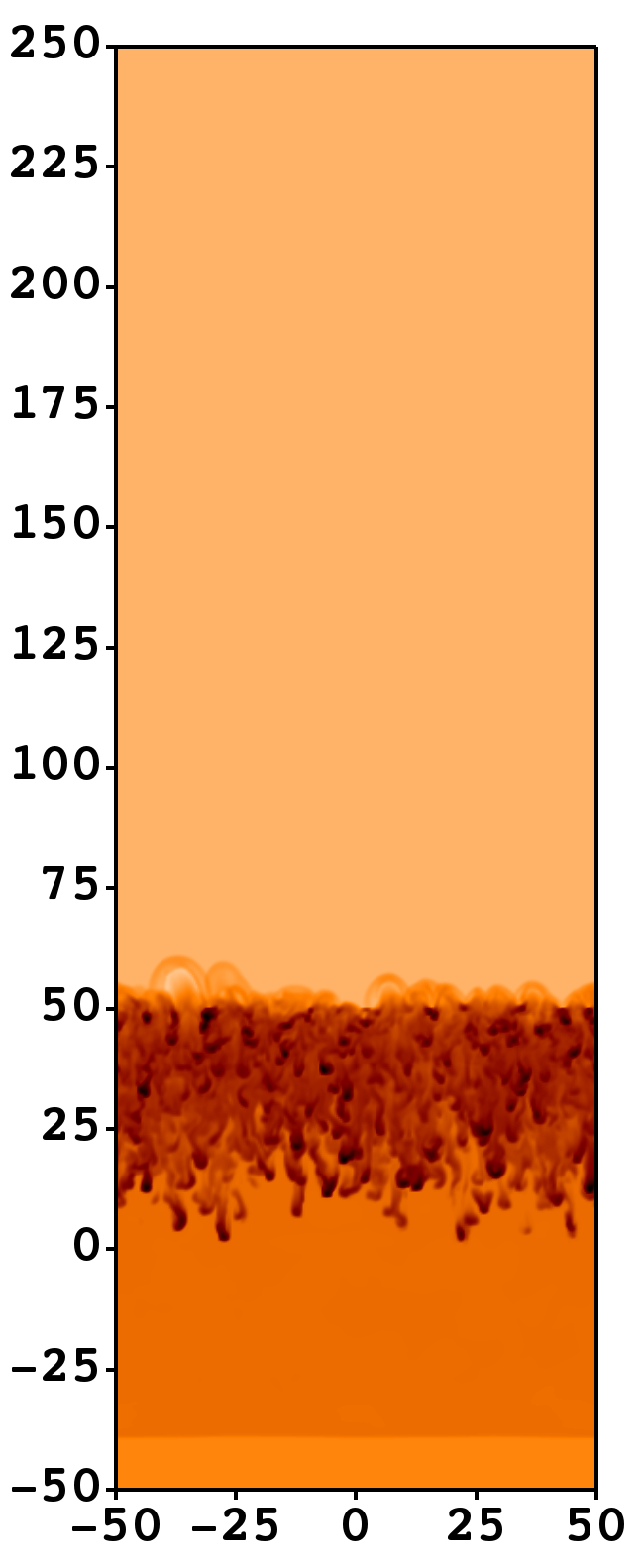}} & \hspace{-0.4cm}\resizebox{27mm}{!}{\includegraphics{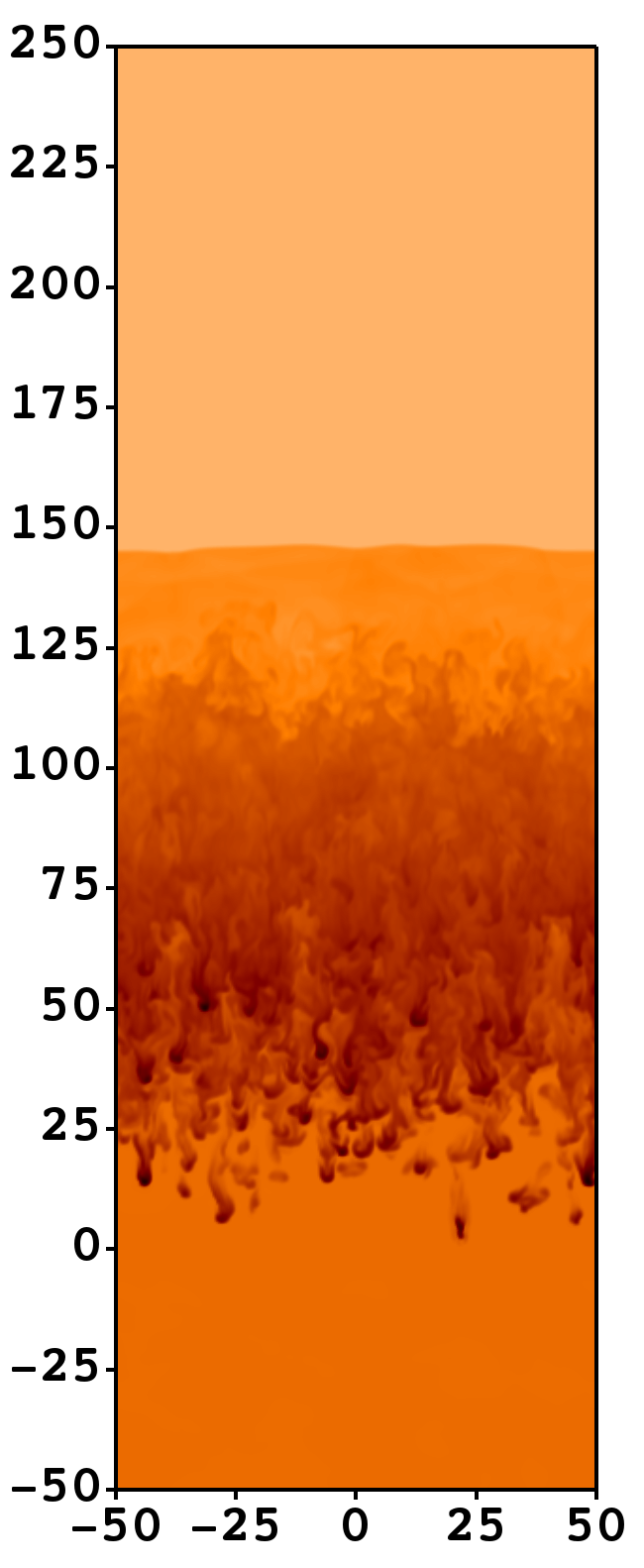}} & \hspace{-0.4cm}\resizebox{27mm}{!}{\includegraphics{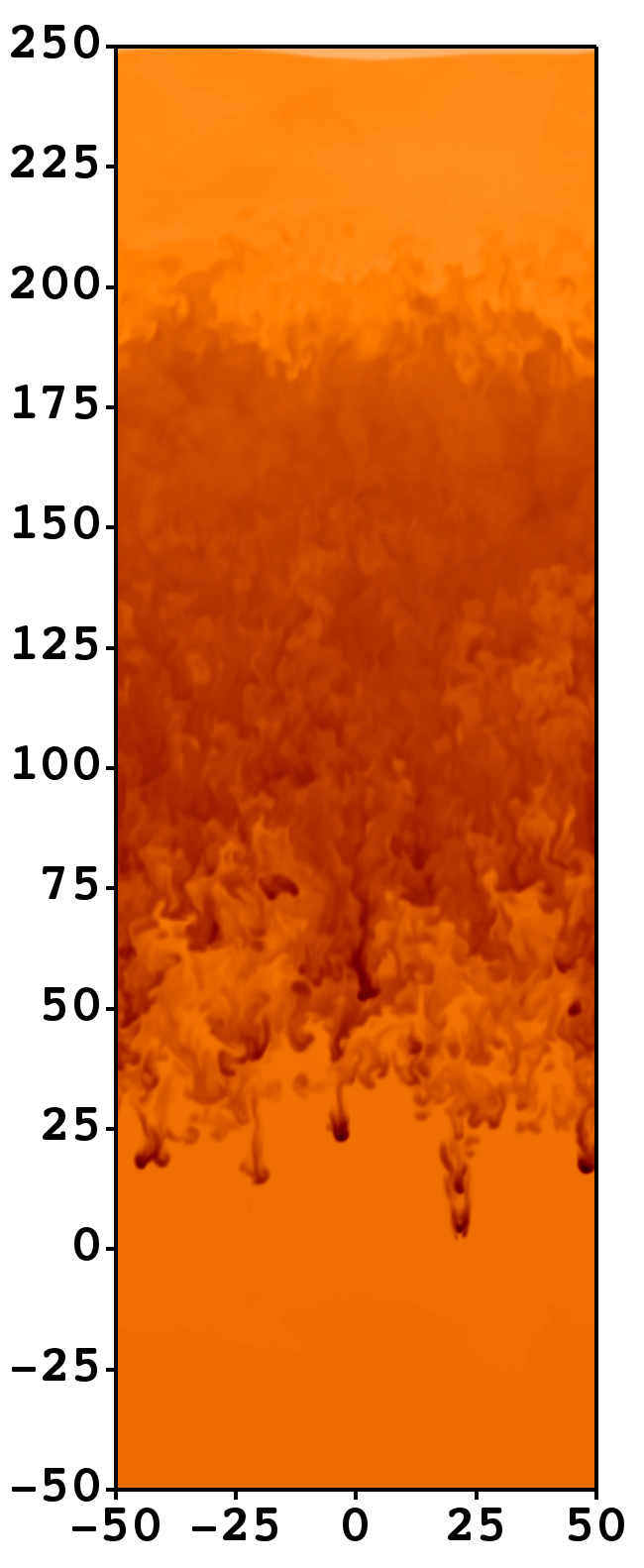}} & \hspace{-0.4cm}\resizebox{27mm}{!}{\includegraphics{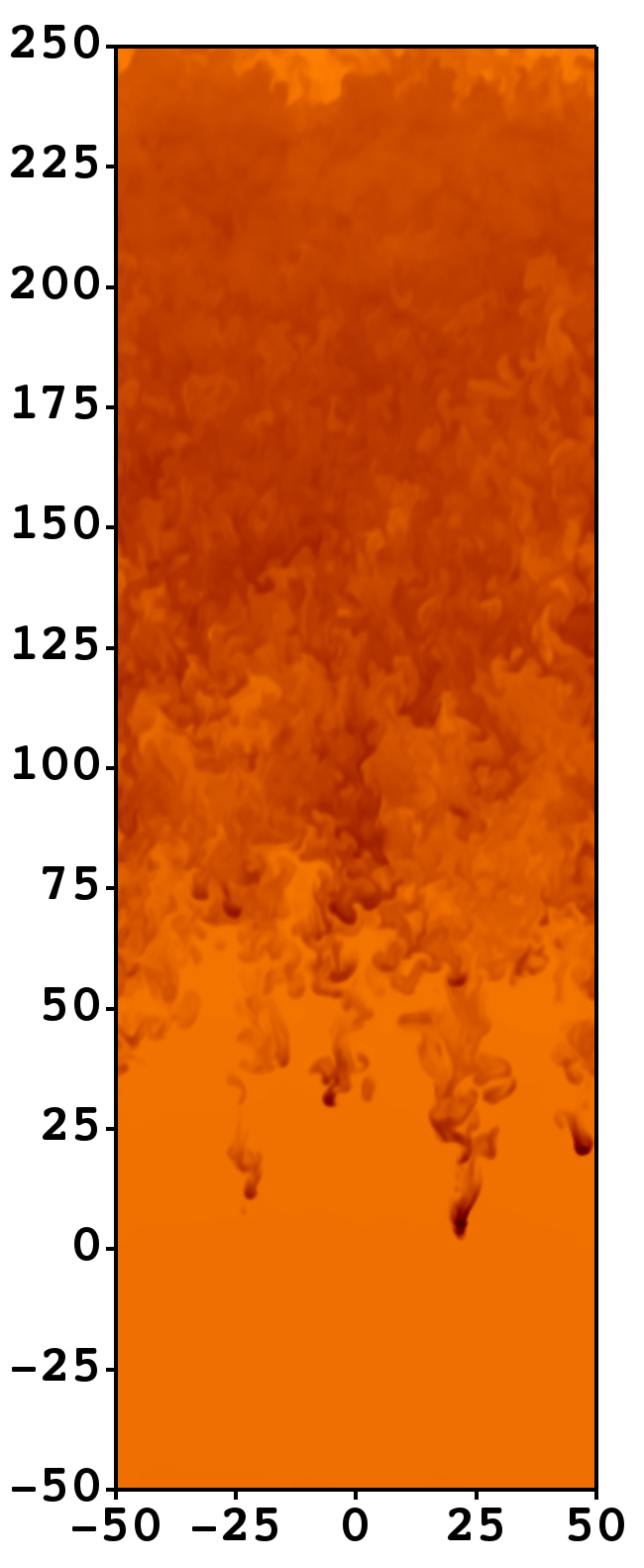}} & \hspace{-0.4cm}\resizebox{27mm}{!}{\includegraphics{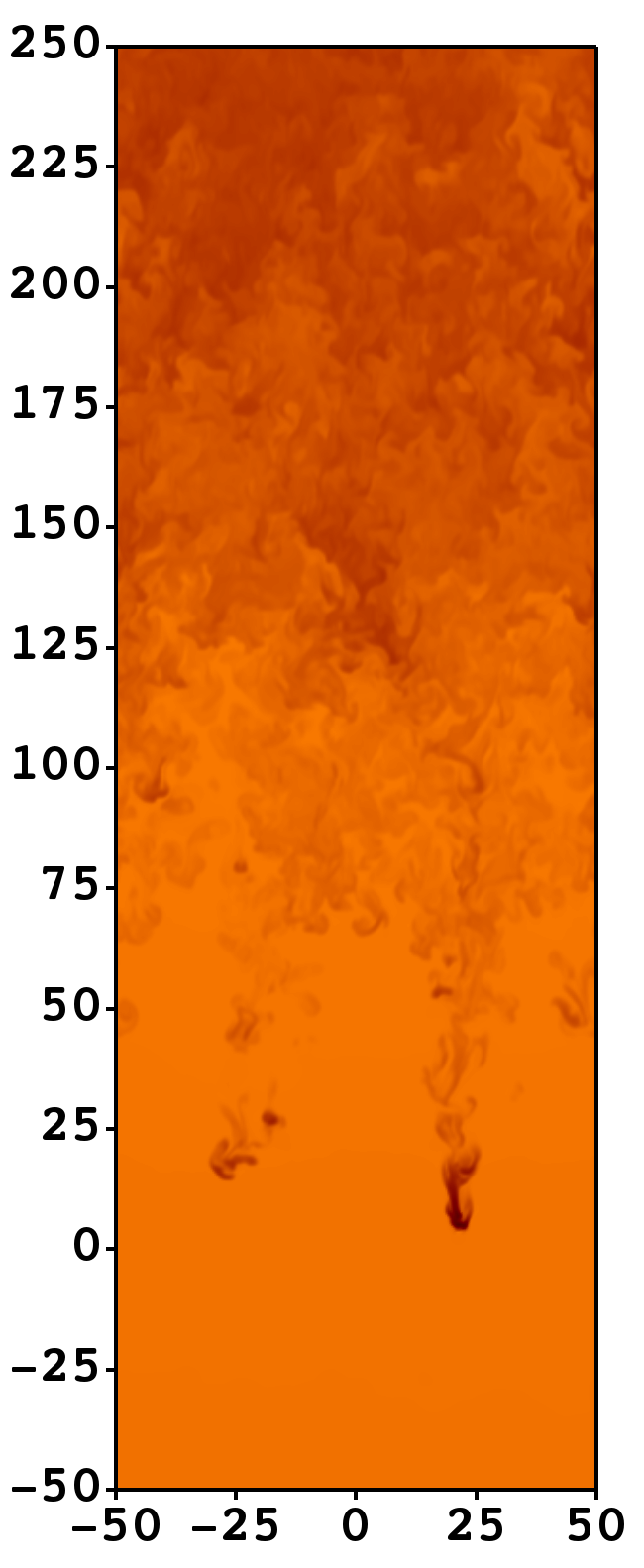}}&
\hspace{-0.2cm}\resizebox{12.8mm}{!}{\includegraphics{bar_vert_2.png}}\\
  \end{tabular}
  \caption{Same as Figure \ref{Figure2}, but here we show the number density slices in three compressive multicloud models, comp-k4-M10 (panel 3a), comp-k8-M10 (panel 3b), and comp-k16-M10, which correspond to fractal multicloud layers with normalised wavenumbers $k=4$, $k=8$, $k=16$, respectively, and the same shock Mach number, ${\cal M_{\rm shock}}=10$. The transmitted shock in compressive models travels faster across diffuse gas in the multicloud region than in solenoidal models. High-density gas in compressive models survives for longer time-scales than in solenoidal models. Within each sample, systems with higher wavenumbers slow down the internal shock and increase the stand-off distance of the reflected shock. The times correspond to $t/t_{\rm sp}\leq3.0$ (i.e., $t\leq0.6\,\rm Myr$), and the spatial ($X,Y$) extent is ($L\times3L$)$\equiv$($2\,L_{\rm mc}\times6\,L_{\rm mc}$), i.e., ($100\,\rm pc\times300\,\rm pc$).} 
  \label{Figure3}
\end{center}
\end{figure*}

\begin{figure}
\begin{center}
  \begin{tabular}{l}
    \hspace{0.55cm}4a) Forward and reverse shock Mach numbers\\
    \hspace{-0.4cm}\resizebox{80mm}{!}{\includegraphics{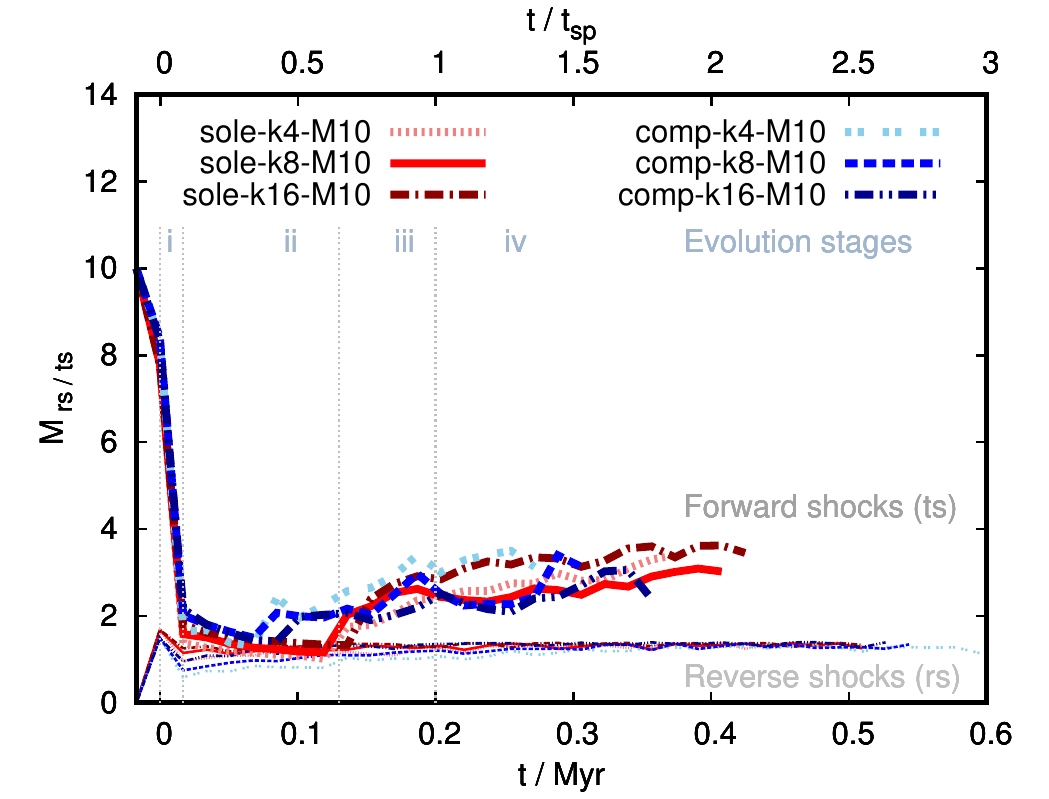}}\\
    \hspace{0.55cm}4b) Thermal pressure in the multicloud system\\
    \hspace{-0.40cm}\resizebox{80mm}{!}{\includegraphics{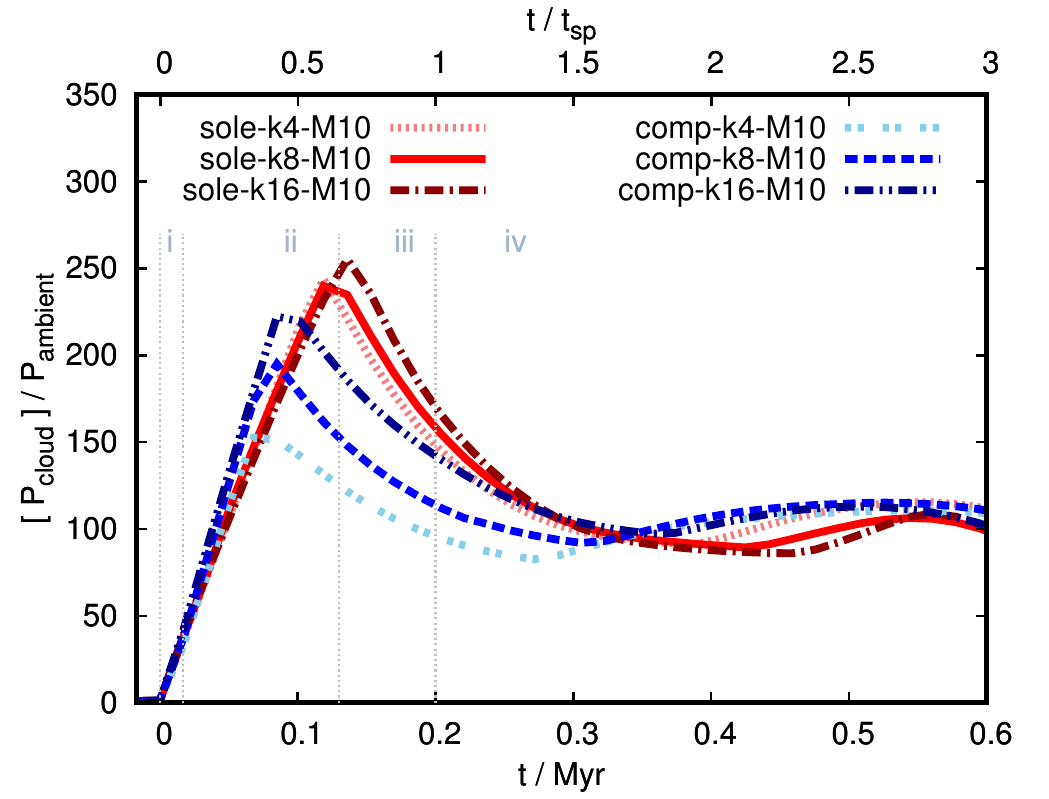}}\\
    \hspace{0.55cm}4c) Cloud volumetric filling factor\\
    \hspace{-0.4cm}\resizebox{80mm}{!}{\includegraphics{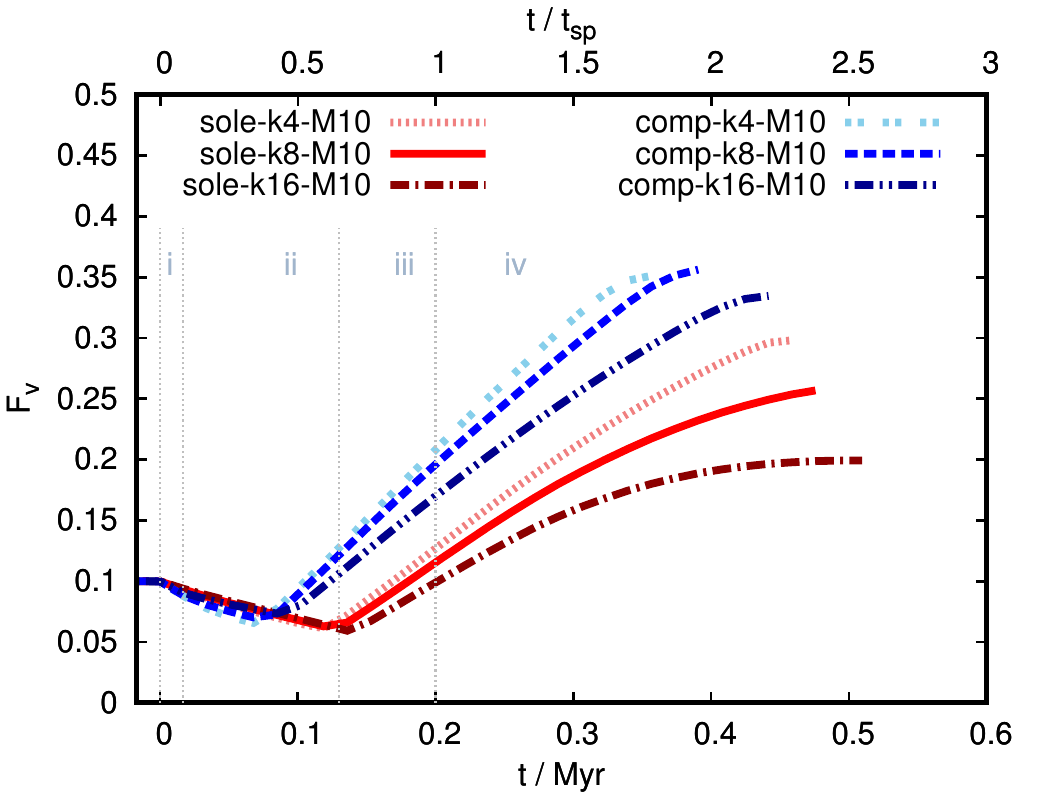}}\\
  \end{tabular}
  \caption{Time evolution of the Mach numbers of forward (thick lines) and reverse (thin lines) shocks (panel 4a), the thermal pressure in the multicloud layer (panel 4b), and the volumetric filling factor of cloud material in the computational domain (panel 4c). The transition between stages (i-iv) for solenoidal models are indicated by grey vertical lines in all panels. The transitions between stages (ii-iv) occur $\sim 0.2\,t_{\rm sp}$ earlier than what the grey lines indicate in compressive models (see Section \ref{subsec:Evolution}). The deceleration, steady crossing, and re-acceleration stages of the forward shock can be seen in the top panel. The compression and expansion phases of the evolution of the shock-multicloud systems are featured in the middle and bottom panels.} 
  \label{Figure4}
\end{center}
\end{figure}

\begin{enumerate}
\item \textbf{Initial contact, shock splitting, and first regime transition.} In the first stage ($t\in[0,0.1]\,t_{\rm sp}$), the impact of the shock on the multicloud systems triggers both reflected and refracted shocks. The reflected shock (reverse shock) propagates upstream while the refracted shock (transmitted forward shock) starts travelling downstream through the multicloud system. The non-uniform density fields in the fractal multicloud systems favour shock splitting (see also \citealt{2005ApJ...633..240P,2019MNRAS.486.4526B}), so the forward shock does not have a single speed, $v_{\rm ts}$, but rather a distribution of speeds around that theoretical value. Both the reflected shock and (especially) the refracted shock are non-planar, display a multi-shock substructure, and cause the post-shock gas to also travel at different speeds inside the multicloud region. The shock/post-shock speeds depend on how steep the local density gradients are. Panel 4a in Figure \ref{Figure4} shows the evolution of both shocks in solenoidal and compressive systems. In all cases, due to momentum conservation, the forward shock (initially hypersonic) decelerates and transitions into a milder supersonic regime, as soon as the initial contact occurs. This phase is short-lived because the Mach number of the transmitted shock quickly readjusts to $M_{\rm ts}\approx M_{\rm shock}\chi^{-0.5}(F_{\rm c1}F_{\rm st})^{\frac{1}{2}}\approx1.73$ (for $M_{\rm shock}=10$). It lasts only $\sim 0.1\,t_{\rm sp}$ ($\approx 0.02\,\rm Myr$ in our fiducial model), after the initial contact. This phase was also identified by \cite{2012MNRAS.425.2212A}, although in their models with multiple uniform clouds the deceleration phase was not as abrupt as in our models because their intercloud gas was more diffuse.
\item \textbf{Cloud layer compression, and shock steady crossing.} In the second stage ($t\in[0.1,0.6]\,t_{\rm sp}$ in solenoidal models, and $t\in[0.1,0.4]\,t_{\rm sp}$ in compressive models), the transmitted forward shock travels across the layer of clouds. While doing so, it compresses the cloud layer, heats up the cloud gas, and advects low-density, intercloud gas downstream. Individual cloudlets (cores within the multicloud system) expand laterally and start to gain momentum. These motions trigger collisions with other surrounding cloudlets and a shell of warm, mixed gas forms inside the multicloud system (also identified in \citealt{2012MNRAS.425.2212A}). The stream-wise length and compactness of the warm gas shell depends on the compactness (or porosity) of the layer as we explain in more detail below. The collisions between dense gas cloudlets re-shape the layer and facilitate the entrainment of gas that is initially placed downstream. In addition, short-wavelength KH instabilities start to grow as a result of vorticity being deposited at the interfaces between the intercloud medium and these cloudlets. Panel 4a in Figure \ref{Figure4} shows that the Mach number of the forward shock remains nearly constant during this phase with values of ${\cal M}_{\rm ts}\sim 1.5-2$. Panels 4b and 4c in Figure \ref{Figure4} show the evolution of the thermal pressure and the volumetric filling factor of cloud material in the computational domain, respectively. The thermal pressure increases by a factor of $\sim 5$ during this phase in all models, while the layer volume contracts by a factor of $\sim 1.7$, consistent with adiabatic compression.
\item \textbf{Cloudlet expansion, shock re-acceleration, and second regime transition.} In the third stage ($t\in[0.6,1.0]\,t_{\rm sp}$ in solenoidal models, and $t\in[0.4,0.8]\,t_{\rm sp}$ in compressive models), the internal forward shock reaches the rear side of the multicloud system, exits the layer of clouds, and enters the downstream ambient gas. The forward shock quickly re-accelerates due to momentum conservation as the ambient density is lower than the mean density in the multicloud system. Panel 4a in Figure \ref{Figure4} shows that the shock exits the multicloud systems at $t\sim 0.6\,\rm t_{sp}$ in solenoidal models and $t\sim 0.4\,\rm t_{sp}$ in compressive models. The forward shock evolves into a supersonic regime characterised by Mach numbers $\sim 2-3$ in all models, regaining a fraction of its initial speed. Low-density gas stripped from the multicloud layer also starts to leave the multicloud region as it comoves with the post-shock flow behind the shock front. Re-acceleration and low-density gas deposition downstream initiates the stream-wise expansion of the multicloud region by promoting the vertical expansion of individual cloudlets whose tails move at higher speeds than their cores. As cloudlets are stretched, filamentary tails emerge and populate the post-shock flow. High-density gas continues mixing with the post-shock flow and some dense cloudlets also gain momentum and leave their original positions at the end of this stage, i.e., $t\sim 1.0\,\rm t_{sp}$ and $t\sim 0.8\,\rm t_{sp}$ for solenoidal and compressive models, respectively. Their relative speed varies in different models. Some cloudlets also merge with others to form a coherent two-phase gas layer composed of dense and diffuse gas. This layer is dominated by vorticity deposited by KH instabilities. Panels 4b and 4c in Figure \ref{Figure4} also show that during this stage there is a factor of $\sim 1.5$ drop in thermal pressure and a factor of $\sim 3$ increase in the volume occupied by cloud material.
\item \textbf{Cloud mixing, turbulence emergence, and comoving post-shock flow.} In the fourth stage of the interaction ($t>1\,t_{\rm sp}$ in solenoidal models and $t>0.8\,t_{\rm sp}$ in compressive models), the forward shock keeps accelerating slowly and eventually moves out of the computational domain. In our simulations, the forward shock never regains its initial velocity as in the multiple cloud models by \cite{2012MNRAS.425.2212A}, but the trend of the curves in panel 4a of Figure \ref{Figure4} suggests that it might reach higher Mach numbers (${\cal M}_{\rm ts}>4$) much further ahead in the downstream flow. The post-shock flow continues to advect mixed gas, injecting momentum into the denser regions of the two-phase filamentary system. The merging of dense cloudlets in the multicloud system also continues, and some dense filaments lose their coherence as a result of shear and RT instabilities. The morphology of the shell is different in solenoidal and compressive models in this phase. While individual cloudlets have lost all coherence in solenoidal models, we find that some cloudlets with a low momentum do survive in compressive models. The emergence of long-wavelength KH instabilities activate a fully turbulent regime, in which warm gas in the shell acquires speeds between $0.6$ and $0.8$ of the post-shock flow. Panel 4b in Figure \ref{Figure4} shows that the thermal pressure in the multicloud gas remains nearly constant after the onset of turbulence (for $t\geq 1.2\,\rm t_{\rm sp}$, i.e., $t\geq 0.25\,\rm Myr$ in our fiducial model), while panel 4c in Figure \ref{Figure4} indicates that the muticloud region continues expanding in the direction of streaming reaching filling factors of $\sim 20-35$ per cent by the time the forward shock leaves the domain.
\end{enumerate}

\subsection{Solenoidal versus compressive multicloud systems}
\label{subsec:SolvsComp}
In agreement with our recent study on single, isolated wind-swept clouds (see \citealt{2019MNRAS.486.4526B}), we find that the dynamics and longevity of individual cloudlets inside fractal, multi-cloud layers overrun by shocks depend on the initial density fields we choose for the cloud layers. In this section, we discuss the differences in the evolution of fractal multicloud systems whose initial density fields are characteristic of two regimes of supersonic turbulence, solenoidal and compressive. Solenoidal multicloud layers have narrow density distributions with low PDF standard deviations ($\sigma_{\rm cloud}=1.9$), while compressive multicloud layers have wide distributions with high PDF standard deviations ($\sigma_{\rm cloud}=5.9$). This difference implies that compressive density fields are more porous as they have higher density cores and larger low-density voids than their solenoidal counterparts, so the growth of KH instabilities, responsible for mixing and turbulence generation, differs in both model samples.\par

Figures \ref{Figure2} and \ref{Figure3} show that the shock can travel more easily across the low-density gas of compressive models, so it reaches the rear side of the multicloud system and the upper side of the computational domain earlier than in solenoidal models. For this reason, the post-shock flow in compressive models is also more efficient in transporting low-density gas downstream than in solenoidal models. On the other hand, the high-density cores in compressive multicloud systems have higher column densities, so they are much harder to disrupt and accelerate than the cores in solenoidal systems. The rightmost panels in Figures \ref{Figure2} and \ref{Figure3} show that a few dense gas cores are able to survive, embedded in the post-shock flow, in compressive multicloud models, while such clumps are totally absent in solenoidal models. These panels also show that solenoidal cloud layers are faster and reach larger distances than their compressive counterparts in $1\,t_{\rm sim}$.\par

Although panel 4a in Figure \ref{Figure4} indicates that the time-dependent propagation of both forward and reverse shocks is very similar in all models, with forward shocks decelerating, crossing, and re-accelerating, and reverse shocks reaching ${\cal M}_{\rm rs}\sim1.2$, the positions of the shock fronts in solenoidal and compressive models differs. For instance, at $t=0.5\,t_{\rm sp}=0.10\,\rm Myr$ the forward shock in solenoidal models is at distances between $y\approx 0.3\,L=+30\,\rm pc$ and $y\approx0.4\,L=+40\,\rm pc$ inside the multicloud layer, while in compressive models it is at distances between $y\approx 0.5\,L=+50\,\rm pc$ and $y\approx0.7\,L=+70\,\rm pc$, i.e., it has exited the multicloud system. Similarly, the stand-off distance of the reverse shock at $t=0.5\,t_{\rm sp}=0.10\,\rm Myr$ also differs. In solenoidal models it is farther upstream (between $y\approx-0.48\,L=-48\,\rm pc$ and $y\approx-0.40\,L=-40\,\rm pc$) than in compressive models where the stand-off distance is between $y\approx-0.35\,L=-35\,\rm pc$ and $y\approx-0.25\,L=-25\,\rm pc$ (see the second column of panels in Figures \ref{Figure2} and \ref{Figure3}).\par

The reason for this behaviour is that solenoidal density fields are more compact (and uniform) and less porous than compressive density fields, so they can more effectively act as a barrier for the upcoming forward shock. Similarly, we find that systems with higher initial wavenumbers result in larger stand-off distances for the reverse shock. This is because high-density cores in high-$k$ models can also "block" the upcoming shock, which is then reflected further away than in low-$k$ models.\par

\subsubsection{Evolution of the density PDFs}
\label{subsubsec:DensityPDF}
\begin{figure}
\begin{center}
  \begin{tabular}{l}
    \hspace{0.55cm}5a) $t=0$\vspace{-0.4cm}\\
    \hspace{-0.40cm}\resizebox{80mm}{!}{\includegraphics{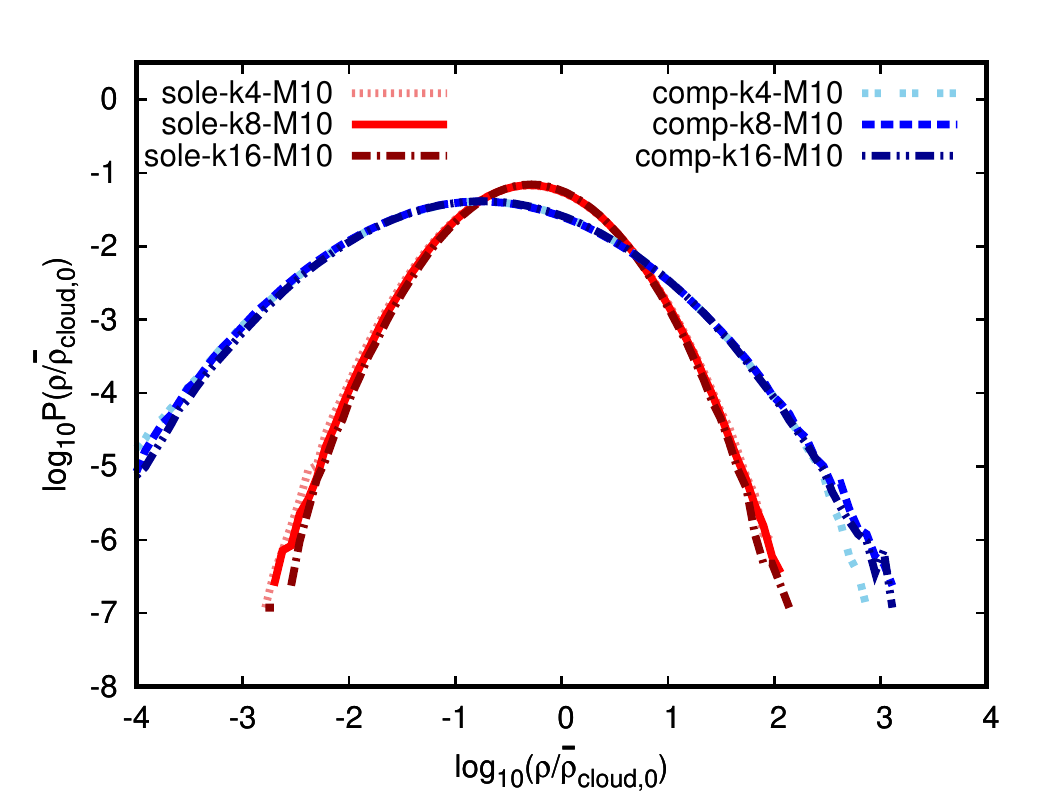}}\\
    \hspace{0.55cm}5b) $t=1.1\,t_{\rm sp}=0.22\,\rm Myr$\vspace{-0.4cm}\\
    \hspace{-0.4cm}\resizebox{80mm}{!}{\includegraphics{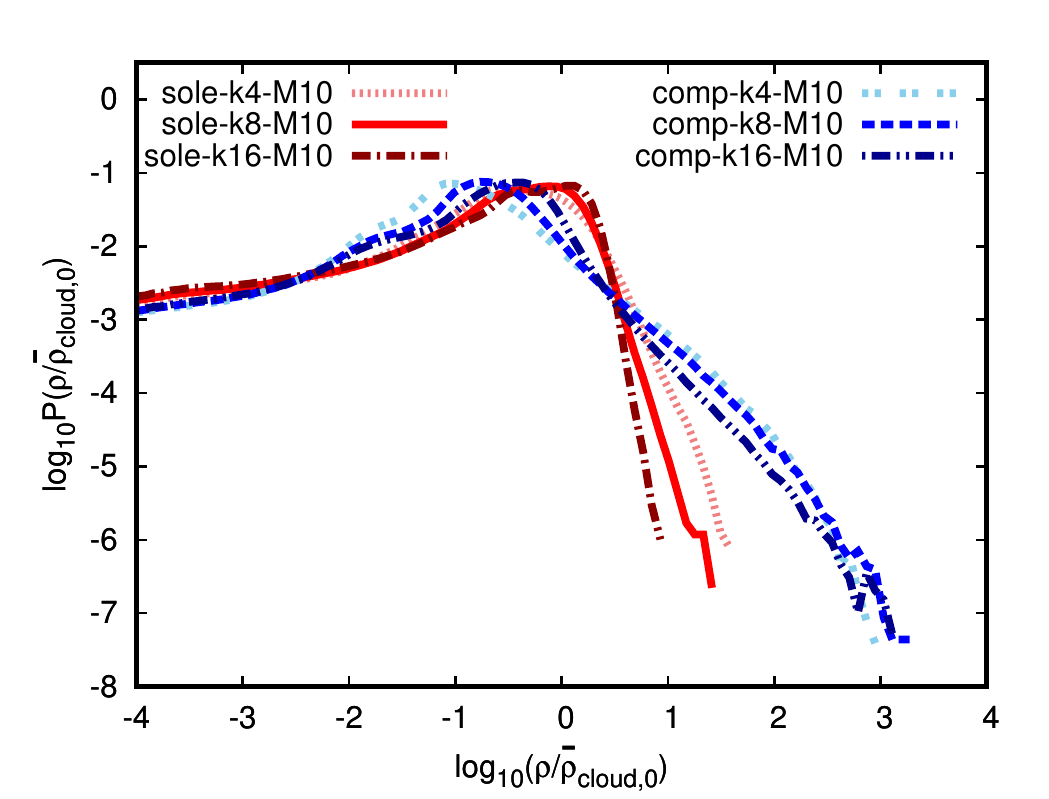}}\\
    \hspace{0.55cm}5c) $t=3.0\,t_{\rm sp}=0.60\,\rm Myr$\vspace{-0.4cm}\\
    \hspace{-0.4cm}\resizebox{80mm}{!}{\includegraphics{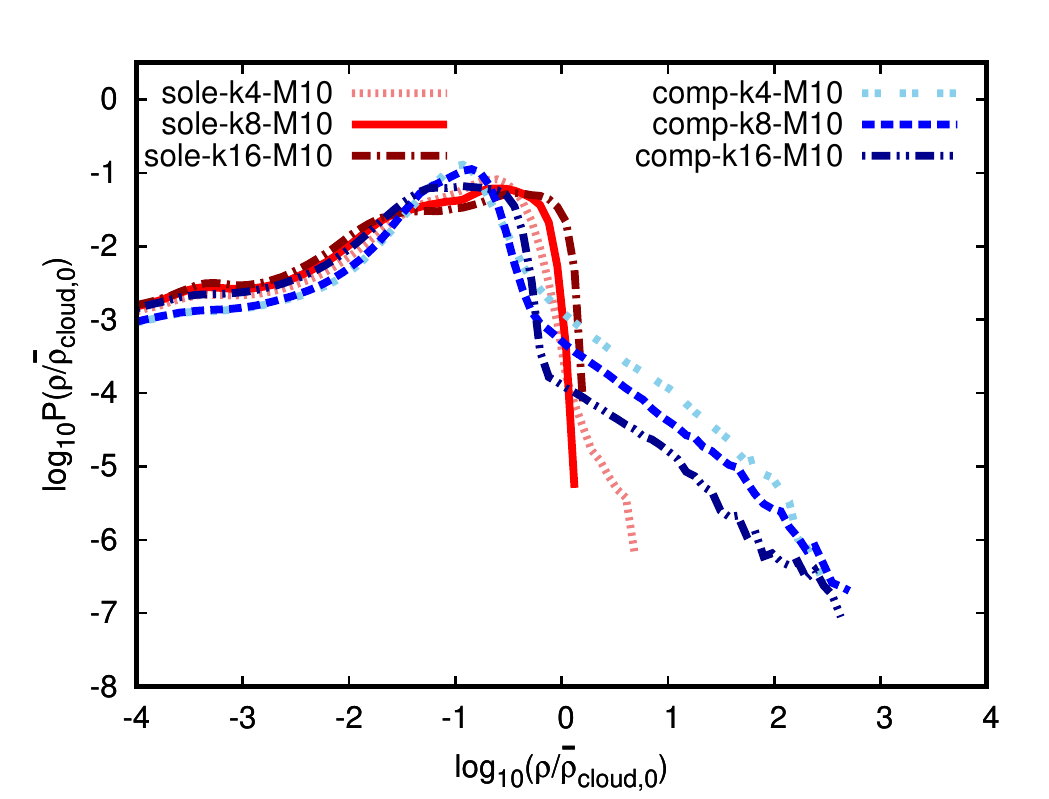}}\\
  \end{tabular}
  \caption{Volume-weighted PDFs of the logarithm of the cloud layer density. The cloud densities are normalised with respect to the initial mean density in the multicloud layer, at three different times: $t=0$ (panel 5a), $t=1.1\,t_{\rm sp}=0.22\,\rm Myr$ (panel 5b), and $t=3.0\,t_{\rm sp}=0.60\,\rm Myr$ (panel 5c). Compressive clouds retain high-density cloudlets/cores until late times of the evolution, regardless of their initial normalised wavenumber.} 
  \label{Figure5}
\end{center}
\end{figure}

The prevalence of high-density cores in the compressive multicloud models can also be seen in Figure \ref{Figure5}, where we compare the density PDFs of solenoidal and compressive models at three different times, $t=0$, $t=1.1\,t_{\rm sp}=0.22\,\rm Myr$, and $t=3.0\,t_{\rm sp}=0.60\,\rm Myr$. The low-density tails of the PDFs evolve similarly in all models. They become flat as time progresses. The high-density tails, on the other hand, show a different behaviour in both regimes. While in solenoidal models they slowly and steadily move towards low-density values, in compressive models they maintain their elongated shape indicating that cores $\sim 10^2-10^3$ times denser than the initial cloud mean density are able to survive until late times. This result is universal to all compressive models regardless of the initial normalised wavenumber assigned to individual multicloud layers.\par

In addition, in solenoidal models most of the cloud gas has densities below the initial mean density at the end of the evolution, which implies that very dense gas has been effectively disrupted and has become part of the mixed-gas shell. On the other hand, in compressive models the PDF develops a sharp peak at low densities and a heavy tail at high densities, which correspond to the wakes (of mixed turbulent gas) and the cores (of dense long-lived gas), respectively, of the filaments seen in Figure \ref{Figure3}. Although radiative cooling and magnetic fields would also influence the density PDFs, this result hints that studying the shapes of the density PDFs of outflowing material in galactic winds might potentially tell us how solenoidal or compressive the cloud population is at the base of the outflows.\par

We also note that the evolution of the low-density tails of the density PDFs in single- and multi-cloud systems is different (c.f., Figure~\ref{Figure5} in this paper with the panels in Figures 3 and 5 in \citealt{2019MNRAS.486.4526B}). In single-cloud systems the low-density mixed gas (i.e., $\rho/\bar{\rho}_{\rm cloud,0}<10^{-2}$) completely dominates the density PDFs, while in multicloud systems the intermediate-density gas (i.e., $10^{-2}<\rho/\bar{\rho}_{\rm cloud,0}<10$) contains most of the cloud mass. This implies that mixing processes in multicloud systems are not only regulated by mass stripping and KH instabilities at the interfaces between cloudlets and the external wind / post-shock flow (i.e., at shear layers), but also by cloud shielding, cloudlet-cloudlet collisions, and cloudlet-intercloudlet vorticity production. Solenoidal cloud layers are more compact and uniform, so cloud mergers occur more efficiently. For the same reason, the relative cloudlet-intercloudlet speeds are lower in solenoidal models than in compressive models. 

\subsubsection{Cloud mixing and turbulence}
\label{subsubsec:Cloud mixing}
Turbulence in these models is generated as a result of dynamical instabilities occurring both at the sides and the front ends of cloudlets inside the multicloud medium. The sides of the cloudlets interact with the fast moving shock and post-shock flow. The difference in velocities causes the local density and pressure gradients to be misaligned and that generates vorticity and KH instabilities (e.g., see \citealt{2006ApJS..164..477N}). The unstable shear layer strips mass from the cloudlets and this leads to mixing between cloudlet, intercloudlet, and ambient gas. In our shock-multicloud models, the growth time-scale of KH instabilities with wavelengths comparable to $r_{\rm cloudlet}$ is $t_{\rm KH}\approx r_{\rm cloudlet}\,\chi_{\rm eff}^{0.5}/(2\pi v_{\rm psh})\lesssim0.2\,t_{\rm cc}$. Similarly, the front ends of the cloudlets are also exposed to RT instabilities, which arise when the post-shock flow pushes through the denser cores of the cloudlets (e.g., see \citealt{2016MNRAS.457.4470P}). The growth time-scale of RT instabilities with wavelengths comparable to $r_{\rm cloudlet}$ is $t_{\rm RT}\approx (r_{\rm cloudlet}/(2\pi a_{\rm eff}))^{0.5}\lesssim0.8\,t_{\rm cc}$, where $a_{\rm eff}\approx 0.4v_{\rm psh}^2/(\chi r_{\rm cloudlet})$ is the effective cloud acceleration. Thus, mixing is regulated by the cloudlet sizes, their effective density contrasts, and the post-shock flow speed. The reader is referred to \cite{2019MNRAS.486.4526B} for a full description on KH and RT instabilities acting on single-cloud systems interacting with supersonic flows.\par

\begin{figure}
\begin{center}
  \begin{tabular}{l}
    \hspace{0.55cm}6a) Mixing fraction\\
    \hspace{-0.40cm}\resizebox{80mm}{!}{\includegraphics{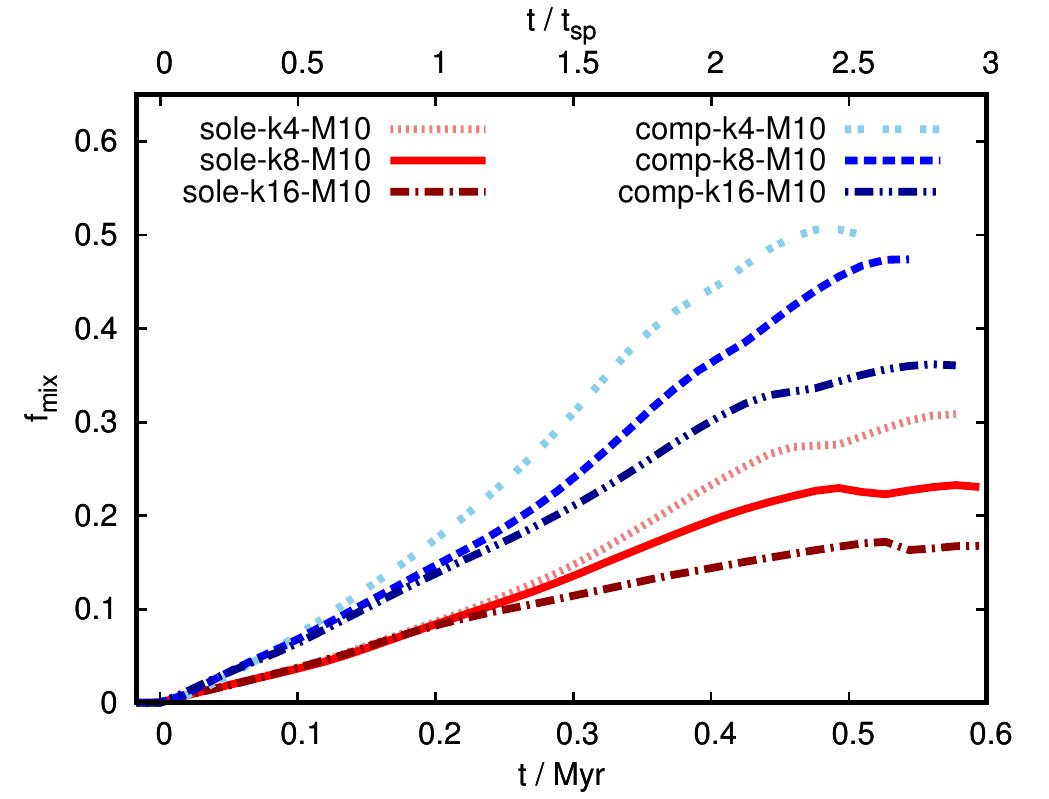}}\\
    \hspace{0.55cm}6b) Velocity dispersion\\
    \hspace{-0.4cm}\resizebox{80mm}{!}{\includegraphics{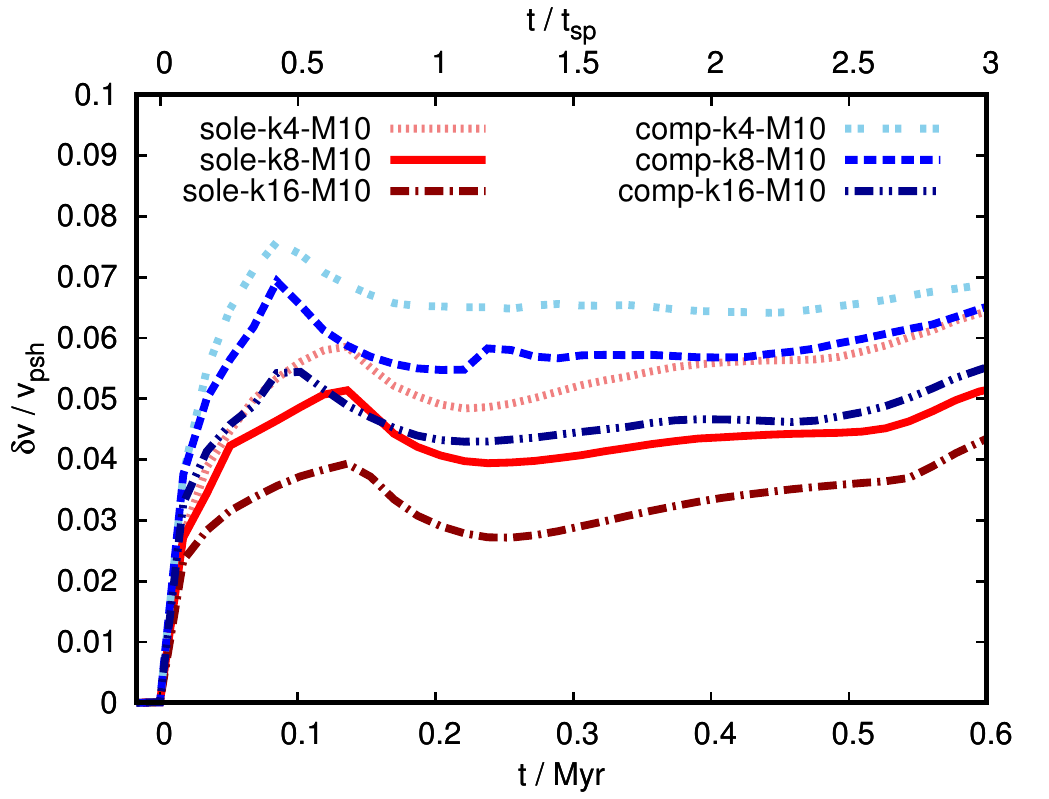}}\\
  \end{tabular}
  \caption{Time evolution of the mixing fraction (panel 6a) and the velocity dispersion of cloud material (panel 6b), normalised to the post-shock flow speed, $v_{\rm psh}\approx 0.75\,v_{\rm shock}$ ($=1080\,\rm km\,s^{-1}$ in our fiducial normalisation), in solenoidal and compressive multicloud models. Both, mixing fractions and velocity dispersions, are higher in compressive models than in their respective solenoidal counterparts. The velocity dispersion curves peak at the time when the shock leaves the multicloud regions, which occurs earlier in compressive models than in solenoidal models. After the end of the re-acceleration stage the velocity dispersion remains nearly constant with only a mild positive slope in all models.}
  \label{Figure6}
\end{center}
\end{figure}

Figure \ref{Figure6} shows the evolution of two parameters, the mixing fraction (panel 6a) and the transverse velocity dispersion (panel 6b) of cloud gas in solenoidal and compressive multicloud systems. The mixing fractions and transverse velocity dispersions in compressive models are in general $\sim1.5$ -- $2.5$ times higher than in solenoidal models. As explained above, this is different to what we found in single-cloud systems with $\chi=10^3$ (see \citealt{2019MNRAS.486.4526B}), where single solenoidal clouds showed more mixing than their compressive counterparts. We attribute this difference to: a) the faster growth of KH instabilities in compressive multicloud models due to higher relative cloud-intercloud gas speeds compared to solenoidal models, b) cloud-cloud interactions (e.g., cloud shielding and cloud-cloud collisions), which are responsible for the formation of a more coherent shell of medium-density warm gas in solenoidal models; and c) the lower density contrasts of $\chi=10^2$ of the models we study in this paper, for which the mixing of medium-density (rather than low-density) gas dominates.\par

The panels in Figure \ref{Figure6} also show that mixing and turbulence generation not only depend on the type of initial fractal density field, but also on the initial normalised wavenumber, $k$, of this field. In other words, the turbulent properties of the post-shock flow and the warm gas shell depend on the number of individual cloudlets/voids in the initial systems. Within each cloud sample, models with more cloudlets (i.e. with higher $k$) display lower mixing fractions and velocity dispersions than models with less clouds (i.e. with lower $k$). For example, at $t=2.5\,t_{\rm sp}=0.50\,\rm Myr$, mixing fractions are $f_{\rm mix}=0.27$ for $k=4$, $f_{\rm mix}=0.22$ for $k=8$, and $f_{\rm mix}=0.17$ for $k=16$ in solenoidal models, and $f_{\rm mix}=0.51$ for $k=4$, $f_{\rm mix}=0.45$ for $k=8$, and $f_{\rm mix}=0.35$ for $k=16$ in compressive models. This implies that upstream clouds can more effectively shield downstream clouds in systems with higher numbers of clouds, and that KH instabilities take longer to develop as the forward shock cannot preclude as readily as in systems with larger voids.\par

The slope of the velocity dispersion curves shown in panel 6b of Figure \ref{Figure6} indicates that the initial contact of the forward shock with the multicloud layer rapidly induces turbulence with internal velocity dispersions increasing to $\sim 4$ per cent of the ambient post-shock flow speed by the end of this phase. During the compression stage, the slope changes and becomes less steep, but turbulent velocities continue rising up owing to the kinetic energy being injected by the forward shock. The turbulence increase stops at the point when the forward shock reaches the rear side of the multicloud system. Typical peak values in normalised and physical units are $\delta_{\rm v}\approx 0.06\,v_{\rm psh}=65\,\rm km\,s^{-1}$ for $k=4$, $\delta_{\rm v}\approx 0.05\,v_{\rm psh}=54\,\rm km\,s^{-1}$ for $k=8$, and $\delta_{\rm v}\approx 0.035\,v_{\rm psh}=38\,\rm km\,s^{-1}$ for $k=16$ in solenoidal models, and $\delta_{\rm v}\approx 0.075\,v_{\rm psh}=81\,\rm km\,s^{-1}$ for $k=4$, $\delta_{\rm v}\approx 0.070\,v_{\rm psh}=76\,\rm km\,s^{-1}$ for $k=8$, and $\delta_{\rm v}\approx 0.055\,v_{\rm psh}=59\,\rm km\,s^{-1}$ for $k=16$ in compressive models. When the shock-multicloud systems transit through the shock re-acceleration phase, velocity dispersions decrease $-0.01\,v_{\rm psh}=-11\,\rm km\,s^{-1}$ from the respective peak values, and remain nearly constant during the fourth stage of the evolution. The mild positive slope present in all models indicates that vorticity deposited by dynamical instabilities dominates turbulence production during this phase.

\subsubsection{Cloud acceleration}
\label{subsec:Acceleration}
The dynamics of cloud layers overrun by shocks depends on how effective the momentum transfer from the shock to individual cloudlets in the layer is. In multicloud systems, pressure gradient forces arising at the leading edges of shock-swept cloudlets and cloud-cloud collisions (if they occur) contribute to momentum transfer. Upstream cloudlets are accelerated by the post-shock flow first, so they can collide with downstream cloudlets and contribute to their acceleration as they move in the direction of streaming. The interaction between cloudlets in a multicloud system is more effective when the systems are more compact and less hollow, so solenoidal clouds can gain more momentum than compressive models.\par

\begin{figure}
\begin{center}
  \begin{tabular}{l}
    \hspace{0.55cm}7a) Mass-weighted velocity\\ 
    \hspace{-0.4cm}\resizebox{80mm}{!}{\includegraphics{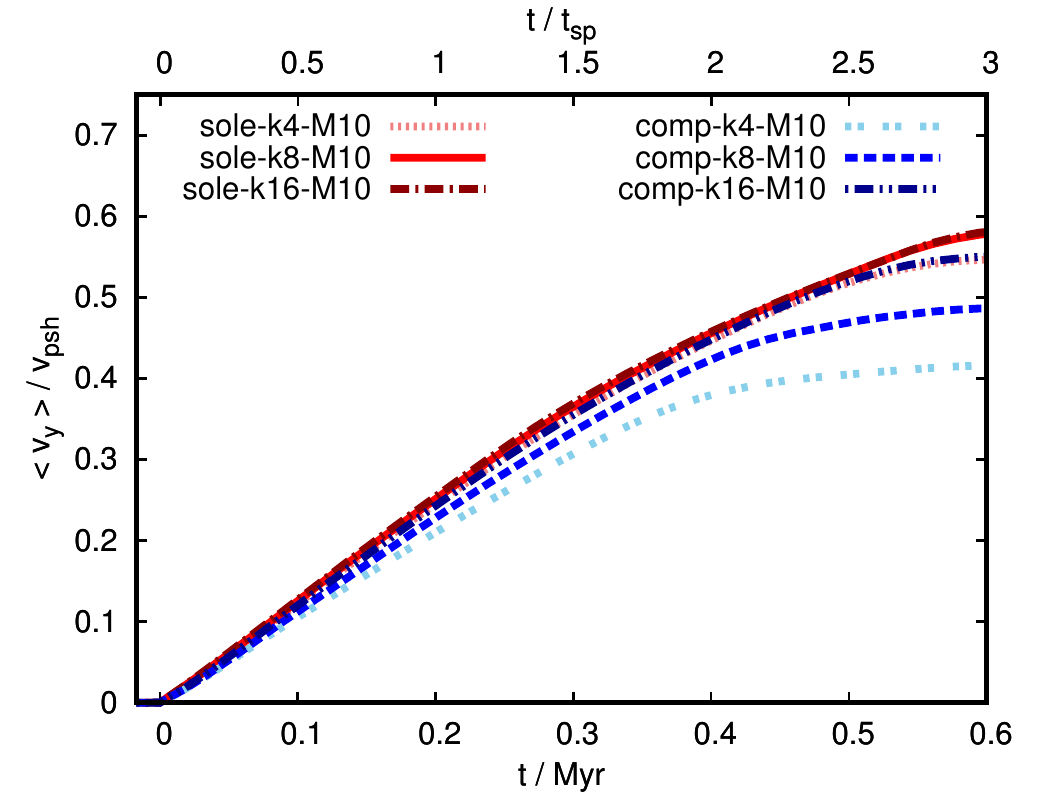}}\\
    \hspace{0.55cm}7b) Travelled distance\\
    \hspace{-0.40cm}\resizebox{80mm}{!}{\includegraphics{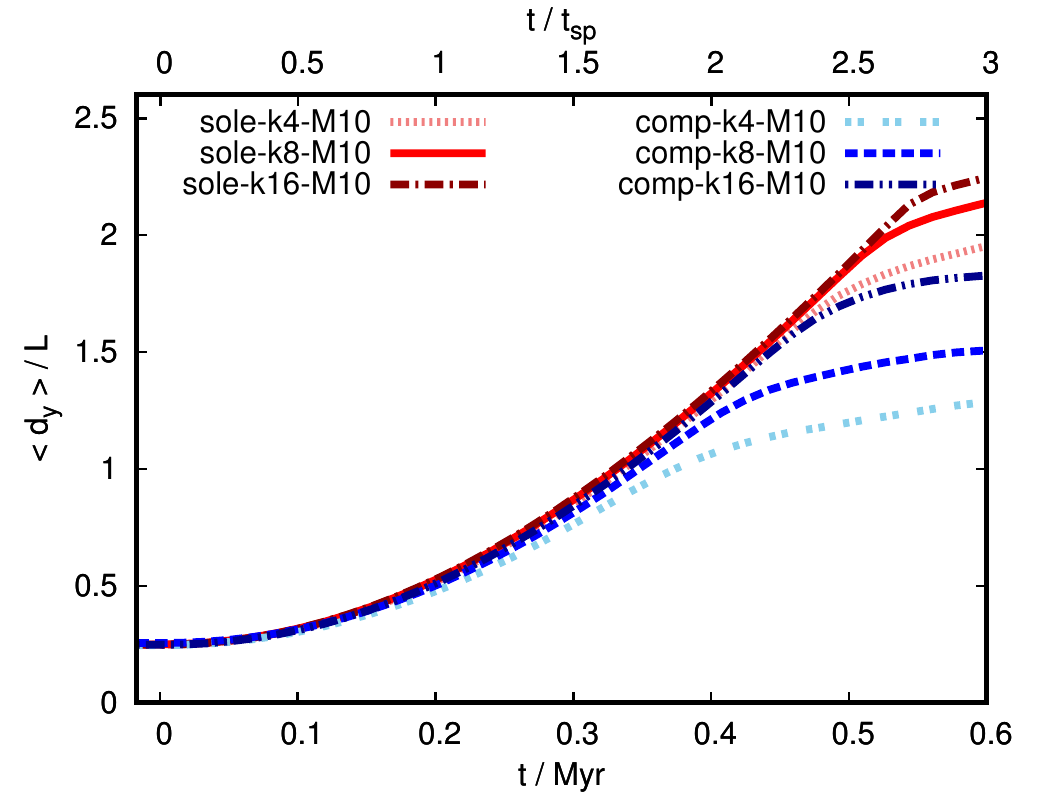}}\\
    \hspace{0.55cm}7c) Mass loss\\
    \hspace{-0.40cm}\resizebox{80mm}{!}{\includegraphics{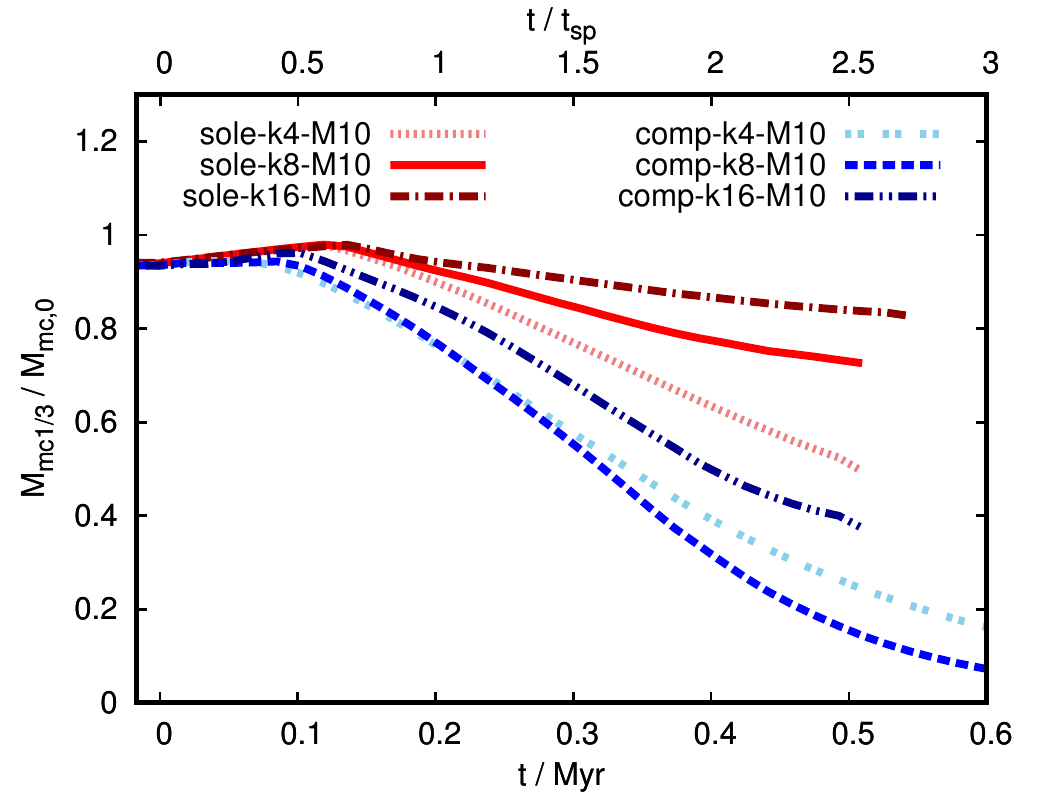}}\\
  \end{tabular}
  \caption{Mass-weighted bulk speed normalised to $v_{\rm psh}$ (panel 7a), distance travelled by the centre of mass normalised to $L$ (panel 7b), and fraction of cloud mass above $\bar{\rho}_{\rm cloud, 0}/3$ of the multicloud layer as a function of time. In general, compressive models are slower than solenoidal models. The dynamics in solenoidal and particularly in compressive models is also sensitive to the initial normalised wavenumber: lower $k$ values reduce momentum transfer from the post-shock flow to the cloud layer. Although individual cloudlets only survive in compressive models, solenoidal cloud layers have higher mass fractions above $\bar{\rho}_{\rm cloud, 0}/3$ than compressive models at all times. Higher $k$ values favour cloud shielding, so in general layers with more cloudlets retain more high-density gas.} 
  \label{Figure7}
\end{center}
\end{figure}

Figure \ref{Figure7} confirms this behaviour and shows that cloud gas in solenoidal models is faster (particularly at late times) and reaches larger distances than in compressive models (which can also be seen in Figures \ref{Figure2} and \ref{Figure3}). By the end of the simulation, solenoidal clouds acquire mass-weighted bulk speeds of $\langle~v_{\rm y}~\rangle\approx0.6\,\rm v_{psh}=648\,\rm km\,s^{-1}$, while compressive clouds are in general slower, reaching speeds of $\langle~v_{\rm y}~\rangle\gtrsim0.4\,\rm v_{psh}=432\,\rm km\,s^{-1}$. Similarly, solenoidal clouds reach distances $\langle~d_{\rm y}~\rangle\gtrsim 1.9\,L=190\,\rm pc$, while compressive clouds are at least 10 per cent behind, reaching distances between $1.3\,L=130\,\rm pc\lesssim\langle~d_{\rm y}~\rangle\lesssim1.8\,L=180\,\rm pc$. This result is in agreement with our earlier study on wind-cloud systems (see Sections 3.3 and 3.6 of \citealt{2019MNRAS.486.4526B}).\par

The dynamics of the multicloud layer also depends on the initial normalised wavenumber, i.e., on the number of cloudlets originally in the cloud layer. Multicloud systems with higher numbers of clouds (i.e., with higher $k$) can gain more momentum than systems with a smaller number of clouds, owing to the contribution from cloudlet-cloudlet collisions. This effect can be best viewed in panel 7b of Figure \ref{Figure7}, which shows that at $t=3.0\,t_{\rm sp}=0.60\,\rm Myr$, travelled distances are $\langle~d_{\rm y}~\rangle\approx 1.9\,L=190\,\rm pc$ for $k=4$, $\langle~d_{\rm y}~\rangle\approx 2.1\,L=210\,\rm pc$ for $k=8$, and $\langle~d_{\rm y}~\rangle\approx 2.3\,L=230\,\rm pc$ for $k=16$ in solenoidal models, and $\langle~d_{\rm y}~\rangle\approx1.3\,L=130\,\rm pc$ for $k=4$, $\langle~d_{\rm y}~\rangle\approx1.5\,L=150\,\rm pc$ for $k=8$, and $\langle~d_{\rm y}\rangle\approx1.8\,L=180\,\rm pc$ for $k=16$ in compressive models.

\subsubsection{Mass loading and dense gas entrainment}
\label{subsec:MassLoading}
Mass loading and dense gas entrainment are important processes for galactic outflows. Both are associated with the disruption of clouds near the base of outflows as wind/shock-cloud interactions can: a) mass load the outflow with warm gas that will later condense back into a cold component via thermal instabilities (e.g., \citealt{2016MNRAS.455.1830T,2018MNRAS.480L.111G}), and b) provide dense gas material directly via entrainment (e.g., \citealt{2015MNRAS.449....2M,2018MNRAS.473.3454B}), understood as the process by which cold gas is advected along the outflowing hot component. To understand how these processes take place in shock-multicloud systems, we quantify the amount of dense gas in the cloud layer at all times.\par

Panel 7c in Figure \ref{Figure7} shows the mass fraction of cloud material with $\rho>\bar{\rho}_{\rm cloud, 0}/3$ (which is the standard threshold; see \citealt{2015ApJ...805..158S}), normalised with respect to the initial cloud mass for models with solenoidal and compressive multicloud systems. The evolution of the mass-loss curves in this panel indicates that solenoidal layers maintain higher amounts of dense gas compared to compressive systems, at all times. Even though individual cloudlets do not survive in solenoidal cloud models, these systems are not destroyed within the time-scale of our simulations (i.e., their $t_{\rm des}>t_{\rm sim }$), but they rather maintain $>40$ per cent of gas with $\rho>\bar{\rho}_{\rm cloud, 0}/3$, mainly contained in a warm, mixed gas shell. Thus, our solenoidal models do not favour a direct dense-gas entrainment scenario, but rather a mass-loading scenario where warm/hot gas is effectively accelerated and very dense gas is fully disrupted. On the other hand, compressive clouds are able to retain some of their cores until the end of the simulations, but the mass-loss curves in panel 7c of Figure \ref{Figure7} show a steady decreasing trend, implying that dense gas is effectively eroded by the post-shock flow in these models, and only survives in a few `islands'. The erosion to which the more porous compressive layers are subjected also makes the warm-gas shell less compact than in solenoidal cases over time (see Figure \ref{Figure3}).\par

The mass-loss curves in Figure \ref{Figure7} also show a general trend with increasing wavenumber. Systems with higher initial $k$ (i.e., with more cloudlets) are able to retain more dense gas than models with lower $k$. The reason for this behaviour is that more clouds allow shielding of downstream cloudlets (by upstream cloudlets) to be more effective (in agreement with \citealt{2019AJ....158..124F}, who showed that hydrodynamical shielding can prolong the lifetime of clouds in gas streams). As a result, more dense gas in the layer is able to survive. For instance, at $t=2.5\,t_{\rm sp}=0.50\,\rm Myr$, $\sim 85$ per cent of gas with $\rho>\bar{\rho}_{\rm cloud, 0}/3$ survives in the $k=16$ solenoidal model, and $\sim 40$ per cent in the $k=16$ compressive model, which are both at least $\sim 20$ per cent higher than in their respective low-$k$ counterparts.\par

\begin{figure}
\begin{center}
  \begin{tabular}{c c}
    \hspace{-0.25cm}8a) sole-k8-10 & \hspace{-0.25cm}8b) comp-k8-M10\\
    at $t=2.4\,t_{\rm sp}=0.48\,\rm Myr$ & at $t=2.4\,t_{\rm sp}=0.48\,\rm Myr$\\
    \resizebox{38mm}{!}{\includegraphics{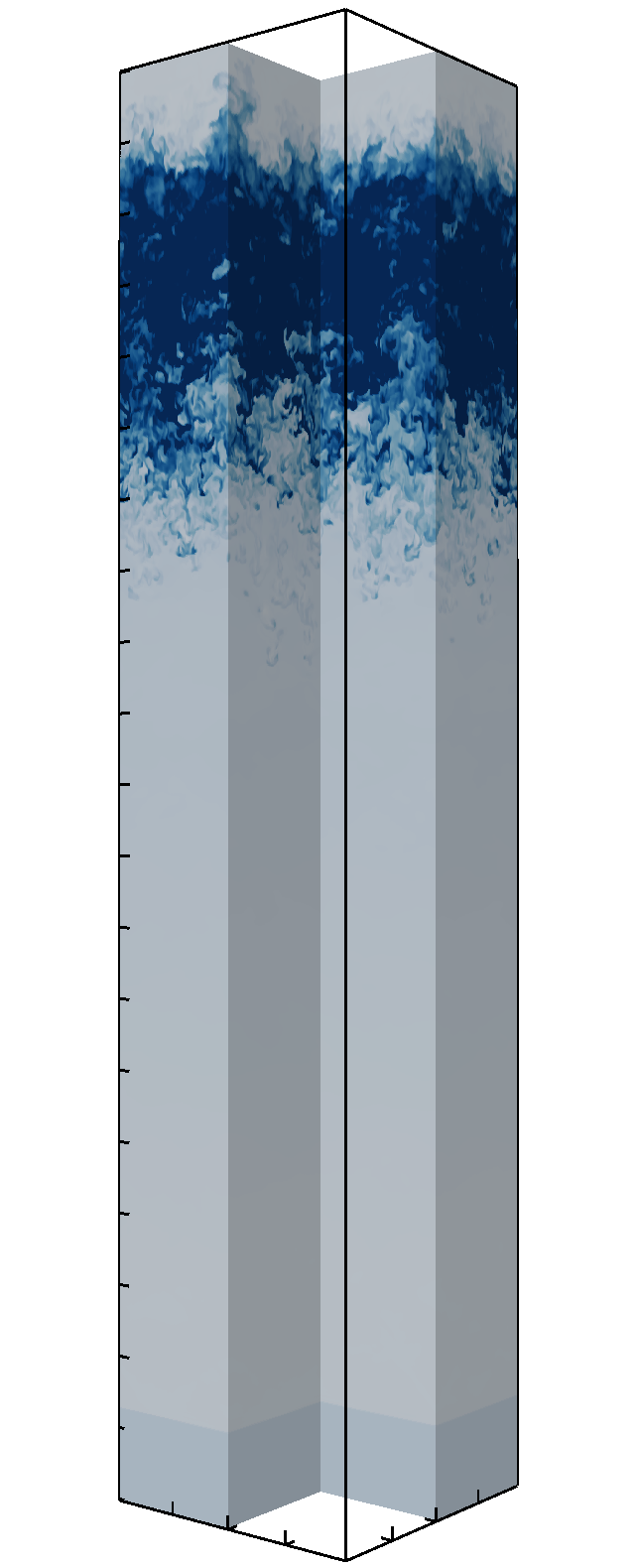}} & \hspace{-0.15cm}\resizebox{38mm}{!}{\includegraphics{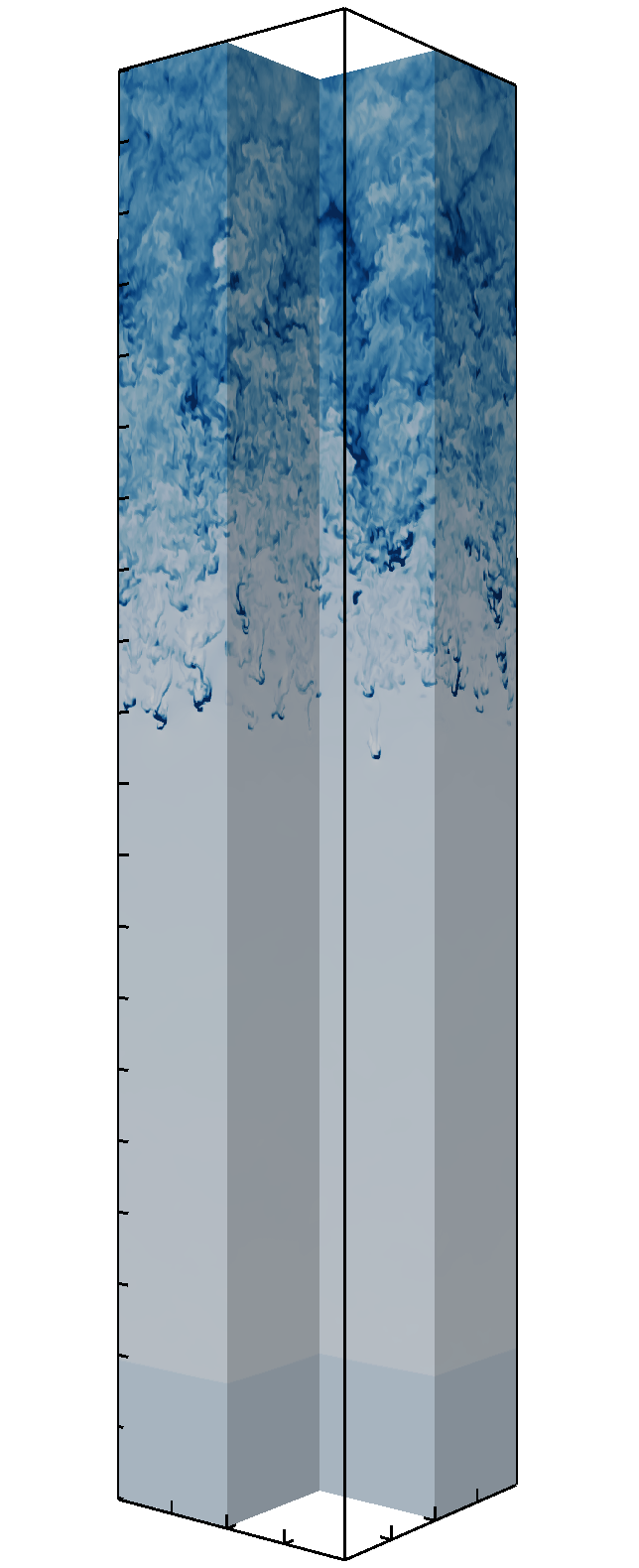}}\\
    \multicolumn{2}{c}{\resizebox{80mm}{!}{\includegraphics{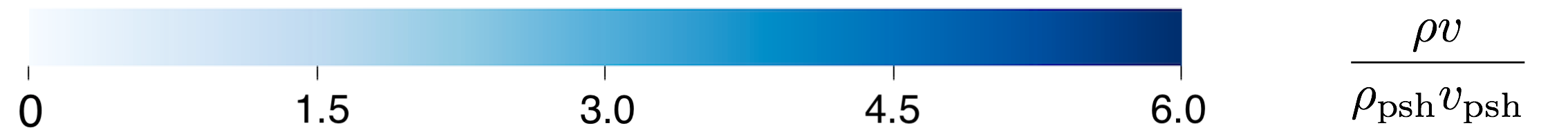}}}\\
    \hspace{-0.25cm}\resizebox{42mm}{!}{\includegraphics{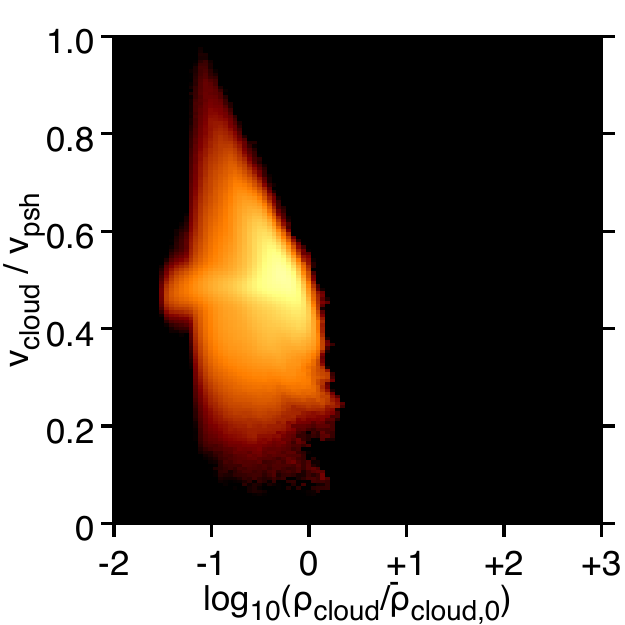}} & \hspace{-0.35cm}\resizebox{42mm}{!}{\includegraphics{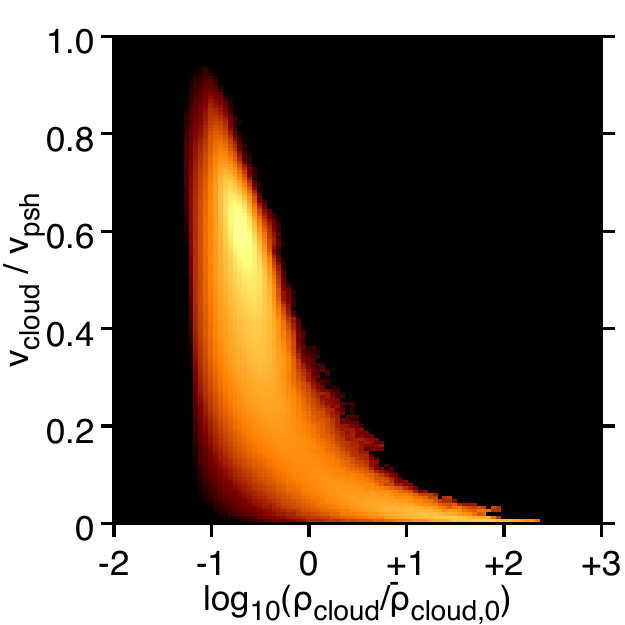}} \\
    \multicolumn{2}{c}{\resizebox{75mm}{!}{\includegraphics{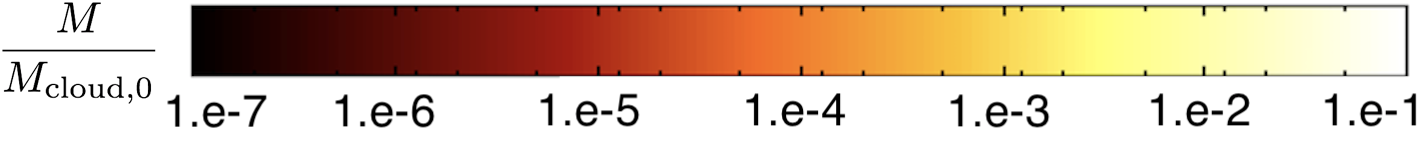}}}\vspace{-0.1cm}\\
  \end{tabular}
  \caption{Gas momentum maps (top panels) and mass-weighted phase diagrams of gas velocity versus density (bottom panels) in the $k=8$ solenoidal model (panel 8a) and the $k=8$ compressive model (panel 8b) at $t=2.4\,t_{\rm sp}=0.48\,\rm Myr$. Mixed gas entrainment is more efficient in the solenoidal model than in the compressive model. Gas with large densities, $\rho>\bar{\rho}_{\rm cloud, 0}$, only survives in compressive models until late times, but it has a very low momentum and it is not entrained in the post-shock flow.} 
  \label{Figure8}
\end{center}
\end{figure}

In terms of gas entrainment, we can separate the analysis in entrainment of warm mixed gas and of cold dense gas. Warm mixed gas in solenoidal models can gain significant momentum owing to its compactness. Panels 8a and 8b of Figure \ref{Figure8} show that mixed gas can gain momenta $>6$ times the momentum of the post-shock flow in solenoidal models, while compressive layers only gain $\sim 3$ times the post-shock momentum over the same time-scale ($t=2.4\,t_{\rm sp}=0.48\,\rm Myr$). On the other hand, dense gas has very low momentum in all cases. The densest cloudlets/cores in multicloud systems are more difficult to accelerate owing to their larger column densities, so they stay behind the most diffuse gas as the cloud layers expand vertically. Since these cores have higher initial densities in compressive models, this effect is accentuated in such models.\par

In Figure \ref{Figure8} we also show mass-weighted phase diagrams of cloud speed versus cloud density for a solenoidal model (bottom panel 8a) and a compressive model (bottom panel 8b) with $k=8$ at $t=2.4\,t_{\rm sp}=0.48\,\rm Myr$. The warm-gas shells stand out as the brightest zones on the 2D histograms, corresponding to densities $0.1\lesssim\rho_{\rm cloud}/\bar{\rho}_{\rm cloud}\lesssim1$ and speeds $0.4\lesssim v_{\rm cloud}/v_{\rm psh}\lesssim0.7$. The overall momentum is dominated by this shell's in solenoidal layers, while low-momentum dense gas also contributes in compressive layers. Panel 8b in Figure \ref{Figure8} confirms that compressive layers can also mass load the outflow with some warm diffuse gas, but dense gas has very low speeds and it is not entrained in the post-shock flow.

\subsection{Dependence on wavenumber, cloud layer thickness, shock Mach number, and numerical resolution}
\label{subsec:dependences}
The above results have important implications for the physics of multi-phase galactic outflows. Our results suggest that the warm phases of observed outflows may contain some imprints of their native environments, i.e., information of the cloud layers in which they originate. In particular, we have shown that the morphology, the volume filling factors, the density PDFs, and the kinematical profiles of mass-loaded gas differ in models that start with different cloud layer distributions. For instance, if the cloud layers at the base of such outflows are more compact and uniform (i.e., solenoidal), our results suggest that the cloud layers evolve into shells of mixed gas that can efficiently gain momentum, while if the cloud layers at the base of outflows are more porous and clumpy (i.e., compressive), some dense gas stays behind and survives, while mixed gas evolves into more vertically-extended filamentary systems. Thus, identifying these morphological signatures in observations of warm atomic, diffuse ionised, and X-ray-emitting gas in galactic outflows can tell us more about the density properties and the cloud distribution at the base of outflows. In fact, our models suggest that differences in the density structure of the gas surrounding star-forming regions may account for the asymmetric morphology of Galactic chimneys (e.g., see \citealt{2003ApJ...590..906T,2008MNRAS.387...31D}), the formation of filamentary shells around stellar-blown super-bubbles (e.g., see \citealt{2006ApJ...638..196M,2011A&A...528A.136S}), and the different filling factors of molecular, atomic, ionised, and shocked gas emission in different large-scale outflows (e.g., \citealt{2009ApJ...701.1636M,2018ApJ...856...97S,2018NatAs...2..901M}). In order to see if our results hold for a wider set of parameters we now discuss invariance in models with different initial conditions.

\subsubsection{Dependence on wavenumber: layer porosity vs. cloudlet population density}
\label{subsubsec:k_dependence}
As mentioned above, the evolution of shock-multicloud systems depends on both the porosity and the cloudlet population density of the cloud layers. Compressive clouds are more porous than solenoidal clouds as the dense gas in them has lower volumetric filling factors. Porous multicloud layers lead to higher mixing fractions and higher velocity dispersions as vorticity can more readily be deposited at cloudlet-intercloudlet interfaces when the internal forward shock moves faster across the intercloud gas. On the other hand, the cloudlet population density can differ in models within each sample (solenoidal and compressive) as this parameter is related to the initial normalised wavenumber, $k$. Models with higher $k$ have more cloudlets clustered in the layer, while models with lower $k$ have less cloudlets. A higher number of cloudlets in a layer results in larger stand-off distances for the shock, lower mixing fractions, lower velocity dispersions, higher accelerations, and milder mass losses. This signifies that the emergence of downstream turbulence is also tied to the initial wavenumber of the multicloud layer.\par

Another aspect that depends on the wavenumber of the initial density distribution is the overall volume filling factor of the cloud layer (see panel 4c in Figure \ref{Figure4}), which also indicates the vertical extent of cloud material (as the transverse cross section is constant in these models). In compressive models, cloud material has higher volume filling factors and is more vertically-extended than in solenoidal models, but there is also a systematic dependence on the cloudlet population density. Models with more cloudlets lead to less vertically-extended outflows with smaller volume filling factors, and vice versa. Thus, in higher $k$ models momentum transfer is more uniformly distributed across gas with different densities inside the layers, while in lower $k$ models diffuse gas gains momentum faster and dense gas is slower.


\subsubsection{Dependence on the cloud layer thickness}
\label{subsubsec:Layer_thickness}
\begin{figure}
\begin{center}
  \begin{tabular}{l}
    \hspace{0.55cm}9a) Normalised volumetric filling factor\vspace{-0.25cm}\\
    \hspace{-0.4cm}\resizebox{80mm}{!}{\includegraphics{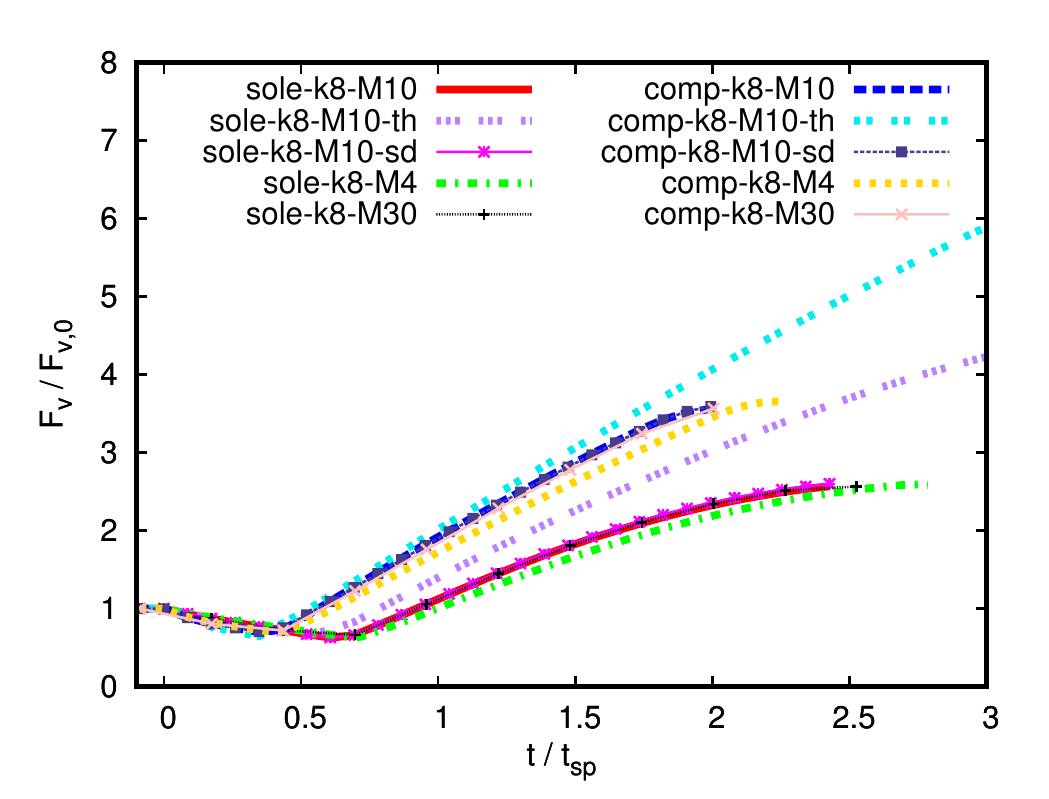}}\\
    \hspace{0.55cm}9b) Mixing fraction\vspace{-0.25cm}\\
    \hspace{-0.4cm}\resizebox{80mm}{!}{\includegraphics{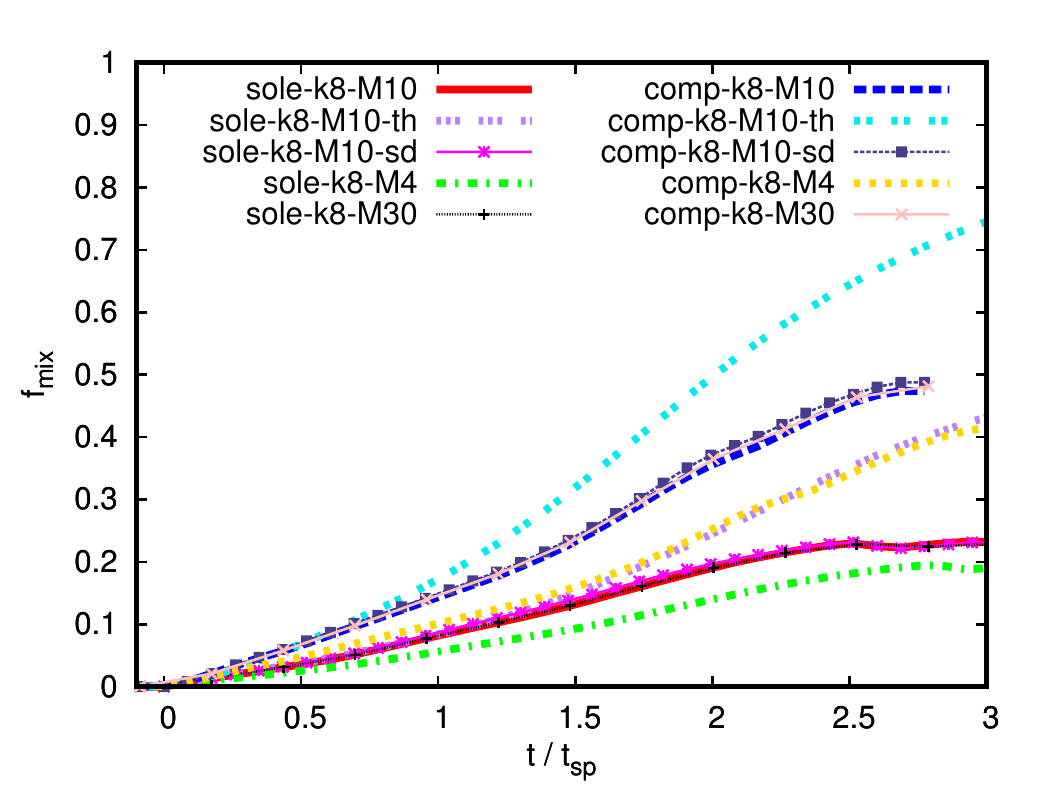}}\\
    \hspace{0.55cm}9c) Bulk speed\vspace{-0.25cm}\\
    \hspace{-0.4cm}\resizebox{80mm}{!}{\includegraphics{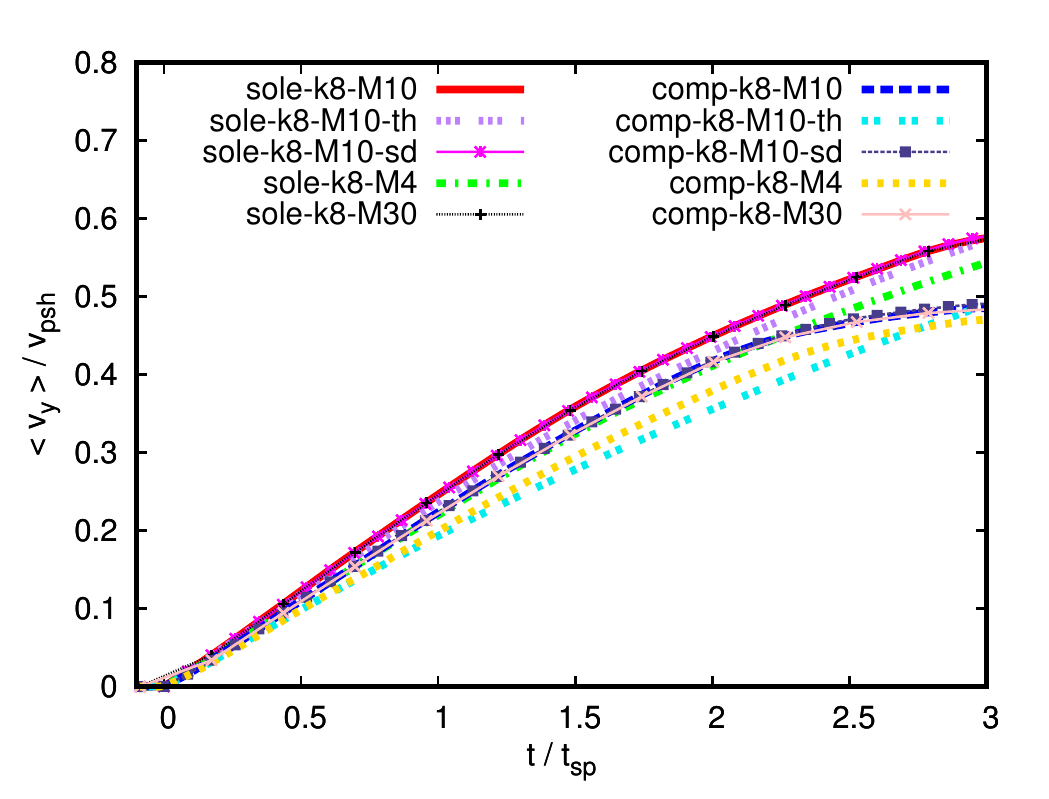}}\\
  \end{tabular}
  \caption{Normalised volumetric filling factor of cloud material (panel 9a), mixing fraction (panel 9b), and mass-weighted bulk speed (panel 9c) in solenoidal and compressive models with different vertical extents, distinct cloud generating seeds, and three different Mach numbers. Models are invariant (i.e., they scale) with respect to the cloud generating seed and Mach numbers $\geq 10$, when times are given in units of the shock-passage time, $t_{\rm sp}$. There is also good, but weaker, scaling in models with thin layers and weaker Mach numbers. Thin-layer models produce higher mixing fractions than their thick-layer counterparts, and weaker Mach numbers result in slightly slower and less turbulent flows. In physical units, slower shocks delay the evolution, while faster shocks and thin-layer systems speed it up, e.g., in our fiducial case, $1\,\rm t_{\rm sp}=0.098\,\rm Myr$, $1\,\rm t_{\rm sp}=0.196\,\rm Myr$, $1\,\rm t_{\rm sp}=0.491\,\rm Myr$, and $1\,\rm t_{\rm sp}=0.065\,\rm Myr$ in thin-layer models, standard thick-layer models, Mach-4 models, and Mach-30 models respectively (see also Appendix \ref{sec:App}).} 
  \label{Figure9}
\end{center}
\end{figure}

The initial vertical extent of the cloud layer, i.e., its streamwise thickness regulates the shock-passage time-scale, which is the most important time-scale to describe multicloud layers. In our thin-layer models, $L_{\rm mc}$ is half the length of the standard (thick-)layer models. When normalised with respect to the shock-passage time, the time-scale on which the forward shock exits the multicloud layer is the same in thin- and thick-layer models, so the four stages of the evolution of the systems occur over similar time-scales (in normalised units). Panel 9a in Figure \ref{Figure9} shows that the change of slope in the curves of the normalised volumetric filling factors (which demarcate the beginning/end of compression and re-expansion phases) occur at the same time in thick- and thin-layer models, thus suggesting there is at least some invariance with respect to the cloud layer thickness.\par

Despite the above, a careful examination of the mixing fractions in panel 9b of Figure \ref{Figure9} indicates that mixing fractions are systematically higher in thin-layer models (by similar factors in solenoidal and compressive models) than in standard thick-layer models. This suggests thin-layer models generate more turbulence than thick-layer models over similar (normalised) time-scales. This also explains why they tend to occupy higher volume filling factors at late times (see panel 9a of Figure \ref{Figure9}) than their thick-layer counterparts. The reason for this is that cloudlet shielding is limited by the smaller $L_{\rm mc}$, so the shock can travel more easily across thin layers owing to their smaller column densities in the streaming direction (see Appendix \ref{sec:App}). The post-shock flow can then more easily deposit vorticity at cloud-intercloud interfaces and low-density gas is rapidly pushed downstream. This prevents the formation of the well-defined, mixed shells seen in standard models, and also reduces (slightly) the effectiveness of momentum transfer. Thus, thin-layer clouds are slightly slower than their thick-layer counterparts (see panel 9c of Figure \ref{Figure9}).\par




\subsubsection{Mach number scaling and seed invariance}
\label{subsubsec:M_invariance}
Figure \ref{Figure9} also shows the effects of changing the shock Mach number from the standard value of ${\cal M}_{\rm shock}=10$ to a weaker case ${\cal M}_{\rm shock}=4$, and to a stronger case, ${\cal M}_{\rm shock}=30$. The physical time-scales over which the different evolutionary stages occur are longer for ${\cal M}_{\rm shock}=4$ and shorter for ${\cal M}_{\rm shock}=30$. However, in normalised time-scales the evolution of all parameters: the volumetric filling factor, the mixing fraction, and the mass-weighted bulk speed is very similar in all cases. Changing the Mach number from ${\cal M}_{\rm shock}=10$ to ${\cal M}_{\rm shock}=4$ does have some minor effects on the parameters, particularly on the mixing fractions. The weaker shock produces less mixing due to the slightly lower velocity difference between the pre- and post-shock ambient media with respect to the standard case. This and the slightly smaller density contrast also reduce momentum transfer with respect to ${\cal M}_{\rm shock}=10$, and while there is a reasonably good scaling, this is weaker compared to the strong-shock scaling. Indeed, changing the Mach number from ${\cal M}_{\rm shock}=10$ to ${\cal M}_{\rm shock}=30$ does not have an effect at all, in normalised time-scales. This implies that the shock-multicloud problem in this adiabatic set is invariant for strong shocks (${\cal M}_{\rm shock}\geq10$), i.e., the evolution patterns hold for high Mach numbers (see also the density slices in Figures \ref{FigureA1}, \ref{FigureA2} and \ref{FigureA3} in the Appendix). Albeit the initial Mach numbers are different, this confirms that there is scaling in the strong-shock regime as discussed in \citealt{1994ApJ...420..213K,2006ApJS..164..477N}.\par

Following \cite{2019MNRAS.486.4526B}, we also check whether or not the aforementioned results depend on the seed we choose to generate the log-normal density distributions for the multicloud systems. For this we generate additional solenoidal and compressive clouds layers with $k=8$ using the pyFC code. Figure \ref{Figure9} shows that the evolution of shock-swept multicloud systems holds for fractal multicloud systems generated with different seeds. We find that there is virtually no difference between the curves, which indicates that multicloud systems have better convergence properties than single-cloud systems, for which we found a slightly higher dependence on the cloud-generating seed.

\subsubsection{Dependence on numerical resolution}
\label{subsubsec:Resolution}
\begin{figure}
\begin{center}
  \begin{tabular}{c c c c}
   \hspace{-0.35cm}10a) sole-k8-M10-lr & sole-k8-M10 & sole-k8-M10-hr \\
    \resizebox{22mm}{!}{\includegraphics{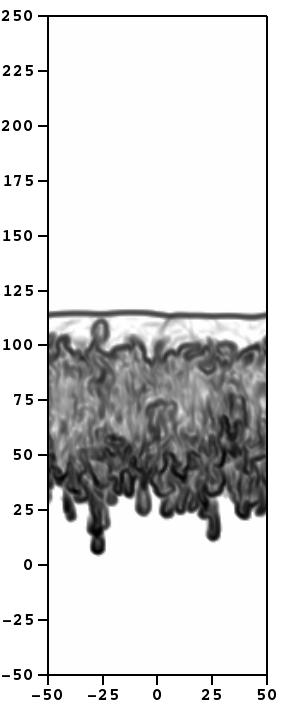}} & \hspace{-0.2cm}\resizebox{22mm}{!}{\includegraphics{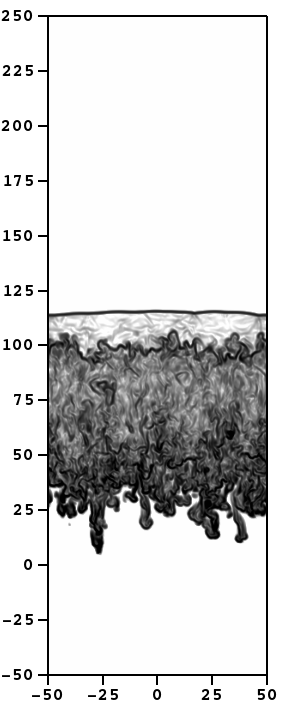}} & \hspace{-0.2cm}\resizebox{22mm}{!}{\includegraphics{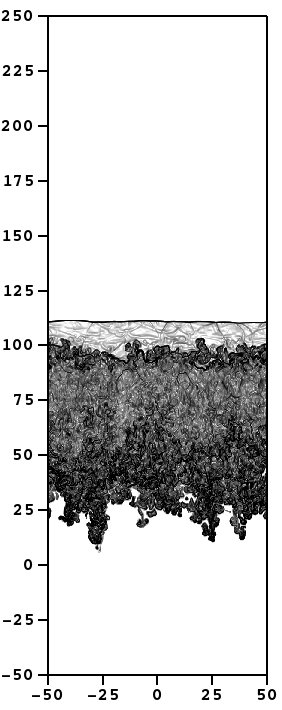}} & \hspace{-0.35cm}\resizebox{12mm}{!}{\includegraphics{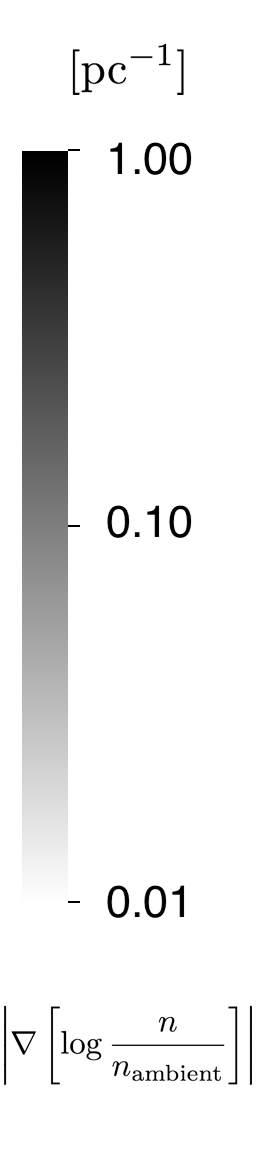}}\\ 
   \hspace{-0.35cm}10b) comp-k8-M10-lr & comp-k8-M10 & comp-k8-M10-hr \\
    \resizebox{22mm}{!}{\includegraphics{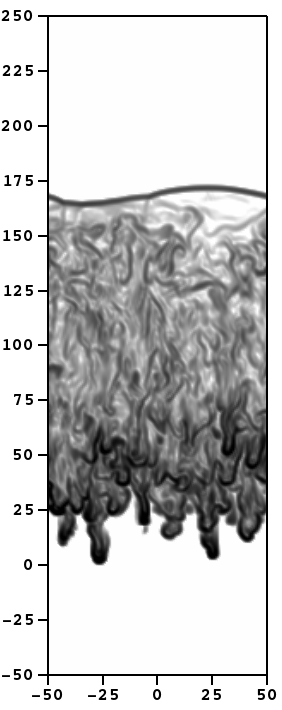}} & \hspace{-0.2cm}\resizebox{22mm}{!}{\includegraphics{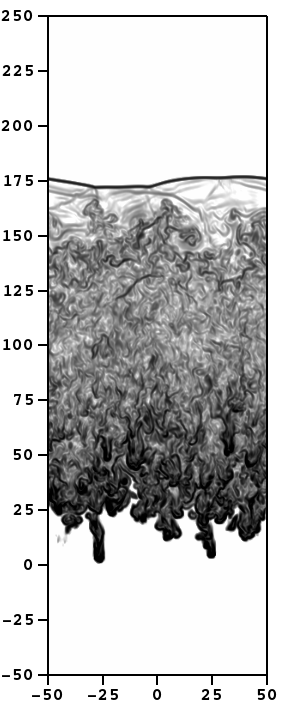}} & \hspace{-0.2cm}\resizebox{22mm}{!}{\includegraphics{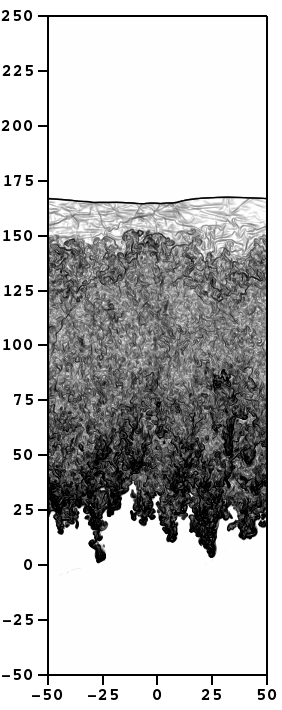}} & \hspace{-0.35cm}\resizebox{12mm}{!}{\includegraphics{bar_grad.png}}\\ 
  \end{tabular}
  \caption{2D slices at $Z=0$ of the gradient of the logarithmic density, normalised to the ambient density, (Schlieren images) of the solenoidal multicloud system, sole-k8-M10 (top panels), and the compressive multicloud system, comp-k8-M10 (bottom panels), at $t=1.1\,t_{\rm sp}=0.22\,\rm Myr$ for three resolutions (8, 16, and 32 cells per cloudlet radius). The figures cover the spatial extent ($L\times3L$)$\equiv$($2\,L_{\rm mc}\times6\,L_{\rm mc}$), i.e., ($100\,\rm pc\times300\,\rm pc$). The forward shock front becomes thinner and the length scales of vortices in the multicloud layer decrease with increasing resolution.} 
  \label{Figure10}
\end{center}
\end{figure}

\begin{figure*}
\begin{center}
  \begin{tabular}{l l}
    \hspace{0.55cm}11a) Thermal pressure & \hspace{0.55cm}11b) Normalised volumetric filling factor\\
    \hspace{-0.4cm}\resizebox{80mm}{!}{\includegraphics{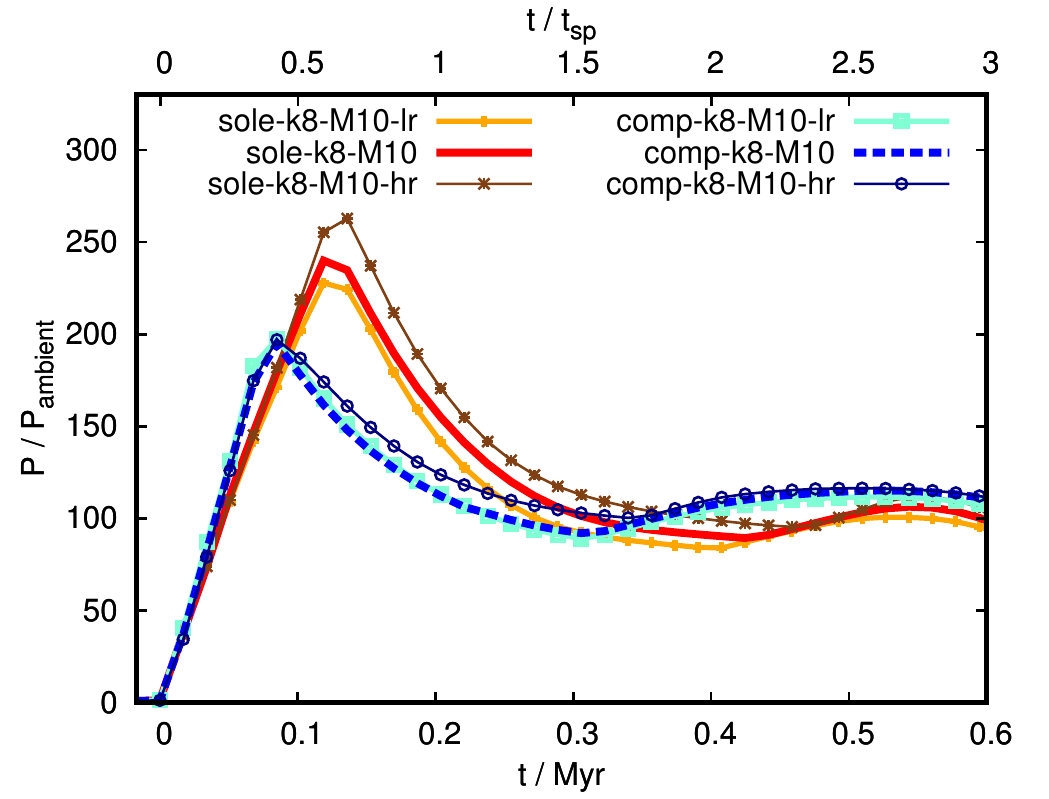}} & \hspace{-0.4cm}\resizebox{80mm}{!}{\includegraphics{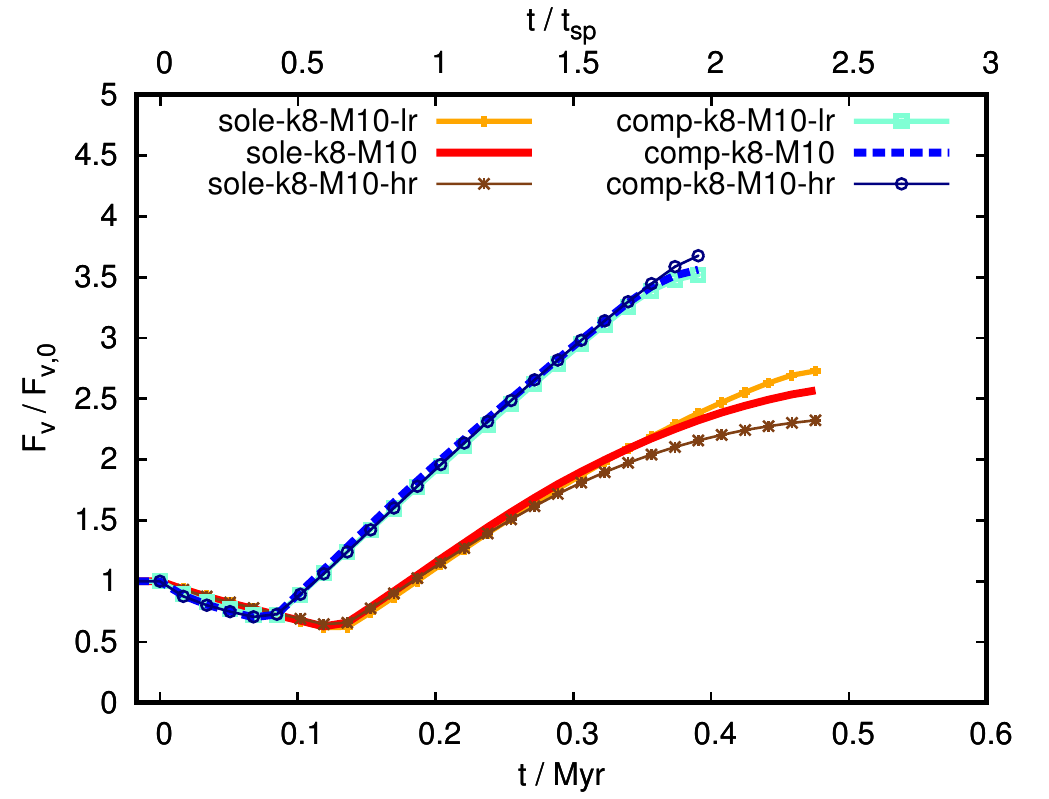}} \\
    \hspace{0.55cm}11c) Mixing fraction & \hspace{0.55cm}11d) Velocity dispersion\\
    \hspace{-0.4cm}\resizebox{80mm}{!}{\includegraphics{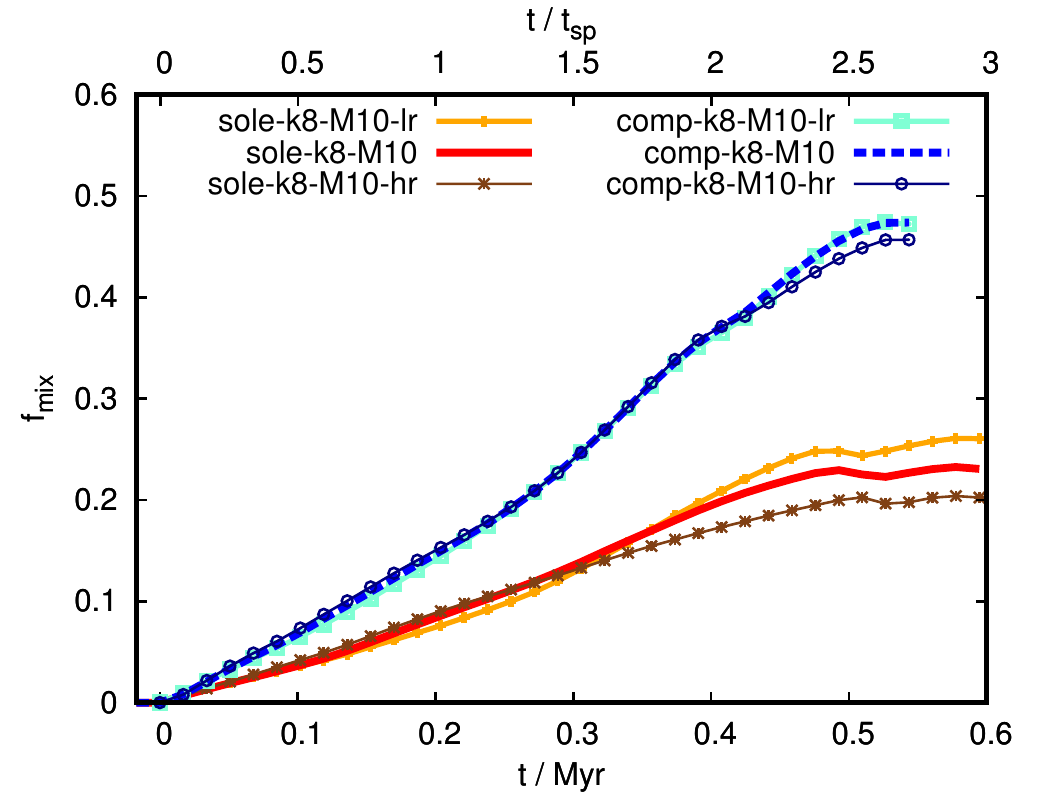}} & \hspace{-0.4cm}\resizebox{80mm}{!}{\includegraphics{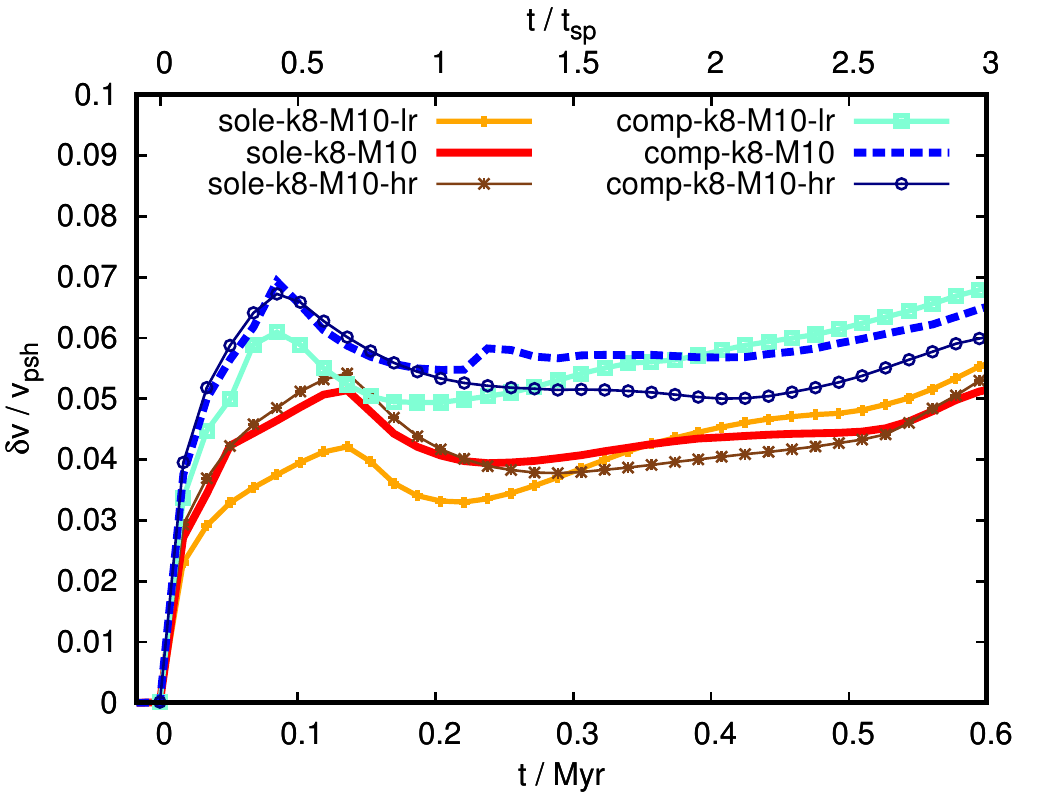}}\\
    \hspace{0.55cm}11e) Bulk speed & \hspace{0.55cm}11f) Mass loss\\
    \hspace{-0.4cm}\resizebox{80mm}{!}{\includegraphics{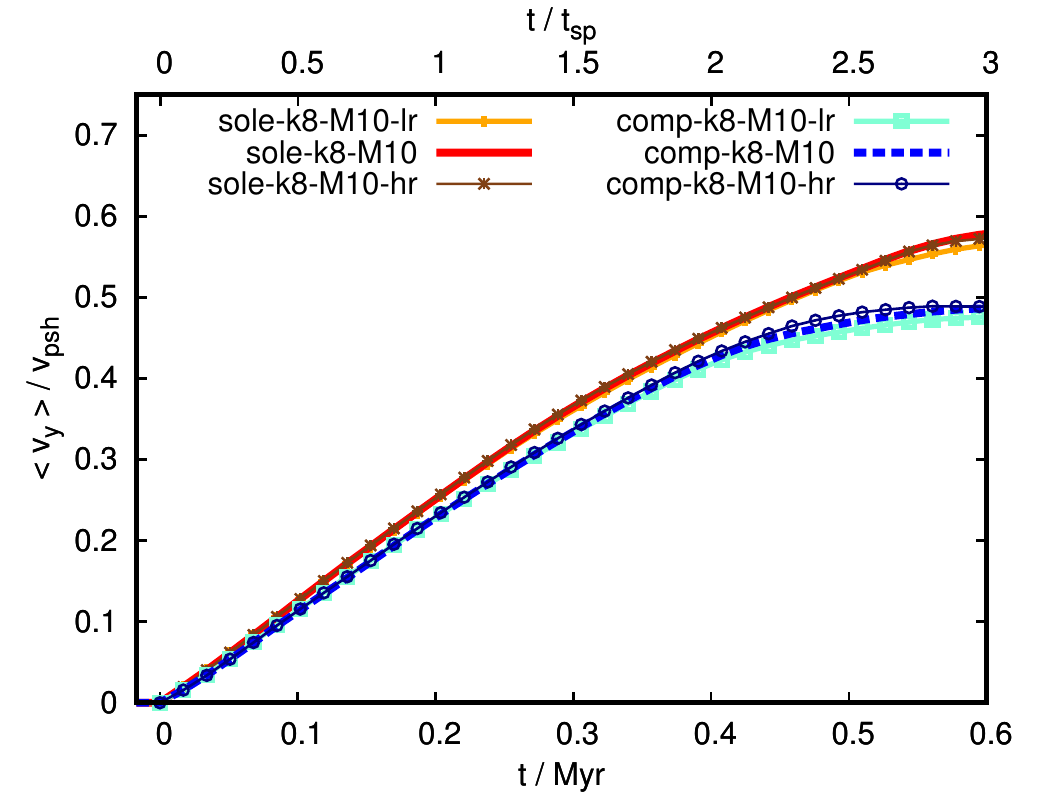}} & \hspace{-0.4cm}\resizebox{80mm}{!}{\includegraphics{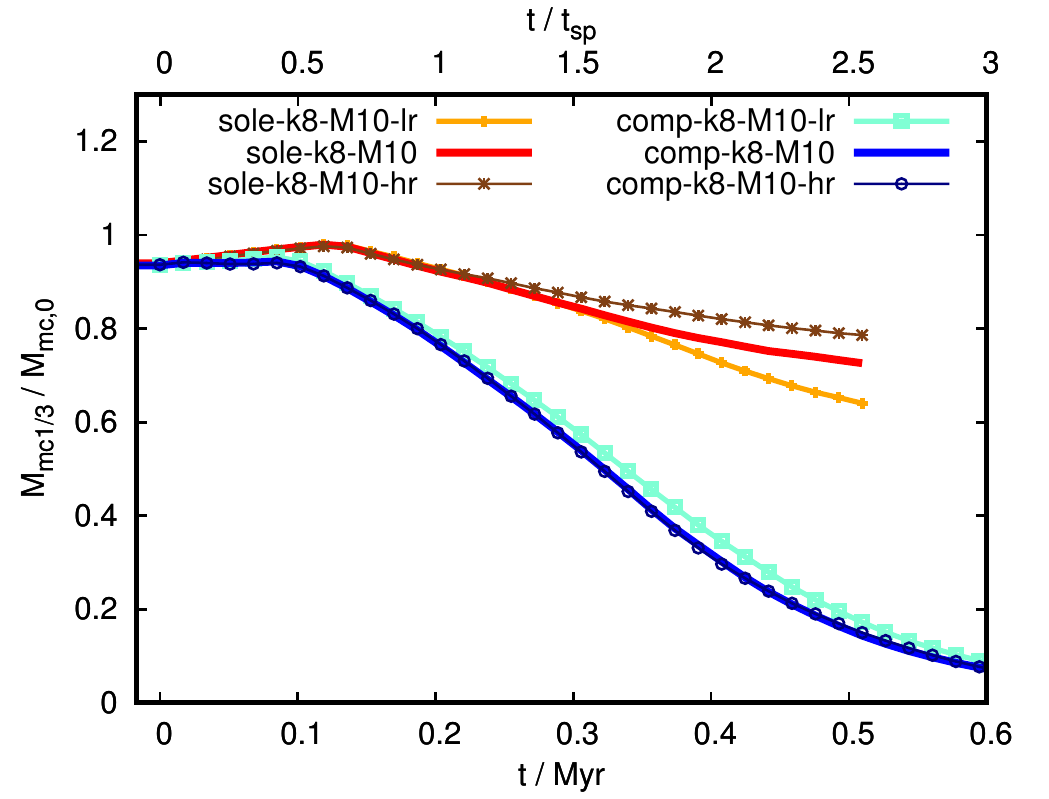}}\\
  \end{tabular}
  \caption{Numerical resolution study showing the time evolution of six diagnostics: the thermal pressure and the normalised volumetric filling factor in the top panels (11a and 11b), the mixing fraction and the velocity dispersion in the middle panels (11c and 11d), and the bulk speed and the mass fraction of dense gas in the bottom panels (11e and 11f), for three resolutions 8, 16, and 32 cells per cloudlet radius. Global dynamical and geometrical quantities show convergence, even at the lowest resolution. Diagnostics that depend on the generation of vorticity, such as the mixing fraction and the velocity dispersion, are the least converged, but the differences due to resolution changes are small compared to those in different models. Our fiducial resolution of 16 cells per cloud radius ($R_{16}$) and even $R_8$ adequately capture the evolution of shock-multicloud systems.}
  \label{Figure11}
\end{center}
\end{figure*}

Finally, we study the effects of changing the numerical resolution of the computational domain on several diagnostics. Figure \ref{Figure10} shows the effect of resolution on the generation of vorticity. The panels show the gradient of the logarithmic mass density (Schlieren images), which indicate that as the resolution increases (from left to right), small-scale turbulence is better captured. How does this affect the evolution of our diagnostics and our conclusions? Figure \ref{Figure11} shows the evolution of six diagnostics. The top panels show the thermal pressure (panel 11a) and the volumetric filling factor (panel 11b) of the cloud layer. Thermal pressures appear to have a very subtle increasing trend when resolutions go up, but the curves are overall well converged. The filling factors show convergence, even at the lowest resolutions we considered, $R_{8}$.\par

The middle panels of Figure \ref{Figure11} show the mixing fraction (panel 11c) and the velocity dispersion (panel 11d) in cloud material. Both quantities depend on how dynamical instabilities grow at gas interfaces and how fast turbulence grows in the multicloud system. These quantities are usually the least converged as they depend on how small-scale vorticities are deposited at shear layers (see \citealt{2018MNRAS.473.3454B,2019MNRAS.486.4526B}). Increasing the resolution leads to higher vortical motions and less mixing (as we study inviscid gases). Despite this, the curves describing these parameters in both sets, solenoidal and compressive, show very good agreement with each other with only very subtle indications of these previously-identified trends. The velocity dispersions are the least converged in our simulations, but our standard-resolution models do capture the overall trend of the high-resolution models.\par

The bottom panels of Figure \ref{Figure11} show the mass-weighted bulk speed (panel 11e) and the mass fraction of dense gas (panel 11f) in our models. The curves corresponding to the bulk speeds show convergence in both model sets. The curves corresponding to the mass fractions of dense gas show better convergence in compressive models than in solenoidal models. In the latter, increasing the resolution slightly reduces the mass-loss rate, which is consistent with the mixing fraction curves reported in panel 11c. Since there is less mixing at higher resolutions, mass loss slows down. Despite this, the differences are very small and occur mostly at late times, so we can conclude that convergence has been achieved.\par

Overall, our resolution tests indicate that diagnostics that depend on the generation of vorticity, such as the mixing fraction and the velocity dispersion show slightly larger differences when increasing the numerical resolution than dynamical diagnostics. However, even in these cases the differences are small compared to differences between the models that we included in our sample. Thus, we also conclude that our fiducial resolution is adequate to capture the turbulent properties of shock-multicloud systems, and also that multi-cloud systems have better convergence properties than single-cloud systems. The intra-system interactions between cloudlets, intercloud gas, and other cloudlets play a more significant role than extra-system interactions in multicloud models; while for single-cloud models, extra-system interactions between the cloud and the post-shock flow become more important as the cloud is directly exposed to fast-flowing material. This also explains why the relevant time-scale to describe multi-cloud systems is not the cloud-crushing time of individual cloudlets, but rather the shock-passage time of the cloud system as a whole.

\section{Limitations and future work}
\label{sec:FutureWork}
The simulations presented here are the first set of a larger sample of models. The purpose of this study is to survey a broad set of the parameter space in order to isolate the effects of changing the density structure and the shock properties of shock-multicloud systems upon the disruption of cloud layers. In this study we ignore several ingredients that are important for the physics of galactic winds, but these will be systematically included in forthcoming studies. We briefly comment on the effects of radiative cooling and magnetic fields below, as they will be the first to be discussed in our next papers in this series.\par


We did not include radiative cooling which is known to extend the lifetimes of clouds (in the strong cooling regime; e.g., see \citealt{2009ApJ...703..330C}). Based on the results we have presented here, mass loading is effective in both cases, solenoidal and compressive, so considering our fiducial physical model we expect cooling to extend the lifetime of dense gas in both cases. We would also expect the clumping factor to be higher in these models as cooling will aid gas condensation, thus creating steeper density contrasts. The forward shock would then be able to travel faster across the intercloud gas, thus modifying the stand-off distance of the reflected shock and also the time-scales for cloud erosion. Similarly, magnetic fields can alter the dynamics of dense gas in shock-cloud systems. Depending on their strength and orientation, uniform fields can, e.g., prevent cloudlet-cloudlet collisions, stretch cloud gas along the field lines, and contribute to momentum transfer (e.g., see \citealt{2020ApJ...892...59C}). Thus, in multicloud systems we would also expect them to reduce the amount of mixing and delay the disruption of cloudlets via draping.


\section{Conclusions}
\label{sec:Conclusions}
We have presented the first part of a comprehensive study of shock-multicloud systems, in which we consider adiabatic fractal clouds embedded in (supersonic) shocks (${\cal M}_{\rm shock}=10,\:4,\:30$). The clouds have initially log-normal density distributions characteristic of supersonic turbulence (${\cal M}_{\rm turb}\approx 5$) driven by solenoidal and compressive modes. The solenoidal density fields have low standard deviations ($\sigma_{\rm cloud,0}=1.9\,\bar{\rho}_{\rm cloud,0}$), while the compressive density fields have high standard deviations ($\sigma_{\rm cloud,0}=5.9\,\bar{\rho}_{\rm cloud,0}$), so solenoidal clouds are more compact and compressive clouds are more porous. Within each sample we vary the cloud population density by changing the minimum wavenumber of the density fields ($k=4,\:8,\:16$), which effectively modifies the number and the size of the cloudlets in the system. In addition, we study cases with thinner cloud layers and with other cloud-generating seeds. Our conclusions are as follows:

\begin{itemize}
 \item In agreement with earlier studies by \cite{2002ApJ...576..832P,2012MNRAS.425.2212A,2014MNRAS.444..971A} on uniform multicloud systems, we find that the evolution of shocked fractal multicloud systems consists of four stages: 1) a shock-splitting phase in which the shock splits into reflected and refracted shocks after the initial contact, 2) a cloud-layer compression phase in which the refracted shock compresses the cloud layer as it travels through it, 3) a shock re-acceleration phase in which the forward shock leaves the cloud layer and enters the downstream medium triggering a rapid expansion of the cloud layer, and 4) a mixing phase in which shear instabilities stir cloud gas and generate turbulence.
 \item The dynamics and disruption of multicloud systems depend on the standard deviation of the density distribution (i.e., on the porosity of the multicloud layer) and on the minimum wavenumber (i.e., on the number of cloudlets in the layer). More compact and uniform multicloud systems (i.e., solenoidal cloud layers) mix less, generate less turbulence, accelerate faster, and form a more coherent, high-momentum shell of mixed gas than porous systems (i.e., compressive cloud layers).
 \item In all the scenarios we studied, dense-gas entrainment is highly inefficient. Dense gas does not survive in solenoidal models. It only survives in compressive clouds, but it has low momentum. The density PDFs of compressive models are wider than in solenoidal cases at all times, as they maintain extended high-density tails. Mass loading of low-density gas is efficient in both solenoidal and compressive models. 
 \item Multicloud systems with high wavenumbers ($k$), i.e., with a higher number of cloudlets, quench mixing via a shielding effect. Upstream cloudlets protect downstream cloudlets, and the high number of cloudlets obstruct the motion of the post-shock flow across the intercloud medium. This reduces the emergence of vorticity and instabilities at cloudlet-intercloudlet boundaries. On the other hand, the low number of cloudlets in models with low wavenumbers ($k$) facilitate the percolation of the post-shock flow and increase the effect of dynamical instabilities.
 \item Compressive models also have larger volume filling factors and vertical extents than solenoidal models. The vertical extent of the layers also depends on the number of cloudlets in the initial distribution. Models with more cloudlets have lower vertical extents, and vice versa.
 \item If the diagnostic variables are normalised with respect to the shock-passage time, the evolution of multicloud systems with the same density contrast and log-normal distribution is invariant with respect to the shock Mach number for $M_{\rm shock}\geq 10$ and the seed used to generate the initial log-normal fractal clouds. We also find weaker scaling for weaker shocks and thinner cloud layers. Weaker shocks produce less mixing, and thin-layer models do not develop the shell-like structures characteristic of their thick-layer counterparts. Thin-layer models also lead to higher mixing fractions and more turbulence than thick multicloud layers over the same normalised time-scales.
 \item Our resolution study suggests that multicloud systems have better convergence properties than single-clouds systems. In terms of cells per cloud radius, resolutions of 8 cells per cloud radius (i.e., $R_{8}$) are sufficient to capture the global dynamics and geometrical aspects of the multicloud layers, while our standard resolution of $R_{16}$ captures small-scale vorticity better and is therefore adequate to describe these systems. In general, diagnostics that depend on small-scale vorticity, such as the mixing fractions and the velocity dispersions, are slightly more resolution-dependent than global dynamical parameters.
\end{itemize}

Overall, the above results suggest that the morphology and properties of some entrained gas in multi-phase outflows contain information on their native environments, e.g., on the original cloud density distribution. If the cloud layers at the base of such outflows are more compact, the layers evolve into a shell of mixed gas that can effectively gain momentum, while if the cloud layers are more porous, dense gas stays behind and survives for long time-scales, while mixed gas develops more coherent filamentary systems. Our study is an attempt to understand how the properties of turbulent gas in star-forming regions relate to those of the outflows generated by their stellar feedback. Our next goal is to study shock-multicloud systems with radiative heating and cooling.


\section*{Acknowledgements}
We thank the anonymous referee for their detailed and constructive report. WBB is supported by the Deutsche Forschungsgemeinschaft (DFG) via grant BR2026125. WBB also thanks for support from the National Secretariat of Higher Education, Science, Technology, and Innovation of Ecuador, SENESCYT. CF acknowledges funding provided by the Australian Research Council (Discovery Project DP170100603 and Future Fellowship FT180100495), and the Australia-Germany Joint Research Cooperation Scheme (UA-DAAD). AYW is partially supported by the Japan Society for the Promotion of Science (JSPS) KAKENHI grant 19K03862. ES was supported by NSF grant AST-1715876. The authors gratefully acknowledge the Gauss Centre for Supercomputing e.V. (\url{www.gauss-centre.eu}) for funding this project (pn34qu) by providing computing time on the GCS Supercomputer SuperMUC-NG at the Leibniz Supercomputing Centre (\url{www.lrz.de}) and on the GCS Supercomputer JUWELS at the J\"ulich Supercomputing Centre (JSC) under projects 16072 and 19590. We further acknowledge computing resources provided by these centres for grants pr32lo, pr48pi, and GCS Large-scale project~10391, and by the Australian National Computational Infrastructure for grant~ek9 in the framework of the National Computational Merit Allocation Scheme and the ANU Merit Allocation Scheme. This work has made use of the VisIt visualisation software (\citealt{HPV:VisIt}), the GDL language (\citealt{2019ASPC..523..365C}), and the gnuplot program (\url{http://www.gnuplot.info}).

\section*{Data availability}
The data underlying this article will be shared on reasonable request to the corresponding author.





\bibliographystyle{mnras}
\bibliography{wlady} 

\begin{thebibliography}{}
\makeatletter
\relax
\def\mn@urlcharsother{\let\do\@makeother \do\$\do\&\do\#\do\^\do\_\do\%\do\~}
\def\mn@doi{\begingroup\mn@urlcharsother \@ifnextchar [ {\mn@doi@}
  {\mn@doi@[]}}
\def\mn@doi@[#1]#2{\def\@tempa{#1}\ifx\@tempa\@empty \href
  {http://dx.doi.org/#2} {doi:#2}\else \href {http://dx.doi.org/#2} {#1}\fi
  \endgroup}
\def\mn@eprint#1#2{\mn@eprint@#1:#2::\@nil}
\def\mn@eprint@arXiv#1{\href {http://arxiv.org/abs/#1} {{\tt arXiv:#1}}}
\def\mn@eprint@dblp#1{\href {http://dblp.uni-trier.de/rec/bibtex/#1.xml}
  {dblp:#1}}
\def\mn@eprint@#1:#2:#3:#4\@nil{\def\@tempa {#1}\def\@tempb {#2}\def\@tempc
  {#3}\ifx \@tempc \@empty \let \@tempc \@tempb \let \@tempb \@tempa \fi \ifx
  \@tempb \@empty \def\@tempb {arXiv}\fi \@ifundefined
  {mn@eprint@\@tempb}{\@tempb:\@tempc}{\expandafter \expandafter \csname
  mn@eprint@\@tempb\endcsname \expandafter{\@tempc}}}

\bibitem[\protect\citeauthoryear{{Al{\= u}zas}, {Pittard}, {Hartquist}, {Falle}
   \& {Langton}}{{Al{\= u}zas} et~al.}{2012}]{2012MNRAS.425.2212A}
{Al{\= u}zas} R.,  {Pittard} J.~M.,  {Hartquist} T.~W.,  {Falle} S.~A.~E.~G.,
  {Langton} R.,  2012, \mn@doi [\mnras] {10.1111/j.1365-2966.2012.21598.x},
  \href {http://adsabs.harvard.edu/abs/2012MNRAS.425.2212A} {425, 2212}

\bibitem[\protect\citeauthoryear{{Al{\= u}zas}, {Pittard}, {Falle}  \&
  {Hartquist}}{{Al{\= u}zas} et~al.}{2014}]{2014MNRAS.444..971A}
{Al{\= u}zas} R.,  {Pittard} J.~M.,  {Falle} S.~A.~E.~G.,   {Hartquist} T.~W.,
  2014, \mn@doi [\mnras] {10.1093/mnras/stu1501}, \href
  {http://adsabs.harvard.edu/abs/2014MNRAS.444..971A} {444, 971}

\bibitem[\protect\citeauthoryear{{Armillotta}, {Fraternali}, {Werk},
  {Prochaska}  \& {Marinacci}}{{Armillotta} et~al.}{2017}]{2017MNRAS.470..114A}
{Armillotta} L.,  {Fraternali} F.,  {Werk} J.~K.,  {Prochaska} J.~X.,
  {Marinacci} F.,  2017, \mn@doi [\mnras] {10.1093/mnras/stx1239}, \href
  {http://adsabs.harvard.edu/abs/2017MNRAS.470..114A} {470, 114}

\bibitem[\protect\citeauthoryear{{Banda-Barrag{\'a}n}, {Parkin}, {Federrath},
  {Crocker}  \& {Bicknell}}{{Banda-Barrag{\'a}n}
  et~al.}{2016}]{2016MNRAS.455.1309B}
{Banda-Barrag{\'a}n} W.~E.,  {Parkin} E.~R.,  {Federrath} C.,  {Crocker} R.~M.,
    {Bicknell} G.~V.,  2016, \mn@doi [\mnras] {10.1093/mnras/stv2405}, \href
  {http://adsabs.harvard.edu/abs/2016MNRAS.455.1309B} {455, 1309}

\bibitem[\protect\citeauthoryear{{Banda-Barrag{\'a}n}, {Federrath}, {Crocker}
  \& {Bicknell}}{{Banda-Barrag{\'a}n} et~al.}{2018}]{2018MNRAS.473.3454B}
{Banda-Barrag{\'a}n} W.~E.,  {Federrath} C.,  {Crocker} R.~M.,   {Bicknell}
  G.~V.,  2018, \mn@doi [\mnras] {10.1093/mnras/stx2541}, \href
  {http://adsabs.harvard.edu/abs/2018MNRAS.473.3454B} {473, 3454}

\bibitem[\protect\citeauthoryear{{Banda-Barrag{\'a}n}, {Zertuche}, {Federrath},
  {Garc{\'\i}a Del Valle}, {Br{\"u}ggen}  \& {Wagner}}{{Banda-Barrag{\'a}n}
  et~al.}{2019}]{2019MNRAS.486.4526B}
{Banda-Barrag{\'a}n} W.~E.,  {Zertuche} F.~J.,  {Federrath} C.,  {Garc{\'\i}a
  Del Valle} J.,  {Br{\"u}ggen} M.,   {Wagner} A.~Y.,  2019, \mn@doi [\mnras]
  {10.1093/mnras/stz1040}, \href
  {https://ui.adsabs.harvard.edu/abs/2019MNRAS.486.4526B} {486, 4526}

\bibitem[\protect\citeauthoryear{{Bieri}, {Dubois}, {Rosdahl}, {Wagner}, {Silk}
   \& {Mamon}}{{Bieri} et~al.}{2017}]{2017MNRAS.464.1854B}
{Bieri} R.,  {Dubois} Y.,  {Rosdahl} J.,  {Wagner} A.,  {Silk} J.,   {Mamon}
  G.~A.,  2017, \mn@doi [\mnras] {10.1093/mnras/stw2380}, \href
  {https://ui.adsabs.harvard.edu/abs/2017MNRAS.464.1854B} {464, 1854}

\bibitem[\protect\citeauthoryear{{Bland-Hawthorn} \& {Cohen}}{{Bland-Hawthorn}
  \& {Cohen}}{2003}]{2003ApJ...582..246B}
{Bland-Hawthorn} J.,  {Cohen} M.,  2003, \mn@doi [\apj] {10.1086/344573}, \href
  {http://adsabs.harvard.edu/abs/2003ApJ...582..246B} {582, 246}

\bibitem[\protect\citeauthoryear{{Br{\"u}ggen} \& {Scannapieco}}{{Br{\"u}ggen}
  \& {Scannapieco}}{2016}]{2016ApJ...822...31B}
{Br{\"u}ggen} M.,  {Scannapieco} E.,  2016, \mn@doi [\apj]
  {10.3847/0004-637X/822/1/31}, \href
  {http://adsabs.harvard.edu/abs/2016ApJ...822...31B} {822, 31}

\bibitem[\protect\citeauthoryear{{Burkhart}, {Lee}, {Murray}  \&
  {Stanimirovi{\'c}}}{{Burkhart} et~al.}{2015}]{2015ApJ...811L..28B}
{Burkhart} B.,  {Lee} M.-Y.,  {Murray} C.~E.,   {Stanimirovi{\'c}} S.,  2015,
  \mn@doi [\apjl] {10.1088/2041-8205/811/2/L28}, \href
  {https://ui.adsabs.harvard.edu/abs/2015ApJ...811L..28B} {811, L28}

\bibitem[\protect\citeauthoryear{Childs et~al.,}{Childs
  et~al.}{2012}]{HPV:VisIt}
Childs H.,  et~al., 2012, in , {High Performance Visualization--Enabling
  Extreme-Scale Scientific Insight}.
pp 357--372

\bibitem[\protect\citeauthoryear{{Cooper}, {Bicknell}, {Sutherland}  \&
  {Bland-Hawthorn}}{{Cooper} et~al.}{2008}]{2008ApJ...674..157C}
{Cooper} J.~L.,  {Bicknell} G.~V.,  {Sutherland} R.~S.,   {Bland-Hawthorn} J.,
  2008, \mn@doi [\apj] {10.1086/524918}, \href
  {http://adsabs.harvard.edu/abs/2008ApJ...674..157C} {674, 157}

\bibitem[\protect\citeauthoryear{{Cooper}, {Bicknell}, {Sutherland}  \&
  {Bland-Hawthorn}}{{Cooper} et~al.}{2009}]{2009ApJ...703..330C}
{Cooper} J.~L.,  {Bicknell} G.~V.,  {Sutherland} R.~S.,   {Bland-Hawthorn} J.,
  2009, \mn@doi [\apj] {10.1088/0004-637X/703/1/330}, \href
  {http://adsabs.harvard.edu/abs/2009ApJ...703..330C} {703, 330}

\bibitem[\protect\citeauthoryear{{Cottle}, {Scannapieco}  \&
  {Br{\"u}ggen}}{{Cottle} et~al.}{2018}]{2018ApJ...864...96C}
{Cottle} J.,  {Scannapieco} E.,   {Br{\"u}ggen} M.,  2018, \mn@doi [\apj]
  {10.3847/1538-4357/aad55c}, \href
  {http://adsabs.harvard.edu/abs/2018ApJ...864...96C} {864, 96}

\bibitem[\protect\citeauthoryear{{Cottle}, {Scannapieco}, {Br{\"u}ggen},
  {Banda-Barrag{\'a}n}  \& {Federrath}}{{Cottle}
  et~al.}{2020}]{2020ApJ...892...59C}
{Cottle} J.,  {Scannapieco} E.,  {Br{\"u}ggen} M.,  {Banda-Barrag{\'a}n} W.,
  {Federrath} C.,  2020, \mn@doi [\apj] {10.3847/1538-4357/ab76d1}, \href
  {https://ui.adsabs.harvard.edu/abs/2020ApJ...892...59C} {892, 59}

\bibitem[\protect\citeauthoryear{{Coulais}}{{Coulais}}{2019}]{2019ASPC..523..365C}
{Coulais} A.,  2019, {GDL - GNU Data Language 0.9.9}.
p.~365

\bibitem[\protect\citeauthoryear{{Dawson}, {Mizuno}, {Onishi},
  {McClure-Griffiths}  \& {Fukui}}{{Dawson} et~al.}{2008}]{2008MNRAS.387...31D}
{Dawson} J.~R.,  {Mizuno} N.,  {Onishi} T.,  {McClure-Griffiths} N.~M.,
  {Fukui} Y.,  2008, \mn@doi [\mnras] {10.1111/j.1365-2966.2008.13152.x}, \href
  {https://ui.adsabs.harvard.edu/abs/2008MNRAS.387...31D} {387, 31}

\bibitem[\protect\citeauthoryear{{Di Teodoro}, {McClure-Griffiths}, {Lockman},
  {Denbo}, {Endsley}, {Ford}  \& {Harrington}}{{Di Teodoro}
  et~al.}{2018}]{2018ApJ...855...33D}
{Di Teodoro} E.~M.,  {McClure-Griffiths} N.~M.,  {Lockman} F.~J.,  {Denbo}
  S.~R.,  {Endsley} R.,  {Ford} H.~A.,   {Harrington} K.,  2018, \mn@doi [\apj]
  {10.3847/1538-4357/aaad6a}, \href
  {https://ui.adsabs.harvard.edu/abs/2018ApJ...855...33D} {855, 33}

\bibitem[\protect\citeauthoryear{{Di Teodoro} et~al.,}{{Di Teodoro}
  et~al.}{2019}]{2019ApJ...885L..32D}
{Di Teodoro} E.~M.,  et~al., 2019, \mn@doi [\apjl] {10.3847/2041-8213/ab4fe9},
  \href {https://ui.adsabs.harvard.edu/abs/2019ApJ...885L..32D} {885, L32}

\bibitem[\protect\citeauthoryear{{Dugan}, {Gaibler}, {Bieri}, {Silk}  \&
  {Rahman}}{{Dugan} et~al.}{2017}]{2017ApJ...839..103D}
{Dugan} Z.,  {Gaibler} V.,  {Bieri} R.,  {Silk} J.,   {Rahman} M.,  2017,
  \mn@doi [\apj] {10.3847/1538-4357/aa6984}, \href
  {http://adsabs.harvard.edu/abs/2017ApJ...839..103D} {839, 103}

\bibitem[\protect\citeauthoryear{{Elmegreen} \& {Scalo}}{{Elmegreen} \&
  {Scalo}}{2004}]{2004ARA&A..42..211E}
{Elmegreen} B.~G.,  {Scalo} J.,  2004, \mn@doi [\araa]
  {10.1146/annurev.astro.41.011802.094859}, \href
  {http://adsabs.harvard.edu/abs/2004ARA%26A..42..211E} {42, 211}

\bibitem[\protect\citeauthoryear{{Federrath}, {Klessen}  \&
  {Schmidt}}{{Federrath} et~al.}{2008}]{2008ApJ...688L..79F}
{Federrath} C.,  {Klessen} R.~S.,   {Schmidt} W.,  2008, \mn@doi [\apjl]
  {10.1086/595280}, \href {http://adsabs.harvard.edu/abs/2008ApJ...688L..79F}
  {688, L79}

\bibitem[\protect\citeauthoryear{{Federrath}, {Klessen}  \&
  {Schmidt}}{{Federrath} et~al.}{2009}]{2009ApJ...692..364F}
{Federrath} C.,  {Klessen} R.~S.,   {Schmidt} W.,  2009, \mn@doi [\apj]
  {10.1088/0004-637X/692/1/364}, \href
  {http://adsabs.harvard.edu/abs/2009ApJ...692..364F} {692, 364}

\bibitem[\protect\citeauthoryear{{Federrath}, {Roman-Duval}, {Klessen},
  {Schmidt}  \& {Mac Low}}{{Federrath} et~al.}{2010}]{2010A&A...512A..81F}
{Federrath} C.,  {Roman-Duval} J.,  {Klessen} R.~S.,  {Schmidt} W.,   {Mac Low}
  M.-M.,  2010, \mn@doi [\aap] {10.1051/0004-6361/200912437}, \href
  {http://adsabs.harvard.edu/abs/2010A%26A...512A..81F} {512, A81}

\bibitem[\protect\citeauthoryear{{Fesen} et~al.,}{{Fesen}
  et~al.}{2006}]{2006ApJ...636..859F}
{Fesen} R.~A.,  et~al., 2006, \mn@doi [\apj] {10.1086/498092}, \href
  {https://ui.adsabs.harvard.edu/abs/2006ApJ...636..859F} {636, 859}

\bibitem[\protect\citeauthoryear{{Forbes} \& {Lin}}{{Forbes} \&
  {Lin}}{2019}]{2019AJ....158..124F}
{Forbes} J.~C.,  {Lin} D. N.~C.,  2019, \mn@doi [\aj]
  {10.3847/1538-3881/ab3230}, \href
  {https://ui.adsabs.harvard.edu/abs/2019AJ....158..124F} {158, 124}

\bibitem[\protect\citeauthoryear{{Goldsmith} \& {Pittard}}{{Goldsmith} \&
  {Pittard}}{2017}]{2017MNRAS.470.2427G}
{Goldsmith} K.~J.~A.,  {Pittard} J.~M.,  2017, \mn@doi [\mnras]
  {10.1093/mnras/stx1431}, \href
  {http://adsabs.harvard.edu/abs/2017MNRAS.470.2427G} {470, 2427}

\bibitem[\protect\citeauthoryear{{Goldsmith} \& {Pittard}}{{Goldsmith} \&
  {Pittard}}{2018}]{2018MNRAS.476.2209G}
{Goldsmith} K.~J.~A.,  {Pittard} J.~M.,  2018, \mn@doi [\mnras]
  {10.1093/mnras/sty401}, \href
  {http://adsabs.harvard.edu/abs/2018MNRAS.476.2209G} {476, 2209}

\bibitem[\protect\citeauthoryear{{Gronke} \& {Oh}}{{Gronke} \&
  {Oh}}{2018}]{2018MNRAS.480L.111G}
{Gronke} M.,  {Oh} S.~P.,  2018, \mn@doi [\mnras] {10.1093/mnrasl/sly131},
  \href {http://adsabs.harvard.edu/abs/2018MNRAS.480L.111G} {480, L111}

\bibitem[\protect\citeauthoryear{{Gr{\o}nnow}, {Tepper-Garc{\'{\i}}a},
  {Bland-Hawthorn}  \& {McClure-Griffiths}}{{Gr{\o}nnow}
  et~al.}{2017}]{2017ApJ...845...69G}
{Gr{\o}nnow} A.,  {Tepper-Garc{\'{\i}}a} T.,  {Bland-Hawthorn} J.,
  {McClure-Griffiths} N.~M.,  2017, \mn@doi [\apj] {10.3847/1538-4357/aa7ed2},
  \href {http://adsabs.harvard.edu/abs/2017ApJ...845...69G} {845, 69}

\bibitem[\protect\citeauthoryear{{Gr{\o}nnow}, {Tepper-Garc{\'{\i}}a}  \&
  {Bland-Hawthorn}}{{Gr{\o}nnow} et~al.}{2018}]{2018ApJ...865...64G}
{Gr{\o}nnow} A.,  {Tepper-Garc{\'{\i}}a} T.,   {Bland-Hawthorn} J.,  2018,
  \mn@doi [\apj] {10.3847/1538-4357/aada0e}, \href
  {http://adsabs.harvard.edu/abs/2018ApJ...865...64G} {865, 64}

\bibitem[\protect\citeauthoryear{{Hill}, {Joung}, {Mac Low}, {Benjamin},
  {Haffner}, {Klingenberg}  \& {Waagan}}{{Hill}
  et~al.}{2012}]{2012ApJ...750..104H}
{Hill} A.~S.,  {Joung} M.~R.,  {Mac Low} M.-M.,  {Benjamin} R.~A.,  {Haffner}
  L.~M.,  {Klingenberg} C.,   {Waagan} K.,  2012, \mn@doi [\apj]
  {10.1088/0004-637X/750/2/104}, \href
  {https://ui.adsabs.harvard.edu/abs/2012ApJ...750..104H} {750, 104}

\bibitem[\protect\citeauthoryear{{J{\'a}chym} et~al.,}{{J{\'a}chym}
  et~al.}{2019}]{2019ApJ...883..145J}
{J{\'a}chym} P.,  et~al., 2019, \mn@doi [\apj] {10.3847/1538-4357/ab3e6c},
  \href {https://ui.adsabs.harvard.edu/abs/2019ApJ...883..145J} {883, 145}

\bibitem[\protect\citeauthoryear{{Jun}, {Jones}  \& {Norman}}{{Jun}
  et~al.}{1996}]{1996ApJ...468L..59J}
{Jun} B.-I.,  {Jones} T.~W.,   {Norman} M.~L.,  1996, \mn@doi [\apjl]
  {10.1086/310224}, \href
  {https://ui.adsabs.harvard.edu/abs/1996ApJ...468L..59J} {468, L59}

\bibitem[\protect\citeauthoryear{{Kim} \& {Ostriker}}{{Kim} \&
  {Ostriker}}{2018}]{2018ApJ...853..173K}
{Kim} C.-G.,  {Ostriker} E.~C.,  2018, \mn@doi [\apj]
  {10.3847/1538-4357/aaa5ff}, \href
  {https://ui.adsabs.harvard.edu/abs/2018ApJ...853..173K} {853, 173}

\bibitem[\protect\citeauthoryear{{Klein}, {McKee}  \& {Colella}}{{Klein}
  et~al.}{1994}]{1994ApJ...420..213K}
{Klein} R.~I.,  {McKee} C.~F.,   {Colella} P.,  1994, \mn@doi [\apj]
  {10.1086/173554}, \href {http://adsabs.harvard.edu/abs/1994ApJ...420..213K}
  {420, 213}

\bibitem[\protect\citeauthoryear{{Krieger} et~al.,}{{Krieger}
  et~al.}{2019}]{2019ApJ...881...43K}
{Krieger} N.,  et~al., 2019, \mn@doi [\apj] {10.3847/1538-4357/ab2d9c}, \href
  {https://ui.adsabs.harvard.edu/abs/2019ApJ...881...43K} {881, 43}

\bibitem[\protect\citeauthoryear{{Krumholz} \& {Federrath}}{{Krumholz} \&
  {Federrath}}{2019}]{2019FrASS...6....7K}
{Krumholz} M.~R.,  {Federrath} C.,  2019, \mn@doi [Frontiers in Astronomy and
  Space Sciences] {10.3389/fspas.2019.00007}, \href
  {https://ui.adsabs.harvard.edu/abs/2019FrASS...6....7K} {6, 7}

\bibitem[\protect\citeauthoryear{{Landau} \& {Lifshitz}}{{Landau} \&
  {Lifshitz}}{1987}]{1987flme.book.....L}
{Landau} L.~D.,  {Lifshitz} E.~M.,  1987, {Fluid Mechanics}

\bibitem[\protect\citeauthoryear{{Leaman} et~al.,}{{Leaman}
  et~al.}{2019}]{2019MNRAS.488.3904L}
{Leaman} R.,  et~al., 2019, \mn@doi [\mnras] {10.1093/mnras/stz1844}, \href
  {https://ui.adsabs.harvard.edu/abs/2019MNRAS.488.3904L} {488, 3904}

\bibitem[\protect\citeauthoryear{{Lehmann}, {Federrath}  \& {Wardle}}{{Lehmann}
  et~al.}{2016}]{2016MNRAS.463.1026L}
{Lehmann} A.,  {Federrath} C.,   {Wardle} M.,  2016, \mn@doi [\mnras]
  {10.1093/mnras/stw2015}, \href
  {https://ui.adsabs.harvard.edu/abs/2016MNRAS.463.1026L} {463, 1026}

\bibitem[\protect\citeauthoryear{{Leroy} et~al.,}{{Leroy}
  et~al.}{2015}]{2015ApJ...814...83L}
{Leroy} A.~K.,  et~al., 2015, \mn@doi [\apj] {10.1088/0004-637X/814/2/83},
  \href {https://ui.adsabs.harvard.edu/abs/2015ApJ...814...83L} {814, 83}

\bibitem[\protect\citeauthoryear{{Li}, {Frank}  \& {Blackman}}{{Li}
  et~al.}{2014}]{2014MNRAS.444.2884L}
{Li} S.,  {Frank} A.,   {Blackman} E.~G.,  2014, \mn@doi [\mnras]
  {10.1093/mnras/stu1571}, \href
  {https://ui.adsabs.harvard.edu/abs/2014MNRAS.444.2884L} {444, 2884}

\bibitem[\protect\citeauthoryear{{Li}, {Hopkins}, {Squire}  \& {Hummels}}{{Li}
  et~al.}{2020}]{2020MNRAS.492.1841L}
{Li} Z.,  {Hopkins} P.~F.,  {Squire} J.,   {Hummels} C.,  2020, \mn@doi
  [\mnras] {10.1093/mnras/stz3567}, \href
  {https://ui.adsabs.harvard.edu/abs/2020MNRAS.492.1841L} {492, 1841}

\bibitem[\protect\citeauthoryear{{Liang} \& {Remming}}{{Liang} \&
  {Remming}}{2020}]{2020MNRAS.491.5056L}
{Liang} C.~J.,  {Remming} I.,  2020, \mn@doi [\mnras] {10.1093/mnras/stz3403},
  \href {https://ui.adsabs.harvard.edu/abs/2020MNRAS.491.5056L} {491, 5056}

\bibitem[\protect\citeauthoryear{{Lockman} \& {McClure-Griffiths}}{{Lockman} \&
  {McClure-Griffiths}}{2016}]{2016ApJ...826..215L}
{Lockman} F.~J.,  {McClure-Griffiths} N.~M.,  2016, \mn@doi [\apj]
  {10.3847/0004-637X/826/2/215}, \href
  {http://adsabs.harvard.edu/abs/2016ApJ...826..215L} {826, 215}

\bibitem[\protect\citeauthoryear{{Lockman}, {Di Teodoro}  \&
  {McClure-Griffiths}}{{Lockman} et~al.}{2020}]{2020ApJ...888...51L}
{Lockman} F.~J.,  {Di Teodoro} E.~M.,   {McClure-Griffiths} N.~M.,  2020,
  \mn@doi [\apj] {10.3847/1538-4357/ab55d8}, \href
  {https://ui.adsabs.harvard.edu/abs/2020ApJ...888...51L} {888, 51}

\bibitem[\protect\citeauthoryear{{Mac Low} \& {Klessen}}{{Mac Low} \&
  {Klessen}}{2004}]{2004RvMP...76..125M}
{Mac Low} M.-M.,  {Klessen} R.~S.,  2004, \mn@doi [Reviews of Modern Physics]
  {10.1103/RevModPhys.76.125}, \href
  {http://adsabs.harvard.edu/abs/2004RvMP...76..125M} {76, 125}

\bibitem[\protect\citeauthoryear{{Mac Low}, {McKee}, {Klein}, {Stone}  \&
  {Norman}}{{Mac Low} et~al.}{1994}]{1994ApJ...433..757M}
{Mac Low} M.-M.,  {McKee} C.~F.,  {Klein} R.~I.,  {Stone} J.~M.,   {Norman}
  M.~L.,  1994, \mn@doi [\apj] {10.1086/174685}, \href
  {http://adsabs.harvard.edu/abs/1994ApJ...433..757M} {433, 757}

\bibitem[\protect\citeauthoryear{{Mandelker}, {van Dokkum}, {Brodie}, {van den
  Bosch}  \& {Ceverino}}{{Mandelker} et~al.}{2018}]{2018ApJ...861..148M}
{Mandelker} N.,  {van Dokkum} P.~G.,  {Brodie} J.~P.,  {van den Bosch} F.~C.,
  {Ceverino} D.,  2018, \mn@doi [\apj] {10.3847/1538-4357/aaca98}, \href
  {https://ui.adsabs.harvard.edu/abs/2018ApJ...861..148M} {861, 148}

\bibitem[\protect\citeauthoryear{{Mandelker}, {Nagai}, {Aung}, {Dekel},
  {Birnboim}  \& {van den Bosch}}{{Mandelker}
  et~al.}{2020}]{2020MNRAS.494.2641M}
{Mandelker} N.,  {Nagai} D.,  {Aung} H.,  {Dekel} A.,  {Birnboim} Y.,   {van
  den Bosch} F.~C.,  2020, \mn@doi [\mnras] {10.1093/mnras/staa812}, \href
  {https://ui.adsabs.harvard.edu/abs/2020MNRAS.494.2641M} {494, 2641}

\bibitem[\protect\citeauthoryear{{Marcolini}, {Strickland}, {D'Ercole},
  {Heckman}  \& {Hoopes}}{{Marcolini} et~al.}{2005}]{2005MNRAS.362..626M}
{Marcolini} A.,  {Strickland} D.~K.,  {D'Ercole} A.,  {Heckman} T.~M.,
  {Hoopes} C.~G.,  2005, \mn@doi [\mnras] {10.1111/j.1365-2966.2005.09343.x},
  \href {http://adsabs.harvard.edu/abs/2005MNRAS.362..626M} {362, 626}

\bibitem[\protect\citeauthoryear{{Martin}, {Dijkstra}, {Henry}, {Soto},
  {Danforth}  \& {Wong}}{{Martin} et~al.}{2015}]{2015ApJ...803....6M}
{Martin} C.~L.,  {Dijkstra} M.,  {Henry} A.,  {Soto} K.~T.,  {Danforth} C.~W.,
   {Wong} J.,  2015, \mn@doi [\apj] {10.1088/0004-637X/803/1/6}, \href
  {https://ui.adsabs.harvard.edu/abs/2015ApJ...803....6M} {803, 6}

\bibitem[\protect\citeauthoryear{{Martin} et~al.,}{{Martin}
  et~al.}{2019}]{2019NatAs...3..822M}
{Martin} D.~C.,  et~al., 2019, \mn@doi [Nature Astronomy]
  {10.1038/s41550-019-0791-2}, \href
  {https://ui.adsabs.harvard.edu/abs/2019NatAs...3..822M} {3, 822}

\bibitem[\protect\citeauthoryear{{Matsubayashi}, {Sugai}, {Hattori}, {Kawai},
  {Ozaki}, {Kosugi}, {Ishigaki}  \& {Shimono}}{{Matsubayashi}
  et~al.}{2009}]{2009ApJ...701.1636M}
{Matsubayashi} K.,  {Sugai} H.,  {Hattori} T.,  {Kawai} A.,  {Ozaki} S.,
  {Kosugi} G.,  {Ishigaki} T.,   {Shimono} A.,  2009, \mn@doi [\apj]
  {10.1088/0004-637X/701/2/1636}, \href
  {http://adsabs.harvard.edu/abs/2009ApJ...701.1636M} {701, 1636}

\bibitem[\protect\citeauthoryear{{McClure-Griffiths}, {Ford}, {Pisano},
  {Gibson}, {Staveley-Smith}, {Calabretta}, {Dedes}  \&
  {Kalberla}}{{McClure-Griffiths} et~al.}{2006}]{2006ApJ...638..196M}
{McClure-Griffiths} N.~M.,  {Ford} A.,  {Pisano} D.~J.,  {Gibson} B.~K.,
  {Staveley-Smith} L.,  {Calabretta} M.~R.,  {Dedes} L.,   {Kalberla} P.~M.~W.,
   2006, \mn@doi [\apj] {10.1086/498706}, \href
  {https://ui.adsabs.harvard.edu/abs/2006ApJ...638..196M} {638, 196}

\bibitem[\protect\citeauthoryear{{McClure-Griffiths}, {Green}, {Hill},
  {Lockman}, {Dickey}, {Gaensler}  \& {Green}}{{McClure-Griffiths}
  et~al.}{2013}]{2013ApJ...770L...4M}
{McClure-Griffiths} N.~M.,  {Green} J.~A.,  {Hill} A.~S.,  {Lockman} F.~J.,
  {Dickey} J.~M.,  {Gaensler} B.~M.,   {Green} A.~J.,  2013, \mn@doi [\apjl]
  {10.1088/2041-8205/770/1/L4}, \href
  {http://adsabs.harvard.edu/abs/2013ApJ...770L...4M} {770, L4}

\bibitem[\protect\citeauthoryear{{McClure-Griffiths}
  et~al.,}{{McClure-Griffiths} et~al.}{2018}]{2018NatAs...2..901M}
{McClure-Griffiths} N.~M.,  et~al., 2018, \mn@doi [Nature Astronomy]
  {10.1038/s41550-018-0608-8}, \href
  {http://adsabs.harvard.edu/abs/2018NatAs...2..901M} {2, 901}

\bibitem[\protect\citeauthoryear{{McCourt}, {O'Leary}, {Madigan}  \&
  {Quataert}}{{McCourt} et~al.}{2015}]{2015MNRAS.449....2M}
{McCourt} M.,  {O'Leary} R.~M.,  {Madigan} A.-M.,   {Quataert} E.,  2015,
  \mn@doi [\mnras] {10.1093/mnras/stv355}, \href
  {http://adsabs.harvard.edu/abs/2015MNRAS.449....2M} {449, 2}

\bibitem[\protect\citeauthoryear{{McCourt}, {Oh}, {O'Leary}  \&
  {Madigan}}{{McCourt} et~al.}{2018}]{2018MNRAS.473.5407M}
{McCourt} M.,  {Oh} S.~P.,  {O'Leary} R.,   {Madigan} A.-M.,  2018, \mn@doi
  [\mnras] {10.1093/mnras/stx2687}, \href
  {http://adsabs.harvard.edu/abs/2018MNRAS.473.5407M} {473, 5407}

\bibitem[\protect\citeauthoryear{{McKee} \& {Ostriker}}{{McKee} \&
  {Ostriker}}{2007}]{2007ARA&A..45..565M}
{McKee} C.~F.,  {Ostriker} E.~C.,  2007, \mn@doi [\araa]
  {10.1146/annurev.astro.45.051806.110602}, \href
  {https://ui.adsabs.harvard.edu/abs/2007ARA&A..45..565M} {45, 565}

\bibitem[\protect\citeauthoryear{{Melioli}, {de Gouveia Dal Pino}  \&
  {Geraissate}}{{Melioli} et~al.}{2013}]{2013MNRAS.430.3235M}
{Melioli} C.,  {de Gouveia Dal Pino} E.~M.,   {Geraissate} F.~G.,  2013,
  \mn@doi [\mnras] {10.1093/mnras/stt126}, \href
  {http://adsabs.harvard.edu/abs/2013MNRAS.430.3235M} {430, 3235}

\bibitem[\protect\citeauthoryear{{Mignone}, {Bodo}, {Massaglia}, {Matsakos},
  {Tesileanu}, {Zanni}  \& {Ferrari}}{{Mignone}
  et~al.}{2007}]{2007ApJS..170..228M}
{Mignone} A.,  {Bodo} G.,  {Massaglia} S.,  {Matsakos} T.,  {Tesileanu} O.,
  {Zanni} C.,   {Ferrari} A.,  2007, \mn@doi [\apjs] {10.1086/513316}, \href
  {http://adsabs.harvard.edu/abs/2007ApJS..170..228M} {170, 228}

\bibitem[\protect\citeauthoryear{{Mukherjee}, {Bicknell}, {Sutherland }  \&
  {Wagner}}{{Mukherjee} et~al.}{2016}]{2016MNRAS.461..967M}
{Mukherjee} D.,  {Bicknell} G.~V.,  {Sutherland } R.,   {Wagner} A.,  2016,
  \mn@doi [\mnras] {10.1093/mnras/stw1368}, \href
  {https://ui.adsabs.harvard.edu/abs/2016MNRAS.461..967M} {461, 967}

\bibitem[\protect\citeauthoryear{{Nakamura}, {McKee}, {Klein}  \&
  {Fisher}}{{Nakamura} et~al.}{2006}]{2006ApJS..164..477N}
{Nakamura} F.,  {McKee} C.~F.,  {Klein} R.~I.,   {Fisher} R.~T.,  2006, \mn@doi
  [\apjs] {10.1086/501530}, \href
  {http://adsabs.harvard.edu/abs/2006ApJS..164..477N} {164, 477}

\bibitem[\protect\citeauthoryear{{Orlando}, {Peres}, {Reale}, {Bocchino},
  {Rosner}, {Plewa}  \& {Siegel}}{{Orlando} et~al.}{2005}]{2005A&A...444..505O}
{Orlando} S.,  {Peres} G.,  {Reale} F.,  {Bocchino} F.,  {Rosner} R.,  {Plewa}
  T.,   {Siegel} A.,  2005, \mn@doi [\aap] {10.1051/0004-6361:20052896}, \href
  {http://adsabs.harvard.edu/abs/2005A%26A...444..505O} {444, 505}

\bibitem[\protect\citeauthoryear{{Padoan}, {Jones}  \& {Nordlund}}{{Padoan}
  et~al.}{1997}]{1997ApJ...474..730P}
{Padoan} P.,  {Jones} B. J.~T.,   {Nordlund} {\r{A}}.~P.,  1997, \mn@doi [\apj]
  {10.1086/303482}, \href
  {https://ui.adsabs.harvard.edu/abs/1997ApJ...474..730P} {474, 730}

\bibitem[\protect\citeauthoryear{{Padoan}, {Federrath}, {Chabrier}, {Evans},
  {Johnstone}, {J{\o}rgensen}, {McKee}  \& {Nordlund}}{{Padoan}
  et~al.}{2014}]{2014prpl.conf...77P}
{Padoan} P.,  {Federrath} C.,  {Chabrier} G.,  {Evans} N.~J. I.,  {Johnstone}
  D.,  {J{\o}rgensen} J.~K.,  {McKee} C.~F.,   {Nordlund} {\r{A}}.,  2014, in
  {Beuther} H.,  {Klessen} R.~S.,  {Dullemond} C.~P.,   {Henning} T.,  eds,
  Protostars and Planets VI. p.~77 (\mn@eprint {arXiv} {1312.5365}),
  \mn@doi{10.2458/azu_uapress_9780816531240-ch004}

\bibitem[\protect\citeauthoryear{{Passot} \& {V{\'a}zquez-Semadeni}}{{Passot}
  \& {V{\'a}zquez-Semadeni}}{1998}]{1998PhRvE..58.4501P}
{Passot} T.,  {V{\'a}zquez-Semadeni} E.,  1998, \mn@doi [\pre]
  {10.1103/PhysRevE.58.4501}, \href
  {http://adsabs.harvard.edu/abs/1998PhRvE..58.4501P} {58, 4501}

\bibitem[\protect\citeauthoryear{{Patnaude} \& {Fesen}}{{Patnaude} \&
  {Fesen}}{2005}]{2005ApJ...633..240P}
{Patnaude} D.~J.,  {Fesen} R.~A.,  2005, \mn@doi [\apj] {10.1086/452627}, \href
  {http://adsabs.harvard.edu/abs/2005ApJ...633..240P} {633, 240}

\bibitem[\protect\citeauthoryear{{Pittard} \& {Parkin}}{{Pittard} \&
  {Parkin}}{2016}]{2016MNRAS.457.4470P}
{Pittard} J.~M.,  {Parkin} E.~R.,  2016, \mn@doi [\mnras]
  {10.1093/mnras/stw025}, \href
  {http://adsabs.harvard.edu/abs/2016MNRAS.457.4470P} {457, 4470}

\bibitem[\protect\citeauthoryear{{Pittard}, {Dyson}, {Falle}  \&
  {Hartquist}}{{Pittard} et~al.}{2005}]{2005MNRAS.361.1077P}
{Pittard} J.~M.,  {Dyson} J.~E.,  {Falle} S.~A.~E.~G.,   {Hartquist} T.~W.,
  2005, \mn@doi [\mnras] {10.1111/j.1365-2966.2005.09268.x}, \href
  {http://adsabs.harvard.edu/abs/2005MNRAS.361.1077P} {361, 1077}

\bibitem[\protect\citeauthoryear{{Pittard}, {Falle}, {Hartquist}  \&
  {Dyson}}{{Pittard} et~al.}{2009}]{2009MNRAS.394.1351P}
{Pittard} J.~M.,  {Falle} S.~A.~E.~G.,  {Hartquist} T.~W.,   {Dyson} J.~E.,
  2009, \mn@doi [\mnras] {10.1111/j.1365-2966.2009.13759.x}, \href
  {https://ui.adsabs.harvard.edu/abs/2009MNRAS.394.1351P} {394, 1351}

\bibitem[\protect\citeauthoryear{{Poludnenko}, {Frank}  \&
  {Blackman}}{{Poludnenko} et~al.}{2002}]{2002ApJ...576..832P}
{Poludnenko} A.~Y.,  {Frank} A.,   {Blackman} E.~G.,  2002, \mn@doi [\apj]
  {10.1086/341886}, \href {http://adsabs.harvard.edu/abs/2002ApJ...576..832P}
  {576, 832}

\bibitem[\protect\citeauthoryear{{Salak}, {Tomiyasu}, {Nakai}, {Kuno},
  {Miyamoto}  \& {Kaneko}}{{Salak} et~al.}{2017}]{2017ApJ...849...90S}
{Salak} D.,  {Tomiyasu} Y.,  {Nakai} N.,  {Kuno} N.,  {Miyamoto} Y.,   {Kaneko}
  H.,  2017, \mn@doi [\apj] {10.3847/1538-4357/aa91cb}, \href
  {https://ui.adsabs.harvard.edu/abs/2017ApJ...849...90S} {849, 90}

\bibitem[\protect\citeauthoryear{{Salak}, {Tomiyasu}, {Nakai}, {Kuno},
  {Miyamoto}  \& {Kaneko}}{{Salak} et~al.}{2018}]{2018ApJ...856...97S}
{Salak} D.,  {Tomiyasu} Y.,  {Nakai} N.,  {Kuno} N.,  {Miyamoto} Y.,   {Kaneko}
  H.,  2018, \mn@doi [\apj] {10.3847/1538-4357/aab2ac}, \href
  {https://ui.adsabs.harvard.edu/abs/2018ApJ...856...97S} {856, 97}

\bibitem[\protect\citeauthoryear{{Sasaki}, {Breitschwerdt}, {Baumgartner}  \&
  {Haberl}}{{Sasaki} et~al.}{2011}]{2011A&A...528A.136S}
{Sasaki} M.,  {Breitschwerdt} D.,  {Baumgartner} V.,   {Haberl} F.,  2011,
  \mn@doi [\aap] {10.1051/0004-6361/201015866}, \href
  {https://ui.adsabs.harvard.edu/abs/2011A&A...528A.136S} {528, A136}

\bibitem[\protect\citeauthoryear{{Scannapieco} \& {Br{\"u}ggen}}{{Scannapieco}
  \& {Br{\"u}ggen}}{2015}]{2015ApJ...805..158S}
{Scannapieco} E.,  {Br{\"u}ggen} M.,  2015, \mn@doi [\apj]
  {10.1088/0004-637X/805/2/158}, \href
  {http://adsabs.harvard.edu/abs/2015ApJ...805..158S} {805, 158}

\bibitem[\protect\citeauthoryear{{Schneider} \& {Robertson}}{{Schneider} \&
  {Robertson}}{2015}]{2015ApJS..217...24S}
{Schneider} E.~E.,  {Robertson} B.~E.,  2015, \mn@doi [\apjs]
  {10.1088/0067-0049/217/2/24}, \href
  {http://adsabs.harvard.edu/abs/2015ApJS..217...24S} {217, 24}

\bibitem[\protect\citeauthoryear{{Schneider} \& {Robertson}}{{Schneider} \&
  {Robertson}}{2017}]{2017ApJ...834..144S}
{Schneider} E.~E.,  {Robertson} B.~E.,  2017, \mn@doi [\apj]
  {10.3847/1538-4357/834/2/144}, \href
  {http://adsabs.harvard.edu/abs/2017ApJ...834..144S} {834, 144}

\bibitem[\protect\citeauthoryear{{Schneider}, {Robertson}  \&
  {Thompson}}{{Schneider} et~al.}{2018}]{2018ApJ...862...56S}
{Schneider} E.~E.,  {Robertson} B.~E.,   {Thompson} T.~A.,  2018, \mn@doi
  [\apj] {10.3847/1538-4357/aacce1}, \href
  {http://adsabs.harvard.edu/abs/2018ApJ...862...56S} {862, 56}

\bibitem[\protect\citeauthoryear{{Shopbell} \& {Bland-Hawthorn}}{{Shopbell} \&
  {Bland-Hawthorn}}{1998}]{1998ApJ...493..129S}
{Shopbell} P.~L.,  {Bland-Hawthorn} J.,  1998, \mn@doi [\apj] {10.1086/305108},
  \href {http://adsabs.harvard.edu/abs/1998ApJ...493..129S} {493, 129}

\bibitem[\protect\citeauthoryear{{Sparre}, {Pfrommer}  \&
  {Vogelsberger}}{{Sparre} et~al.}{2019}]{2019MNRAS.482.5401S}
{Sparre} M.,  {Pfrommer} C.,   {Vogelsberger} M.,  2019, \mn@doi [\mnras]
  {10.1093/mnras/sty3063}, \href
  {https://ui.adsabs.harvard.edu/abs/2019MNRAS.482.5401S} {482, 5401}

\bibitem[\protect\citeauthoryear{{Strickland} \& {Heckman}}{{Strickland} \&
  {Heckman}}{2009}]{2009ApJ...697.2030S}
{Strickland} D.~K.,  {Heckman} T.~M.,  2009, \mn@doi [\apj]
  {10.1088/0004-637X/697/2/2030}, \href
  {http://adsabs.harvard.edu/abs/2009ApJ...697.2030S} {697, 2030}

\bibitem[\protect\citeauthoryear{{Sutherland} \& {Bicknell}}{{Sutherland} \&
  {Bicknell}}{2007}]{2007ApJS..173...37S}
{Sutherland} R.~S.,  {Bicknell} G.~V.,  2007, \mn@doi [\apjs] {10.1086/520640},
  \href {http://adsabs.harvard.edu/abs/2007ApJS..173...37S} {173, 37}

\bibitem[\protect\citeauthoryear{{Terebey}, {Fich}, {Taylor}, {Cao}  \&
  {Hancock}}{{Terebey} et~al.}{2003}]{2003ApJ...590..906T}
{Terebey} S.,  {Fich} M.,  {Taylor} R.,  {Cao} Y.,   {Hancock} T.,  2003,
  \mn@doi [\apj] {10.1086/375160}, \href
  {https://ui.adsabs.harvard.edu/abs/2003ApJ...590..906T} {590, 906}

\bibitem[\protect\citeauthoryear{{Thompson}, {Quataert}, {Zhang}  \&
  {Weinberg}}{{Thompson} et~al.}{2016}]{2016MNRAS.455.1830T}
{Thompson} T.~A.,  {Quataert} E.,  {Zhang} D.,   {Weinberg} D.~H.,  2016,
  \mn@doi [\mnras] {10.1093/mnras/stv2428}, \href
  {https://ui.adsabs.harvard.edu/abs/2016MNRAS.455.1830T} {455, 1830}

\bibitem[\protect\citeauthoryear{{Toro}, {Spruce}  \& {Speares}}{{Toro}
  et~al.}{1994}]{Toro:1994}
{Toro} E.~F.,  {Spruce} M.,   {Speares} W.,  1994, \mn@doi [Shock Waves]
  {10.1007/BF01414629}, \href
  {http://adsabs.harvard.edu/abs/1994ShWav...4...25T} {4, 25}

\bibitem[\protect\citeauthoryear{{Tremblin} et~al.,}{{Tremblin}
  et~al.}{2014}]{2014A&A...564A.106T}
{Tremblin} P.,  et~al., 2014, \mn@doi [\aap] {10.1051/0004-6361/201322700},
  \href {https://ui.adsabs.harvard.edu/abs/2014A&A...564A.106T} {564, A106}

\bibitem[\protect\citeauthoryear{{Tripp} et~al.,}{{Tripp}
  et~al.}{2011}]{2011Sci...334..952T}
{Tripp} T.~M.,  et~al., 2011, \mn@doi [Science] {10.1126/science.1209850},
  \href {https://ui.adsabs.harvard.edu/abs/2011Sci...334..952T} {334, 952}

\bibitem[\protect\citeauthoryear{{Vazza}, {Dolag}, {Ryu}, {Brunetti},
  {Gheller}, {Kang}  \& {Pfrommer}}{{Vazza} et~al.}{2011}]{2011MNRAS.418..960V}
{Vazza} F.,  {Dolag} K.,  {Ryu} D.,  {Brunetti} G.,  {Gheller} C.,  {Kang} H.,
   {Pfrommer} C.,  2011, \mn@doi [\mnras] {10.1111/j.1365-2966.2011.19546.x},
  \href {https://ui.adsabs.harvard.edu/abs/2011MNRAS.418..960V} {418, 960}

\bibitem[\protect\citeauthoryear{{Veilleux}, {Cecil}  \&
  {Bland-Hawthorn}}{{Veilleux} et~al.}{2005}]{2005ARA&A..43..769V}
{Veilleux} S.,  {Cecil} G.,   {Bland-Hawthorn} J.,  2005, \mn@doi [\araa]
  {10.1146/annurev.astro.43.072103.150610}, \href
  {http://adsabs.harvard.edu/abs/2005ARA%26A..43..769V} {43, 769}

\bibitem[\protect\citeauthoryear{{Villagran}, {Vel{\'a}zquez}, {G{\'o}mez}  \&
  {Giacani}}{{Villagran} et~al.}{2020}]{2020MNRAS.491.2855V}
{Villagran} M.~A.,  {Vel{\'a}zquez} P.~F.,  {G{\'o}mez} D.~O.,   {Giacani}
  E.~B.,  2020, \mn@doi [\mnras] {10.1093/mnras/stz2811}, \href
  {https://ui.adsabs.harvard.edu/abs/2020MNRAS.491.2855V} {491, 2855}

\bibitem[\protect\citeauthoryear{{Wagner}, {Bicknell}  \& {Umemura}}{{Wagner}
  et~al.}{2012}]{2012ApJ...757..136W}
{Wagner} A.~Y.,  {Bicknell} G.~V.,   {Umemura} M.,  2012, \mn@doi [\apj]
  {10.1088/0004-637X/757/2/136}, \href
  {http://adsabs.harvard.edu/abs/2012ApJ...757..136W} {757, 136}

\bibitem[\protect\citeauthoryear{{Wagner}, {Umemura}  \& {Bicknell}}{{Wagner}
  et~al.}{2013}]{2013ApJ...763L..18W}
{Wagner} A.~Y.,  {Umemura} M.,   {Bicknell} G.~V.,  2013, \mn@doi [\apjl]
  {10.1088/2041-8205/763/1/L18}, \href
  {https://ui.adsabs.harvard.edu/abs/2013ApJ...763L..18W} {763, L18}

\bibitem[\protect\citeauthoryear{{Walter} et~al.,}{{Walter}
  et~al.}{2017}]{2017ApJ...835..265W}
{Walter} F.,  et~al., 2017, \mn@doi [\apj] {10.3847/1538-4357/835/2/265}, \href
  {https://ui.adsabs.harvard.edu/abs/2017ApJ...835..265W} {835, 265}

\bibitem[\protect\citeauthoryear{{Werk}, {Prochaska}, {Thom}, {Tumlinson},
  {Tripp}, {O'Meara}  \& {Peeples}}{{Werk} et~al.}{2013}]{2013ApJS..204...17W}
{Werk} J.~K.,  {Prochaska} J.~X.,  {Thom} C.,  {Tumlinson} J.,  {Tripp} T.~M.,
  {O'Meara} J.~M.,   {Peeples} M.~S.,  2013, \mn@doi [\apjs]
  {10.1088/0067-0049/204/2/17}, \href
  {https://ui.adsabs.harvard.edu/abs/2013ApJS..204...17W} {204, 17}

\bibitem[\protect\citeauthoryear{{Werk} et~al.,}{{Werk}
  et~al.}{2016}]{2016ApJ...833...54W}
{Werk} J.~K.,  et~al., 2016, \mn@doi [\apj] {10.3847/1538-4357/833/1/54}, \href
  {https://ui.adsabs.harvard.edu/abs/2016ApJ...833...54W} {833, 54}

\bibitem[\protect\citeauthoryear{{Xu} \& {Stone}}{{Xu} \&
  {Stone}}{1995}]{1995ApJ...454..172X}
{Xu} J.,  {Stone} J.~M.,  1995, \mn@doi [\apj] {10.1086/176475}, \href
  {http://adsabs.harvard.edu/abs/1995ApJ...454..172X} {454, 172}

\bibitem[\protect\citeauthoryear{{Yirak}, {Frank}  \& {Cunningham}}{{Yirak}
  et~al.}{2010}]{2010ApJ...722..412Y}
{Yirak} K.,  {Frank} A.,   {Cunningham} A.~J.,  2010, \mn@doi [\apj]
  {10.1088/0004-637X/722/1/412}, \href
  {http://adsabs.harvard.edu/abs/2010ApJ...722..412Y} {722, 412}

\bibitem[\protect\citeauthoryear{{Zhang}}{{Zhang}}{2018}]{2018Galax...6..114Z}
{Zhang} D.,  2018, \mn@doi [Galaxies] {10.3390/galaxies6040114}, \href
  {http://adsabs.harvard.edu/abs/2018Galax...6..114Z} {6, 114}

\bibitem[\protect\citeauthoryear{{Zhang}, {Thompson}, {Quataert}  \&
  {Murray}}{{Zhang} et~al.}{2017}]{2017MNRAS.468.4801Z}
{Zhang} D.,  {Thompson} T.~A.,  {Quataert} E.,   {Murray} N.,  2017, \mn@doi
  [\mnras] {10.1093/mnras/stx822}, \href
  {http://adsabs.harvard.edu/abs/2017MNRAS.468.4801Z} {468, 4801}

\makeatother
\end{thebibliography}




\appendix
\section{Appendix}
\label{sec:App}
\begin{figure*}
\begin{center}
  \begin{tabular}{c c c c c c c}
       \multicolumn{1}{l}{\hspace{-2mm}a) sole-k8-M10-th \hspace{+3mm}$t_0$} & \multicolumn{1}{c}{$0.5\,t_{\rm sp}=0.05\,\rm Myr$} & \multicolumn{1}{c}{$1.1\,t_{\rm sp}=0.11\,\rm Myr$} & \multicolumn{1}{c}{$1.8\,t_{\rm sp}=0.18\,\rm Myr$} & \multicolumn{1}{c}{$2.4\,t_{\rm sp}=0.24\,\rm Myr$} & \multicolumn{1}{c}{$3.0\,t_{\rm sp}=0.30\,\rm Myr$} & $\frac{n}{n_{\rm ambient}}$\\    
       \hspace{-0.00cm}\resizebox{27mm}{!}{\includegraphics{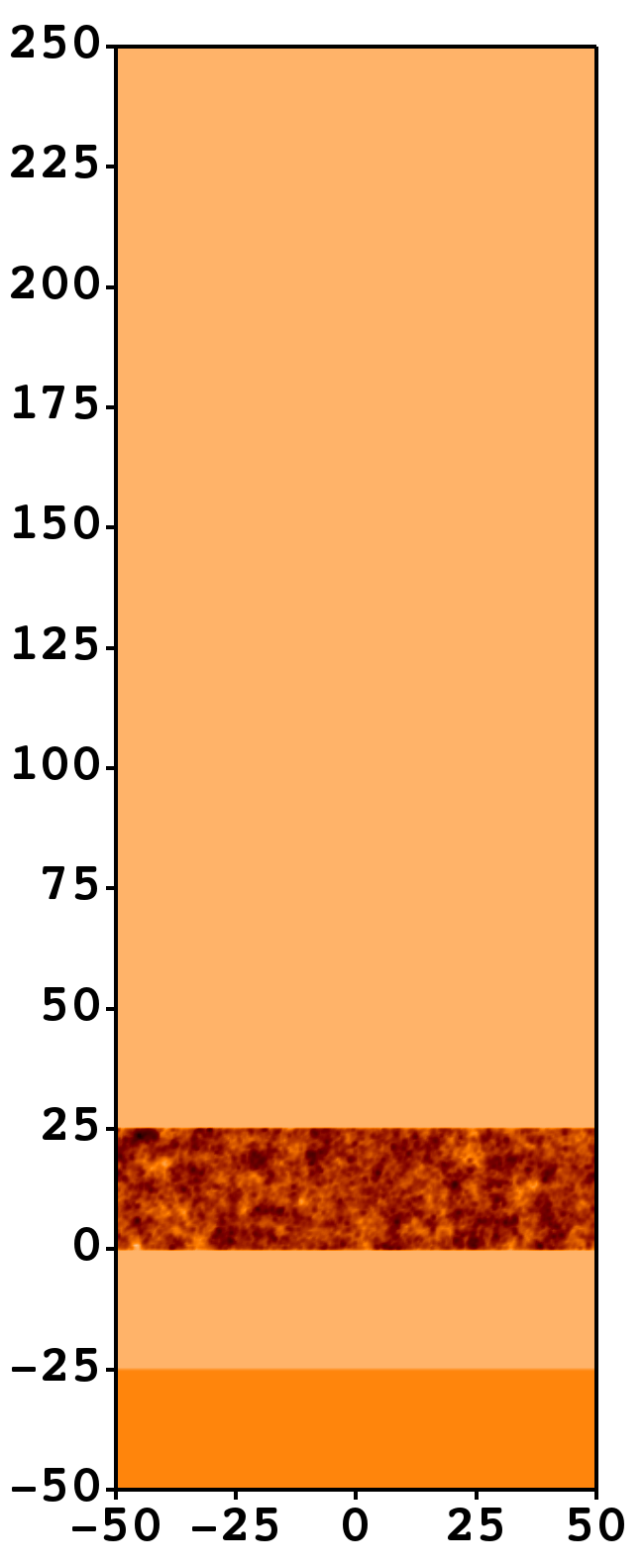}} & \hspace{-0.4cm}\resizebox{27mm}{!}{\includegraphics{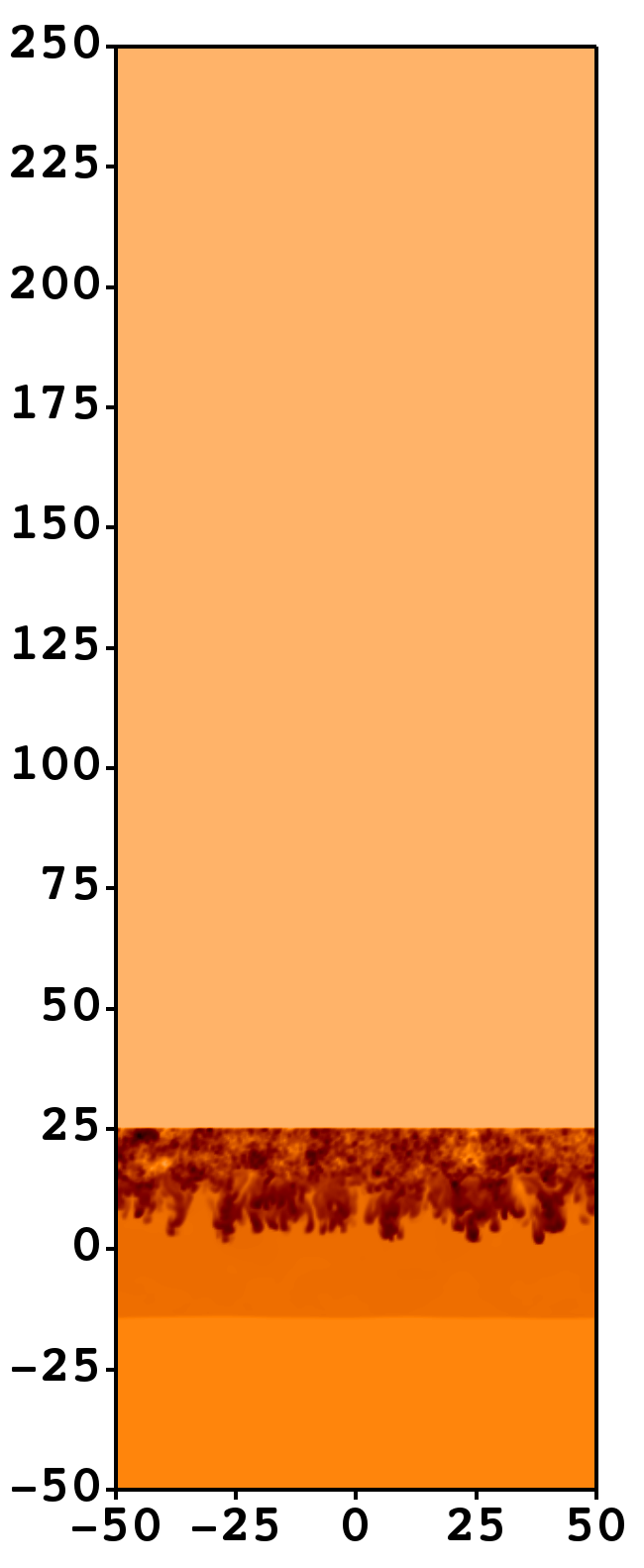}} & \hspace{-0.4cm}\resizebox{27mm}{!}{\includegraphics{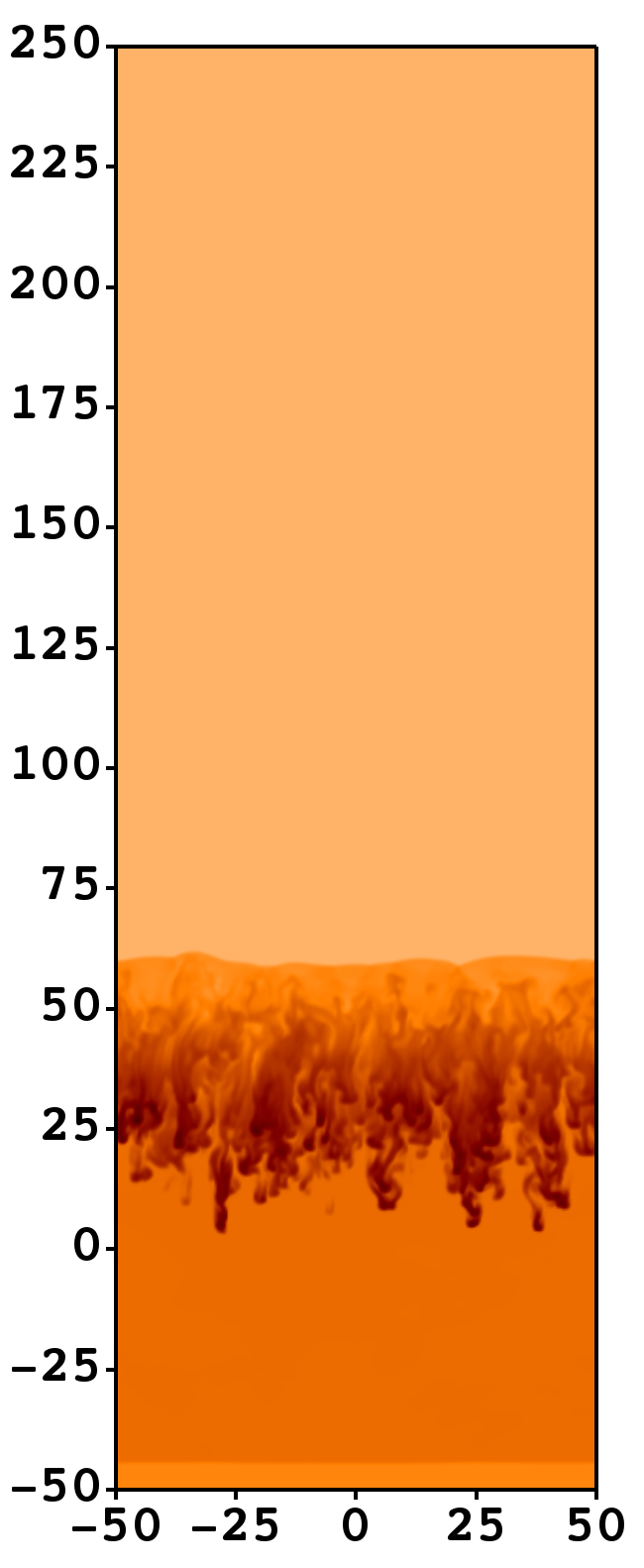}} & \hspace{-0.4cm}\resizebox{27mm}{!}{\includegraphics{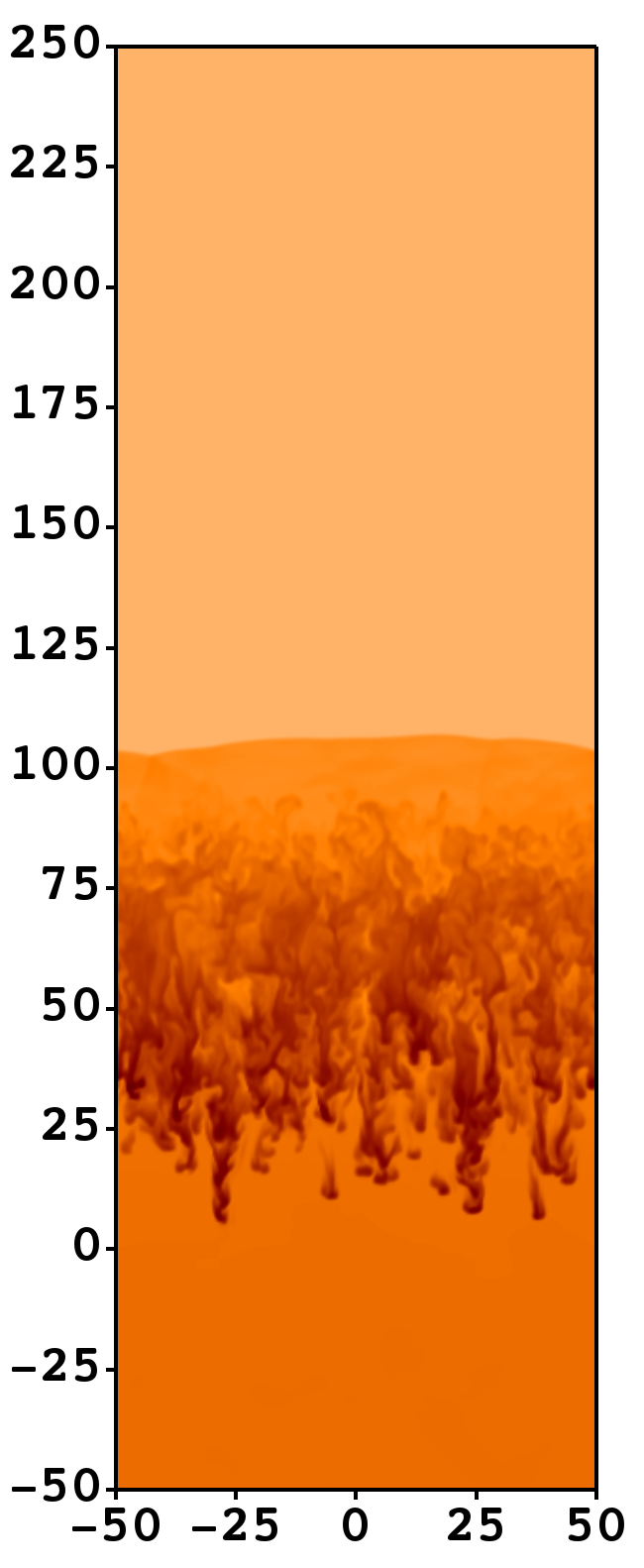}} & \hspace{-0.4cm}\resizebox{27mm}{!}{\includegraphics{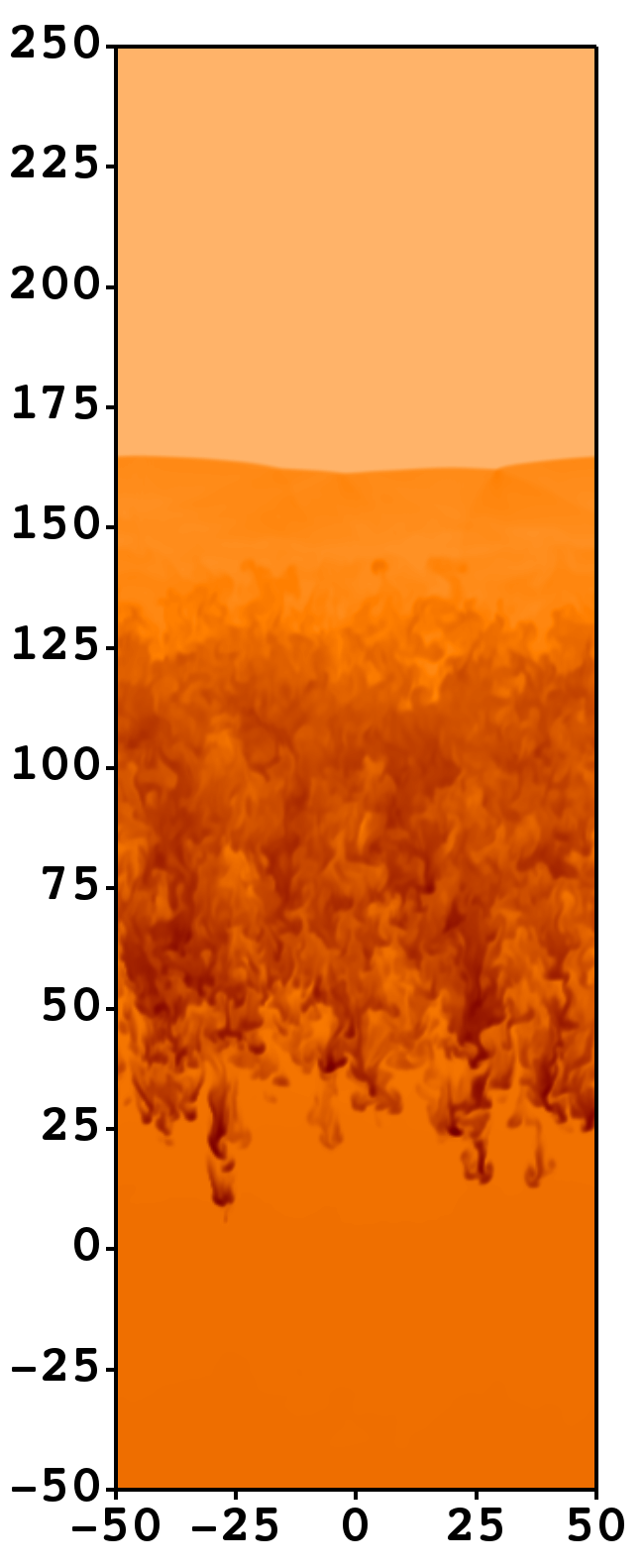}} & \hspace{-0.4cm}\resizebox{27mm}{!}{\includegraphics{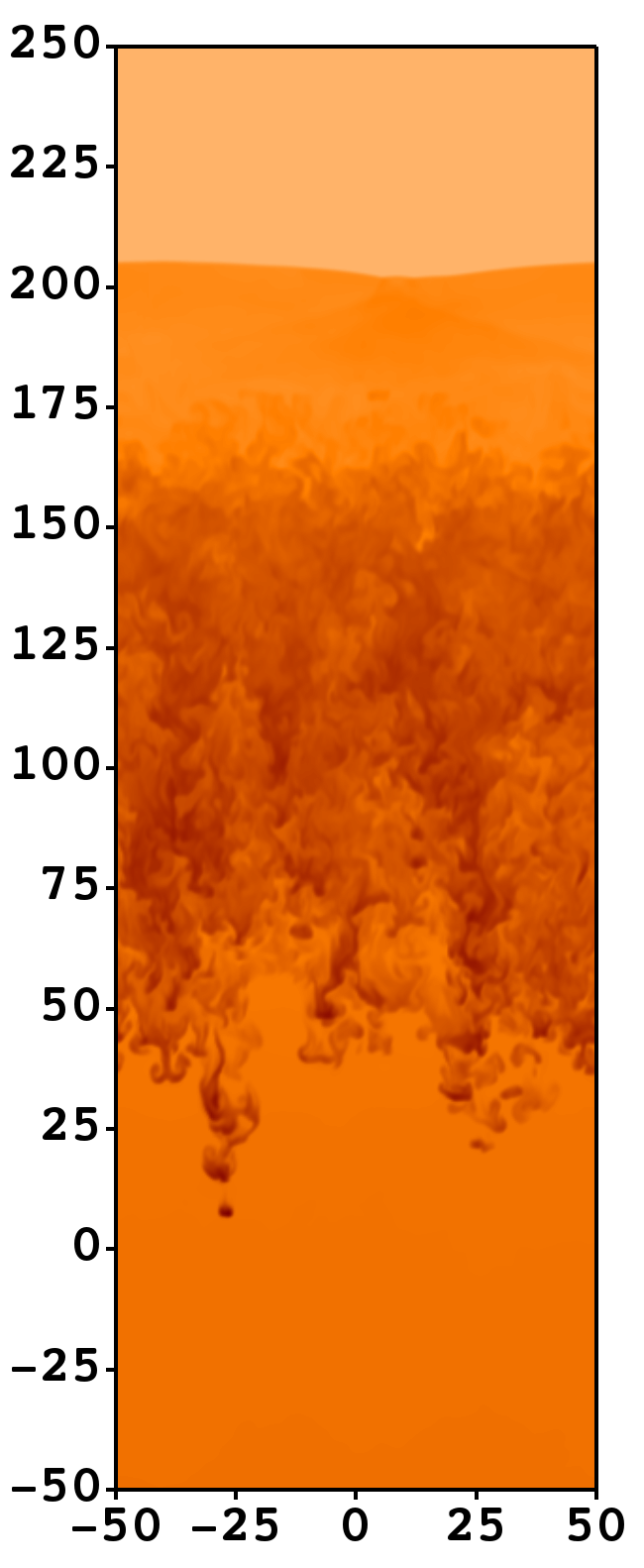}} &
\hspace{-0.2cm}\resizebox{12.8mm}{!}{\includegraphics{bar_vert_2.png}}\\
       \multicolumn{1}{l}{\hspace{-2mm}b) sole-k8-M4 \hspace{+7.2mm}$t_0$} & \multicolumn{1}{c}{$0.5\,t_{\rm sp}=0.25\,\rm Myr$} & \multicolumn{1}{c}{$1.1\,t_{\rm sp}=0.55\,\rm Myr$} & \multicolumn{1}{c}{$1.8\,t_{\rm sp}=0.90\,\rm Myr$} & \multicolumn{1}{c}{$2.4\,t_{\rm sp}=1.20\,\rm Myr$} & \multicolumn{1}{c}{$3.0\,t_{\rm sp}=1.50\,\rm Myr$} & $\frac{n}{n_{\rm ambient}}$\\      
       \hspace{-0.00cm}\resizebox{27mm}{!}{\includegraphics{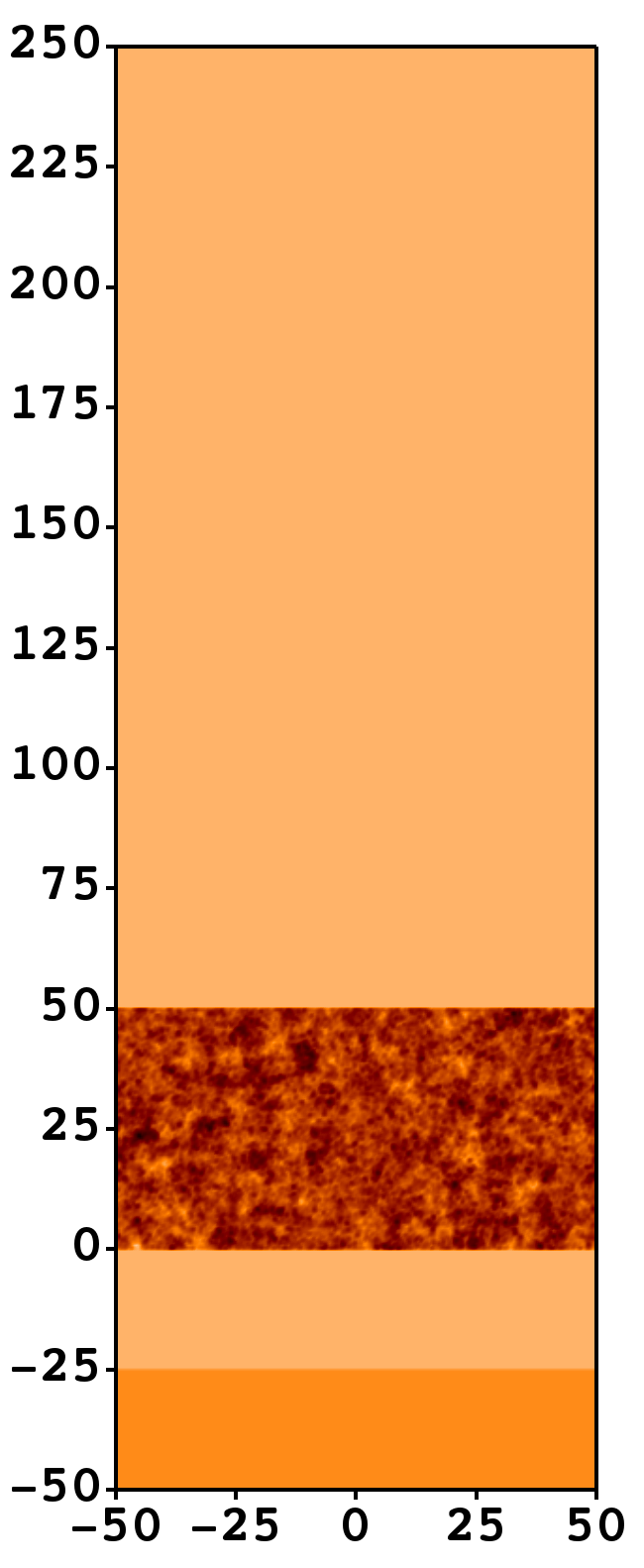}} & \hspace{-0.4cm}\resizebox{27mm}{!}{\includegraphics{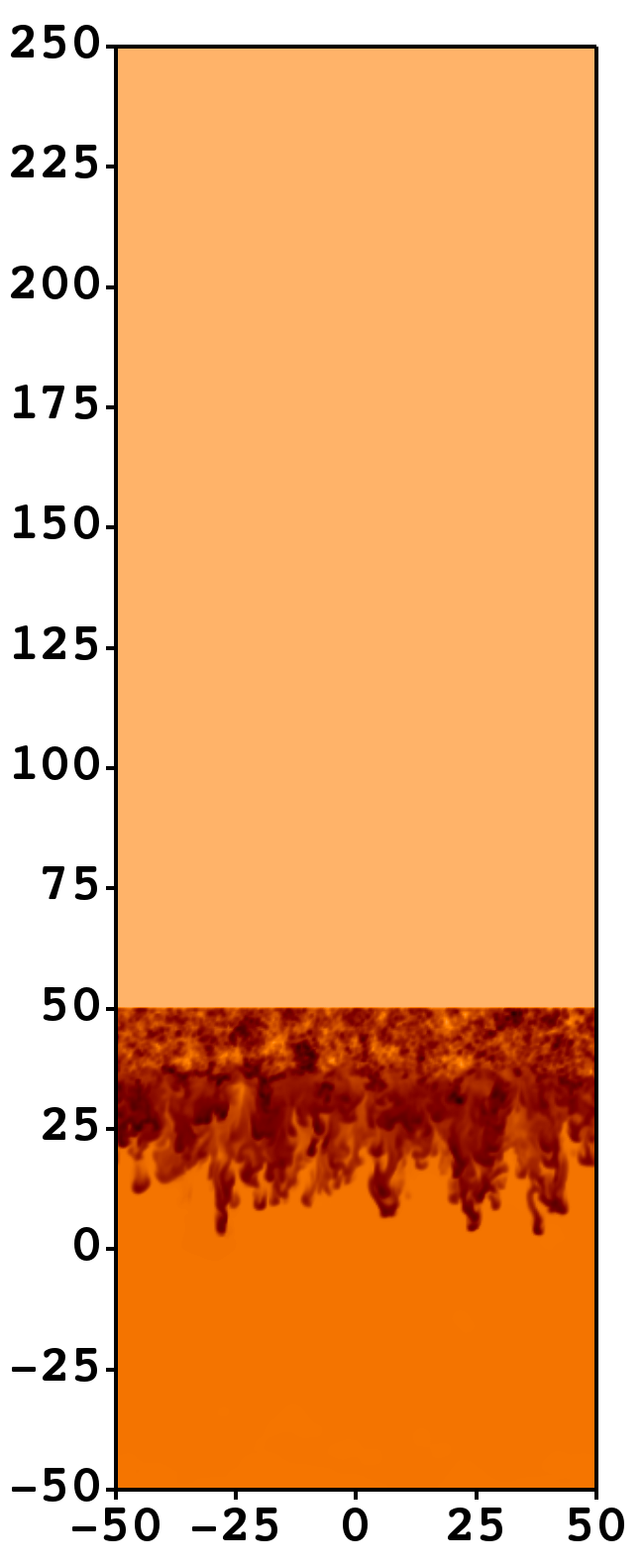}} & \hspace{-0.4cm}\resizebox{27mm}{!}{\includegraphics{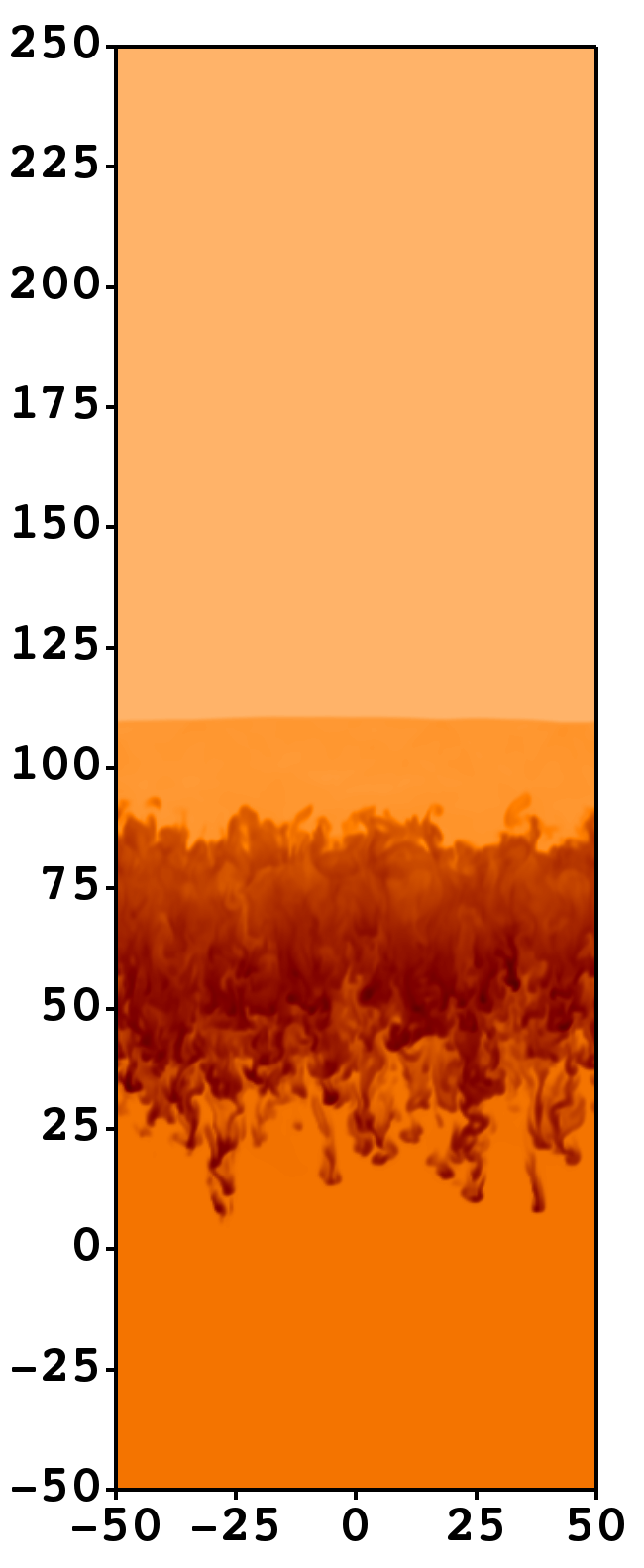}} & \hspace{-0.4cm}\resizebox{27mm}{!}{\includegraphics{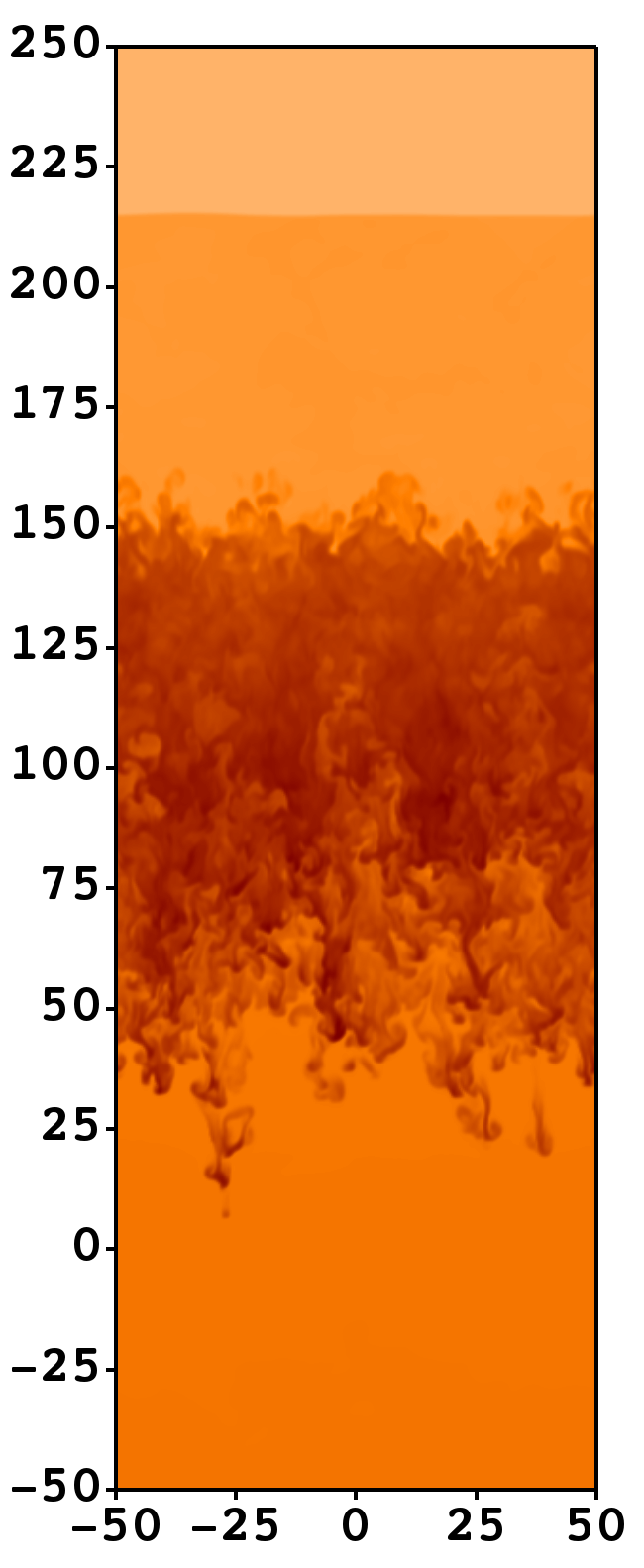}} & \hspace{-0.4cm}\resizebox{27mm}{!}{\includegraphics{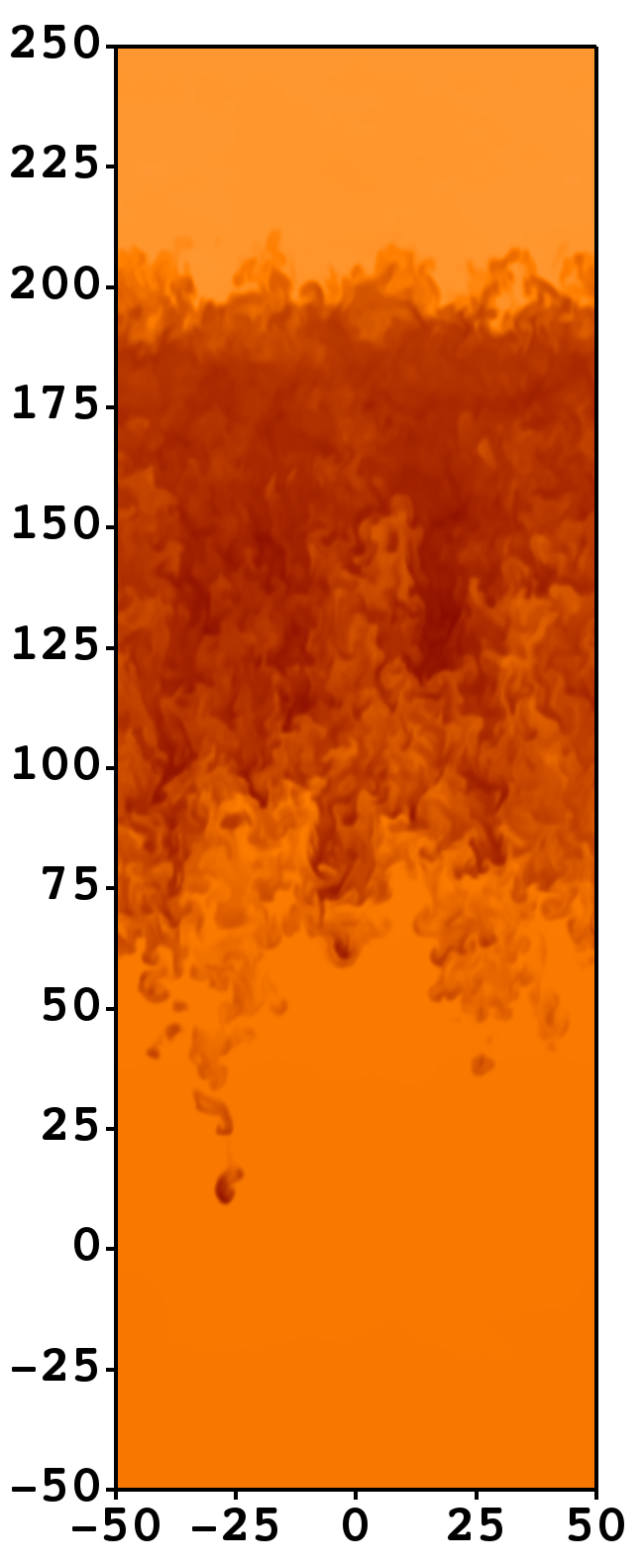}}  & \hspace{-0.4cm}\resizebox{27mm}{!}{\includegraphics{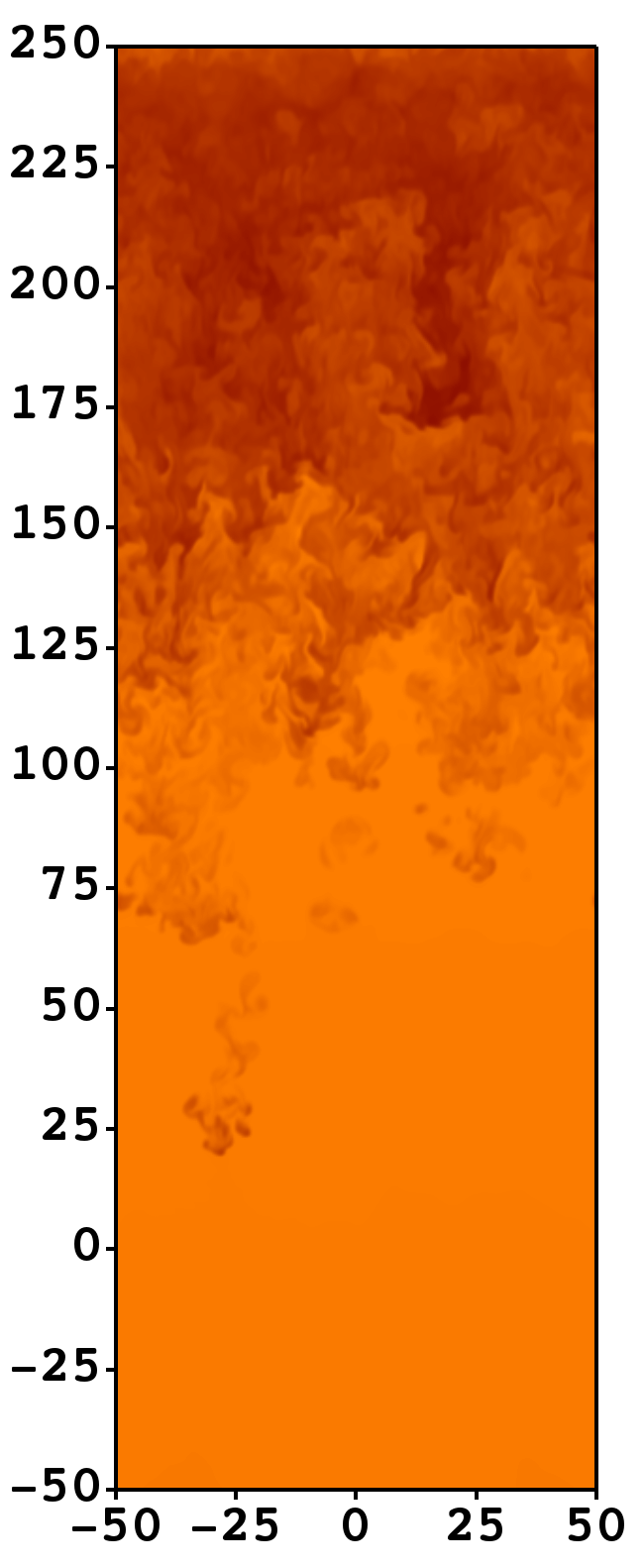}} &
\hspace{-0.2cm}\resizebox{12.8mm}{!}{\includegraphics{bar_vert_2.png}}\\
       \multicolumn{1}{l}{\hspace{-2mm}c) sole-k8-M10-sd \hspace{+2.5mm}$t_0$} & \multicolumn{1}{c}{$0.5\,t_{\rm sp}=0.10\,\rm Myr$} & \multicolumn{1}{c}{$1.1\,t_{\rm sp}=0.22\,\rm Myr$} & \multicolumn{1}{c}{$1.8\,t_{\rm sp}=0.36\,\rm Myr$} & \multicolumn{1}{c}{$2.4\,t_{\rm sp}=0.48\,\rm Myr$} & \multicolumn{1}{c}{$3.0\,t_{\rm sp}=0.60\,\rm Myr$} & $\frac{n}{n_{\rm ambient}}$\\ 
        \hspace{-0.00cm}\resizebox{27mm}{!}{\includegraphics{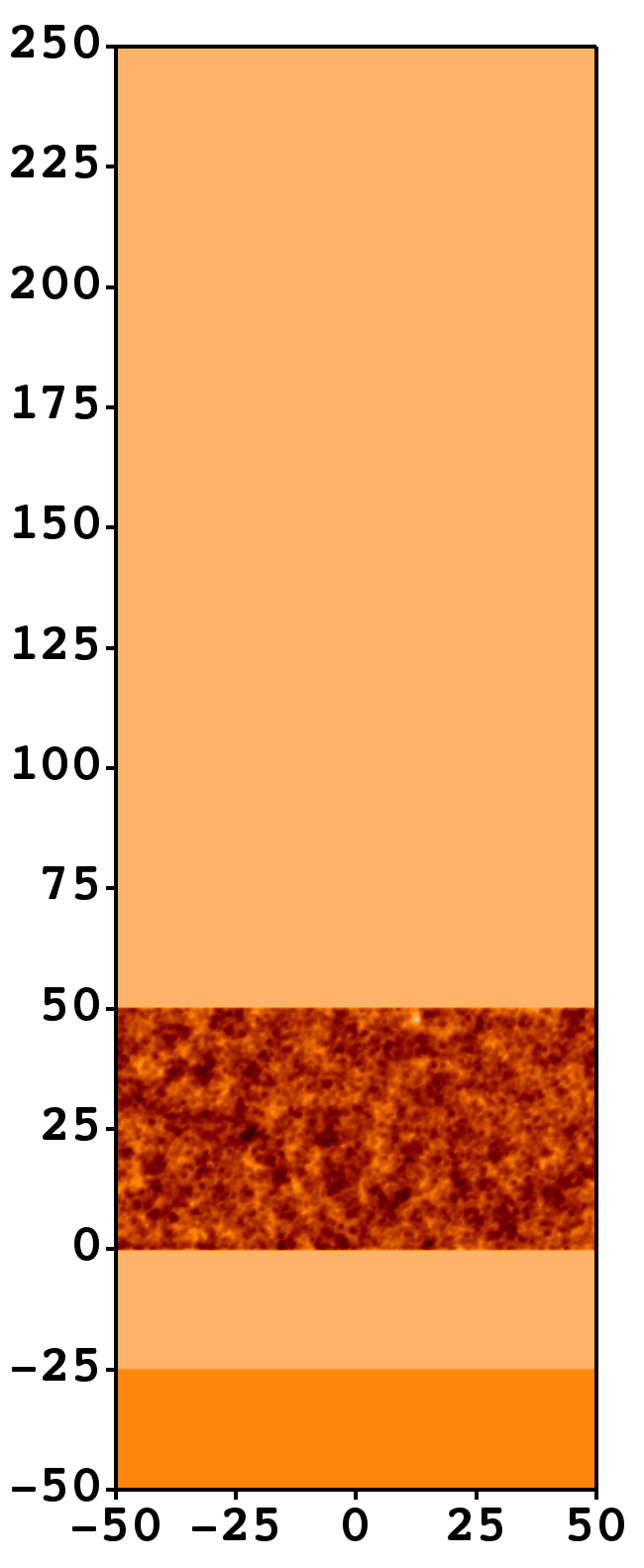}} & \hspace{-0.4cm}\resizebox{27mm}{!}{\includegraphics{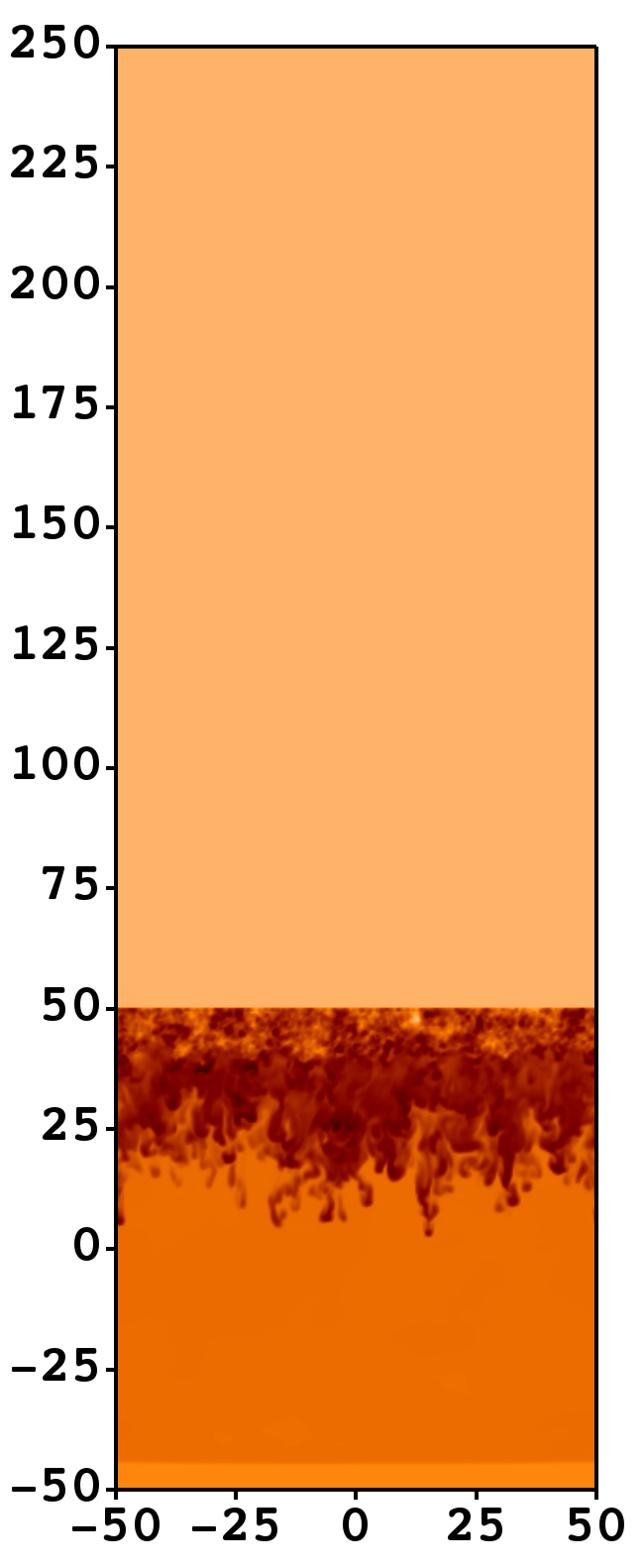}} & \hspace{-0.4cm}\resizebox{27mm}{!}{\includegraphics{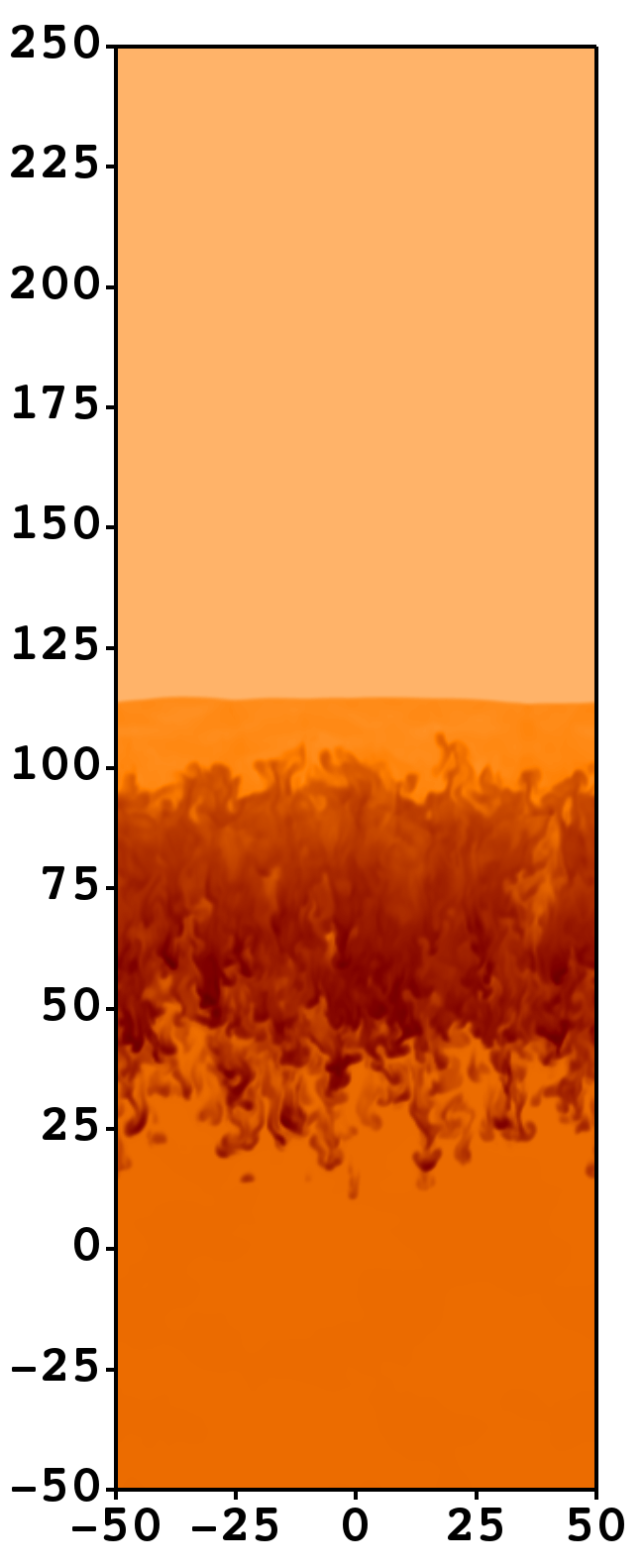}} & \hspace{-0.4cm}\resizebox{27mm}{!}{\includegraphics{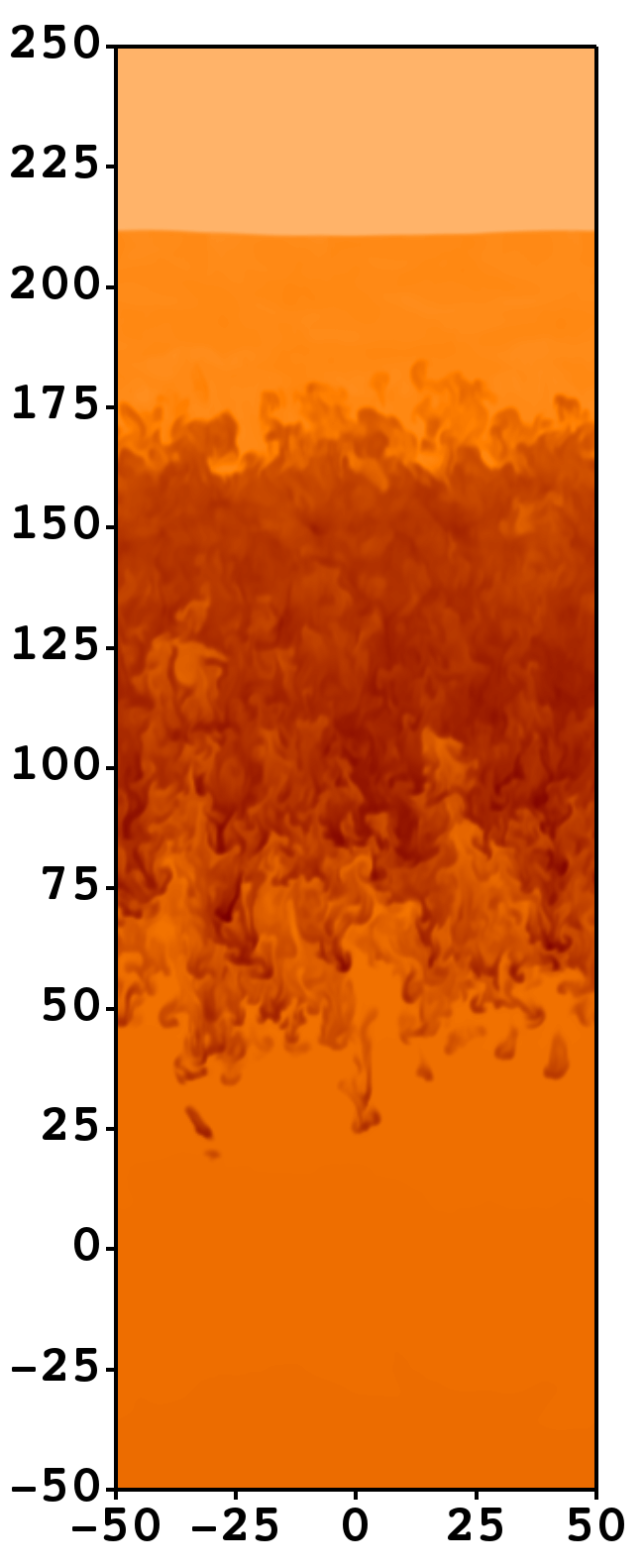}} & \hspace{-0.4cm}\resizebox{27mm}{!}{\includegraphics{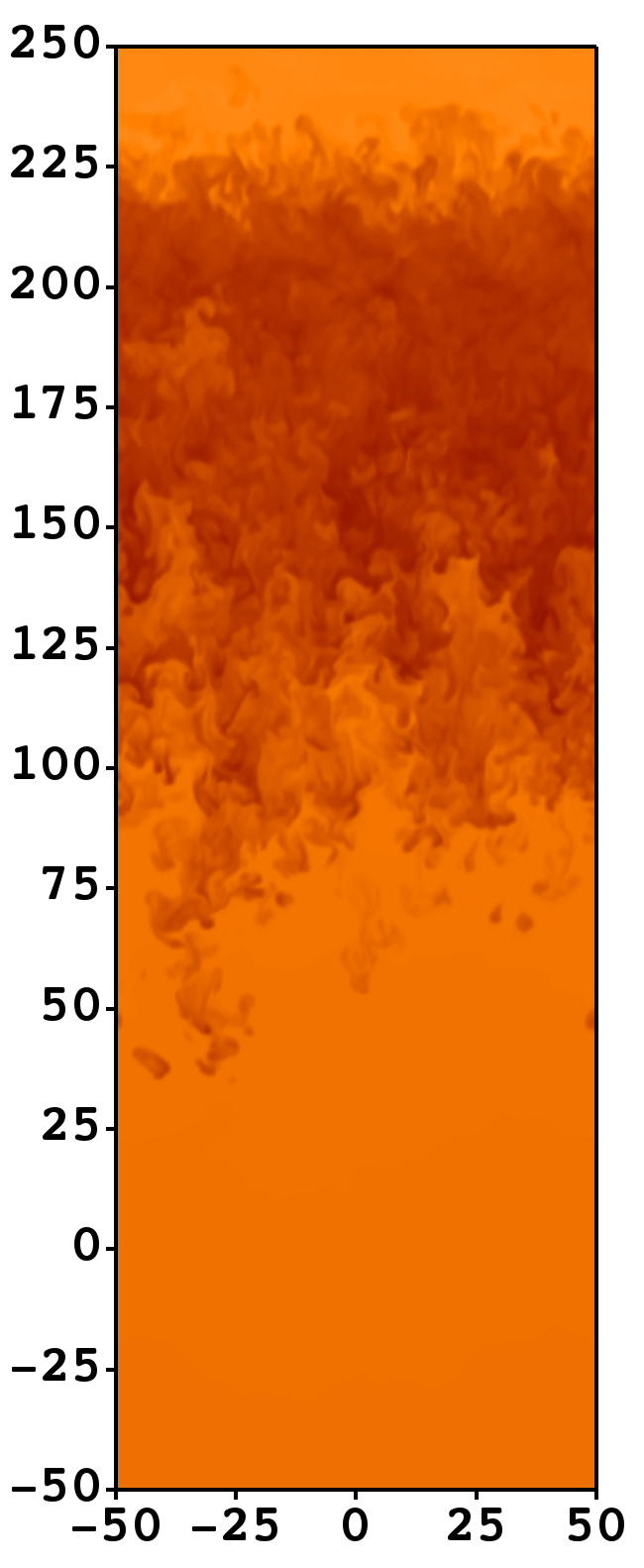}} & \hspace{-0.4cm}\resizebox{27mm}{!}{\includegraphics{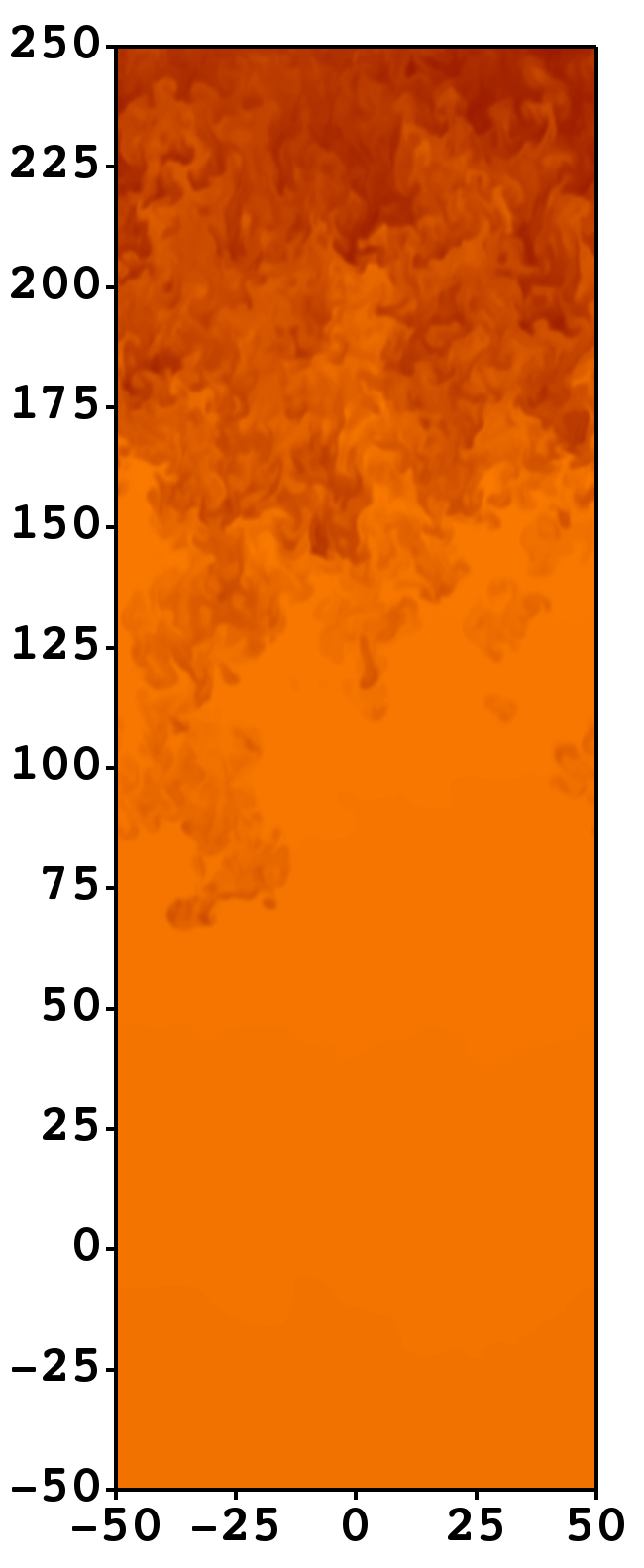}} &
\hspace{-0.2cm}\resizebox{12.8mm}{!}{\includegraphics{bar_vert_2.png}}\\
  \end{tabular}
  \caption{Same as Figure \ref{Figure2}, but here we show the number density slices in three solenoidal multicloud models, sole-k8-M10-th (panel a), sole-k8-M4 (panel b), and sole-k8-M10-sd, which correspond to the thin-layer model, the run with ${\cal M_{\rm shock}}=4$, and the model with a different seed. The spatial ($X,Y$) extent is ($L\times3L$)$\equiv$($4L_{\rm mc}\times12L_{\rm mc}$) in sole-k8-M10-th, and $\equiv$($2L_{\rm mc}\times6L_{\rm mc}$) in the other models. The $X$ and $Y$ axes are given in $\rm pc$, so they cover a spatial extent of ($100\,\rm pc\times300\,\rm pc$) in all models. Time-scales in physical units are also different as $1\,\rm t_{\rm sp}=0.098\,\rm Myr$, $1\,\rm t_{\rm sp}=0.196\,\rm Myr$, $1\,\rm t_{\rm sp}=0.491\,\rm Myr$, in thin-layer models, standard thick-layer models, and Mach-4 models, respectively.} 
  \label{FigureA1}
\end{center}
\end{figure*}

\begin{figure*}
\begin{center}
  \begin{tabular}{c c c c c c c}
       \multicolumn{1}{l}{\hspace{-3mm}a) comp-k8-M10-th \hspace{+1mm}$t_0$} & \multicolumn{1}{c}{$0.5\,t_{\rm sp}=0.05\,\rm Myr$} & \multicolumn{1}{c}{$1.1\,t_{\rm sp}=0.11\,\rm Myr$} & \multicolumn{1}{c}{$1.8\,t_{\rm sp}=0.18\,\rm Myr$} & \multicolumn{1}{c}{$2.4\,t_{\rm sp}=0.24\,\rm Myr$} & \multicolumn{1}{c}{$3.0\,t_{\rm sp}=0.30\,\rm Myr$} & $\frac{n}{n_{\rm ambient}}$\\      
       \hspace{-0.25cm}\resizebox{27mm}{!}{\includegraphics{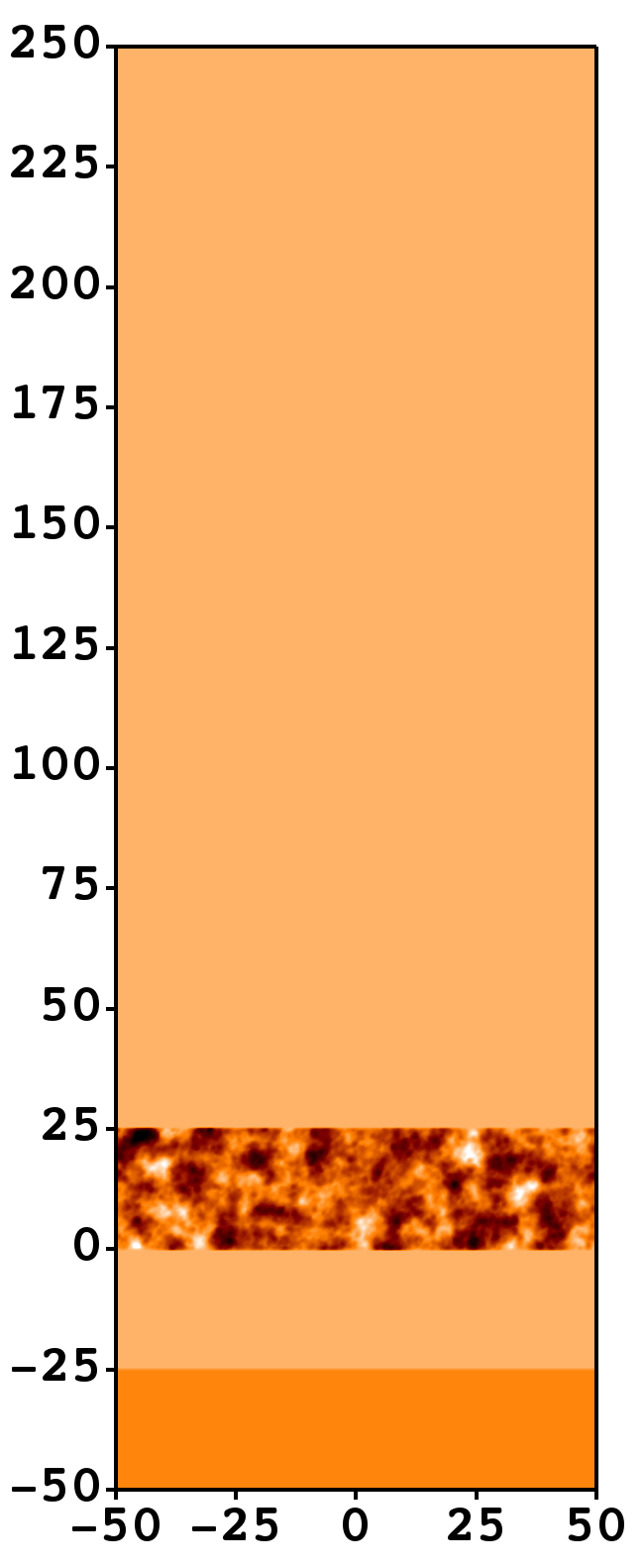}} & \hspace{-0.4cm}\resizebox{27mm}{!}{\includegraphics{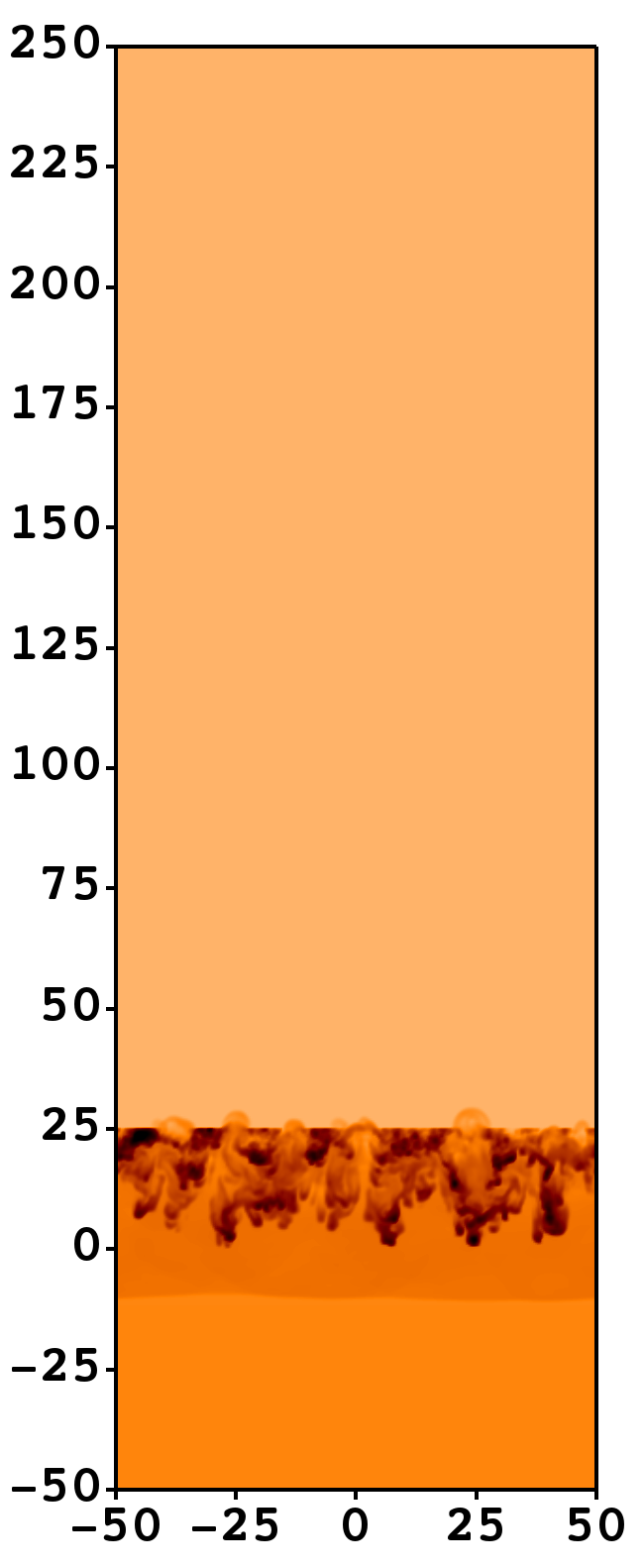}} & \hspace{-0.4cm}\resizebox{27mm}{!}{\includegraphics{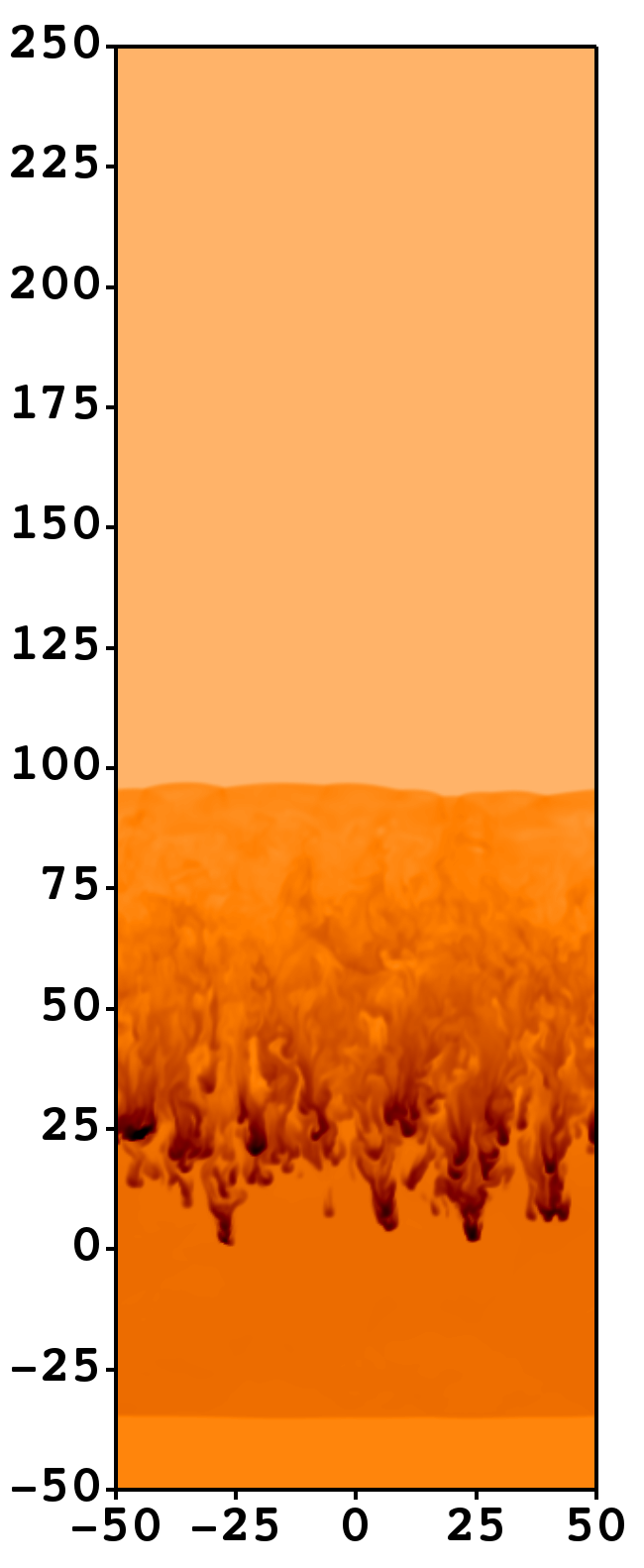}} & \hspace{-0.4cm}\resizebox{27mm}{!}{\includegraphics{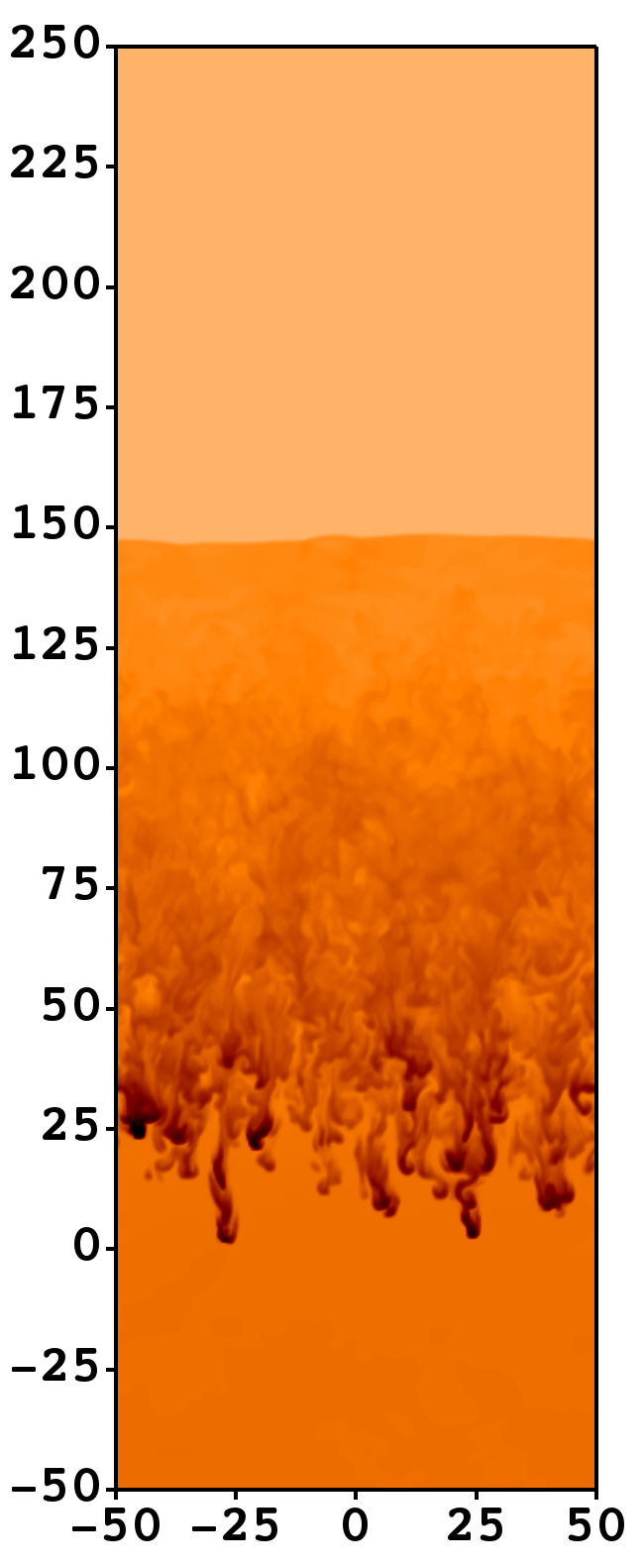}} & \hspace{-0.4cm}\resizebox{27mm}{!}{\includegraphics{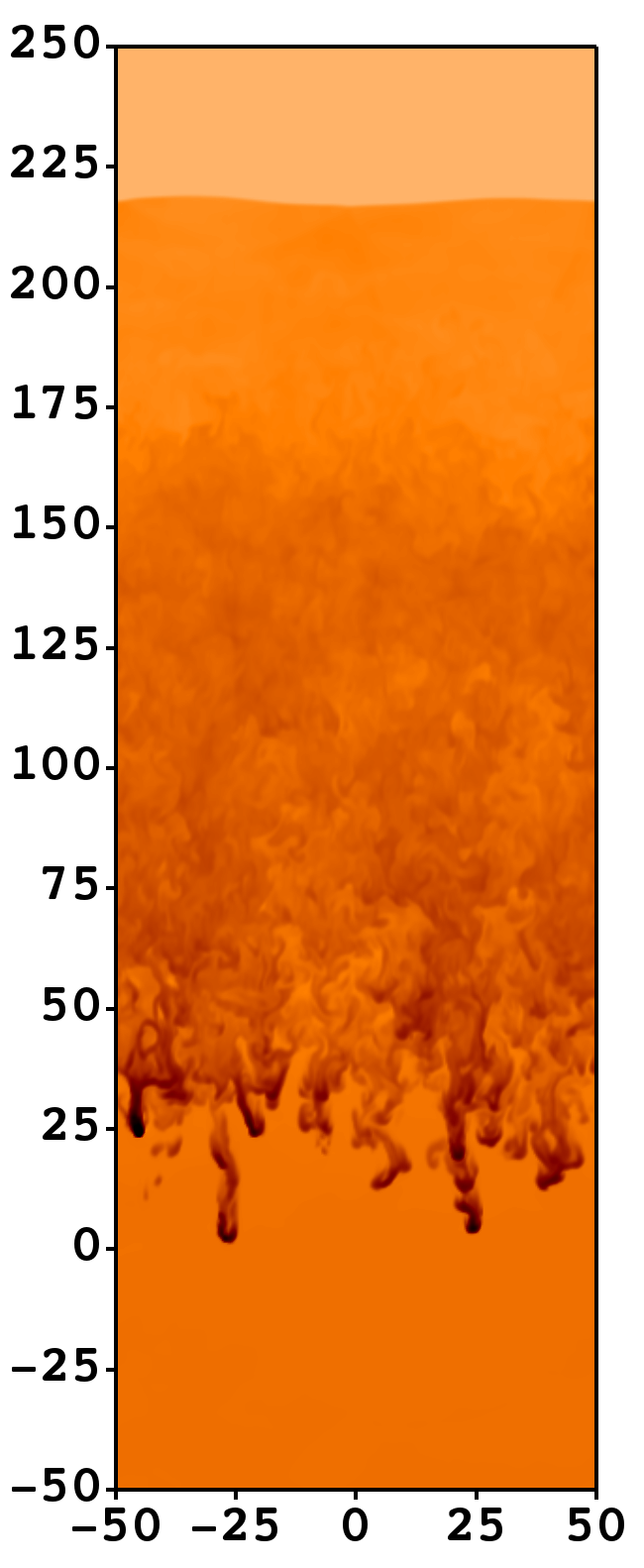}}  & \hspace{-0.4cm}\resizebox{27mm}{!}{\includegraphics{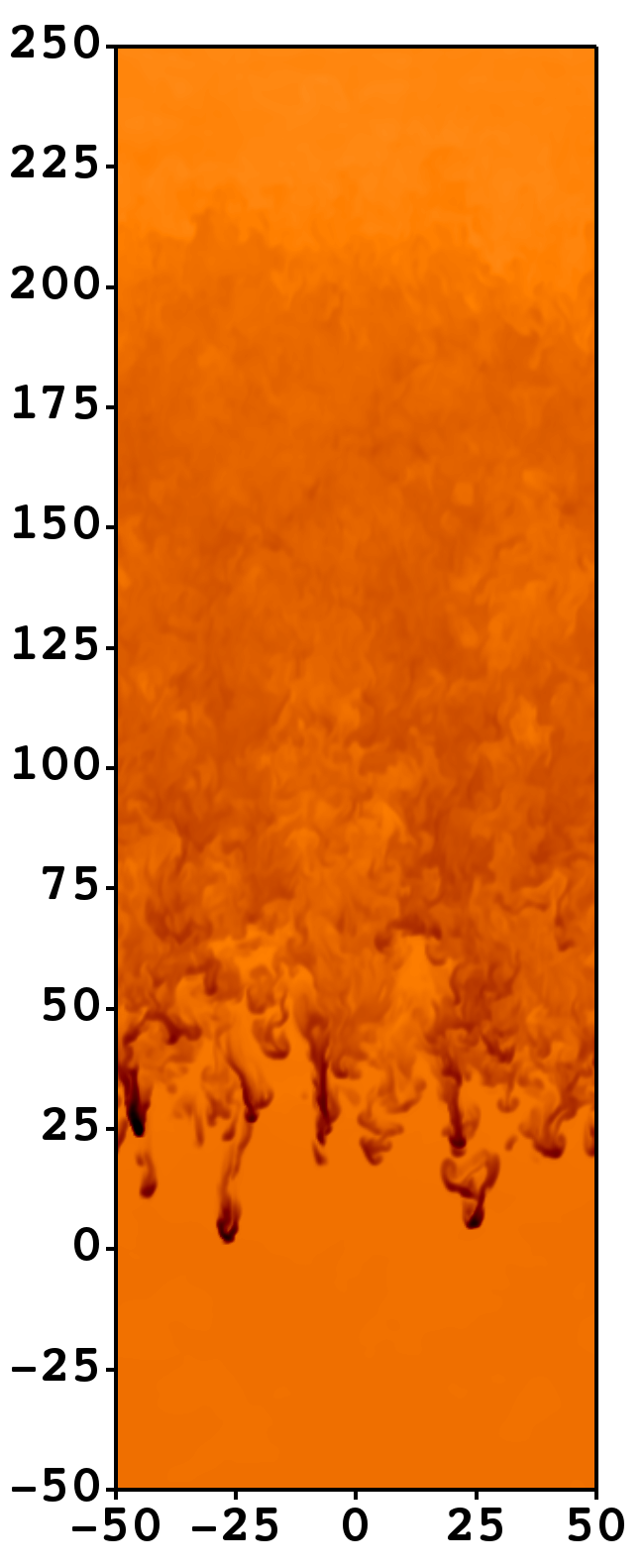}} &
\hspace{-0.2cm}\resizebox{12.8mm}{!}{\includegraphics{bar_vert_2.png}}\\
       \multicolumn{1}{l}{\hspace{-3mm}b) comp-k8-M4 \hspace{+2mm}$t_0$} & \multicolumn{1}{c}{$0.5\,t_{\rm sp}=0.25\,\rm Myr$} & \multicolumn{1}{c}{$1.1\,t_{\rm sp}=0.55\,\rm Myr$} & \multicolumn{1}{c}{$1.8\,t_{\rm sp}=0.90\,\rm Myr$} & \multicolumn{1}{c}{$2.4\,t_{\rm sp}=1.20\,\rm Myr$} & \multicolumn{1}{c}{$3.0\,t_{\rm sp}=1.50\,\rm Myr$} & $\frac{n}{n_{\rm ambient}}$\\   
       \hspace{-0.25cm}\resizebox{27mm}{!}{\includegraphics{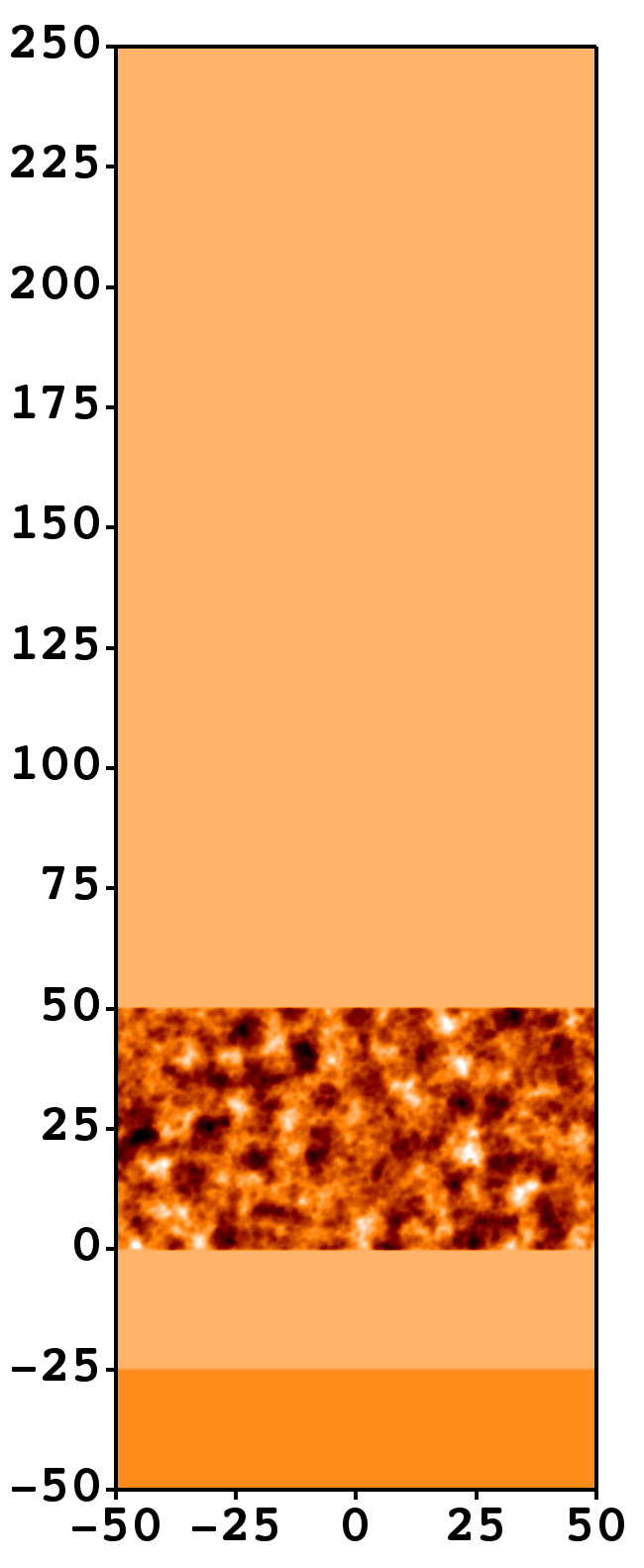}} & \hspace{-0.4cm}\resizebox{27mm}{!}{\includegraphics{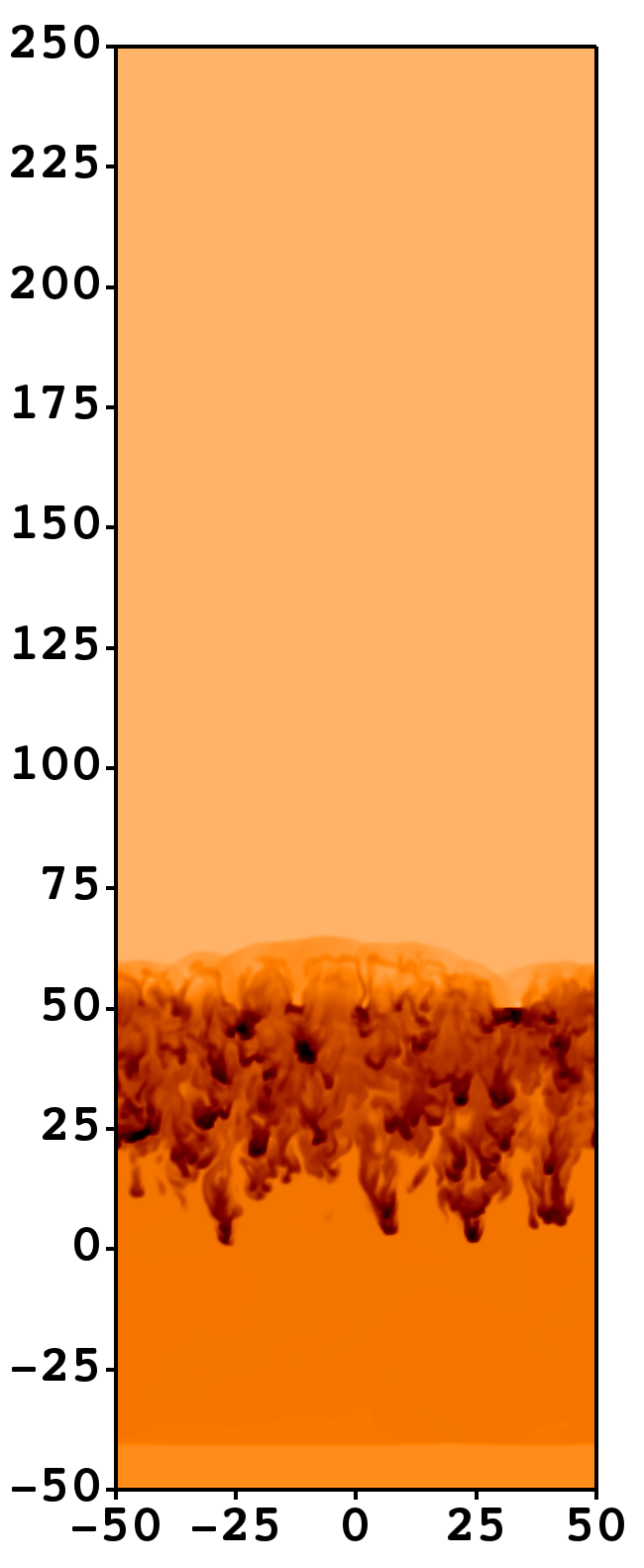}} & \hspace{-0.4cm}\resizebox{27mm}{!}{\includegraphics{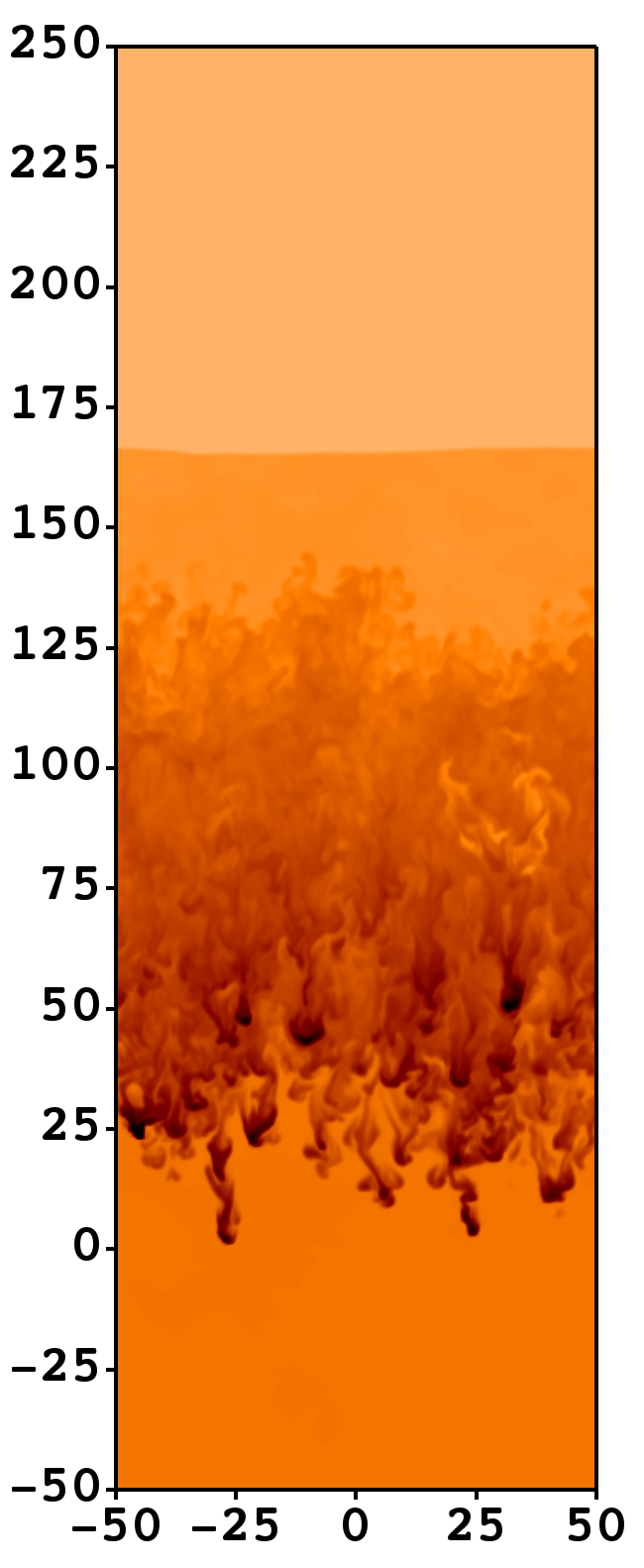}} & \hspace{-0.4cm}\resizebox{27mm}{!}{\includegraphics{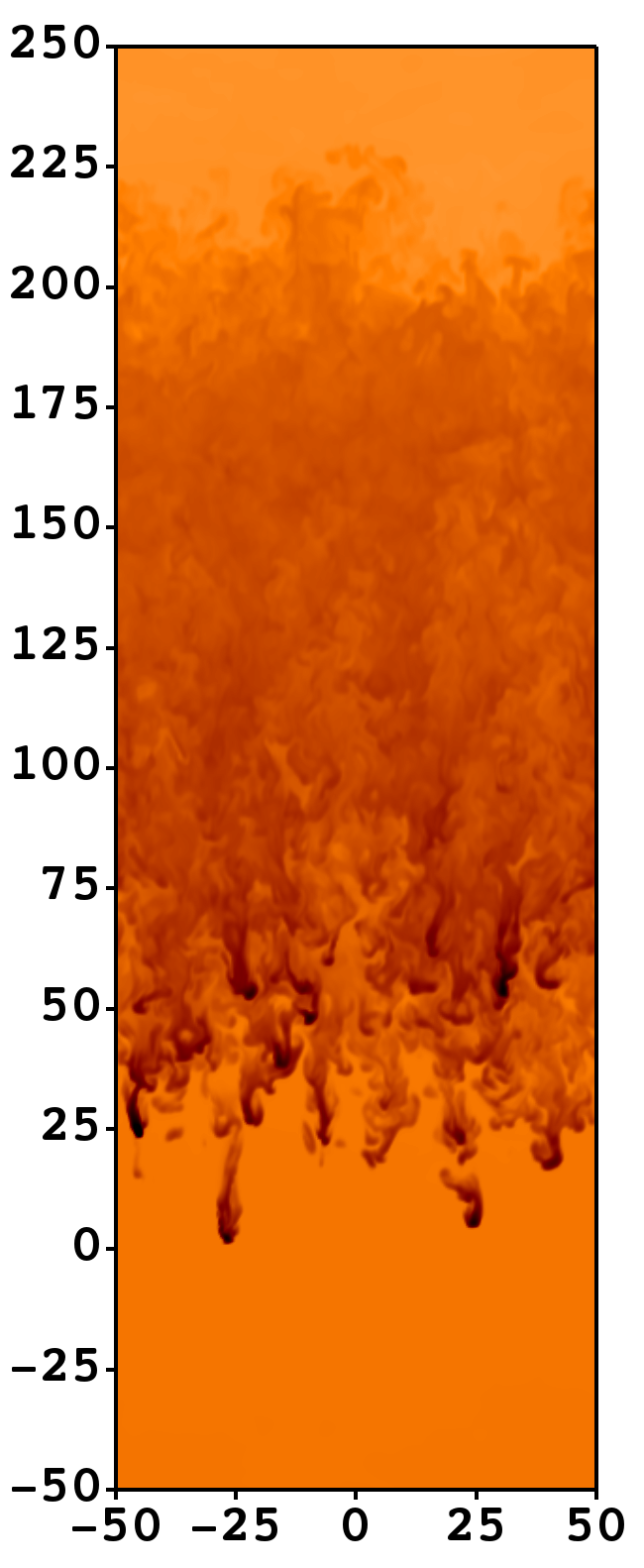}} & \hspace{-0.4cm}\resizebox{27mm}{!}{\includegraphics{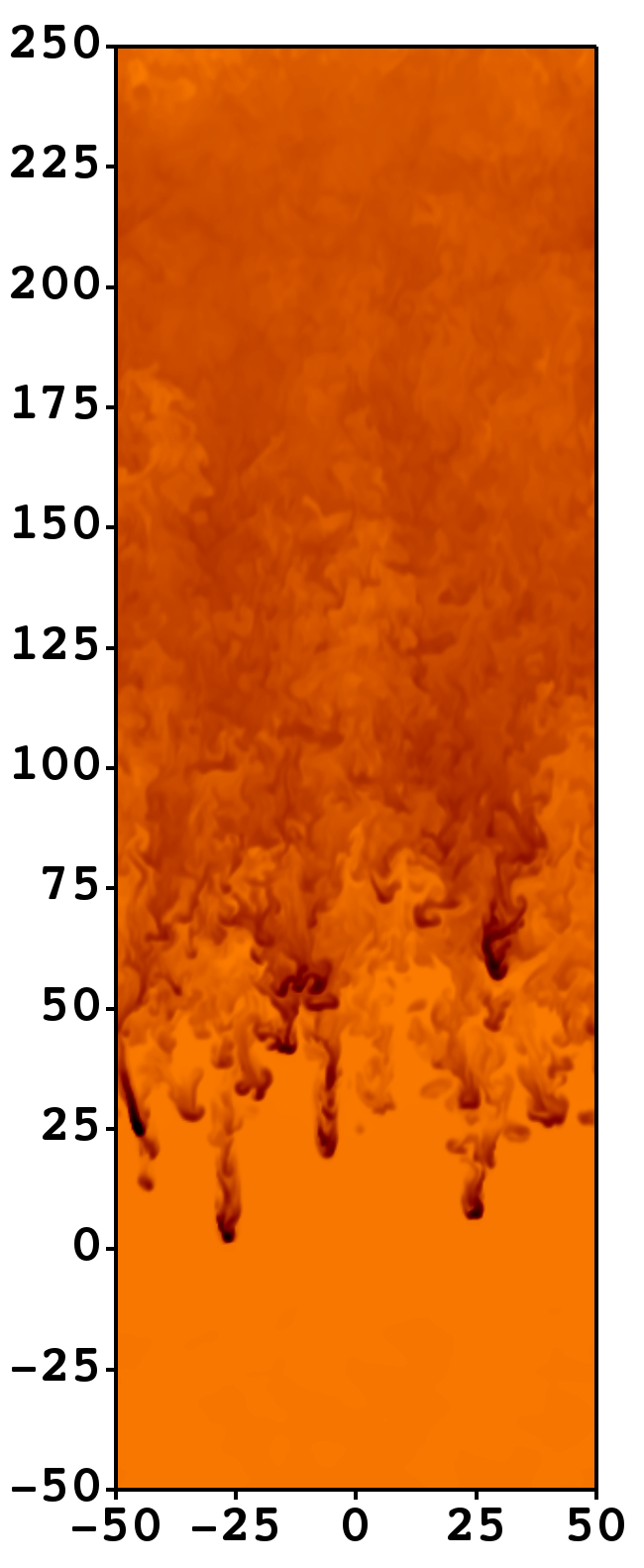}}  & \hspace{-0.4cm}\resizebox{27mm}{!}{\includegraphics{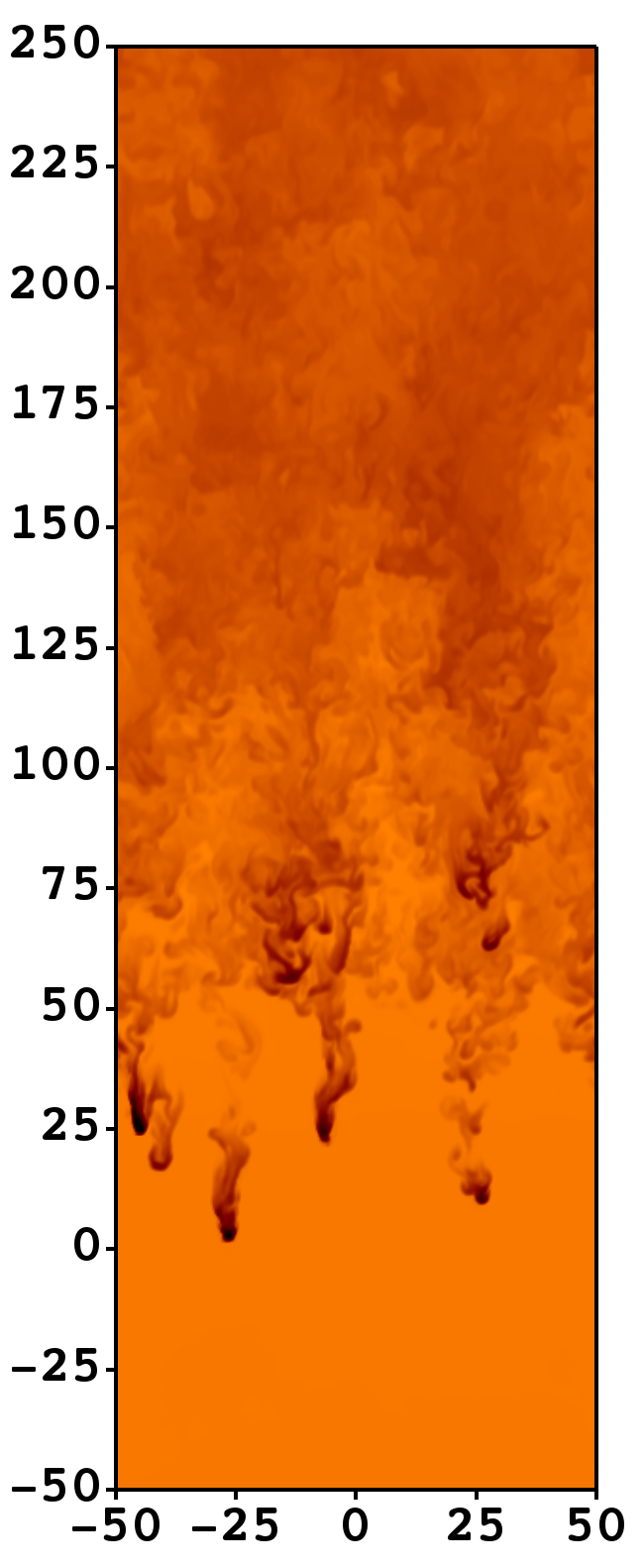}}&
\hspace{-0.2cm}\resizebox{12.8mm}{!}{\includegraphics{bar_vert_2.png}}\\
       \multicolumn{1}{l}{\hspace{-3mm}c) comp-k8-M10-sd \hspace{+1.1mm}$t_0$} & \multicolumn{1}{c}{$0.5\,t_{\rm sp}=0.10\,\rm Myr$} & \multicolumn{1}{c}{$1.1\,t_{\rm sp}=0.22\,\rm Myr$} & \multicolumn{1}{c}{$1.8\,t_{\rm sp}=0.36\,\rm Myr$} & \multicolumn{1}{c}{$2.4\,t_{\rm sp}=0.48\,\rm Myr$} & \multicolumn{1}{c}{$3.0\,t_{\rm sp}=0.60\,\rm Myr$} & $\frac{n}{n_{\rm ambient}}$\\    
       \hspace{-0.25cm}\resizebox{27mm}{!}{\includegraphics{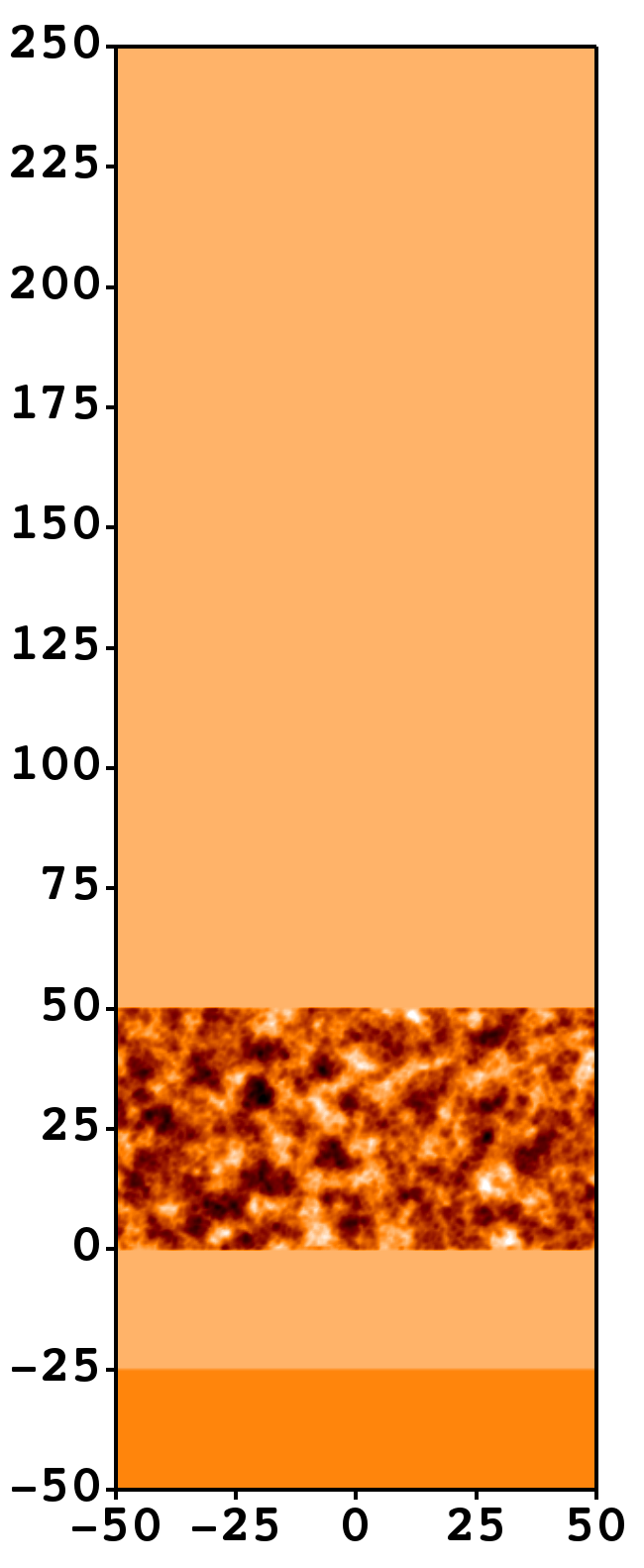}} & \hspace{-0.4cm}\resizebox{27mm}{!}{\includegraphics{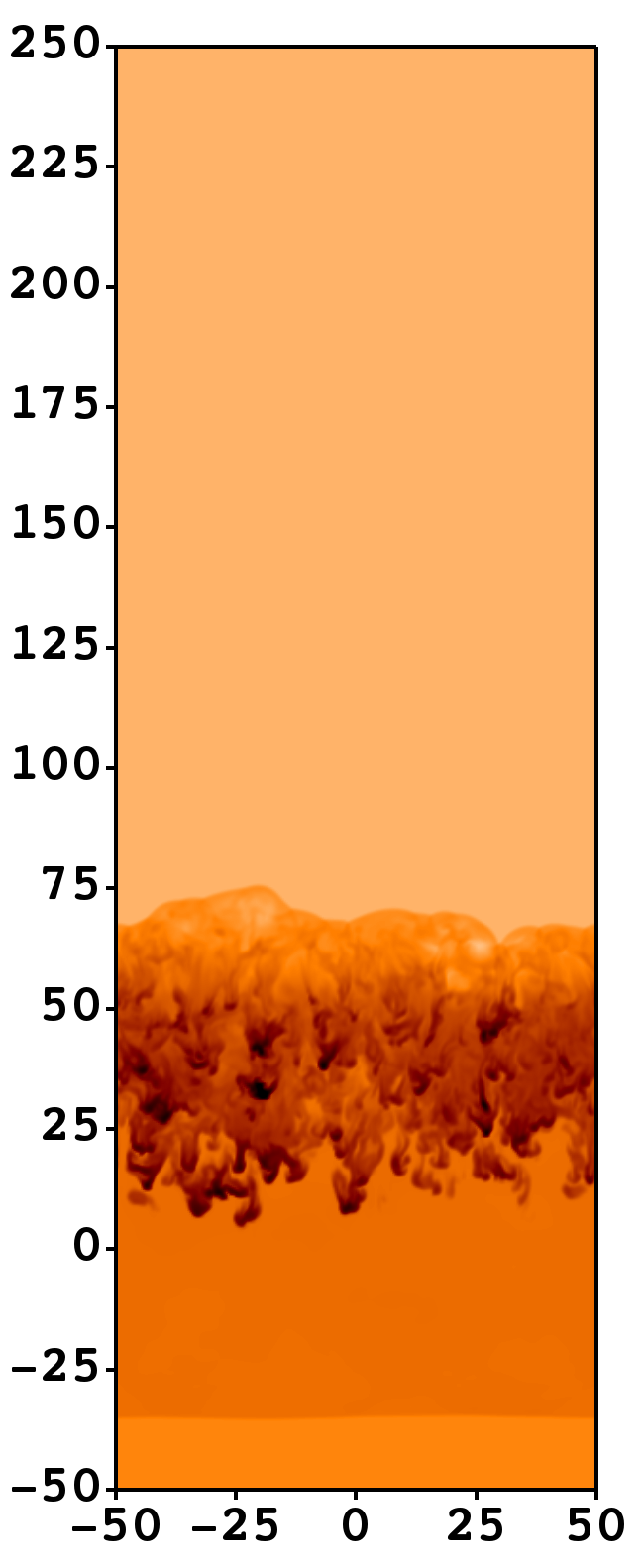}} & \hspace{-0.4cm}\resizebox{27mm}{!}{\includegraphics{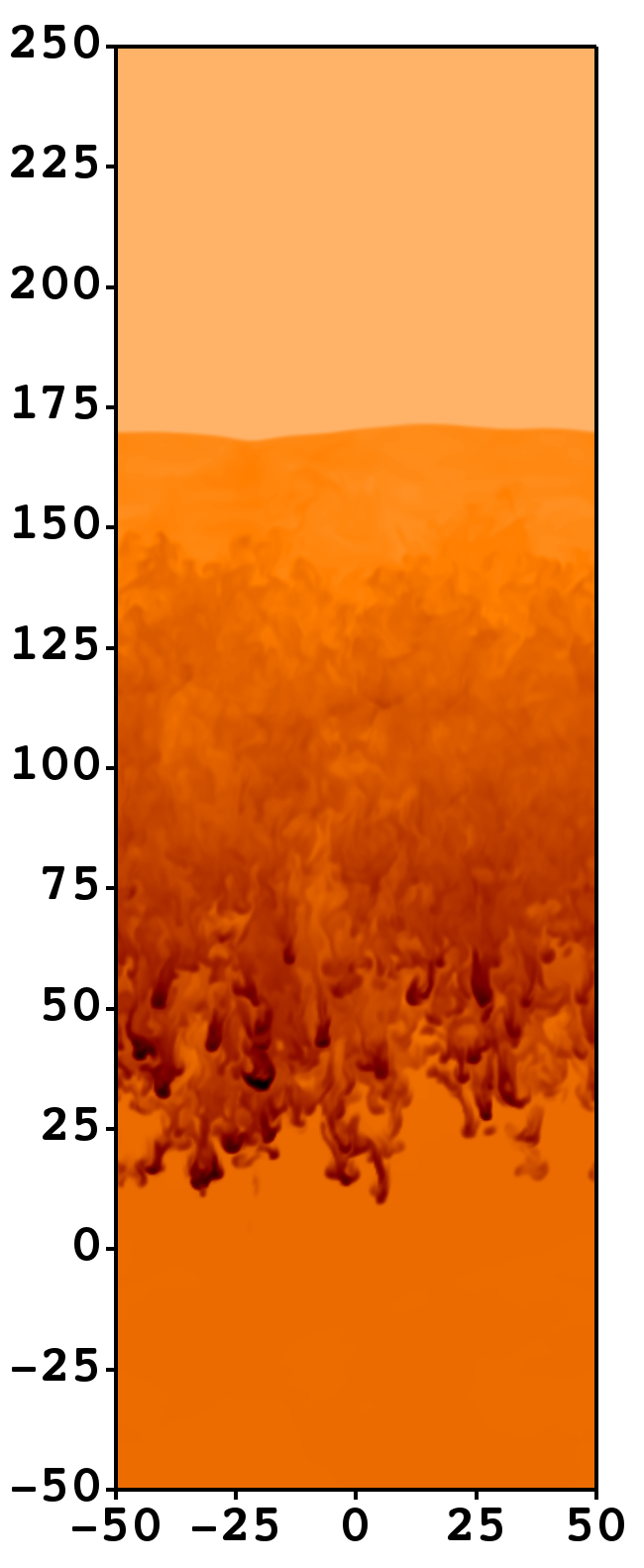}} & \hspace{-0.4cm}\resizebox{27mm}{!}{\includegraphics{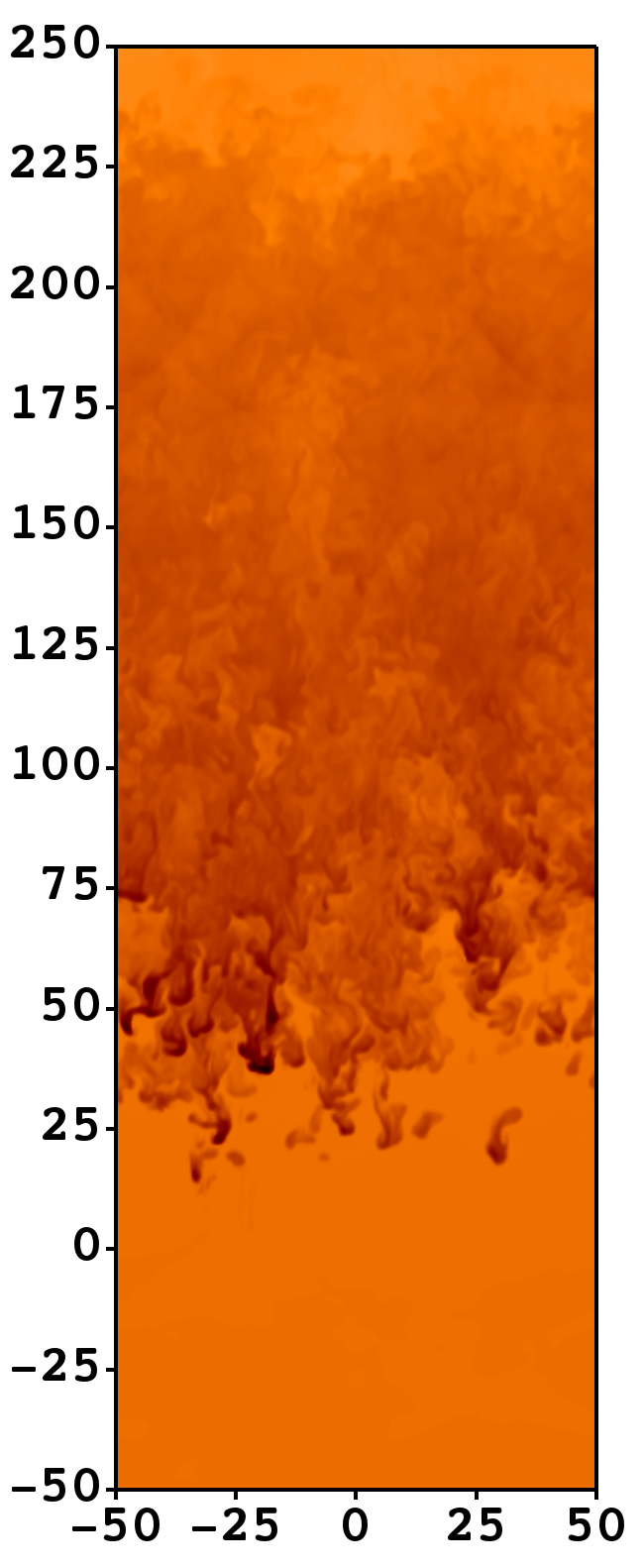}} & \hspace{-0.4cm}\resizebox{27mm}{!}{\includegraphics{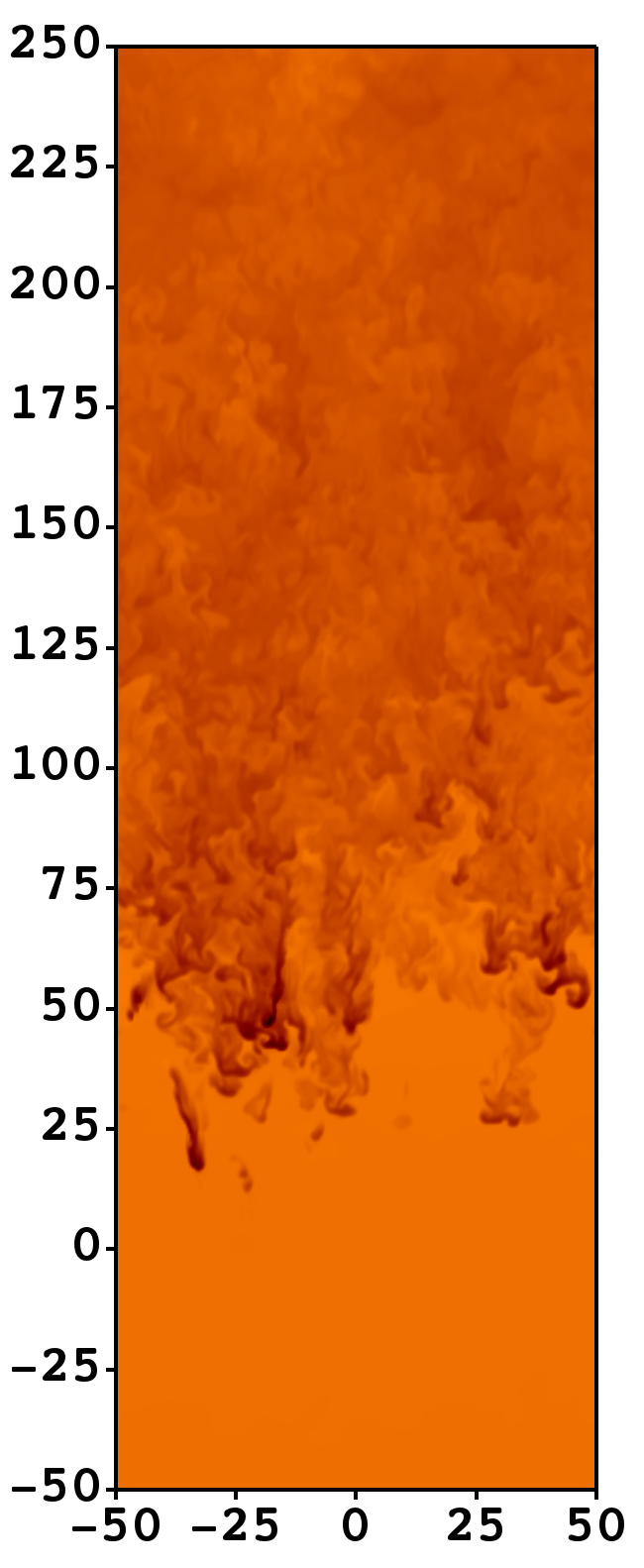}} & \hspace{-0.4cm}\resizebox{27mm}{!}{\includegraphics{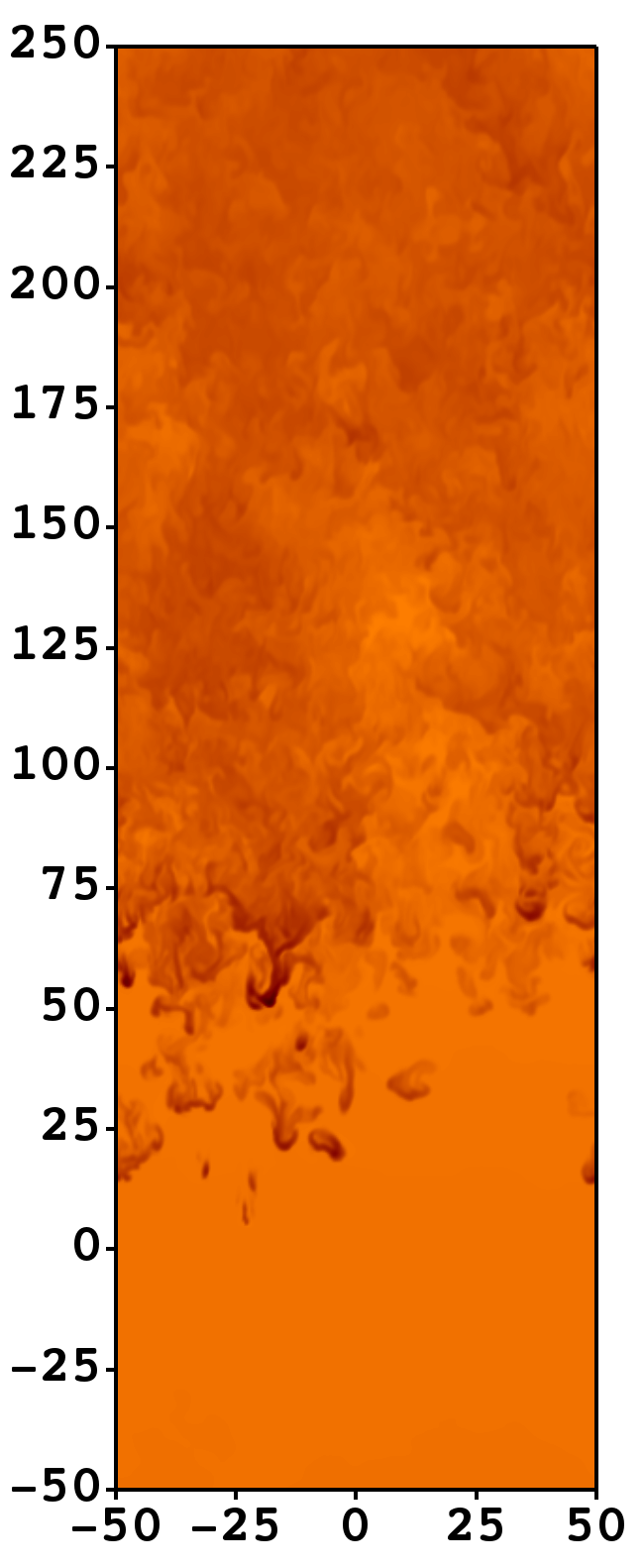}}&
\hspace{-0.2cm}\resizebox{12.8mm}{!}{\includegraphics{bar_vert_2.png}}\\
  \end{tabular}
  \caption{Same as Figure \ref{Figure3}, but here we show the number density slices in three compressive multicloud models, comp-k8-M10-th (panel a), comp-k8-M4 (panel b), and comp-k8-M10-sd, which correspond to the thin-layer model, the run with ${\cal M_{\rm shock}}=4$, and the model with a different seed. The spatial ($X,Y$) extent is ($L\times3L$)$\equiv$($4L_{\rm mc}\times12L_{\rm mc}$) in comp-k8-M10-th, and $\equiv$($2L_{\rm mc}\times6L_{\rm mc}$) in the other models. The $X$ and $Y$ axes are given in $\rm pc$, so they cover a spatial extent of ($100\,\rm pc\times300\,\rm pc$) in all models. Time-scales in physical units are also different as $1\,\rm t_{\rm sp}=0.098\,\rm Myr$, $1\,\rm t_{\rm sp}=0.196\,\rm Myr$, $1\,\rm t_{\rm sp}=0.491\,\rm Myr$, in thin-layer models, standard thick-layer models, and Mach-4 models, respectively.} 
  \label{FigureA2}
\end{center}
\end{figure*}

\begin{figure*}
\begin{center}
  \begin{tabular}{c c c c c c c}
       \multicolumn{1}{l}{\hspace{-3mm}a) sole-k8-M30 \hspace{+3.2mm}$t_0$} & \multicolumn{1}{c}{$0.5\,t_{\rm sp}=0.03\,\rm Myr$} & \multicolumn{1}{c}{$1.1\,t_{\rm sp}=0.07\,\rm Myr$} & \multicolumn{1}{c}{$1.8\,t_{\rm sp}=0.12\,\rm Myr$} & \multicolumn{1}{c}{$2.4\,t_{\rm sp}=0.16\,\rm Myr$} & \multicolumn{1}{c}{$3.0\,t_{\rm sp}=0.20\,\rm Myr$} & $\frac{n}{n_{\rm ambient}}$\\    
       \hspace{-0.25cm}\resizebox{27mm}{!}{\includegraphics{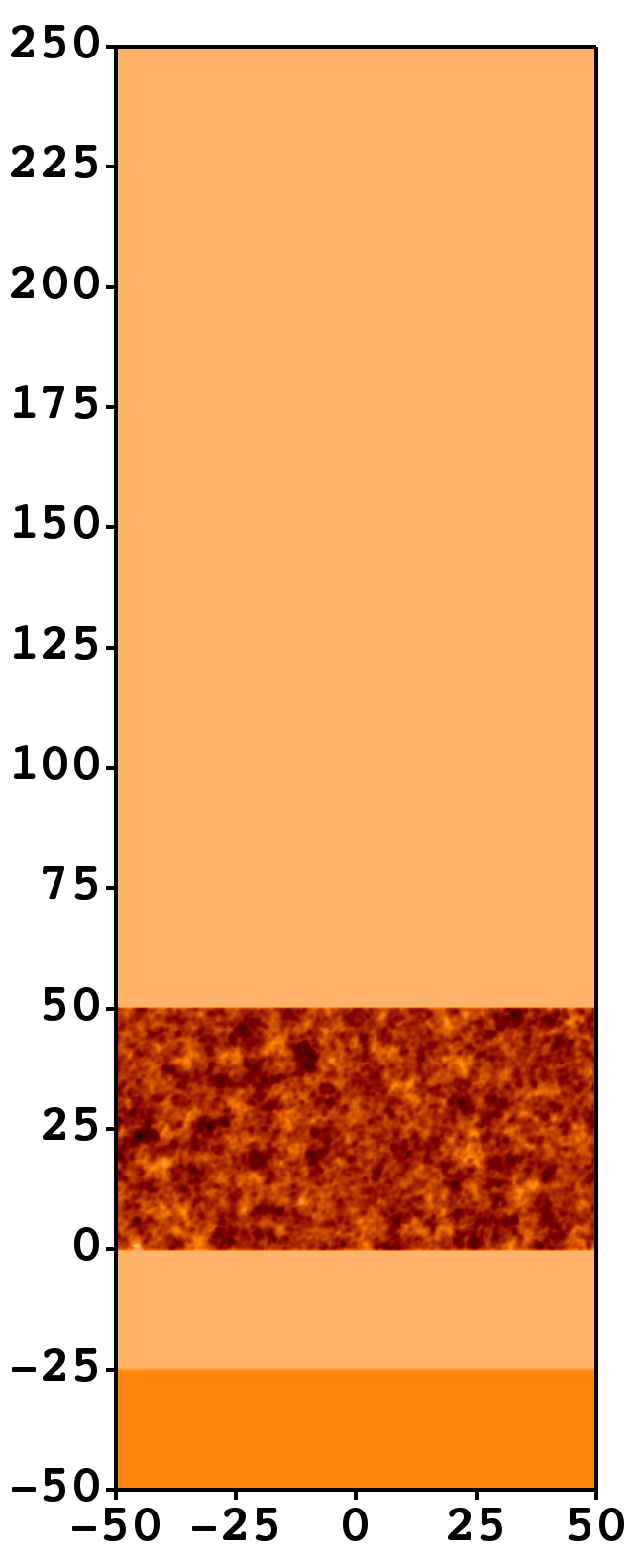}} & \hspace{-0.4cm}\resizebox{27mm}{!}{\includegraphics{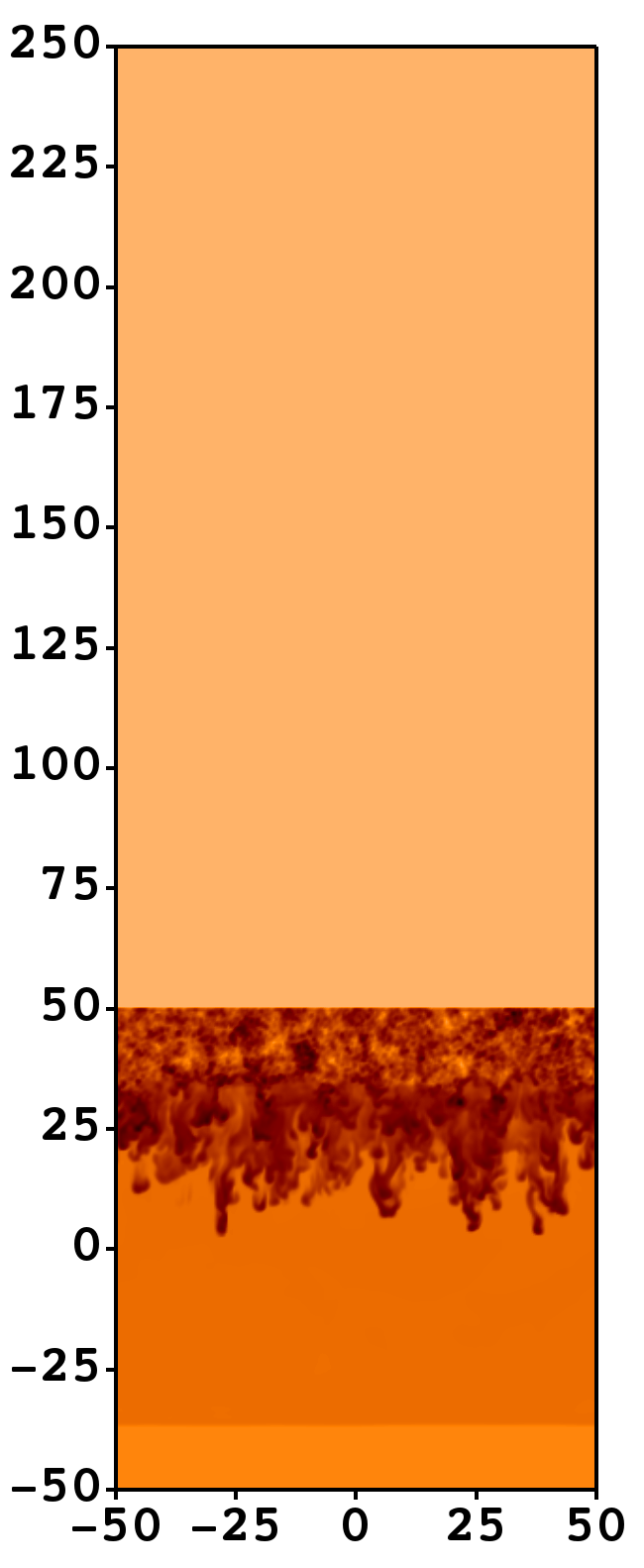}} & \hspace{-0.4cm}\resizebox{27mm}{!}{\includegraphics{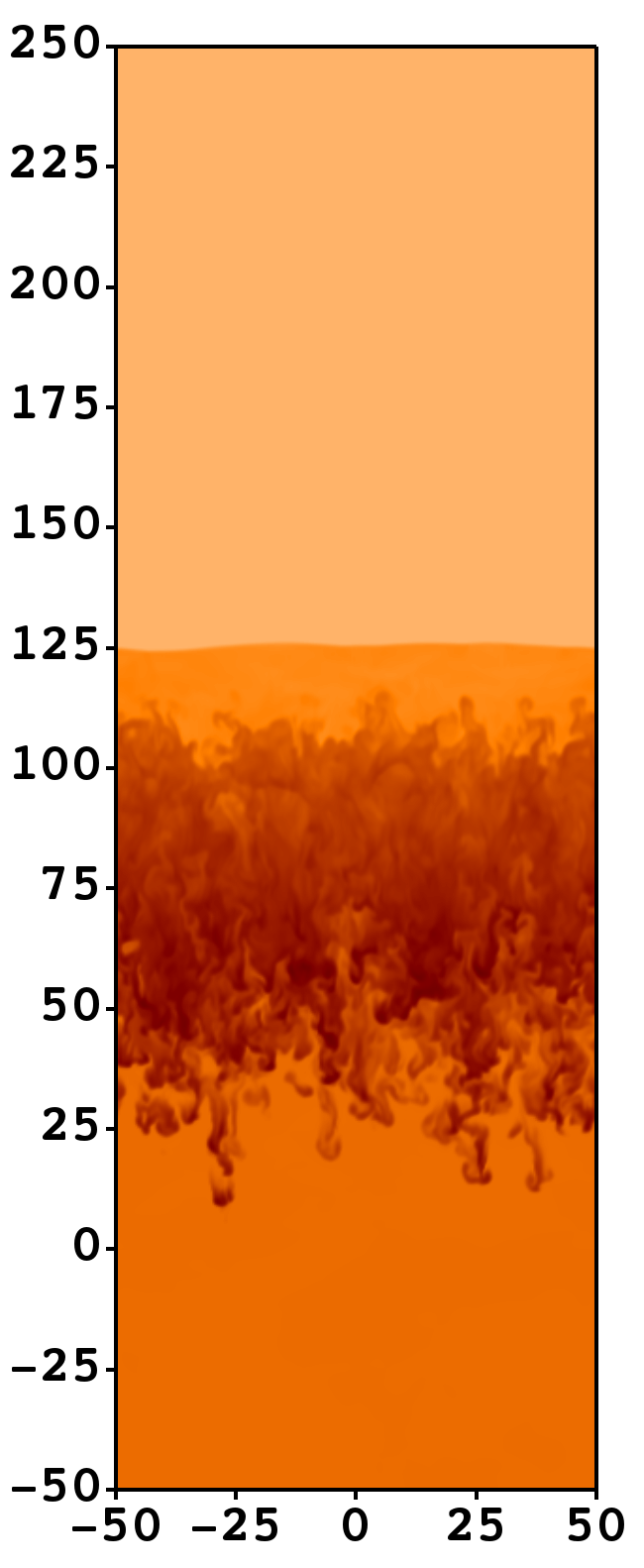}} & \hspace{-0.4cm}\resizebox{27mm}{!}{\includegraphics{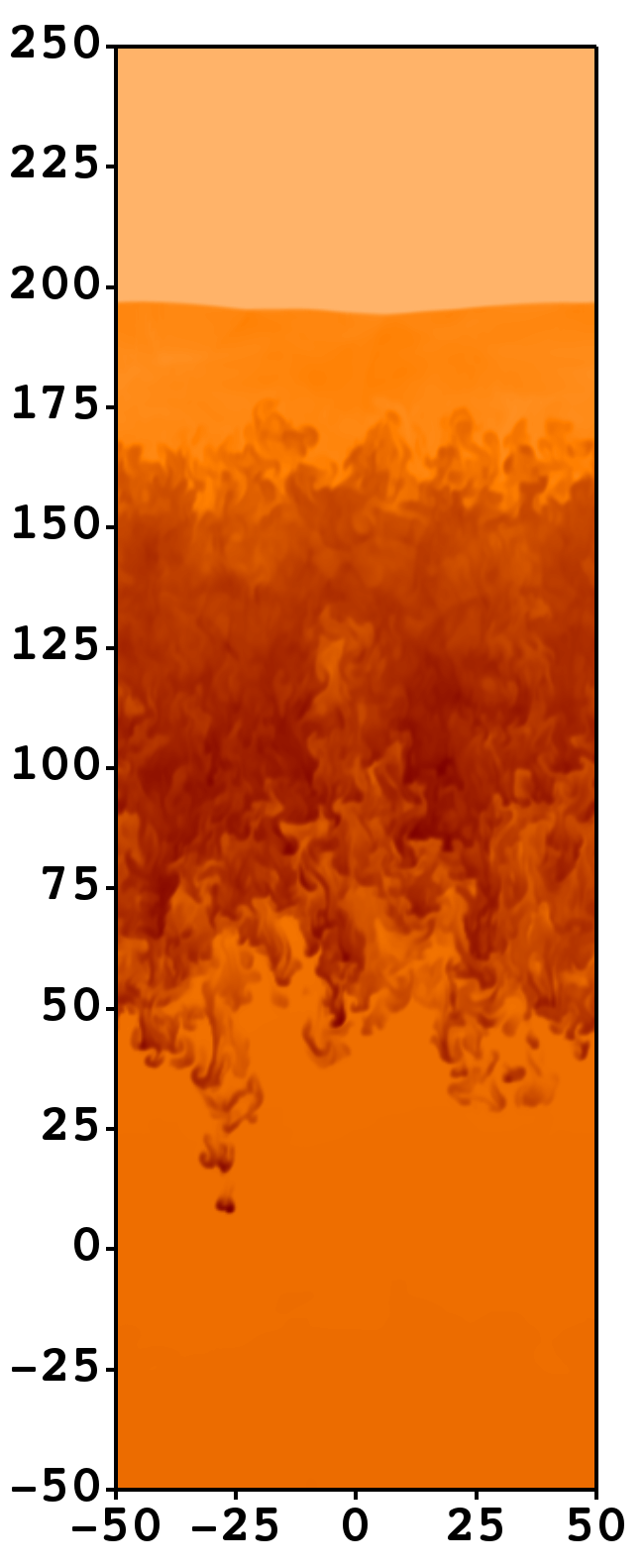}} & \hspace{-0.4cm}\resizebox{27mm}{!}{\includegraphics{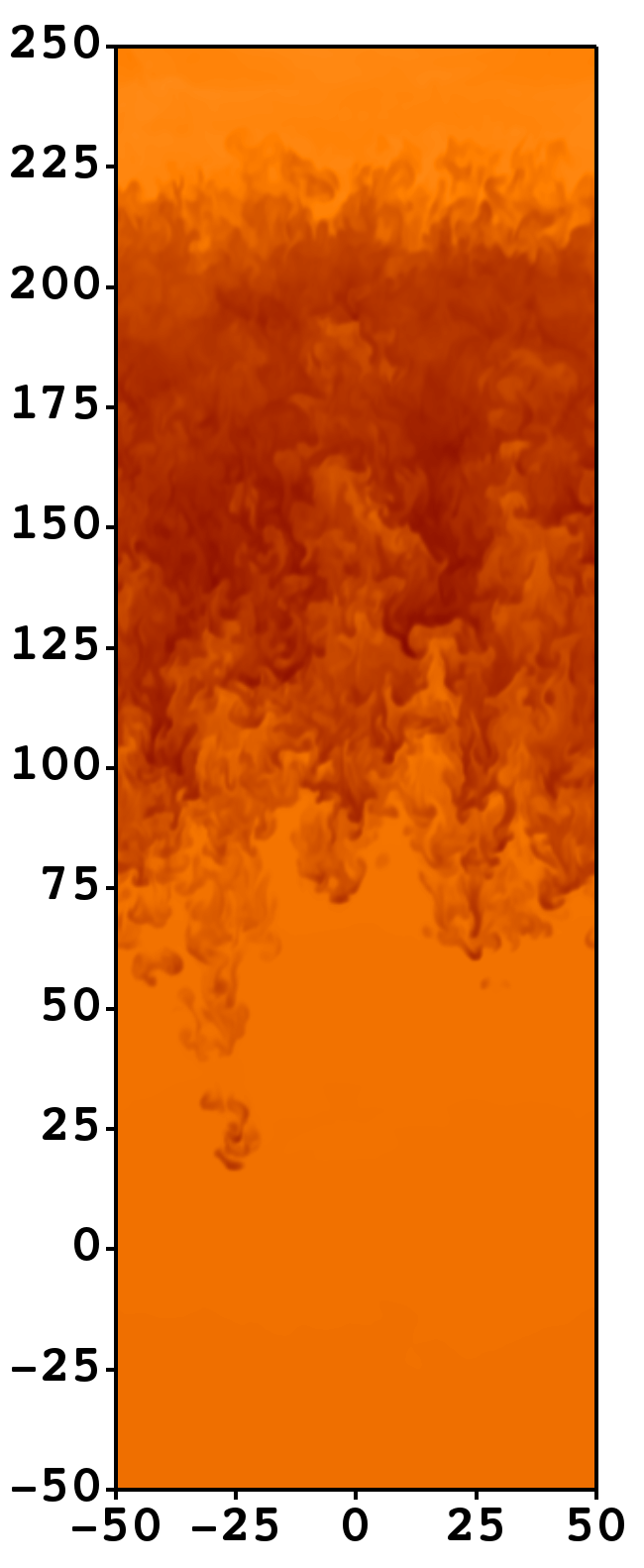}}  & \hspace{-0.4cm}\resizebox{27mm}{!}{\includegraphics{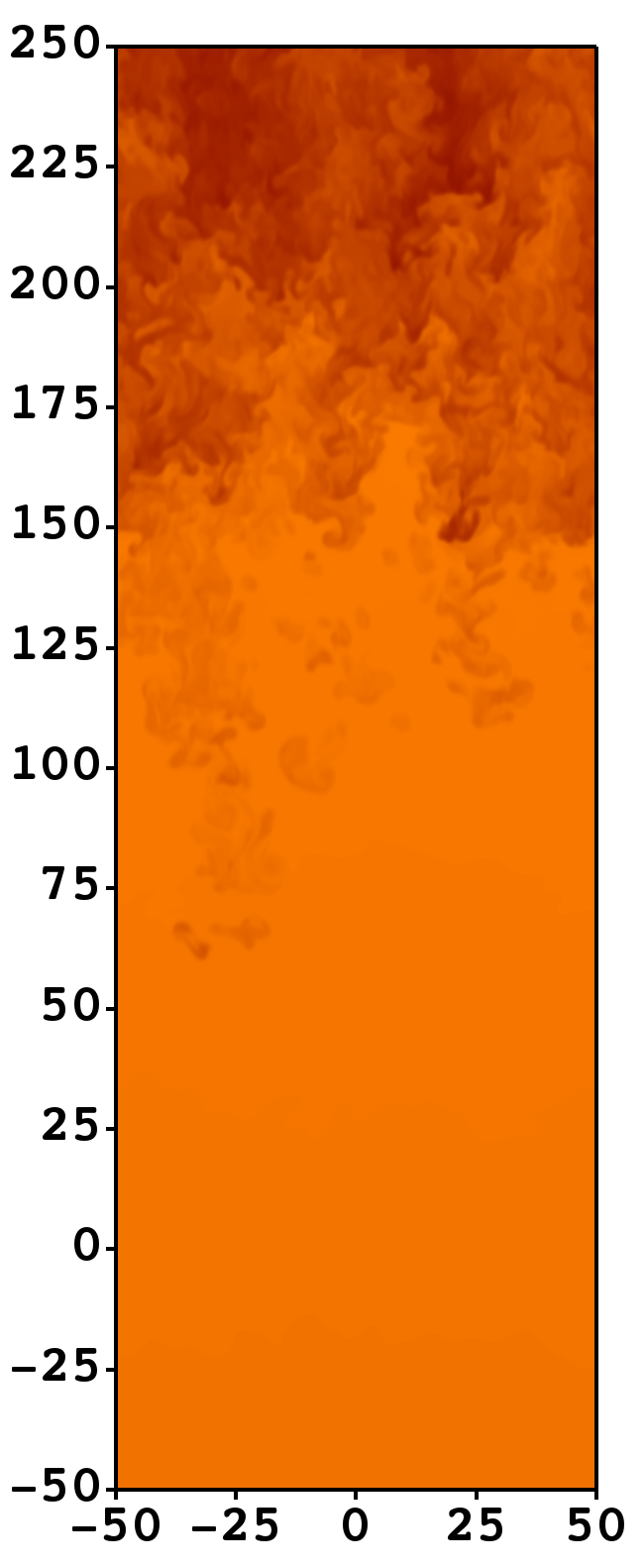}} &
\hspace{-0.2cm}\resizebox{12.8mm}{!}{\includegraphics{bar_vert_2.png}}\\
       \multicolumn{1}{l}{\hspace{-3mm}b) comp-k8-M30\hspace{+3.2mm}$t_0$} & \multicolumn{1}{c}{$0.5\,t_{\rm sp}=0.03\,\rm Myr$} & \multicolumn{1}{c}{$1.1\,t_{\rm sp}=0.07\,\rm Myr$} & \multicolumn{1}{c}{$1.8\,t_{\rm sp}=0.12\,\rm Myr$} & \multicolumn{1}{c}{$2.4\,t_{\rm sp}=0.16\,\rm Myr$} & \multicolumn{1}{c}{$3.0\,t_{\rm sp}=0.20\,\rm Myr$} & $\frac{n}{n_{\rm ambient}}$\\     
       \hspace{-0.25cm}\resizebox{27mm}{!}{\includegraphics{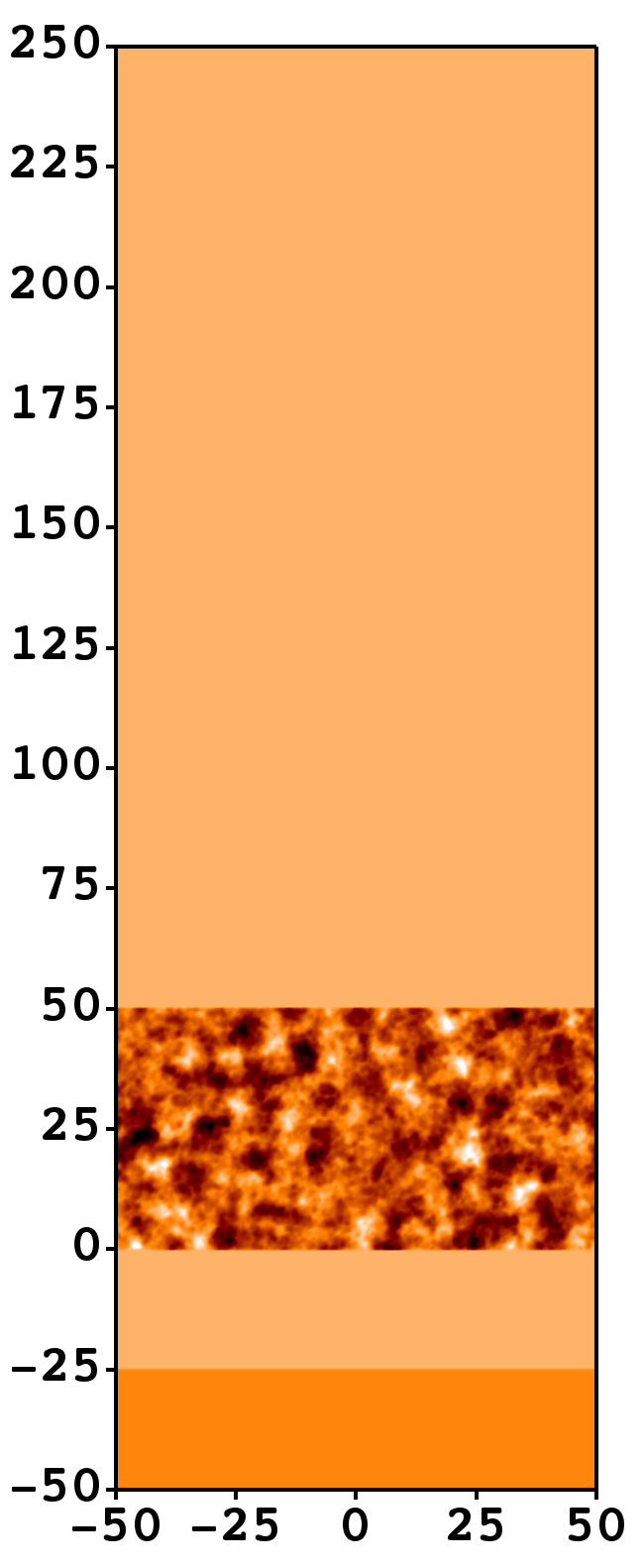}} & \hspace{-0.4cm}\resizebox{27mm}{!}{\includegraphics{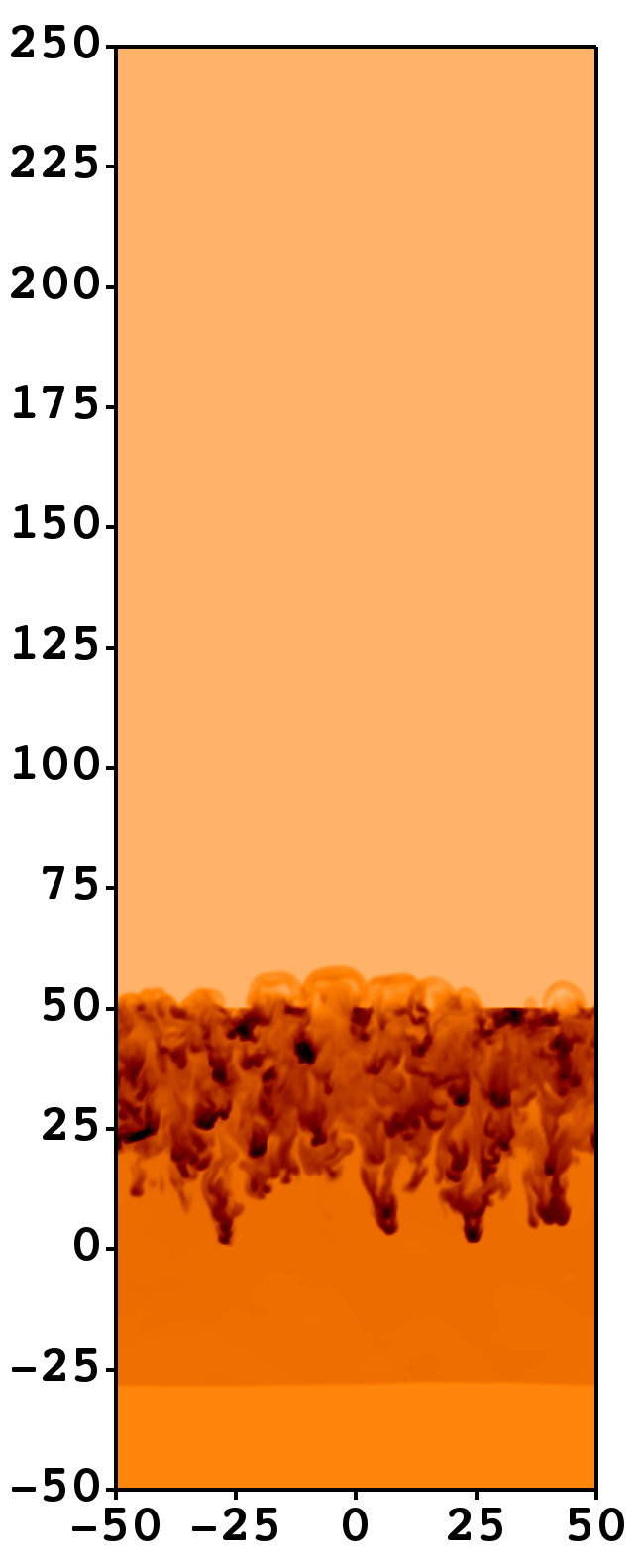}} & \hspace{-0.4cm}\resizebox{27mm}{!}{\includegraphics{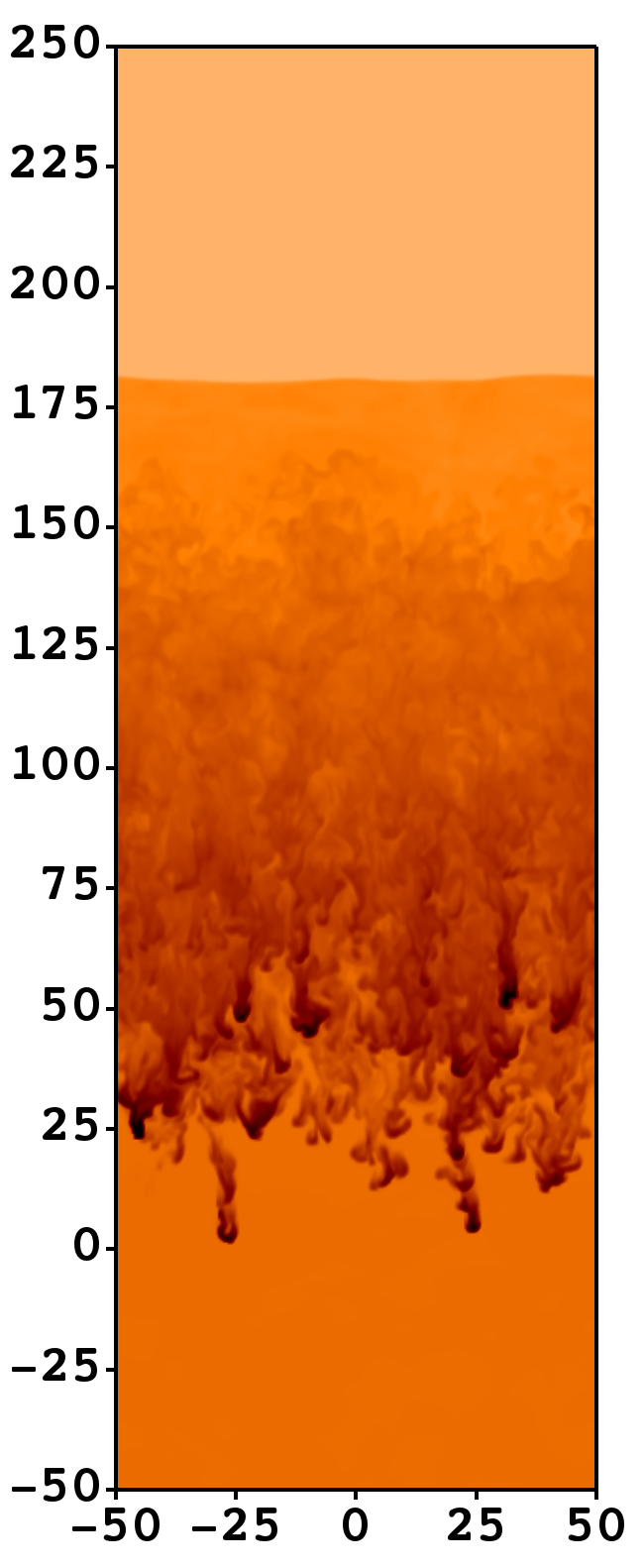}} & \hspace{-0.4cm}\resizebox{27mm}{!}{\includegraphics{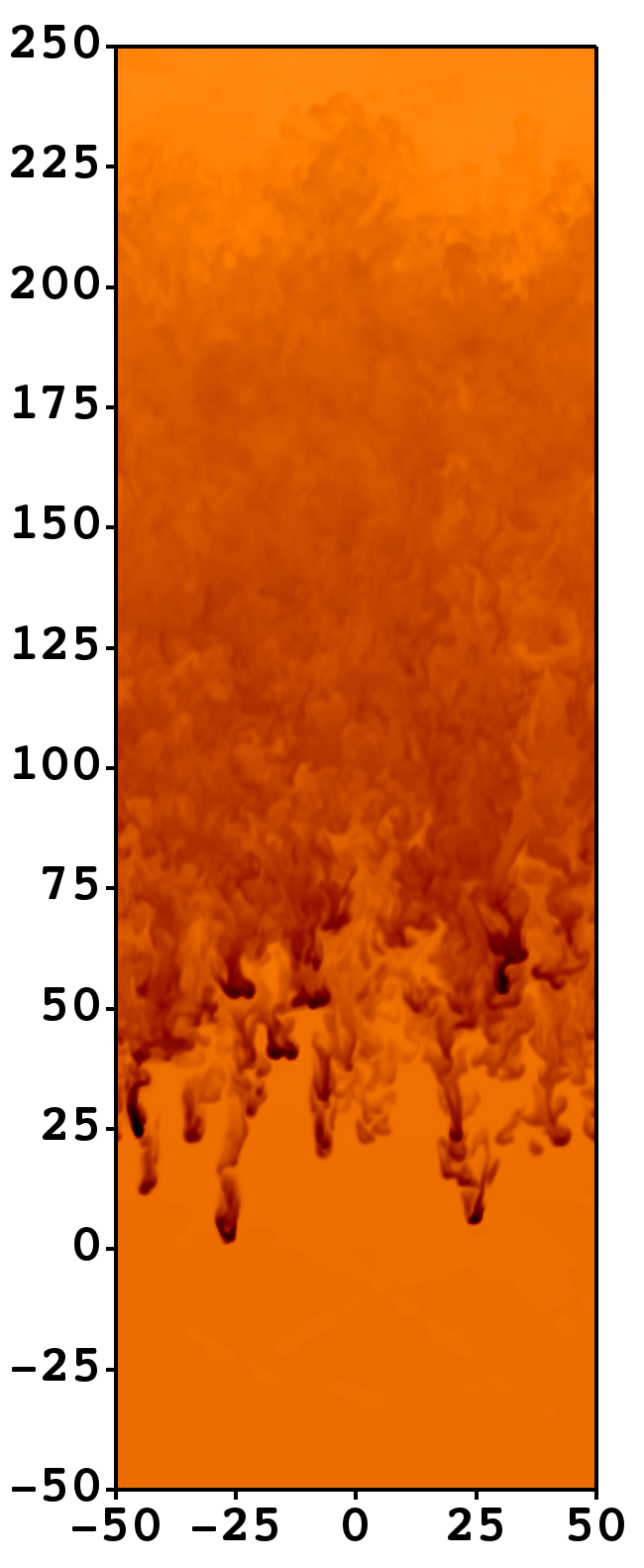}} & \hspace{-0.4cm}\resizebox{27mm}{!}{\includegraphics{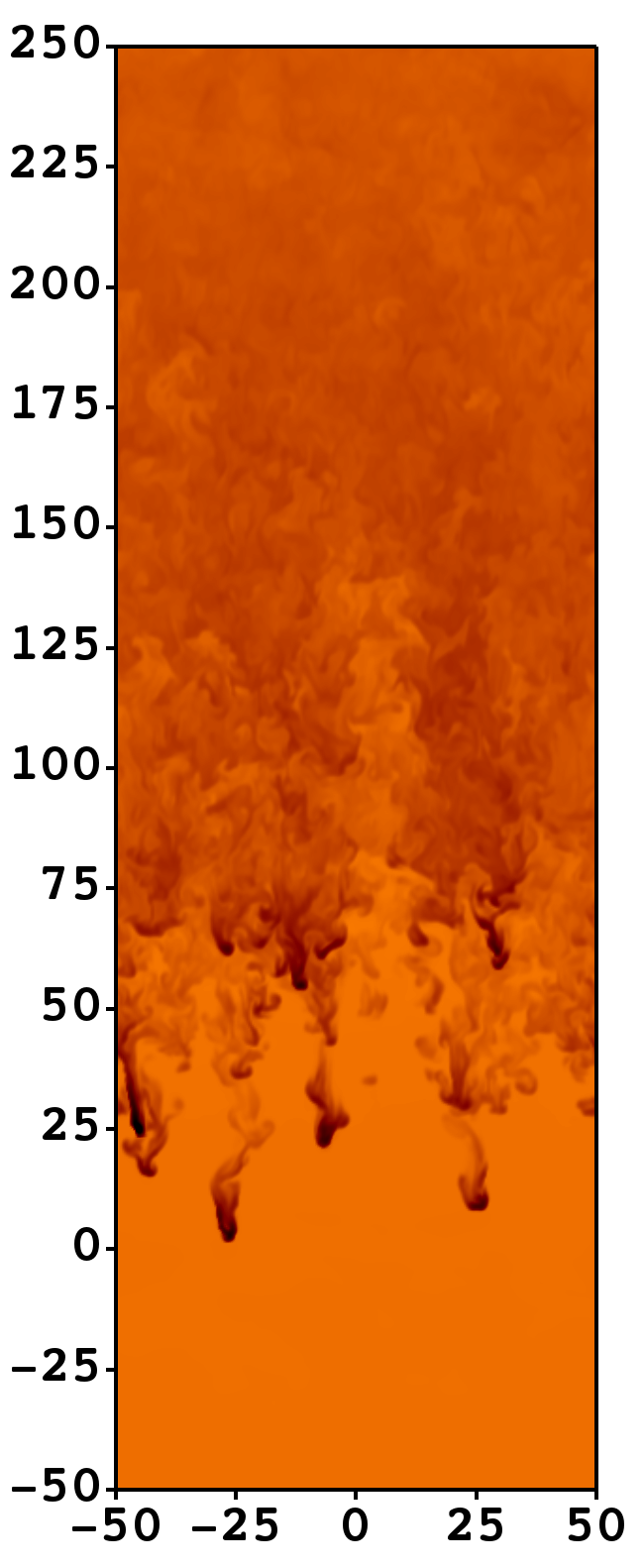}}  & \hspace{-0.4cm}\resizebox{27mm}{!}{\includegraphics{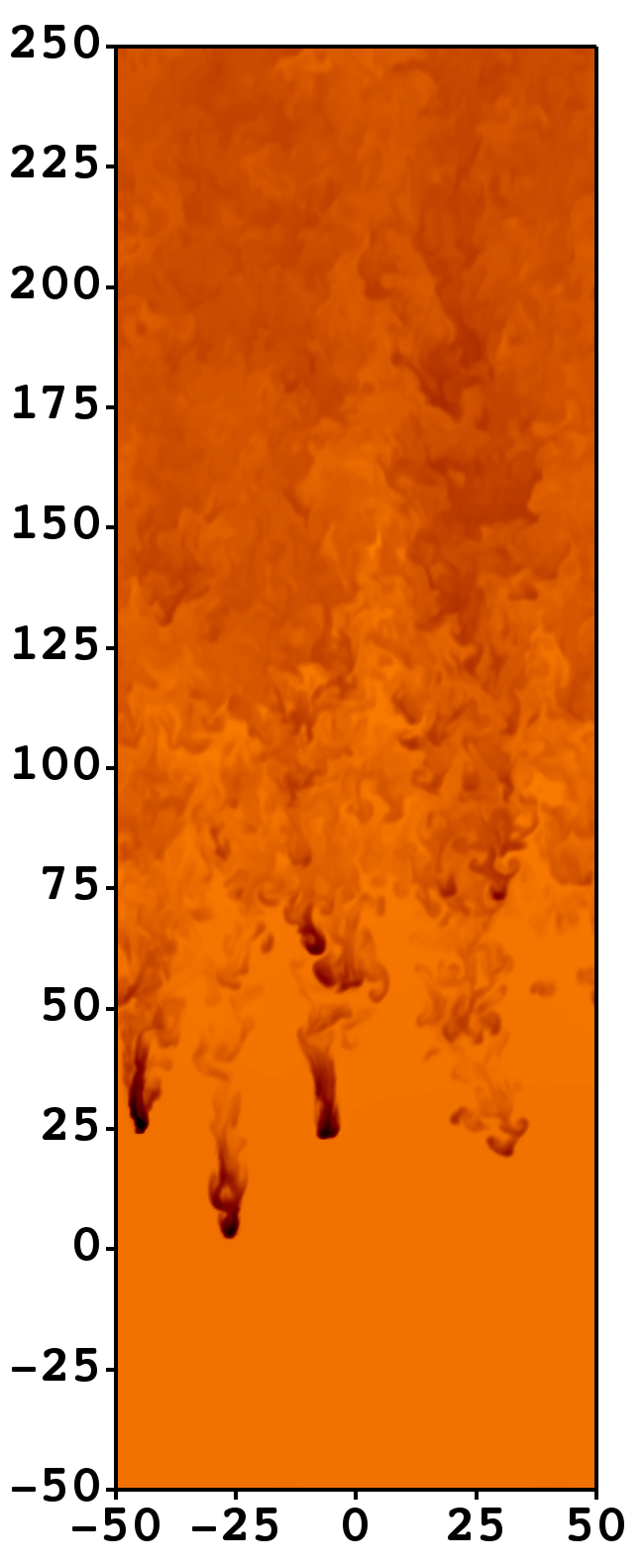}}&
\hspace{-0.2cm}\resizebox{12.8mm}{!}{\includegraphics{bar_vert_2.png}}\\
  \end{tabular}
  \caption{Same as Figures \ref{Figure2} and \ref{Figure3}, but here we show the number density slices in a solenoidal model, sole-k8-M30 (panel a), and a compressive model, comp-k8-M30 (panel b), which correspond to the runs with ${\cal M_{\rm shock}}=30$. The spatial ($X,Y$) extent is ($L\times3L$)$\equiv$($2L_{\rm mc}\times6L_{\rm mc}$), i.e., ($100\,\rm pc\times300\,\rm pc$) in physical units. Time-scales in physical units are different than in the standard Mach-10 models as $1\,\rm t_{\rm sp}=0.065\,\rm Myr$ in Mach-30 models.} 
  \label{FigureA3}
\end{center}
\end{figure*}

Figures~\ref{FigureA1}, ~\ref{FigureA2}, and ~\ref{FigureA3} show the normalised gas number density in the thin-layer, Mach-4, seed-2, and Mach-30 solenoidal and compressive multicloud models discussed in the text.

\bsp	
\label{lastpage}
\end{document}